\documentclass[a4paper,twocolumn,11pt,accepted=2020-07-03]{quantumarticle}
\pdfoutput=1
\usepackage[utf8]{inputenc}
\usepackage[T1]{fontenc}
\usepackage{graphicx,tikz}
\input{tikzit.sty}

\tikzstyle{rectangle}=[fill=white, draw=black, shape=rectangle]
\tikzstyle{red}=[fill=red, draw=black, shape=circle]
\tikzstyle{system_label}=[fill=white, draw=white, shape=circle]
\tikzstyle{prep}=[fill=white, draw=black, shape=rectangle, tikzit shape=rectangle, new atom]
\tikzstyle{four wire box}=[fill=white, draw=black, shape=rectangle, minimum width=0.70 cm, minimum height=3  cm]

\tikzstyle{lambda}=[-, draw={rgb,255: red,191; green,191; blue,191}]
\tikzstyle{state}=[<-]
\tikzstyle{new edge style 0}=[-]
\tikzstyle{new edge style 1}=[-]

\usepackage{changepage}
\usepackage{amsmath,amssymb,mathtools,amsthm,mathrsfs,enumitem}
\usepackage[backref=none
]{hyperref}
\usepackage[matrix,frame,arrow]{xypic}
\usepackage[matrix,frame,arrow]{xy}
\usepackage{amsmath}

\newcommand{\qw}[1][-1]{\ar @{-} [0,#1]}

\newcommand{\gate}[1]{*{\xy *+<.6em>{#1};p\save+LU;+RU **\dir{-}\restore\save+RU;+RD **\dir{-}\restore\save+RD;+LD **\dir{-}\restore\POS+LD;+LU **\dir{-}\endxy} \qw}

\newcommand{\measureD}[1]{*{\xy*+=+<.5em>{\vphantom{\rule{0em}{.1em}#1}}*\cir{r_l};p\save*!R{#1} \restore\save+UC;+UC-<.5em,0em>*!R{\hphantom{#1}}+L **\dir{-} \restore\save+DC;+DC-<.5em,0em>*!R{\hphantom{#1}}+L **\dir{-} \restore\POS+UC-<.5em,0em>*!R{\hphantom{#1}}+L;+DC-<.5em,0em>*!R{\hphantom{#1}}+L **\dir{-} \endxy} \qw}

\newcommand{\multimeasureD}[2]{*+<1em,.9em>{\hphantom{#2}}\save[0,0].[#1,0];p\save !C *{#2},p+LU+<0em,0em>;+RU+<-.8em,0em> **\dir{-}\restore\save +LD;+LU **\dir{-}\restore\save +LD;+RD-<.8em,0em> **\dir{-} \restore\save +RD+<0em,.8em>;+RU-<0em,.8em> **\dir{-} \restore \POS !UR*!UR{\cir<.9em>{r_d}};!DR*!DR{\cir<.9em>{d_l}}\restore \qw}

\newcommand{\multigate}[2]{*+<1em,.9em>{\hphantom{#2}} \qw \POS[0,0].[#1,0];p !C *{#2},p \save+LU;+RU **\dir{-}\restore\save+RU;+RD **\dir{-}\restore\save+RD;+LD **\dir{-}\restore\save+LD;+LU **\dir{-}\restore}

\newcommand{\ghost}[1]{*+<1em,.9em>{\hphantom{#1}} \qw}

\newcommand{\ustick}[1]{*!D!<0em,-.2em>=<0em>{{\scriptstyle #1}}}

\newcommand{\Qcircuit}[1][0em]{\xymatrix @*=<#1>} 
\newcommand{\node}[2][]{{\begin{array}{c} \ _{#1}\  \\ {#2} \\ \ \end{array}}\drop\frm{o} }

\newcommand{\pureghost}[1]{*+<1em,.9em>{\hphantom{#1}}}
\newcommand{\multiprepareC}[2]{*+<1em,.9em>{\hphantom{#2}}\save[0,0].[#1,0];p\save !C
  *{#2},p+RU+<0em,0em>;+LU+<+.8em,0em> **\dir{-}\restore\save +RD;+RU **\dir{-}\restore\save
  +RD;+LD+<.8em,0em> **\dir{-} \restore\save +LD+<0em,.8em>;+LU-<0em,.8em> **\dir{-} \restore \POS
  !UL*!UL{\cir<.9em>{u_r}};!DL*!DL{\cir<.9em>{l_u}}\restore}
\newcommand{\prepareC}[1]{*{\xy*+=+<.5em>{\vphantom{#1\rule{0em}{.1em}}}*\cir{l^r};p\save*!L{#1} \restore\save+UC;+UC+<.5em,0em>*!L{\hphantom{#1}}+R **\dir{-} \restore\save+DC;+DC+<.5em,0em>*!L{\hphantom{#1}}+R **\dir{-} \restore\POS+UC+<.5em,0em>*!L{\hphantom{#1}}+R;+DC+<.5em,0em>*!L{\hphantom{#1}}+R **\dir{-} \endxy}}

\newtheorem{lemma}{Lemma} \newtheorem{proposition}{Proposition}
\newtheorem{corollary}{Corollary} \newtheorem{theorem}{Theorem}
\newtheorem{principle}{Principle} \newtheorem{definition}{Definition}
\newtheorem{assumption}{Assumption}
\newtheorem{remark}{Remark}

\def\>{\rangle}
\def\<{\langle}
 
\def\trnsfrm#1{\mathscr #1}
 
	\def\rA{{\rm A}}
	\def\rB{{\rm B}}
	\def\rC{{\rm C}}
	\def\rD{{\rm D}} 
	\def\rE{{\rm E}} 
	\def\rF{{\rm F}}

    \def\rI{{\rm I}} \def\rR{{\rm R}}

\def\camp#1{\mathsf{#1}}
    \def\cX{\camp X}
    \def\cY{\camp Y}
    \def\cZ{\camp Z}
    \def\cW{\camp W}
\def\sH{{\sf H}} 
 
\def\sK{{\sf K}} 
\def\tA{\trnsfrm A}
\def\tB{\trnsfrm B}
\def\tC{\trnsfrm C}
\def\tD{\trnsfrm D}
\def\tG{\trnsfrm G}

\def\tI{\trnsfrm I}
\def\tT{\trnsfrm T}
\def\tU{\trnsfrm U}
\def\tV{\trnsfrm V} 
\def\tW{\trnsfrm W}

\def\tX{\trnsfrm X} 
\def\tY{\trnsfrm Y}
\def\tZ{\trnsfrm Z}
\def\tF{\trnsfrm F}

\def\tE{\trnsfrm E}
\def\tJ{\trnsfrm J}
\def\tR{\trnsfrm R}
   
\def\tS{\trnsfrm S}
\def\Elety{\mathsf{Sys}}
\def\Testun{\mathsf{Test}}
\def\Evun{\mathsf{Ev}}
\def\test#1{\mathsf{#1}}
\def\teA{\test{A}}
\def\teB{\test{B}}
\def\teD{\test{D}}
\def\teE{\test{E}}
\def\teP{\test{P}}
\def\teQ{\test{Q}}
\def\teR{\test{R}}
\def\teS{\test{S}}
\def\teT{\test{T}} 
\def\teI{\test{I}}
\def\teW{\test{W}} 

\def\Cntset#1{[\![\bar#1]\!]}
\def\Cntsetcomp#1#2{[\![\bar#1\bar#2]\!]}
 \def\Span{\mathsf{Span}}
\def\Stset#1{[\![ #1]\!]} 
\def\Trnset#1{[\![#1]\!]}
\def\Testset#1{\<\!\<#1\>\!\>}

  \def\Tr{{\rm Tr}}
 
\def\Comps{{\mathbb C}}
\def\Reals{{\mathbb R}}
\def\Nats{{\mathbb N}}
\def\Intg{{\mathbb Z}}
\def\bvec#1{\mathbf{#1}}

\def\set#1{\mathsf{#1}}
\def\normop#1{\|{#1}\|_\mathrm{op}}
\def\normsup#1{\|{#1}\|_\mathrm{sup}}
\def\normloc#1{\|{#1}\|_{B,\rho_0}}
\def\normq#1{\|{#1}\|_{Q}}
\def\normst#1{\|{#1}\|_{*}}
\def\reg#1{\set R^{(#1)}}
\def\regpart{\set R^{(G,B)}}
\def\infreg#1{\overline{\set R}^{(#1)}}
\def\infregpart{\overline{\set R}^{(G,B)}}

\begin{document}

\title{Cellular automata in operational probabilistic theories}

\author{Paolo Perinotti} 
\affiliation{{\em QUIT Group}, Dipartimento di Fisica, Universit\`a degli studi di Pavia, and INFN sezione di Pavia, via Bassi 6, 27100 Pavia, Italy}
\homepage{http://www.qubit.it}
\email{paolo.perinotti@unipv.it}
\orcid{0000-0003-4825-4264}

\maketitle

\begin{abstract}
The theory of cellular automata in operational probabilistic theories is developed. We start introducing the composition of infinitely many elementary systems, and then use this notion to define update rules for such infinite composite systems. The notion of causal influence is introduced, and its relation with the usual property of signalling is discussed. We then introduce homogeneity, namely the property of an update rule to evolve every system in the same way, and prove that systems evolving by a homogeneous rule always correspond to vertices of a Cayley graph. Next, we define the notion of locality for update rules. Cellular automata are then defined as homogeneous and local update rules. Finally, we prove a general version of the wrapping lemma, that connects CA on different Cayley graphs sharing some small-scale structure of neighbourhoods.
\end{abstract}

\tableofcontents
\section{Introduction}
In the last two decades a new approach to quantum 
foundations arose, grounded on 
the quantum information 
experience~\cite{hardy1999disentangling,hardy2001quantum,Fuchs:2002p89,Brassard:2005aa}.
This line of research on the fundamental aspects of quantum 
theory benefits from new ideas, concepts and methods \cite{PhysRevA.75.032304,PhilosophyQuantumInformationEntanglement2,d2010testing} 
that lead  to a wealth of remarkable results 
\cite{PhysRevA.81.062348,chiribella2011informational,masanes2011derivation,dakic2009quantum}. 
In particular, quantum theory can be now understood as a 
theory of 
information processing, that is selected among a universe of 
alternate 
theories \cite{Rastall1985,Popescu1994,hardy1999disentangling,Spekkens:2007aa} by operational principles about the possibility or impossibility 
to perform specific information processing 
tasks~\cite{chiribella2011informational,DAriano:2017aa}. 

The scenario of alternate theories
among which the principles select quantum theory is called the framework of Operational
Probabilistic Theories (OPTs)~\cite{PhysRevA.81.062348,DAriano:2017aa}, and has  
connections with the less structured concept of Generalized Probabilistic Theory
(GPT)~\cite{PhysRevA.75.032304,barnum2008teleportation,BARNUM20113}, as well as the diagrammatic category theoretical approach
often referred to as quantum picturialism~\cite{Coecke:2006fd,coecke_kissinger_2017}. 
Inspiration for the framework came also from  
quantum logic~\cite{ludwig1985foundations,Wilce2000}. 

Quantum theory, namely the theory of Hilbert spaces, density matrices, 
completely positive maps and POVMs (in particular we refer to the elegant 
exposition of Ref.~\cite{nielsen_chuang_2010}),
can thus be re-formulated as a special theory of information processing. Besides the 
many advantages of this result, one has to face a main issue: the theory as such is
devoid of its physical content. Elementary systems are thought of as elementary
information carriers---brutally speaking, memory cells---rather than elementary particles 
or fields in space-time. 
While this framework is satisfactory for an effective, empirical
description of physical experiments, when it comes to provide a theoretical foundation
for the physics of elementary systems, the informational approach at this stage calls for
a way to re-embrace mechanical notions such as mass, energy, position, 
space-time, and complete the picture encompassing the dynamics of quantum systems.

A recent proposal for this endeavour is based on the idea that physical laws 
have to be ultimately understood as algorithms, that make systems evolve, 
changing their state, exactly as the memory cells of a computer are updated by 
the run of an algorithm~\cite{bisio2013dirac,Bisio2015244,PhysRevA.90.062106}. Such a program already achieved successful results in the reconstruction of Weyl's, Dirac's and Maxwell's equations (for a comprehensive review see Ref.~\cite{Bisio2015}). The most natural candidate algorithm for describing a physical law 
in this context is a cellular automaton. The theory of cellular automata is a wide and
established branch of computer science. The notion of a cellular automaton 
for quantum systems was first devised as the quantum version of its classical counterpart, 
e.g.~in Refs.~\cite{feynman1982simulating,vanDam:tn,492583,Gruska:QKjNtRg3}, but turned out 
to give rise to a rather independent theory, developed starting
from Ref.~\cite{schumacher2004reversible}: the theory of Quantum Cellular Automata (QCAs).
The latter counts presently various important results---see 
e.g.~Refs.~\cite{arrighi2011unitarity,Gross2012}, just to mention a few. We stress that,
most commonly, QCAs are defined to be reversible algorithms, and most results in the 
literature are proved with this hypothesis. It is known, however, that many desirable 
preoperties fail to hold in the irreversible case (see 
e.g.~Ref.~\cite{10.1007/978-3-642-21875-0_1}). In the present work we will not consider 
irreversible cellular automata, and leave this subject for further studies.

In order to use cellular automata as candidate physical laws in the foundational perspective
based on OPTs, one has two choices at hand. The first one is to start treating automata 
within a definite theory, and this is the approach adopted so far, in particular within Fermionic theory~\cite{Bravyi2002210,D_Ariano_2014,doi:10.1142/S0217751X14300257}. In the
linear case, Fermionic cellular automata reduce to Quantum Walks, and this brings in 
the picture all the tools from such widely studied topic~\cite{aharonov1993quantum,aharonov2001quantum,knight2004propagating,konno2005new,carteret2003three,ambainis2001one}. The non-linear case is far less studied~\cite{PhysRevA.97.032132,PhysRevA.98.052337}, and does not offer as many results for the analysis. 
Needless to say, this approach faces difficulties that are specific of the theory at hand, and
prevents a comparison of different theories on a ground that is genuinely physical.

The second approach is initiated in the present paper, and consists in defining 
cellular automata in the general context of OPTs. This perspective offers the possibility of
extracting the essential features of the theory of quantum or Fermionic cellular automata,
those that are not specific of the theory but are well suited in any theory of 
information processing. As a consequence, one can generalise some results, and figure out
why and how others fail to extend to the broader scenario. Moving a much less structured 
mathematical context, this approach can use only few tools, but provides results 
that have the widest applicability range.

\section{Detailed outlook}

In this section we provide a short, non technical discussion of the main results.
The purpose of this work is to define cellular automata in OPTs, and prove some 
general results that will help applying the theory to special cases of interest.

A cellular automaton is an algorithm that updates the information stored in an array of 
memory cells in discrete steps, in such a way that one needs to read the content of a few 
neighbouring cells to determine the state of a given cell at the next step. Cellular 
automata in the literature are often, but not always, defined to be homogeneous: in this 
case the local rule for the update of the cell is the same for every cell. Here we will 
adopt homogeneity, but the subject is presented so that the generalisation of definitions 
to inhomogeneous CA is straightforward.

Typically, the interesting case of a cellular automaton is the one involving an infinite 
memory array. The first challenge we have to face is then to extend the theory 
of OPTs from finite, arbitrarily large composite systems to actually infinite ones. This 
piece of theory has an interest per se, for many reasons ranging from the possibility to 
introduce thermodynamic limits to the extension of the theory of C$^*$ and von Neumann algebras. 

We then start with a review of the framework of OPTs, and build the necessary tools to 
define infinite composite systems. The starting point is the construction of mathematical 
objects that describe measurements on finitely many systems within an infinite array---
the OPT counterpart of the space of local effects of quantum theory. Effects for the 
infinite system are then defined as limits of Cauchy sequences of local effects. Since
the introduction of a suitable topology for the definition of limits is needed, we open 
the paper with section~\ref{sec:OPTs}, where a review of OPTs is provided, along with a 
few new results that will be useful, and a rather consistent part of the section will be
dedicated to the introduction and discussion of norms that will provide the necessary
topological framework for a consistent definition of limits.

The subject of the subsequent Section~\ref{sec:vonn} is then the construction of the 
Banach space of effects for the infinite composite system, and consequently the 
construction of the space of states as suitable linear functionals on effects. An 
important subsection will be dedicated to the construction of the algebra of quasi-local
transformations, that allows for the description of transformations on an infinite system 
that can be arbitrarily well approximated by local operations on finitely many subsystems.
Indeed, an important part of the theory of CAs in OPTs is built by ruling the way in 
which quasi-local transformations are transformed by the CA. In particular, the way in
which the CA propagates the effects of a local transformation on surrounding subsystems
will be the key to the definition of the neighbourhood of the subsystem, a concept that
is central to the theory of CAs.

Once this is done, the next step consists in defining update rules, and their 
admissibility conditions, which are the subject of Section~\ref{sec:UR}, along with causal 
influence and a block-decomposition theorem. Update rules are defined in the first place
as automorphisms $\tV$ of the space of quasi-local effects, but an important request they 
must abide is that when they act on a quasi-local transformation $\tA$ by conjugation as
$\tV\tA\tV^{-1}$, the obtained transformation is again quasi-local.

The next step is taken in section~\ref{sec:homog}, and consists in defining the property 
of homogeneity. The latter presents with some difficulties,
stemming from the fact that, as we mentioned above, we are defining update rules
{\em prior} to any mechanical notion. This implies that
even space-time is not available at this fundamental stage, and without geometry 
we cannot define homogeneity as translational invariance under the 
group corresponding to a given space-time. On the contrary, 
we will define homogeneity by formalising the idea that 
every single cell has to be treated equally by the update rule, and this notion will be 
defined operationally, requiring that no experiment made of local operations will allow 
establishing any difference between cells. As a consequence of homogeneity, one can prove 
that every homogeneous update rule is underpinned by the Cayley graph of some group, 
generalising a result of Refs.~\cite{PhysRevA.90.062106,Ariano2016}. The mathematical
theory known as {\em geometric group theory} then tells us that the Cayley graph 
representing the causal connections of cells in the memory array, being a metric space,
uniquely identifies an equivalence class of metric spaces, that captures both the 
algebraic and geometric essential features of the group. Astonishingly, for Cayley graphs
of finitely presented groups---which is the case for cellular automata as we define 
them here---the equivalence class always contains a smooth manifold 
of dimension at most four (see Ref.~\cite{de2000topics}, pag.~90). This result means that
we can always think of a cellular automaton as if it was embedded in a Riemannian 
manifold, in such a way that the Riemannian distance between nodes is almost the same as
the distance between nodes given in terms of steps along the graph edges.

Also the notion of locality, presented in section~\ref{sec:local}, 
comes with its own difficulties, that can be overcome by proving
a generalisation of the result known as ``unitarity plus causality implies 
localizability"~\cite{arrighi2011unitarity}. This is where the theory, which is inspired 
by that of QCAs, deeply differs from the theory of classical cellular automata. In 
particular, considering the collection of transformations $\tV\tA\tV^{-1}$ for all $\tA$ 
acting on a given system $g$, we will define the neighbourhood of $g$ as the set of 
systems on which transformations $\tV\tA\tV^{-1}$ act non trivially.

Once the theory of cellular automata is fully developed, some results are proved 
in Section~\ref{sec:cas}. In particular, a very useful theorem is the wrapping lemma, 
which under very wide hypotheses---though not universal---allows for 
the classification of automata on a given infinite graph by classifying automata on any 
suitably ``wrapped" finite version of the same graph.

In Section~\ref{sec:exa}, a few examples are reviewed. In particular, using the general 
notion of locality, we apply the definition of causal influence and the neighbourhood 
scheme to the case of classical cellular automata. With the above definitions at hand, 
we show that allegedly local automata are actually non-local. This solves a long standing 
puzzle about the connection between classical and quantum automata---the so-called 
quantisation of classical automata~\cite{schumacher2004reversible,inomizo05}. 

The paper is concluded by Section~\ref{sec:conc} with a summary of the results and some closing remarks.

\section{Operational Probabilistic Theories}\label{sec:OPTs}

In this section we will provide a thorough introduction to OPTs, which is partly a review
of the literature, and partly presentation of new results that will be used in the remainder. We provide here a sketchy introduction before going through formal definitions,
and use Quantum Theory as an illustrative example for the main notions in the framework.

Quantum theory is about system types $\rA$~
\footnote{We remark that for the purpose of information processing the type of a system is captured
by the dimension of the corresponding Hilbert space, e.g.~the electron spin is of 
the same type as the photon polarisation. This is slightly different than the usual
notion of system type in physics, where the two types in the above example are different.}
---namely complex Hilbert spaces $\sH_\rA$ that are classified by their dimension 
$d_\rA=\dim(\sH_\rA)$---and transformations that can occur on systems as a consequence of 
undergoing a {\em test}. Tests are represented by quantum instruments 
$\teE_\cX=\{\tE_i\}_{i\in\cX}$, i.e.~a collection of Completely Positive (CP) maps $\tE_i$ 
that sum 
to a trace-preserving one (a channel): $\tE=\sum_{i\in\cX}\tE_i$. The quantum operation 
$\tE_i$---a CP trace non-increasing 
map---represents the change in the system occurring upon the event of a specific outcome 
$i\in\cX$ in the test.  In the diagrammatic 
language of OPTs systems are represented as labelled wires, and instruments or quantum 
operations as boxes with an input and output wire, e.g.
\begin{align*}
\begin{aligned}
    \Qcircuit @C=1em @R=.7em @! R {&\ustick{\rA}\qw&\gate{\teE_\cX}&\ustick{\rB}\qw&\qw}
\end{aligned}\ ,\quad
\begin{aligned}
    \Qcircuit @C=1em @R=.7em @! R {&\ustick{\rA}\qw&\gate{\tE_i}&\ustick{\rB}\qw&\qw}
\end{aligned}\ ,
\end{align*}
respecitvely. States can be considered as special transformations where the input system 
is $\rI$, having $\sH_\rI=\Comps$, i.e.~$d_\rI=1$. Notice that the a {\em preparation 
test}, i.e.~the most general test from $\rI$ to $\rA$, represents a {\em probabilistic} 
preparation procedure where some state in the collection $\teP_\cX=\{\rho_i\}_{i\in\cX}$ 
is prepared, with the sub-normalised density matrix $\rho_i$ occurring upon reading the 
outcome $i\in\cX$. The probability of occurrence of $\rho_i$ is $\Tr[\rho_i]$, and 
$\rho=\sum_{i\in\cX}\rho_i$ has unit trace. Preparation tests or states for system $\rA$ 
are represented by special diagrams
\begin{align*}
\begin{aligned}
    \Qcircuit @C=1em @R=.7em @! R {&\prepareC{\teP_\cX}&\ustick{\rA}\qw&\qw}
\end{aligned}\ ,\quad
\begin{aligned}
    \Qcircuit @C=1em @R=.7em @! R {&\prepareC{\rho_i}&\ustick{\rA}\qw&\qw}
\end{aligned}\ ,
\end{align*}
respectively. The space $\mathcal H_\rA$ of Hermitian operators 
on $\sH_\rA$, whose dimension is $D_\rA=d_\rA^2$, is the real span of states of 
system $\rA$. 

A POVM $\teQ_\cY=\{Q_j\}_{j\in\cY}$ is a collection of {\em effects} $0\leq Q_j\leq I_\rA$ 
that sum to the identity operator: $\sum_{j\in\cY}Q_j=I_\rA$. This kind of test can be 
seen as a quantum instrument from $\rA$ to $\rI$, namely a collection of linear 
functionals on the real space spanned by density matrices, where the functionals 
$a_j(\rho)$ are defined as $a_j(\rho)\coloneqq\Tr[\rho Q_j]$. The diagrammatic 
representation of POVMs and effects is the following
\begin{align*}
\begin{aligned}
    \Qcircuit @C=1em @R=.7em @! R {&\ustick{\rA}\qw&\measureD{\teQ_\cY}}
\end{aligned}\ ,\quad
\begin{aligned}
    \Qcircuit @C=1em @R=.7em @! R {&\ustick{\rA}\qw&\measureD{Q_j}}
\end{aligned}\ ,
\end{align*}
respectively.

Composite systems, such as $\rA\rB$, correspond to to the Hilbert 
space $\sH_\rA\otimes\sH_\rB$ of dimension $d_{\rA\rB}=d_\rA d_\rB$. 
Tests $\{\tE_i\}_{i\in\cX}$ from $\rA$ to $\rB$ and $\{\tF_j\}_{j\in\cY}$ from $\rB$ to 
$\rC$ can be run in sequence, obtaining a new test: $\{\tD_{ij}\}_{(i,j)\in\cX\times\cY}$ 
where $\tD_{ij}\coloneqq\tF_j\tE_i$. On the other hand, tests $\{\tE_i\}_{i\in\cX}$ from 
$\rA$ to $\rB$ and $\{\tE_j\}_{j\in\cY}$ from $\rB$ to $\rB'$ on subsystems of a composite 
system $\rA\rA'$ can be run in parallel, obtaining the test 
$\{\tD_{ij}\}_{(i,j)\in\cX\times\cY}$ with $\tD_{ij}\coloneqq\tE_i\otimes\tF_j$. In 
diagrams, we draw for quantum operations
\begin{align*}
&\begin{aligned}
    \Qcircuit @C=1em @R=.7em @! R {&\ustick{\rA}\qw&\gate{\tF_j\tE_i}&\ustick{\rC}\qw&\qw}
\end{aligned}\ =\ 
\begin{aligned}
    \Qcircuit @C=1em @R=.7em @! R {&\ustick{\rA}\qw&\gate{\tE_i}&\ustick{\rB}\qw&\gate{\tF_j}&\ustick{\rC}\qw&\qw}
\end{aligned}\ ,\\ \\
&\begin{aligned}
    \Qcircuit @C=1em @R=.7em @! R {&\ustick{\rA\rA'}\qw&\gate{\tE_i\otimes\tF_j}&\ustick{\rB\rB'}\qw&\qw}
\end{aligned}\ =\ 
\begin{aligned}
    \Qcircuit @C=1em @R=.7em @! R {&\ustick{\rA}\qw&\gate{\tE_i}&\ustick{\rB}\qw&\qw\\
    &\ustick{\rA'}\qw&\gate{\tF_j}&\ustick{\rB'}\qw&\qw}
\end{aligned}\ ,
\end{align*}
and analogously for instruments.

Very relevant structures in the theory are the real space $\mathcal H_\rA$ of Hermitian 
operators on $\sH_\rA$, spanned by quantum states, the cone of positive operators in 
$\mathcal P_\rA\subseteq\mathcal H_\rA$, the convex set of states $\Stset{\rA}$, obtained 
by intersecting the positive cone with the half-space $\Tr[X]\leq1$, and the
convex set of deterministic states $\Stset{\rA}_1$, obtained by intersecting the positive 
cone with the affine hyperplane $\Tr[X]=1$. Similar structures can be generalised to the 
space spanned by quantum operations. In the general framework of OPTs, we will 
systematically refer to the generalisation of the above concepts.

\subsection{Formal framework}

The framework of OPTs is meant to capture the main traits of Quantum Theory 
(shared e.g.~by Classical Theory, or Fermionic Theory, etc.) summarised above, and use 
them as defining properties of a family of abstract theories that might be candidates for 
an alternative representation of elementary physical systems and their transformations. In 
the remainder of this section we provide a brief review of the framework of OPTs (for 
reference see 
e.g.~\cite{DAriano:2017aa,PhysRevA.81.062348,chiribella2011informational}). Some of the most relevant differences with respect to the quantum case stem 
from the fact that systems might not compose with the tensor product rule.

We warn the reader that this presentation has some elements that are slightly different 
from other reviews in the literature, and is tailored to ease the presentation of 
subsequent material. Most of the results in the present sections are new, and their proof 
will be given. Those theorems that are not original are not proved, and due reference is 
provided.

An operational theory $\Theta$ consists in i) a collection $\Testun(\Theta)$ of tests 
$\teT^{\rA\to\rB}_\cX$, each labelled by input and output letters from a collection 
$\Elety(\Theta)$ denoting system types, e.g.~$\rA\to\rB$---that will be systematically 
omitted---and by a finite set of outcomes $\cX$; for every pair of types 
$\rA,\rB\in\Elety(\Theta)$ the set of tests of type $\rA\to\rB$ is denoted 
$\Testset{\rA\to\rB}$; ii) a very basic associative rule for sequential composition: the test 
$\teT_\cX\in\Testset{\rA\to\rB}$ can be followed by the test $\teS_\cY\in\Testset{\rA'\to\rB'}$ 
if $\rA'\equiv\rB$, thus obtaining the sequential composition 
$\teS\teT_{\cX\times\cY}\in\Testset{\rA\to\rB'}$; iii) a rule $\otimes:(\rA,\rB)\mapsto\rA\rB$ 
for composing labels in parallel, and a corresponding rule for tests 
$\otimes:(\teS_\cX,\teT_\cY)\mapsto(\teS\otimes\teT)_{\cX\times\cY}$, with the following 
properties
\begin{enumerate}
\item Associativity: $(\rA\rB)\rC=\rA(\rB\rC)$. 
\item For every $\teS_\cX\in\Testset{\rA\to\rB}$ and $\teT_\cY\in\Testset{\rC\to\rD}$, one has $\teS_\cX\otimes\teT_\cY\in\Testset{\rA\rC\to\rB\rD}$. Associativity of $\otimes$ holds: 
\begin{align*}
(\teS_\cX\otimes\teT_\cY)\otimes\teW_\cZ=\teS_\cX\otimes(\teT_\cY\otimes\teW_\cZ). 
\end{align*}
\item Unit: there is a label $\rI$ such that $\rI\rA=\rA\rI=\rA$ for every $\rA\in\Elety(\Theta)$.
\item Identity: for every $\rA\in\Elety(\Theta)$, a test $\teI_\rA\in\Testset{\rA\to\rA}$ such that $\teI_\rB\teS_\cX=\teS_\cX\teI_\rA=\teS_\cX$, for every $\teS_\cX\in\Testset{\rA\to\rB}$.
\item For every $\teA_\cX\in\Testset{\rA\to\rB}$, $\teB_\cY\in\Testset{\rB\to\rC}$, $\teD_\cZ\in\Testset{\rD\to\rE}$, $\teE_\cW\in\Testset{\rE\to\rF}$, one has 
\begin{align}
(\teB_\cY\otimes\teE_\cW)(\teA_\cX\otimes\teD_\cZ)=(\teB_\cY\teA_\cX)\otimes(\teE_\cW\teD_\cZ).
\label{eq:natur}
\end{align}
\item Braiding: for every pair of system types $\rA,\rB$, there exist tests $\teS_{\rA\rB},\teS^*_{\rA\rB}\in\Testset{\rA\rB\to\rB\rA}$ such that $\teS^*_{\rB\rA}\teS_{\rA\rB}=\teS_{\rB\rA}\teS^*_{\rA\rB}=\teI_{\rA\rB}$, and $\teS_{\rA\rB}(\teA_\cX\otimes\teB_\cY)=(\teB_\cY\otimes\teA_\cX)\teS_{\rA\rB}$. Moreover, 
\begin{align*}
&\teS_{(\rA\rB)\rC}=(\teI_\rA\otimes\teS_{\rB\rC})(\teS_{\rA\rC}\otimes\teI_\rB),\\
&\teS_{\rA(\rB\rC)}=(\teS_{\rA\rB}\otimes\teI_\rC)(\teI_\rB\otimes\teS_{\rA\rC}).
\end{align*}
When $\teS^*_{\rA\rB}\equiv\teS_{\rA\rB}$, the theory is {\em symmetric}. 
\end{enumerate}
All the theories developed so far are symmetric. 

All tests of an operational theory are (finite) collections of {\em events}: $\Testset{\rA\to\rB}\ni\teR_\cX=\{\tR_i\}_{i\in\cX}$. If $\Testset{\rA\to\rB}\ni\teR_\cX=\{\tR_i\}_{i\in\cX}$ and $\Testset{\rB\to\rC}\ni\teT_\cY=\{\tT_j\}_{j\in\cY}$, then 
\begin{align*}
\Testset{\rA\to\rC}\ni(\teT\teR)_{\cX\times\cY}\coloneqq \{\tT_j\tR_i\}_{(i,j)\in\cX\times\cY}. 
\end{align*}
Similarly, for $\teR_\cX\in\Testset{\rA\to\rB}$ and $\teT_\cY\in\Testset{\rC\to\rD}$,
\begin{align*}
(\teR\otimes\teT)_{\cX\times\cY}\coloneqq \{\tR_i\otimes\tT_j\}_{(i,j)\in\cX\times\cY}.
\end{align*}
The set of events of tests in $\Testset{\rA\to\rB}$ is denoted by $\Trnset{\rA\to\rB}$.
By the properties of sequential and parallel composition of tests, one can easily derive associativity of sequential and parallel composition of events, as well as the analogue of Eq.~\eqref{eq:natur}. For every test $\teT_\cX\in\Testset{\rA\to\rB}$ with $\teT_\cX=\{\tT_i\}_{i\in\cX}$, and every disjoint partition $\{\cX_j\}_{j\in\cY}$ of $\cX=\bigcup_{j\in\cY}\cX_j$, one has a {\em coarse graining} operation that maps $\teT_\cX$ to $\teT'_\cY\in\Testset{\rA\to\rB}$, with $\teT'_\cY=\{\tT'_j\}_{j\in\cY}$. We define $\tT_{\cX_j}\coloneqq \tT'_j$. The parallel and sequential compositions distribute over coarse graining: 
\begin{align*}
&\tT_{\cX_j}\otimes\tR_k=(\tT\otimes\tR)_{\cX_j\times \{k\}},\\
&\tA_l\tT_{\cX_j}\tB_k=(\tA\tT\tB)_{\{l\}\times\cX_j\times \{k\}}.
\end{align*}
Notice that for every test $\teT_\cX\in\Testset{\rA\to\rB}$ there exists the singleton test $\teT'_*\coloneqq \{\tT_\cX\}$. One can easily prove that the identity test $\teI_{\rA}$ is a singleton: $\teI_{\rA}=\{\tI_{\rA}\}$, and $\tI_\rB\tT=\tT\tI_\rA$ for every event $\tT\in\Trnset{\rA\to\rB}$. Similarly, for $\teS_{\rA\rB}=\{\tS_{\rA\rB}\}$ and $\teS^*_{\rA\rB}=\{\tS^*_{\rA\rB}\}$ we have $\tS^*_{\rB\rA}\tS_{\rA\rB}=\tS_{\rB\rA}\tS^*_{\rA\rB}=\tI_{\rA\rB}$. The collection of events of an operational theory $\Theta$ will be denoted by $\Evun(\Theta)$. The above requirements make the collections $\Testun(\Theta)$ and $\Evun(\Theta)$ the families of morphisms of two braided monoidal categories with the same objects---system types $\Elety(\Theta)$.

An operational theory is an OPT if the tests $\Testset{\rI\to\rI}$ are probability distributions: $1\geq\tT_i=p_i\geq0$, so that $\sum_{i\in\cX}p_i=1$, and given two tests $\teS_\cX,\teT_\cY\in\Testset{\rI\to\rI}$ with $\tS_i=p_i$ and $\tT_i=q_i$, the following identities hold
\begin{align*}
&\tS_i\otimes \tT_j=\tS_i\tT_j\coloneqq p_iq_j,\\
&\tT_{\cX_j}\coloneqq \sum_{i\in\cX_j}p_i,
\end{align*}
meaning that events in the same test are mutually exclusive and events in different tests 
of system $\rI$ are independent. While it is immediate that $1\in\Trnset{\rI\to\rI}$---since $\{1\}_*=\{\tI_\rI\}$ is the only singleton test---we will assume that 
$0\in\Trnset{\rI\to\rI}$. This means that we can 
consider e.g.~tests of the form $\{1,0,0\}$.

Events in $\Trnset{\rA\to\rB}$ are called {\em transformations}. 
As a consequence of the above definitions, 
every set $\Stset{\rA}\coloneqq \Trnset{\rI\to\rA}$ can be viewed as a set of functionals on 
$\Cntset{\rA}$. As such, it can be viewed as a spanning subset of the real vector space 
$\Stset{\rA}_\Reals$ of linear funcitonals on $\Cntset{\rA}$. On the other hand 
$\Cntset{\rA}\coloneqq \Trnset{\rA\to\rI}$ is a separating set of 
positive linear functionals on $\Stset{\rA}$, which then spans the dual space 
$\Stset{\rA}_{\Reals}^*=:\Cntset{\rA}_{\Reals}$. The dimension $D_\rA$ of $\Stset{\rA}_\Reals$ 
(which is the same as that of $\Cntset{\rA}_\Reals$) is called {\em size} of system $\rA$.  One 
can easily prove that in any OPT $\Theta$, $\rI$ is the unique system with unit size $D_\rI=1$.
Using the properties of parallel composition, one can also prove that $D_{\rA\rB}\geq D_\rA D_\rB$. Events in $\Stset{\rA}$ are called {\em states}, and denoted by lower-case greek letters, e.g.~$\rho$, while events in $\Cntset{\rA}$ are called {\em effects}, and denoted by lower-case latin letters, e.g.~$a$. When it is appropriate, we will use the symbol $|\rho)$ to denote a state, and $(a|$ to denote an effect.
We will also use the circuit notation, where we denote states, transformations and effects by the symbols
\begin{align*}
\begin{aligned}
    \Qcircuit @C=1em @R=.7em @! R {&\prepareC{\rho}&\ustick{\rA}\qw&\qw}
\end{aligned}\ ,
\begin{aligned}
    \Qcircuit @C=1em @R=.7em @! R {&\ustick{\rA}\qw&\gate{\tA}&\ustick{\rB}\qw&\qw}
\end{aligned}\ ,
\begin{aligned}
    \Qcircuit @C=1em @R=.7em @! R {&\ustick{\rA}\qw&\measureD{a}}
\end{aligned}\ ,
\end{align*}
respectively. 
Sequential composition of $\tA\in\Trnset{\rA\to\rB}$  and $\tB\in\Trnset{\rB\to\rC}$ is 
denoted by the diagram
\begin{align*}
\begin{aligned}
    \Qcircuit @C=1em @R=.7em @! R {&\ustick{\rA}\qw&\gate{\tB\tA}&\ustick{\rC}\qw&\qw}
\end{aligned}\ =\ 
\begin{aligned}
    \Qcircuit @C=1em @R=.7em @! R {&\ustick{\rA}\qw&\gate{\tA}&\ustick{\rB}\qw&\gate{\tB}&\ustick{\rC}\qw&\qw}
\end{aligned}\ .
\end{align*}
For composite systems we use diagrams with multiple wires, e.g.
\begin{align*}
\begin{aligned}
    \Qcircuit @C=1em @R=.7em @! R {&\ustick{\rA}\qw&\multigate{1}{\tA}&\ustick{\rB}\qw&\qw\\
    &\ustick{\rC}\qw&\ghost{\tA}&\ustick{\rD}\qw&\qw}
\end{aligned}.
\end{align*}
The identity  will be omitted: $\Qcircuit @C=1em @R=.7em @! R {&\ustick{\rA}\qw&\gate{\tI}&\ustick{\rA}\qw&\qw}=    \Qcircuit @C=1em @R=.7em @! R {&\ustick{\rA}\qw&\qw}$. 
The swap $\tS_{\rA\rB}$ and its inverse will be denoted as follows
\begin{align*}
&
\begin{aligned}
    \Qcircuit @C=1em @R=.7em @! R {&\ustick{\rA}\qw&\multigate{1}{\tS}&\ustick{\rB}\qw&\qw\\
    &\ustick{\rB}\qw&\ghost{\tS}&\ustick{\rA}\qw&\qw}
\end{aligned}=
\begin{tikzpicture}
	\begin{pgfonlayer}{nodelayer}
		\node [style=none] (20) at (1.5, 1) {$\scriptstyle\rB$};
		\node [style=none] (19) at (-1.5, -0.5) {$\scriptstyle\rB$};
		\node [style=none] (0) at (-1.5, -1) {};
		\node [style=none] (1) at (-1.5, 0.5) {};
		\node [style=none] (3) at (1.5, -1) {};
		\node [style=none] (5) at (-1.5, -1) {};
		\node [style=none] (22) at (-0.25, -0.5) {};
		\node [style=none] (23) at (1.5, 0.5) {};
		\node [style=none] (24) at (0.25, 0) {};
		\node [style=none] (25) at (-1.5, 1) {$\scriptstyle\rA$};
		\node [style=none] (26) at (1.5, -0.5) {$\scriptstyle\rA$};
	\end{pgfonlayer}
	\begin{pgfonlayer}{edgelayer}
		\draw [in=-180, out=0, looseness=0.75] (1.center) to (3.center);
		\draw [in=-135, out=0] (5.center) to (22.center);
		\draw [in=45, out=-180] (23.center) to (24.center);
	\end{pgfonlayer}
\end{tikzpicture}\\
\\
&
\begin{aligned}
    \Qcircuit @C=1em @R=.7em @! R {&\ustick{\rA}\qw&\multigate{1}{\tS^*}&\ustick{\rB}\qw&\qw\\
    &\ustick{\rB}\qw&\ghost{\tS^*}&\ustick{\rA}\qw&\qw}
\end{aligned}=
\begin{tikzpicture}
	\begin{pgfonlayer}{nodelayer}
		\node [style=none] (20) at (1.5, 1) {$\scriptstyle\rB $};
		\node [style=none] (19) at (-1.5, -0.5) {$\scriptstyle\rB$};
		\node [style=none] (1) at (-1.5, -1) {};
		\node [style=none] (3) at (1.5, 0.5) {};
		\node [style=none] (5) at (-1.5, 0.5) {};
		\node [style=none] (22) at (-0.25, 0) {};
		\node [style=none] (23) at (1.5, -1) {};
		\node [style=none] (24) at (0.25, -0.5) {};
		\node [style=none] (25) at (-1.5, 1) {$\scriptstyle\rA$};
		\node [style=none] (26) at (1.5, -0.5) {$\scriptstyle\rA$};
	\end{pgfonlayer}
	\begin{pgfonlayer}{edgelayer}
		\draw [in=-180, out=0, looseness=0.75] (1.center) to (3.center);
		\draw [in=135, out=0] (5.center) to (22.center);
		\draw [in=-45, out=180, looseness=0.75] (23.center) to (24.center);
	\end{pgfonlayer}
\end{tikzpicture}
\end{align*}
In the present paper we will always assume that the theory under consideration is 
symmetric, however all the results will be straightforwardly generalisable. We will
consequently draw the swap $\tS_{\rA\rB}$ as
\begin{align*}
&
\begin{aligned}
    \Qcircuit @C=1em @R=.7em @! R {&\ustick{\rA}\qw&\multigate{1}{\tS}&\ustick{\rB}\qw&\qw\\
    &\ustick{\rB}\qw&\ghost{\tS}&\ustick{\rA}\qw&\qw}
\end{aligned}=\ 
\begin{aligned}
    \Qcircuit @C=1em @R=.7em @! R {&\ustick{\rA}\qw&\multigate{1}{\tS^*}&\ustick{\rB}\qw&\qw\\
    &\ustick{\rB}\qw&\ghost{\tS^*}&\ustick{\rA}\qw&\qw}
\end{aligned}=
\begin{tikzpicture}
	\begin{pgfonlayer}{nodelayer}
		\node [style=none] (20) at (1.5, 1) {$\scriptstyle\rB $};
		\node [style=none] (19) at (-1.5, -0.5) {$\scriptstyle\rB$};
		\node [style=none] (0) at (-1.5, -1) {};
		\node [style=none] (1) at (-1.5, 0.5) {};
		\node [style=none] (3) at (1.5, -1) {};
		\node [style=none] (5) at (-1.5, -1) {};
		\node [style=none] (22) at (1.5, 0.5) {};
		\node [style=none] (25) at (-1.5, 1) {$\scriptstyle\rA$};
		\node [style=none] (26) at (1.5, -0.5) {$\scriptstyle\rA$};
	\end{pgfonlayer}
	\begin{pgfonlayer}{edgelayer}
		\draw [in=-180, out=0, looseness=0.75] (1.center) to (3.center);
		\draw [in=-180, out=0, looseness=0.75] (5.center) to (22.center);
	\end{pgfonlayer}
\end{tikzpicture}
\end{align*}

An OPT $\Theta$ is specified by the collections of systems and tests, along with the parallel composition rule $\otimes$ 
\begin{align*}
\Theta\equiv(\Testun(\Theta),\Elety(\Theta),\otimes). 
\end{align*}
\begin{definition}[Equal transformations]\label{def:eqtr}
Let $\tA,\tB\in\Trnset{\rA\to\rB}$. Then we define $\tA=\tB$ if for every system $\rC$ and every $\rho\in\Stset{\rA\rC}$ and $a\in\Cntsetcomp{\rB}{\rC}$, one has
\begin{align*}
\begin{aligned}
    \Qcircuit @C=1em @R=.7em @! R {&\multiprepareC{1}{\rho}&\ustick{\rA}\qw&\gate{\tA}&\ustick{\rB}\qw&\multimeasureD{1}{a}\\
    &\pureghost{\rho}&\qw&\ustick{\rC}\qw&\qw&\ghost{a}}
\end{aligned}\ =\ 
\begin{aligned} 
    \Qcircuit @C=1em @R=.7em @! R {&\multiprepareC{1}{\rho}&\ustick{\rA}\qw&\gate{\tB}&\ustick{\rB}\qw&\multimeasureD{1}{a}\\
    &\pureghost{\rho}&\qw&\ustick{\rC}\qw&\qw&\ghost{a}}
\end{aligned}
\end{align*}
\end{definition}
Notice that, since states separate effects and viceversa effects separate states, the above definition is equivalent to the two following equality criteria.
\begin{lemma}\label{lem:eqcrit}
Let $\tA,\tB\in\Trnset{\rA\to\rB}$. Then the following conditions are equivalent.
\begin{enumerate}
\item $\tA=\tB$.
\item For every $\rC$ and every $\rho\in\Stset{\rA\rC}$, one has 
\begin{align}
\begin{aligned}
    \Qcircuit @C=1em @R=.7em @! R {&\multiprepareC{1}{\rho}&\ustick{\rA}\qw&\gate{\tA}&\ustick{\rB}\qw&\\
    &\pureghost{\rho}&\qw&\ustick{\rC}\qw&\qw&}
\end{aligned}\ =\ 
\begin{aligned} 
    \Qcircuit @C=1em @R=.7em @! R {&\multiprepareC{1}{\rho}&\ustick{\rA}\qw&\gate{\tB}&\ustick{\rB}\qw& \\
    &\pureghost{\rho}&\qw&\ustick{\rC}\qw&\qw&}
\end{aligned}\ .
\end{align}
\item For every $\rC$ and every $a\in\Cntsetcomp{\rB}{\rC}$, one has 
\begin{align}
\begin{aligned}
    \Qcircuit @C=1em @R=.7em @! R {&\ustick{\rA}\qw&\gate{\tA}&\ustick{\rB}\qw&\multimeasureD{1}{a}\\
    &\qw&\ustick{\rC}\qw&\qw&\ghost{a}}
\end{aligned}\ =\ 
\begin{aligned} 
    \Qcircuit @C=1em @R=.7em @! R {&\ustick{\rA}\qw&\gate{\tB}&\ustick{\rB}\qw&\multimeasureD{1}{a}\\
    &\qw&\ustick{\rC}\qw&\qw&\ghost{a}}
\end{aligned}\ .
\label{eq:eqcrit}
\end{align}
\end{enumerate}
\end{lemma}

One can show~\cite{PhysRevA.81.062348} that events $\tT\in\Trnset{\rA\to\rB}$ can be 
identified with a family of linear maps, one for every $\rC$, that characterize the action 
of $\tT\otimes\tI_\rC$ on $\Stset{\rA\rC}_{\Reals}$ as a linear map to 
$\Stset{\rB\rC}_{\Reals}$. As anticipated in the introductory paragraph, 
we remind the reader that some of the difficulties that we will be faced with in the 
remainder originate from the fact that in a general theory it 
is not true that $\Stset{\rA\rB}_\Reals=\Stset{\rA}_\Reals\otimes\Stset{\rB}_\Reals$, but
only $\Stset{\rA}_\Reals\otimes\Stset{\rB}_\Reals\subseteq \Stset{\rA\rB}_\Reals$. As a 
consequence, the linear map representing $\tT$ on $\Stset{\rA}_\Reals$ is not sufficient 
to determine the linear map representing $\tT\otimes\tI_\rC$ on $\Stset{\rA\rC}_\Reals$.

One can easily prove that $\Trnset{\rA\to\rB}$ spans a real vector space $\Trnset{\rA\to\rB}_{\Reals}$. Being $\Trnset{\rA\to\rB}$ spanning for $\Trnset{\rA\to\rB}_\Reals$, the criteria of definition~\ref{def:eqtr} and lemma~\ref{lem:eqcrit} hold for $\tA,\tB\in\Trnset{\rA\to\rB}_\Reals$. Elements of $\Trnset{\rA\to\rB}_\Reals$ are called {\em generalized events}. Every space $\Trnset{\rA\to\rB}_{\Reals}$ has a {\em zero event} $0_{\rA\to\rB}\coloneqq 0_{\rI\to\rI}\otimes\tT\in\Trnset{\rA\to\rB}$, where $\tT$ is an arbitrary event in $\Trnset{\rA\to\rB}$. As a consequence of the coarse graining rule for tests on $\Trnset{\rI\to\rI}$, one can easily show that coarse graining of two or more transformations is represented by their sum. Precisely, given a test $\{\tT_i\}_{i\in\cX}\subseteq\Trnset{\rA\to\rB}$, for $\cX_0=\{0,1\}\subseteq \cX$ one has $\tT'_0=\tT_0+\tT_1$, namely for every system $\rC$ it is $\tT'_0\otimes\tI_\rC=\tT_0\otimes\tI_\rC+\tT_1\otimes\tI_\rC$. Finally, every set $\Trnset{\rA\to\rB}$ has a subset consisting in singleton events, that we denote by $\Trnset{\rA\to\rB}_1$, and call {\em deterministic}. For $\rA,\rB\neq\rI$, a deterministic event is called {\em channel}. A channel $\tU\in\Trnset{\rA\to\rB}_1$ is {\em reversible} if there exists a channel $\tV\in\Trnset{\rB\to\rA}_1$ such that $\tV\tU=\tI_\rA$, $\tU\tV=\tI_\rB$.

Let us now define the cones
\begin{align*}
\Trnset{\rA\to\rB}_+\coloneqq \{\lambda\tT\mid\lambda\geq0,\ \tT\in\Trnset{\rA\to\rB}\}.
\end{align*}

We will often write $\tA\triangleright0$ as a shorthand for $\tA\in\Trnset{\rA\to\rB}_+$. The cone $\Trnset{\rA\to\rB}_+$ introduces a partial ordering in $\Trnset{\rA\to\rB}_\Reals$, defined by
\begin{align*}
\tA\triangleright\tB\quad\Leftrightarrow\quad(\tA-\tB)\triangleright0.
\end{align*}
OPTs are assumed to have all sets $\Trnset{\rA\to\rB}$ (and thus also cones $\Trnset{\rA\to\rB}_+$) closed in the operational norm.

Two systems may be operationally equivalent if they can be mapped one onto the other via a 
reversible transformation. Clearly, in this case every processing of the first system is
perfectly simulated by a processing of the other, and viceversa.We define operationally 
equivalent systems as follows.
\begin{definition}\label{def:opeqsys}
Let $\rA$, $\rB$ be two systems. We say that $\rA$ and $\rB$ are {\em operationally equivalent}, in formula $\rA\cong\rB$, if there exists a reversible transformation $\tU\in\Trnset{\rA\to\rB}_1$.
\end{definition}

\begin{definition}
If $\rA_1$ and $\rA_2$ are operationally equivalent through $\tU$ and $\rB_1$ and $\rB_2$ through $\tV$, then, for every system $\rC$, $\tA_1\in\Trnset{\rA_{1}\rC\to\rB_{1}\rC}_{\Reals}$ and $\tA_2\in\Trnset{\rA_{2}\rC\to\rB_{2}\rC}_{\Reals}$ are operationally equivalent if $\tA_2=(\tV\otimes\tI_\rC)\tA_1(\tU^{-1}\otimes\tI_\rC)$. In particular, $\rho_1\in\Stset{\rA_1\rC}_{\Reals}$ and $\rho_2\in\Stset{\rA_2\rC}_{\Reals}$ are operationally equivalent if $\rho_2=(\tU\otimes\tI_\rC)\rho_1$, and $a_1\in\Cntsetcomp{\rA_1}{\rC}_{\Reals}$ and $a_2\in\Cntsetcomp{\rA_2}{\rC}_{\Reals}$ are operationally equivalent if $a_2=a_1(\tU^{-1}\otimes\tI_\rC)$. 
\end{definition}

\begin{lemma}\label{lem:compi}
Let $\rA\rB$ be operationally equivalent to $\rA$. Then $\rB$ must be operationally equivalent to the trivial 
system $\rI$.
\end{lemma}
\begin{proof}
First of all, since $D_{\rA\rB}\geq D_\rA D_\rB\geq D_\rA$, if $\rA\rB\cong\rA$ the chain of inequalities is 
saturated, and thus $D_\rB=1$. This implies that every state of system $\rA\rB$ is of the form 
$\tau\otimes 1_\rB$, 
where $\tau\in\Stset{\rA}$ and $1_\rB$ is the unique deterministic state of system $\rB$. The same result holds for
states of $\rA\rB\rC$ for arbitrary $\rC$. Now, let $\tU\in\Trnset{\rA\rB\to\rA}_1$ be the reversible channel 
implementing operational equivalence. Let $\eta\in\Stset{\rA\rC}$, and 
$(\tU\otimes\tI_\rC)(1_\rB\otimes\eta)=\tilde\eta\in\Stset{\rA\rC}$. 
One can choose $\tU$ as to identify $\tilde\eta=\eta$. Indeed, there
is a reversible transformation $\tV\in\Trnset{\rA\to\rA}$ such that $(\tV\otimes\tI_\rC)\tilde\eta=\eta$. In 
particular, one can choose $\tV\coloneqq 1_\rB\tU^{-1}$, where now $1$ represents the unique deterministic effect 
of $\rB$. Right-reversibility of $\tV$ is trivial. On the other hand, let $\sigma\in\Stset{\rA\rC}$, and let 
$1_\rB\otimes\tau\coloneqq(\tU^{-1}\otimes\tI_\rC)\sigma\in\Stset{\rA\rB\rC}$. Now, after applying the 
deterministic effect $1_\rB$ one has $(\tV\otimes\tI_\rC)\sigma=\tau$. One can reverse $\tV$ by 
composing $\tau$ with $1_\rB$, obtaining $1_\rB\otimes\tau$, and then applying $\tU\otimes\tI_\rC$, finally 
obtaining $(\tU\otimes\tI_\rC)(1_\rB\otimes\tau)=\sigma$. With a suitable choice of $\tU$, one then has 
$(\tU\otimes\tI_\rC)(1_\rB\otimes\eta)=\eta$. Let now $1_\rB\otimes\eta=\tau_{\rB\rC}\otimes\nu_\rA$, with 
$\nu\in\Stset{\rA}_1$. Discarding system $\rA$ on both sides via some deterministic effect 
$e_\rA\in\Cntset{\rA}_1$, one obtains $\tau_{\rB\rC}=1_\rB\otimes\theta$, with 
$\theta=(e_\rA\otimes\tI_\rC)\eta$. Finally, this implies that every state of system $\rB\rC$ is of the form
$\tau_{\rB\rC}=1_\rB\otimes\theta$, with $\theta\in\Stset{\rC}$. Once we fix $\eta\in\Stset{\rA}_1$, we can then
define the reversible channel $\tW\in\Trnset{\rB\rC\to\rC}$ given by 
$\tW\coloneqq e_\rA(\tU\otimes\tI_\rC)(\eta\otimes\tI_{\rB\rC})$. Clearly, $\tW(1_\rB\otimes\theta)=\theta$, and 
the inverse map $\tW^{-1}\in\Trnset{\rC\to\rB\rC}$ is obtained as 
$\tW^{-1}\coloneqq e_\rA(\tU^{-1}\eta\otimes\tI_\rC)$. We then have $\rB\rC\simeq\rC$ for every $\rC$, 
i.e.~$\rB\simeq\rI$.
\end{proof}

\subsection{Causality and the no-restriction hypothesis}

In the remainder of the paper we will focus on {\em causal} theories.  The causality property, that we define right away, characterizes those theories where signals can propagate only in the direction defined by input and output of processes, within a cone of causal influence determined by interactions between systems. These are the only theories where one can consistently use information acquired in a set of tests to condition the choice of subsequent tests, where the partial ordering we are referring to is that determined by the input/output direction.
\begin{definition}[Causal theories]
A theory $(\tT,\rA)$ is {\em causal} if for every test $\{\tT_i\}_{i\in\cX}$ and every collection of tests $\{\tS^{i}_j\}_{j\in\cY}$ labelled by $i\in\cY$, the generalized test $\{\tC_{i,j}\}_{(i,j)\in\cX\times\cY}$ with
\begin{align*}
\begin{aligned}
    \Qcircuit @C=1em @R=.7em @! R {&\ustick{\rA}\qw&\gate{\tC_{i,j}}&\ustick{\rC}\qw&\qw}
\end{aligned}\ \coloneqq \ 
\begin{aligned}
    \Qcircuit @C=1em @R=.7em @! R {&\ustick{\rA}\qw&\gate{\tT_i}&\ustick{\rB}\qw&\gate{\tS^{i}_j}&\ustick{\rC}\qw&\qw}
\end{aligned}\ ,
\end{align*}
is a test of the theory.
\end{definition}
Notice that this notion of causality is strictly stronger than the one usually adopted in the literature about OPTs, that can be summarised as {\em uniqueness of the deterministic effect} (see e.g.~Refs.~\cite{PhysRevA.81.062348,chiribella2011informational,DAriano:2017aa}). Indeed, a first result of crucial importance 
derives uniqueness of the deterministic effect for every system type in causal theories.
\begin{theorem}[\cite{PhysRevA.81.062348,DAriano:2017aa}]
In a causal theory, for every system type $\rA$ the set of deterministic effects $\Cntset{\rA}_1$ is the singleton $\{e_\rA\}$.
\end{theorem}
A proof of the above theorem, that proceeds by contradiction, can be found in the mentioned 
references. We remark that, being the notion of causality in these references different form the 
one defined here, the theorem has a slightly different statement. In the same references, 
one can find the equivalence of uniqueness of the deterministic effect with non-signalling from 
output to input.
\begin{theorem}[\cite{PhysRevA.81.062348,DAriano:2017aa}]
In an OPT, every system type $\rA$ has a unique deterministic effect if and only if the marginal 
probabilities of preparation tests cannot depend on the choice of subsequent observation tests.
\end{theorem}

As a consequence of our notion of causality, in a non-deterministic theory 
(i.e.~a theory where $\Stset{\rI}\neq\{0,1\}$), every convex combination of tests is a test (see 
e.g.~\cite{DAriano:2017aa}). This implies that the sets $\Stset{\rA}_*$, $\Cntset{\rA}_*$, 
$\Trnset{\rA\to\rB}_*$ are convex, where $*$ can be replaced by nothing, $1$ or $+$. Moreover, 
the sets with $*=+$ are convex cones.

Another important consequence of causality is that a transformation $\tC\in\Trnset{\rA\to\rB}$ is deterministic if and only if it maps the deterministic effect $e_\rB$ to the deterministic effect $e_\rA$.
\begin{theorem}[\cite{PhysRevA.81.062348,DAriano:2017aa}]\label{th:charchan}
In a causal theory, a transformation $\tC\in\Trnset{\rA\to\rB}$ is a channel iff
\begin{align*}
\begin{aligned}
    \Qcircuit @C=1em @R=.7em @! R {&\ustick{\rA}\qw&\gate{\tC}&\ustick{\rB}\qw&\measureD{e}}
\end{aligned}\ =
\  
\begin{aligned}
    \Qcircuit @C=1em @R=.7em @! R {&\ustick{\rA}\qw&\measureD{e}}
\end{aligned}\ .
\end{align*}
\end{theorem}

In causal theories, one can always assume that every state is proportional to a deterministic one. Indeed, including in the theory every state $\rho\in\Stset{\rA}_+$ such that $(e_\rA|\rho)=1$ does not introduce any inconsistency with the set of transformations, as we now prove. Let $\Theta$ be a causal OPT, and define the theory $\Theta'$ through the bijection $\kappa:\Elety(\Theta)\to\Elety(\Theta')::\rA\mapsto\rA'$, with
\begin{align*}
&\Trnset{\rA'\to\rB'}_\Reals\equiv\Trnset{\rA\to\rB}_\Reals,\\
&\Trnset{\rA'\to\rB'}_1\coloneqq\{\tT\triangleright0\mid(e_\rB|\tT=(e_\rA|\},\\
&\Trnset{\rA'\to\rB'}\coloneqq\{\tT\triangleright0\mid\exists\tC\in\Trnset{\rA'\to\rB'}_1,\ \tC\triangleright\tT\},\\
&\Testset{\rA'\to\rB'}\coloneqq\\
&\quad\{\teT\subseteq\Trnset{\rA'\to\rB'}\mid\sum_{\tT\in\teT}\tT\in\Trnset{\rA'\to\rB'}_1\}.
\end{align*}
where $\triangleright$ denotes the ordering induced by the cones of the theory $\Theta$, and the cardinality of sets $\teT$ in the last definition is implicitly assumed to be finite.

\begin{theorem}\label{th:cauclo}
Let $\Theta$ be a causal OPT, and consider the theory $\Theta'$ defined above. Then $\Theta'$ is an OPT with a unique deterministic effect $e_{\rA'}$ for every $\rA'$.
\end{theorem}
\begin{proof}
All the compositional structures of $\Theta$ are inherited by $\Theta'$. 
The defining conditions in the special case of $\Stset{\rA'}=\Trnset{\rI'\to\rA'}$ give
\begin{align*}
\Stset{\rA'}=\{\rho\in\Stset{\rA}_+\mid(e_\rA|\rho)\leq1\},
\end{align*}
while for $\Cntset{\rA'}=\Trnset{\rA'\to\rI'}$ they give
\begin{align*}
\Cntset{\rA'}_1=\{e_\rA\}.
\end{align*}
What remains to be proved is that the set of transformations is closed under sequential and parallel composition. First of all, we observe that by theorem~\ref{th:charchan} $\Trnset{\rA\to\rB}_1\subseteq\Trnset{\rA'\to\rB'}_1$, thus $\Trnset{\rA\to\rB}\subseteq\Trnset{\rA'\to\rB'}$. This makes $\Trnset{\rA\to\rB}_+\subseteq\Trnset{\rA'\to\rB'}_+$. On the other hand, since by definition it is also $\Trnset{\rA'\to\rB'}\subseteq\Trnset{\rA\to\rB}_+$, we have $\Trnset{\rA'\to\rB'}_+\subseteq\Trnset{\rA\to\rB}_+$, which makes $\Trnset{\rA'\to\rB'}_+\equiv\Trnset{\rA\to\rB}_+$. Thus the ordering $\triangleright$ is the same for the two theories. As a consequence, the preservation of cones under the two compositions is inherited by $\Theta'$ from the same property in $\Theta$. Now, if $\tA\in\Trnset{\rA'\to\rB'}_1$ and $\tB\in\Trnset{\rB'\to\rC'}_1$, then 
\begin{align*}
(e_\rC|\tB\tA=(e_\rB|\tA=(e_\rA|,
\end{align*}
thus $\tB\tA\in\Trnset{\rA'\to\rC'}_1$. Moreover, thanks to causality of the theory $\Theta$ one has $e_{\rA\rB}=e_\rA\otimes e_\rB$, thus for $\tA\in\Trnset{\rA'\to\rB'}_1$ and $\tB\in\Trnset{\rC'\to\rD'}_1$,
\begin{align*}
(e_{\rB\rD}|\tA\otimes\tB&=(e_{\rB}|\tA\otimes(e_\rD|\tB\\
&=(e_\rA|\otimes(e_\rC|=(e_{\rA\rC}|,
\end{align*}
and consequently $\tA\otimes\tB\in\Trnset{\rA'\rC'\to\rB'\rD'}_1$. Now, given events $\tA,\tB$ in the theory $\Theta'$, by definition we have $\tC,\tD$ deterministic in $\Theta'$ such that $\tC\triangleright\tA$ and $\tD\triangleright\tB$, thus
\begin{align*}
\tF\coloneqq &(\tD-\tB)(\tC-\tA)+(\tD-\tB)\tA\\
&+\tB(\tC-\tA)\triangleright0,\\
\tG\coloneqq &(\tC-\tA)\otimes(\tD-\tB)\\
&+(\tC-\tA)\otimes\tB+\tA\otimes(\tD-\tB)\triangleright0.
\end{align*}
Finally, $\tF+\tB\tA\triangleright\tB\tA$ and $\tG+\tA\otimes\tB\triangleright\tA\otimes\tB$ are channels, and thus $\tB\tA$ and $\tA\otimes\tB$ are events.
\end{proof}

One can now show that the new theory $\Theta'$ is causal.
%
%

\begin{corollary}\label{cor:cauclo}
Let $\Theta$ be a causal OPT, and $\Theta'$ as in theorem~\ref{th:cauclo}.
Then $\Theta'$ is a causal OPT.
\end{corollary}
\begin{proof}
Let $\{\tT_i\}_{i\in\cX}$ be a test in $\Trnset{\rA'\to\rB'}$, and $\{\tS^i_j\}_{j\in\cY}$ be tests in $\Trnset{\rB'\to\rC'}$ for every $i\in\cX$. Then
\begin{align*}
\forall i\in\cX\ \sum_{j\in\cY}(e_{\rC'}|\tS^i_j=(e_{\rB'}|,\quad\sum_{i\in\cX}(e_{\rB'}|\tT_i=(e_{\rA'}|.
\end{align*}
This implies that
\begin{align*}
&\sum_{i\in\cX}\sum_{j\in\cY}(e_{\rC'}|\tW_{i,j}=(e_{\rA'}|,\\
&\tW_{i,j}\coloneqq \tS^i_j\tT_i\in\Trnset{\rA'\to\rC'}\ \forall i\in\cX,j\in\cY.
\end{align*}
As a consequence, every conditional test $\tW_{i,j}\coloneqq \tS^i_j\tT_i\in\Trnset{\rA'\to\rC'}$ is admitted in the theory $\Theta'$.
\end{proof}

\begin{remark}\label{rem:norestcau}
In the remainder, we will focus on theories satisfying causality and the further requirements 
\begin{align*}
&\Trnset{\rA\to\rB}_1=\{\tT\triangleright0\mid(e_\rB|\tT=(e_\rA|\},\\
&\Trnset{\rA\to\rB}=\{\tT\triangleright0\mid\exists\tC\in\Trnset{\rA\to\rB}_1,\,\tC\triangleright\tT\},\\
&\Testset{\rA\to\rB}=\{\teT\subseteq\Trnset{\rA\to\rB}\mid\sum_{\tT\in\teT}\tT\in\Trnset{\rA\to\rB}_1\}. 
\end{align*}
In particular, this implies that $\Stset{\rA}=\{\rho\in\Stset{\rA}_+\mid(e_\rA|\rho)\leq1\}$. Thanks to corollary~\ref{cor:cauclo}, this is not a significant restriction.
\end{remark}

We now narrow down focus on theories that satisfy the {\em no-restriction 
hypothesis}. Let us consider the set of preparation-tests for every system of the 
theory $\Theta$. Then we will complete the set of tests allowing for all those tests
that transform preparation tests to preparation tests, even when applied locally. This 
requirement is the generalisation of the assumption made in quantum theory that a map is a 
transformation if and only if it is completely positive and trace non increasing, and a 
collection of transformations is a test if and only if the sum of its elements is a 
channel, i.e.~completely positive and trace-preserving.
\begin{assumption}
Let $\tA\in\Trnset{\rA\to\rB}_\Reals$. The no-restriction hypothesis consists in the requirement that 
$\teT\in\Testset{\rA\to\rB}$ if and only if for every $\rC$ and every $\teP\in\Testset{\rI\to\rA\rC}$, one has $(\teT\otimes\teI_\rC)\teP\in\Testset{\rI\to\rB\rC}$.
\label{ass:nore}
\end{assumption}

We will write $\tA\succeq0$ if for every $\rC$ and every $P\in\Stset{\rA\rC}_+$, one has $(\tA\otimes\tI_\rC)P\in\Stset{\rB\rC}_+$. The set of transformations $\tA\succeq0$ is a cone, that we denote by 
\begin{align*}
\sK(\rA\to\rB)\coloneqq \{\tA\in\Trnset{\rA\to\rB}_\Reals\mid\tA\succeq0\}.
\end{align*}
The above defined cone introduces a (partial) ordering in $\Trnset{\rA\to\rB}_\Reals$, defined by
\begin{align}
\tA\succeq\tB\quad\Leftrightarrow\ \tA-\tB\succeq0.
\end{align}

Notice that, under the no-restriction hypothesis, the following identity holds
\begin{align*}
\sK(\rA\to\rB)=\Trnset{\rA\to\rB}_+,
\end{align*}
and thus for every $\tA,\tB\in\Trnset{\rA\to\rB}_\Reals$,
\begin{align}
\tA\succeq\tB\ \Leftrightarrow\ \tA\triangleright\tB.
\end{align}

We remark that, under assumption~\ref{ass:nore}, the hypothesis of theorem~\ref{th:charchan} can be relaxed to $\tC\in\Trnset{\rA\to\rB}_+$, since the latter implies that $\tC=\lambda\tC_0$ for some $\tC_0\in\Trnset{\rA\to\rB}$ and $\lambda\geq0$. Then, $\tC\succeq0$. Moreover, since $\tC$ sends $e_\rB$ to $e_\rA$, one has $(\tC\otimes\tI_\rC)\Stset{\rA\rC}_1\subseteq\Stset{\rB\rC}_1$, and thus $(\tC\otimes\tI_\rC)\Stset{\rA\rC}\subseteq\Stset{\rB\rC}$. By the no-restriction hypothesis, this is tantamount to $\tC\in\Trnset{\rA\to\rB}$.


We can now prove a theorem that is a very important consequence of the no-restriction hypothesis, along with causality.

\begin{theorem}\label{th:norest}
In a theory satisfying the no-restriction hypothesis, let $\tT\in\Trnset{\rA\to\rB}_\Reals$. Then $\tT,\tC-\tT\succeq0$ for some channel $\tC\in\Trnset{\rA\to\rB}_1$ iff $\tT\in\Trnset{\rA\to\rB}$.
\end{theorem}
\begin{proof}
The hypothesis $\tT,\tC-\tT\succeq0$ is equivalent to $\tT,\tC-\tT\in\Trnset{\rA\to\rB}_+$. Equivalently, for every $\rC$ and every $P\in\Stset{\rA\rC}_+$, one has $(\tX\otimes\tI_\rC)P\in\Stset{\rB\rC}_+$ for $\tX=\tT,\tC-\tT$. Let now $P\in\Stset{\rA\rC}$. Then one has
\begin{align*}
(e_{\rB\rC}|[(\tC-\tT)\otimes\tI_\rC]|P)\geq0,
\end{align*}
and thus
\begin{align*}
0\leq(e_{\rB\rC}|(\tT\otimes\tI_\rC)|P)\leq(e_{\rA\rC}|P)\leq1.
\end{align*}
This implies that $(\tT\otimes\tI_\rC)P\in\Stset{\rA\rC}$. Finally, by the no-restriction hypothesis, $\tT\in\Trnset{\rA\to\rB}$. The converse statement is trivial.
\end{proof}

In the literature~\cite{masanes2011derivation,dakic2009quantum} one can often find a different requirement under the name ``no-restriction hypothesis'', namely that for every system $\rA$ the set $\Cntset{\rA}$ coincides with the set of functionals $a$ on $\Stset{\rA}_\Reals$ that satisfy $0\leq(a|\rho)\leq1$ for every $\rho\in\Stset{\rA}$. 
The last condition may be neither necessary nor sufficient for the no-restriction hypothesis as we state it, despite counterexamples are still unknown for both cases. Indeed, the no-restriction hypothesis for effects imposes the requirement that for every $\rC$ and every state $P\in\Stset{\rA\rC}$, one has
\begin{align}
\begin{aligned}
    \Qcircuit @C=1em @R=.7em @! R {\multiprepareC{1}{P}&\ustick{\rA}\qw&\measureD{a}\\
    \pureghost{P}&\ustick{\rC}\qw&\qw}
\end{aligned}\ \in\Stset{\rC}\ .
\end{align}

\subsection{Norms}

As the first step in the present work is to construct infinite composite systems, we will
need a thorough notion of sequences and limits. From a topological point of view the 
vector spaces that we constructed so far have no special structure, however the 
operational procedures for discrimination of processes provide a natural definition of 
distance between events. Such a distance is related to the success probability in 
discriminating events. Making the space of events into a metric space, the operational  
distance immediately provides a topological structure through the induced operational 
norm. However, in order to make the space of events of infinite 
composite systems into a Banach algebra, a stronger norm is needed, that is introduced 
here: the sup-norm. In the quantum and classical case the two norms coincide.

The operational norm $\normop{\cdot}$ for states is defined as 
follows~\cite{DAriano:2017aa}.
\begin{definition}
The operational norm on $\Stset{\rA}_\Reals$ is
\begin{align}
\normop{\rho}\coloneqq \sup_{a\in \Cntset{\rA}}(2a-e|\rho)=\sup_{\substack{a_0,a_1\in\Cntset{\rA}\\a_0+a_1=e_\rA}}(a_0-a_1|\rho).
\end{align}
\end{definition}

The operational norm $\normop{\cdot}$ for transformations $\tA\in\Trnset{\rA\to\rB}_\Reals$ on finite systems is then defined as
\begin{align*}
\normop{\tA}&\coloneqq \sup_{\rC,\Psi\in\Stset{\rA\rC}_1}\normop{(\tA\otimes\tI_\rC)\Psi}\nonumber\\
&=\sup_{\rC,\Psi\in\Stset{\rA\rC}}\sup_{a\in\Cntsetcomp{\rB}{\rC}}(2a-e|\tilde\tA\otimes\tI_\rC|\Psi).
\end{align*}
In the special case of effects $a\in\Cntset{\rA}_\Reals$ we have
\begin{align*}
\normop{a}&\coloneqq \sup_{\rC,\Psi\in\Stset{\rA\rC}}\normop{(a\otimes\tI_\rC)\Psi}\nonumber\\
&=\sup_{\rC,\Psi\in\Stset{\rA\rC}}\sup_{b\in\Cntset{\rC}}(a\otimes[2b-e_\rC]|\Psi).
\end{align*}
As a consequence,
\begin{align*}
\normop{a}&=\sup_{\substack{\rho_0,\rho_1\in\Stset{\rA}\\\rho_0+\rho_1\in\Stset{\rA}}}(a|\rho_0-\rho_1)\\
&=\sup_{p\in[0,1]}\{p\sup_{\rho\in\Stset{\rA}}(a|\rho)-(1-p)\inf_{\sigma\in\Stset{\rA}}(a|\sigma)\}\\
&=\max\{\sup_{\rho\in\Stset{\rA}}(a|\rho),\inf_{\sigma\in\Stset{\rA}}(a|\sigma)\}\\
&=\sup_{\rho\in\Stset{\rA}}|(a|\rho)|.
\end{align*}
Here we only prove one new result about the operational norm, that we will use in the following.
\begin{lemma}\label{lem:optens}
Let $\rho\in\Stset{\rA}_\Reals$ and $\sigma\in\Stset{\rB}_1$. Then
\begin{align}
\normop{\rho\otimes\sigma}=\normop\rho.
\end{align}
\end{lemma}
\begin{proof}
By definition it is
\begin{align*}
\normop{\rho\otimes\sigma}&=\sup_{(a_0,a_1)}(a_0-a_1|{\rho\otimes\sigma})\\
&=\sup_{(b_0,b_1)\in M}(b_0-b_1|\rho),
\end{align*}
where $M\coloneqq \{(b_0,b_1)\mid b_i=a_i(\tI_\rA\otimes \sigma)\}$. Thus
\begin{align*}
\normop{\rho\otimes\sigma}\leq\normop{\rho}.
\end{align*}
On the other hand, for every binary observation-test $(a_0,a_1)$ in $\Cntset{\rA}$ one has
\begin{align*}
a_i=(a_i\otimes e_{\rB})(\tI_{\rA}\otimes\sigma),
\end{align*}
and thus $M$ actually contains every possible binary observation-test on $\rA$. Finally, this implies that
\begin{align*}
\normop{\rho\otimes\sigma}=\normop{\rho}. &\qedhere
\end{align*}
\end{proof}
For more details on $\normop{\cdot}$ see Refs.~\cite{PhysRevA.81.062348,DAriano:2017aa}. 

We now proceed to define the sup-norm. 

\begin{definition}\label{def:normsuptr}
The \emph{sup-norm} $\normsup{\tA}$ of $\tA\in\Trnset{\rA\to\rB}_{\Reals}$ is defined as
\begin{align*}
&\normsup{\tA}\coloneqq \inf J(\tA),\\ 
&J(\tB)\coloneqq \{\lambda\mid\exists\tC\in\Trnset{\rA\to \rB}_1,\ \lambda\tC\succeq\tB\succeq-\lambda\tC\}.
\end{align*}
\end{definition}

We now show that $\normsup{\cdot}$ actually defines a norm. 

\begin{proposition}\label{prop:normsup}
The sup-norm on $\Trnset{\rA\to\rB}_\Reals$ is well defined:
\begin{enumerate}
\item $\normsup{\tA}$ is non-negative, and it is null iff $\tA=0$,
\item for $\mu\in\Reals$ one has $\normsup{\mu \tA}=|\mu|\normsup{\tA}$,
\item $\normsup{\tA+\tB}\leq\normsup{\tA}+\normsup{\tB}$.
\end{enumerate}
\end{proposition}
\begin{proof}1. Let $\tA\in\Trnset{\rA\to\rB}_{\Reals}$. Suppose that $j\coloneqq \inf J(\tA)<0$. Then there exists  $0<\varepsilon<|j|$ such that $j+\varepsilon\in J(\tA)$, namely
\begin{align*}
&-(|j|-\varepsilon)\tC\succeq \tA\succeq(|j|-\varepsilon)\tC\\
& \Rightarrow\ -2(|j|-\varepsilon)\tC\succeq0,
\end{align*}
for some $\tC\in\Trnset{\rA\to\rB}_1$, which is absurd. Then $\inf J(\tA)\geq0$. Suppose now that $J(\tA)=0$. This implies that for every $n\in\Nats$ there exists $\tC_n\in\Trnset{\rA\to\rB}_1$ such that $\frac1n \tC_n\succeq \tA\succeq-\frac1n \tC_n$. Now, by the closure of $\Trnset{\rA}_+$ in the operational norm, and considering that the sequences $(\frac1n \tC_n\pm \tA)$ converge to $\pm \tA$, it must be $\pm \tA\in\Trnset{\rA}_+$. This is possible if and only if $\tA=0$.
2. The proof is trivial for $\mu=0$. Let then $\mu\neq0$. If $x\in J(\tA)$, then $x \tC\pm \tA\succeq0$ for some $\tC\in\Trnset{\rA\to\rB}_1$, and thus $|\mu|(x\tC\pm \tA)\succeq0$, namely $|\mu| x\in J(\mu \tA)$. Thus $\normsup{\mu \tA}\leq|\mu|\normsup{\tA}$. For the same reason,  $\normsup \tA\leq(1/{|\mu|})\normsup{\mu \tA}$, and finally $\normsup{\mu \tA}=|\mu|\normsup \tA$.
3. Let now $x\in J( \tA)$ and $y\in J( \tB)$. Then $x\tC\pm \tA\succeq0$, and $y \tD\pm \tB\succeq0$, for $\tC,\tD\in\Trnset{\rA\to\rB}_1$. Thus $(x+y)\tF\pm(\tA+\tB)\succeq0$, where $\tF\coloneqq x/(x+y)\tC+y/(x+y)\tD\in\Trnset{\rA\to\rB}_1$. Thus, $x+y\in J(\tA+\tB)$, and then $\normsup{\tA+\tB}\leq\normsup \tA+\normsup \tB$.\qedhere
\end{proof}
The following property makes $(\Trnset{\rA\to\rA}_\Reals,\normsup\cdot)$ a Banach algebra.
\begin{proposition}\label{prop:banachsupnorm}
For $\tA\in\Trnset{\rB\to\rC}_\Reals$ and $\tB\in\Trnset{\rA\to\rB}_\Reals$, $\normsup{\tA\tB}\leq\normsup\tA\normsup\tB$.
\end{proposition}
\begin{proof} Let $x\in J(\tA)$, and $y\in J(\tB)$. Then there are $\tC,\tD$ such that $x\tC\pm\tA\succeq0$, $y\tD\pm\tB\succeq0$. Now, we have
\begin{align*}
&\frac12[(x\tC+\tA)(y\tD-\tB)+(x\tC-\tA)(y\tD+\tB)\\
&\quad=xy\tC\tD-\tA\tB\succeq0\\
&\frac12[(x\tC+\tA)(y\tD+\tB)+(x\tC-\tA)(y\tD-\tB)\\
=&xy\tC\tD+\tA\tB\succeq0,
\end{align*}
and then $xy\in J(\tA\tB)$. Thus, $\normsup{\tA\tB}\leq\normsup\tA\normsup\tB$.\end{proof}

\begin{lemma}\label{lem:prodchan}
Let $\tA\in\Trnset{\rA\to\rB}_\Reals$, and for an arbitrary $\rC$, let $\tD\in\Trnset{\rC\to\rD}_1$. Then $\normsup{\tA\otimes\tD}=\normsup\tA$.
\end{lemma}
\begin{proof} Let $x\in J(\tA)$. Then by definition there exists $\tC\in\Trnset{\rA\to\rB}_1$ such that $x\tC\pm\tA\succeq0$. This implies that
\begin{align*}
x\tC\otimes\tD\pm\tA\otimes\tD\succeq0,
\end{align*}
namely $x\in J(\tA\otimes\tD)$, and then $J(\tA)\subseteq J(\tA\otimes\tD)$. Now, let $y\in J(\tA\otimes\tD)$. Then there exists $\tC'\in\Trnset{\rA\rC\to\rB\rD}_1$ such that $y\tC'\pm\tA\otimes\tD\succeq0$. Composing the l.h.s.~of the latter relation with $\psi\in\Stset{\rC}_1$ and $e_\rD$, we obtain $y\tC''\pm\tA\succeq0$, where $\tC''\coloneqq (e_\rD|\tC'|\psi)$ is a channel, and then $y\in J(\tA)$. Thus, $J(\tA\otimes\tD)\subseteq J(\tA)$. Finally, since $J(\tA\otimes\tD)=J(\tA)$, we have $\normsup{\tA\otimes\tD}=\normsup{\tA}$.
\end{proof}

\begin{corollary}\label{cor:supnormtid}
For $\tA\in\Trnset{\rA\to\rB}_\Reals$, and for arbitrary $\rC$, it is $\normsup{\tA\otimes\tI_\rC}=\normsup{\tA}$.
\end{corollary}

\begin{corollary}\label{cor:productsupnorm}
For $\tA\in\Trnset{\rA\to\rB}_\Reals$ and $\tB\in\Trnset{\rC\to\rD}_\Reals$, $\normsup{\tA\otimes\tB}\leq\normsup\tA\normsup\tB$.
\end{corollary}
\begin{proof} 
The result follows straightforwardly from proposition~\ref{prop:banachsupnorm} and corollary~\ref{cor:supnormtid}.
\end{proof}

An important result regarding the sup-norm is provided by the following proposition.
\begin{proposition}\label{prop:opleqsupop}
For $\tA\in\Trnset{\rB\to\rC}_\Reals$ and $\tB\in\Trnset{\rA\to\rB}_\Reals$, $\normop{\tA\tB}\leq\normsup\tA\normop\tB$.
\end{proposition}
\begin{proof} By definition we have
\begin{align*}
&\normop{\tA\tB}\\
&=\sup_{\rD,\Psi\in\Stset{\rA\rD}_1}\sup_{a\in\Cntsetcomp{\rC}{\rD}}(2a-e_{\rC\rD}|\tA\tB\otimes\tI_\rD|\Psi).
\end{align*}
Now, for every $a\in\Cntsetcomp{\rC}{\rD}$, and $\lambda\in J(\tA)$, upon defining $a_0\coloneqq a$, $a_1\coloneqq e_{\rC\rD}-a$, we have
\begin{align*}
&(a_0-a_1|\tA\otimes\tI_\rC\\
&=(a_0|[\tA\otimes\tI_\rD]-(a_1|[\tA\otimes\tI_\rD]\\
&=\lambda[(\tilde a_0|-(\tilde a_1|],
\end{align*}
where $(\tilde a_i|\coloneqq \lambda^{-1}\{(a_i|[\tA\otimes\tI_\rD]+\tfrac12(e|[(\lambda\tC-\tA)\otimes\tI_\rD]\}$, with $\tC:\Trnset{\rB\to\rC}_1$ such that $\lambda\tC\pm\tA\succeq0$. Clearly, 
\begin{align*}
&\tilde a_0,\tilde a_1\in\Cntsetcomp{\rB}{\rD},&&\tilde a_0+\tilde a_1=e_{\rB\rD},
\end{align*}
thus
\begin{align*}
&(a_0-a_1|\tA\tB\otimes\tI_\rD|\Psi)\\
&=\lambda[(\tilde a_0|-(\tilde a_1|]\tB\otimes\tI_\rD|\Psi)\\
&\leq\lambda\sup_{b\in\Cntsetcomp{\rB}{\rD}}(2b-e_{\rB\rD}|\tB\otimes\tI_{\rD}|\Psi)
\end{align*}
This implies that $\normop{\tA\tB}\leq\lambda\sup_{b\in\Cntsetcomp{\rB}{\rD}}(2b-e_{\rB\rD}|\tB\otimes\tI_{\rD}|\Psi)$, and then for every $\lambda\in J(\tA)$
\begin{align*}
&\normop{\tA\tB}\\
&\leq\lambda\sup_{\rD,\Psi\in\Stset{\rA\rD}_1}\sup_{b\in\Cntsetcomp{\rB}{\rD}}(2b-e_{\rB\rD}|\tB\otimes\tI_\rD|\Psi)\\
&=\lambda\normop{\tB}.
\end{align*}
Finally, taking the infimum over $\lambda\in J(\tA)$ we get
\begin{align*}
\normop{\tA\tB}\leq\normsup{\tA}\normop{\tB}.&\qedhere
\end{align*}
\end{proof}

\begin{corollary}\label{cor:supstrongerop}
The sup-norm is stronger than the operational norm.
\end{corollary}
\begin{proof} 
It is sufficient to observe that $\normop{\tA}=\normop{\tA\tI}\leq\normsup{\tA}\normop{\tI}=\normsup{\tA}$.
\end{proof}

\begin{lemma}\label{lem:supnormchan}
Let $\tA\in\Trnset{\rA\to\rB}_1$ be a channel. Then $\normsup{\tA}=1$.
\end{lemma}
\begin{proof}
Clearly, $\tA\succeq\tA\succeq-\tA$, thus $1\in J(\tA)$, and $\inf J(\tA)\leq1$. On the other hand, suppose that there exists $1>\lambda\in J(\tA)$.
This implies that there exists a channel $\tC\in\Trnset{\rA\to\rB}_1$ such that $\tD\coloneqq \lambda\tC-\tA\succeq0$. However, this implies that $(e|_\rB\tD=-(1-\lambda)(e|_\rA\succeq0$. This is absurd, and then $\lambda\in J(\tA)$ must be $\lambda\geq 1$. Thus, $\inf J(\tA)\geq1$. Finally, this implies $\normsup{\tA}=1$.
\end{proof}

\begin{corollary}\label{cor:normid}
The sup-norm of the identity channel $\tI\in\Trnset{\rA\to\rA}_1$ is 1.
\end{corollary}

The sup-norm for effects is just the special case of the sup-norm of transformations with the output system equal to $\rI$.
\begin{definition}\label{def:supnormeff}
Let $a\in\Cntset{\rA}_\Reals$, and let us define the half-line
\begin{align*}
J(a)\coloneqq \{\lambda\in\Reals_+\mid\lambda e_\rA\succeq a\succeq-\lambda e_\rA\}. 
\end{align*}
The sup-norm $\normsup a$ is defined as 
\begin{align*}
&\normsup a\coloneqq \inf J(a).
\end{align*}
\end{definition}
\begin{proposition}\label{prop:normsupeff}
The sup-norm on $\Cntset{\rA}_\Reals$ is well defined:
\begin{enumerate}
\item $\normsup{a}$ is non-negative, and it is null iff $a=0$,
\item for $\mu\in\Reals$ one has $\normsup{\mu a}=|\mu|\normsup{a}$,
\item $\normsup{a+b}\leq\normsup{a}+\normsup{b}$.
\end{enumerate}
\end{proposition}
\begin{proof}
The case of effects is just a special case of the result of Proposition~\ref{prop:normsup}.
\end{proof}

As an immediate consequence of proposition~\ref{prop:opleqsupop}, we have the following result.

\begin{corollary} The sup-norm is stronger than the operational norm on $\Cntset{\rA}_\Reals$.
\end{corollary}

In the special case of $\Trnset{\rI\to\rI}_\Reals=\Reals$, one has $\Trnset{\rI\to\rI}_1=\{1\}$, and $\Trnset{\rI\to\rI}_+=\Reals_+$, thus $J(x)=\{\lambda\in\Reals\mid\lambda\pm x\geq0\}$. Clearly, $\normsup x=\inf J(x)=|x|$.
We now need the following lemma.
\begin{lemma}\label{lem:normotimese}
Let $a\in\Cntset{\rA}_\Reals$. Then $\normop{a\otimes e_\rB}=\normop a$ and $\normsup{a\otimes e_\rB}=\normsup a$.
\end{lemma}
\begin{proof}
For $\normop{\cdot}$, the equality trivially follows from  
the definition. For $\normsup\cdot$, it is a special case of lemma~\ref{lem:prodchan}.
\end{proof}

From lemma~\ref{lem:supnormchan}, the following consequence follows.

\begin{corollary}
For the deterministic effect $e_\rA$ one has $\normsup{e_\rA}=\normop{e_\rA}=1$.
\end{corollary}

We now prove a result that will be very useful later.
\begin{lemma}\label{lem:***}
Let $\tA\in\Trnset{\rA\to\rB}_+$, and $(a|\coloneqq (e_\rB|\tA\in\Cntset{\rA}_+$. Then $\normsup{\tA}=\normsup{a}$.
\end{lemma}
\begin{proof}
First of all, by proposition~\ref{prop:banachsupnorm}, one has $\normsup a\leq\normsup{e_\rA}\normsup\tA=\normsup\tA$. On the other hand, let $\lambda\in J(a)$. This implies that
\begin{align*}
b\coloneqq \lambda e_\rA-a\succeq0.
\end{align*}
Let us then define $\tB\coloneqq |\rho)(b|\in\Trnset{\rA\to\rB}_+$, for some $\rho\in\Stset{\rB}_1$. By theorem~\ref{th:charchan}, it is easy to check that $\tA+\tB=\lambda\tC$ for $\tC\in\Trnset{\rA\to\rB}_1$. Then,
\begin{align}
\lambda\tC-\tA=\tB\succeq0,
\end{align}
which means that $\lambda\in J(\tA)$. Then, $J(a)\subseteq J(\tA)$, and finally $\normsup{\tA}\leq\normsup a$. 
\end{proof}

In the case of Classical and Quantum Theory, phrasing the definitions of sup- and 
operational norm in terms of semidefinite programming problems, one can conclude that they 
are strongly dual. As a consequence, in these special cases they define the same norm.


\section{The quasi-local algebra in OPTs}\label{sec:vonn}

Following the definition of Ref.~ \cite{schumacher2004reversible}, we define a cellular automaton in a general OPT as a triple $(G,\rA,\tV)$, where $G$ is a denumerable set of labels for the systems that compose the automaton---addresses of the memory cells---, $\rA_G$ is a (possibly infinite) composite system corresponding to the collection of systems $\rA_g$ labelled by elements $g\in G$---i.e.~the memory array---, and $\tV$ is a reversible transformation on $\Cntset{\rA_G}$, such that $\tV\cdot\tV^{-1}$ is an automorphism of $\Trnset{\rA_G\to\rA_G}$. However, this is definition is incomplete, for many reasons. In the first place, most of the objects mentioned above are not thoroughly defined. The purpose of the present section is to set the ground for the rigorous definition of a cellular automaton.

We start fixing some notation. For every $R\subseteq G$ let 
\begin{align*}
\rA_R\coloneqq \bigotimes_{g\in R}\rA_g, 
\end{align*}
with the convention that $\rA_\emptyset\coloneqq \rI$. The full 
system is then $\rA_G=\bigotimes_{g\in G}\rA_g$. 
This purely formal notion will now be thoroughly 
substantiated.

With a slight abuse of notation, we will often use $R=g$ instead of $R=\{g\}$, dropping the braces. The set of finite regions of $G$ will be denoted as
\begin{align*}
\reg G \coloneqq \{R\subseteq G\mid|R|<\infty\},
\end{align*}
while the set of arbitrary regions of $G$ will be denoted as 
\begin{align*}
\infreg G \coloneqq \{R\subseteq G\}. 
\end{align*}
Clearly $\reg G \subseteq\infreg G $. In the remainder, when we write e.g.~$e_R$ or $\tI_R$ for $R\in\reg G$, we mean the deterministic effect or the identity transformation for the system $\rA_R$, respectively.

The following construction of $\bigotimes_{g\in G}\rA_g$ is inspired by that of Refs.~\cite{vonNeumann:1939vh,10.2307/1968693}, however with significant differences.

\subsection{Quasi-local effects}\label{sub:qle}

In this subsection we start the mathematical construction of the system $\rA_G$, the 
parallel composition of infinitely many finite systems. The first object that we will 
define is the space of generalised effects, along with its convex cone of positive 
effects, and the convex set of effects. The construction is based on the notion of a local 
effect, that is an effect which acts non trivially only on finitely many systems labelled 
by $g\in R$, with $R\in\reg G$. We will say that such an effect acts on the finite region 
$R$. We will then introduce a real vector space structure over the set of local effects,
and a norm that is induced by the sup norm for local effects. The last step is then the 
topological closure of the space of local effects in the sup-norm. The Banach space thus
obtained is the space of generalised quasi-local effects, and the positive cone along with 
the convex set of effects contain the limits of sequences of elements of local cones or 
convex sets of effects, respectively.

As we will see in the next section, the definition of the space of quasi-local states is much more involved than that of quasi-local effects. The latter is particularly simple, thanks to causality, that provides us with a preferential local effect, the unique deterministic one, without the need of introducing any arbitrary choice of a reference local effect. Moreover, while the construction of the space of quasi-local states is not logically necessary, as we will find them as a subspace of bounded functionals on quasi-local effects, the same is not true of effects. 

If we defined effects as the dual space of quasi-local states, we would end up with far more effects than needed, while lacking sectors of the state space, that are usually reached by the evolution of a quasi-local state through a cellular automaton. These are the main reasons why our construction begins with effects. 

The first notion we introduce is that of a local effect. 
Let $\rA_G$ denote the formal composition of countably 
many systems from a general OPT: 
$\rA_G\coloneqq \bigotimes_{g\in G}\rA_g$. 
Intuitively speaking, a local effect of $\rA_G$ is an event within a 
test that discards all the systems in $G$ but the finite 
region $R$, where a non-trivial measurement is performed. We 
then define the {\em local effect} $(a,R)$ as a pair made of 
an effect $a\in\Cntset{\rA_R}$ and the region $R$. To 
lighten the notation, in the remainder of the paper we will 
use the symbol $a_R$ instead of $(a,R)$.
The set of local effects is denoted by 
\begin{align*}
&\set{Pre}\Cntset{\rA_G}_{L}\\
&\coloneqq \bigsqcup_{R\in\reg G}\Cntset{\rA_R}=\{a_R\mid R\in\reg G,\, a\in\Cntset{\rA_R}\}.
\end{align*}
The above definition can be widened encompassing generalised effects:
\begin{align*}
\set{Pre}\Cntset{\rA_G}_{L\Reals}\coloneqq \bigsqcup_{R\in\reg G}\Cntset{\rA_R}_\Reals.
\end{align*}

Let us now consider a partition of the region $R$ into two disjoint regions $S_0\cup S_1=R$. Since in the definition of a local effect $a_R$ we did not set constraints on the effect $a\in\Cntset{\rA_R}_\Reals$, it might be that $a=e_{{S_0}}\otimes a'$, with $a'\in\Cntset{\rA_{S_1}}_{\Reals}$. It is clear from our intuitive notion of a local effect that $a_R$ and $a'_{S_1}$ should represent the same effect. This observation leads us to the equivalence relation defined as follows.
\begin{definition}[Equivalent local effects]
We say that the effects $a_R$ and $a'_S$ in $\set{Pre}\Cntset{\rA_G}_{L\Reals}$ are equivalent, and denote this as $a_R\sim a'_S$, if there exists $a_0\in\Cntset{\rA_{R\cap S}}_\Reals$ such that the following identities hold
\begin{align}
\left\{
\begin{aligned}
&a=a_0\otimes e_{S\setminus R},\\
&a'=a_0\otimes e_{R\setminus S}.
\end{aligned}
\right.\label{eq:equiveff}
\end{align}
\end{definition}

It is clear that the real notion of a local effect is 
captured by the equivalence classes modulo the above 
equivalence relation.
We thus quotient $\set{Pre}\Cntset{\rA_G}_{\Reals}$ and define the obtained set as the set of local effects of $\rA_G$.

\begin{definition}
A {\em generalised local effect} is an equivalence class $[a_R]_\sim$.
The {\em set of generalised local effects of} $\rA_G$ is
\begin{align*}
\Cntset{\rA_G}_{L\Reals}\coloneqq \set{Pre}\Cntset{\rA_G}_{L\Reals}/\sim.
\end{align*} 
\end{definition}
With a slight abuse of notation, in the following we will write $a_R$ instead of $[a_R]_\sim$, unless the context requires explicit distinction of the two symbols.
One can easily prove the following result

\begin{lemma}\label{lem:extef}
Let $a_R\in\set{Pre}\Cntset{\rA_G}_{L\Reals}$. Then for every finite region $H\in\reg G $ such that $H\cap R=\emptyset$, one has $(a\otimes e_{H})_{R\cup H}\in\set{Pre}\Cntset{\rA_G}_{L\Reals}$, and $(a\otimes e_{H})_{R\cup H}\sim a_R$.
\end{lemma}

The proof of the above lemma is straightforward, and we do not report it here.

Let us now come back to our initial goal, that is to define a local effect as an event that acts non-trivially only on a finite region $R$. Intuition leads again to figure out what is the preferred representative of the class of a local effect: within the equivalence class, it is the element defined on the smallest region. We now provide a formal definition of such a minimal representative, and show that it is well posed.

\begin{definition}
The \emph{minimal representative} of the equivalence class $a_R$, denoted as ${\tilde a}_{R_a}$, is defined through
\begin{align}
&R_a\coloneqq \bigcap_{S\in\set R_{(a,R)}}S,\quad{\tilde a}_{R_a}\sim a_R,
\label{eq:minimalreploceff}
\end{align}
where $\set R_{(a,R)}$ is the set of all those finite regions $S\in\reg G$ for which there exists $b\in\Cntset{\rA_S}_{\Reals}$ such that $b_S\sim a_R$.
\end{definition}

\begin{lemma}\label{lem:exunmineff}
The minimal representative exists and is unique. 
\end{lemma}
\begin{proof} As to existence, we remark that, by Eq.~\eqref{eq:minimalreploceff}, $R_a$ is the set of all $g\in G$ such that for all regions $S\in\set R_{(a,R)}$ one has $g\in S$. Thus, if $h\not\in R_a$, there must exist $S\in\set R_{(a,R)}$ such that $h\not\in S$. This implies that i) by definition there exists $f_S\sim a_R$, and ii) by lemma~\ref{lem:extef} $c_{S\cup h}\sim a_R$, with
\begin{align}
c=f\otimes e_{h}. 
\label{eq:nu0i}
\end{align}
As a consequence, if $c_T\sim a_R$ and $h\in T$, but $h\not\in R_a$, by definition~\eqref{eq:equiveff} $c_T$ must be
of the form of Eq.~\eqref{eq:nu0i}, for some $f\in\Cntset{\rA_{T\setminus h}}_\Reals$. Clearly, $R_a\in \reg G $, since for any $S\in R_{(a,R)}$ one has $R_a\subseteq S$. Now, let $S\in\set R_{(a,R)}$, and $c_S\sim a_R$. One has $S=R_a\cup S'$, where $S'\coloneqq (S\setminus R_a)\in\reg G$, thus $R_a\cap S'=\emptyset$. By Eq.~\eqref{eq:nu0i}, we then have 
\begin{align}
c=b\otimes e_{S'},\quad b\in\Cntset{\rA_{R_a}}_\Reals. 
\label{eq:nu0S}
\end{align}
We now define $\tilde a_{R_a}\coloneqq b_{R_a}$, and finally, since Eq.~\ref{eq:nu0S} holds for any $c_S\sim a_R$, one can easily verify that $\tilde a_{R_a}\sim a_R$.

As to uniqueness, we remark that the region $R_a$ is uniquely defined. Now, suppose that there were two different $b,c\in\Cntset{\rA_{R_a}}_\Reals$ such that $b_{R_a},c_{R_a}\sim a_R$. Then by Eq.~\eqref{eq:equiveff} it must be $b=c$.
\end{proof}

We now make local effects into a vector space.

\begin{definition}\label{def:equiveff}
Let $a_R,b_S\in\Cntset{\rA_G}_{L\Reals}$, and $h\in\Reals$. Then we define
\begin{align*}
&h a_R\coloneqq \left\{
\begin{aligned}
&(ha)_R&h\neq0,\\
&0_\emptyset&h=0,
\end{aligned}
\right.
\\
&a_R+b_S\coloneqq c_{R\cup S}\\
&c\coloneqq a\otimes e_{S\setminus R}+b\otimes e_{R\setminus S}.
\end{align*}
\end{definition}

Notice that it is not always true that $R_{c}=R_a\cup R_b$. As an example, consider $a=f_{g_1}\otimes f_{g_2}$ and $b=f_{g_1}\otimes(e-f_{g_2})$, with $f_{g_1}\neq e_{g_1}$, $0\neq f_{g_2}\neq e_{g_2}$, and $R_a=R_b=\{g_1,g_2\}$. Then $c= a+b=f_{g_1}\otimes e_{g_2}$, and clearly $R_{c}=\{g_1\}$, which is strictly included in $R_a=R_b=R_a\cup R_b$.

It is easy to check that $\Cntset{\rA_G}_{L\Reals}$ is a real vector space with null element given by the equivalence class of $0_\emptyset$.


We now equip the real vector space of local effects with a norm, and we then close it to obtain the Banach space of {\em quasi-local} effects. The definition is based on a norm for effects of finite systems, as every local effect $a_R\in\Cntset{\rA_G}_{L\Reals}$ reduces to the effect $\tilde a\in\Cntset{\rA_{R_a}}_\Reals$. A natural norm one might think of is then the {\em operational norm}, that induces the following definition.

\begin{definition}\label{def:opnormeff}
The \emph{operational norm} $\normop{a_R}$ of $a_R\in\Cntset{\rA_G}_{L\Reals}$ is defined by the following expression
\begin{align}
&\normop{a_R}\coloneqq \normop{\tilde a},
\end{align}
where $\tilde a\in\Cntset{\rA_{R_a}}_\Reals$.
\end{definition}

Unfortunately, if one completes the space of local effects in the operational norm, in general the dual norm on the space of bounded linear functionals---which will be our state space---does not coincide with the operational norm on the state space, for those states that can be interpreted as quasi-local preparations. We will then choose a different norm on our space of effects. The new norm will be referred to as {\em sup-norm}, as it is the extension of the sup-norm to the infinite case. 

As far as finite-dimensional systems are concerned, this choice does not represent a problem, as all norms are equivalent in finite-dimensional vector spaces. However, the sup-norm is stronger than the operational one---see corollary~\ref{cor:supstrongerop}---, and thus the space that we construct, completing our normed vector space of local effects with sup-norm Cauchy sequences, might contain distinct limits that are operationally equivalent. This point is a delicate one, and to avoid an unreasonable construction where there exist different effects that are operationally equivalent, we impose a sufficient constraint for the operational norm and the sup-norm to be equivalent also for infinite systems: we restrict attention to those theories where the following property holds: there exists a finite constant $k$ such that for every system $\rA$ and every $a\in\Cntset{\rA}_\Reals$
\begin{align}
\normsup{a}\leq k\normop a.
\label{eq:equivnor}
\end{align}
This implies that not only the bound~\eqref{eq:equivnor} holds for a fixed system $\rA$, but it holds with a fixed constant independent of the system $\rA$.
In turn, a sufficient condition for~\eqref{eq:equivnor} is that for every system $\rA$
\begin{align}
\Stset{\rA}_+^*\equiv\Cntset{\rA}_+,
\label{eq:assfullcone}
\end{align}
namely every positive functional on the cone of states is proportional to an effect by a positive constant, and viceversa. In all the presently known theories the latter condition is satisfied. The two above conditions in Eqs.~\eqref{eq:equivnor} and~\eqref{eq:assfullcone} are discussed in detail in Appendix~\ref{app:equinor}, where we prove that condition~\eqref{eq:assfullcone} implies condition~\eqref{eq:equivnor}.

Our new norm is an order-unit norm. As we will discuss in subsection~\ref{sub:extst}, its dual coincides with the operational norm on quasi-local states. 



Let us now see the definition of the sup-norm in detail. 
\begin{definition}\label{def:normsup}
The \emph{sup-norm} $\normsup{a_R}$ of $a_R\in\Cntset{\rA_G}_{L\Reals}$ is defined by the following expression
\begin{align}
&\normsup{a_R}\coloneqq \normsup{\tilde a},
\end{align}
where $\tilde a\in\Cntset{\rA_{R_a}}_\Reals$.
\end{definition}

We now want to prove that the operational norm and the sup-norm are well-defined norms on $\Cntset{\rA_G}_{L\Reals}$. 

\begin{proposition}\label{prop:eqnormop}
Let $a_R\sim\tilde a_{R_a}$. Then $\normop{a_R}=\normop{a}$, and $\normsup{a_R}=\normsup{a}$.
\end{proposition}
\begin{proof}
Let $a_R\sim\tilde a_{R_a}$, and $R=R_a\cup R'$, with $R_a\cap R'=\emptyset$. Then by Eq.~\eqref{eq:equiveff} we have
\begin{align*}
a=\tilde a\otimes e_{{R'}},
\end{align*}
and thus by definitions~\ref{def:opnormeff} and~\ref{def:normsup} and lemma~\ref{lem:normotimese}, it is
\begin{align*}
&\normop{a_R}=\normop{\tilde a}=\normop{\tilde a\otimes e_{R'}}=\normop a,\\
&\normsup{a_R}=\normsup{\tilde a}=\normsup{\tilde a\otimes e_{R'}}=\normsup a.&&\qedhere
\end{align*}
\end{proof}

We can then prove the desired result.
\begin{proposition}
The functionals $\normop\cdot$ and $\normsup\cdot$ are norms on $\Cntset{\rA_G}_{L\Reals}$.
\end{proposition}
\begin{proof}
1. For every $a_R\in\Cntset{\rA_G}_{L\Reals}$ it is clear that $\normop{a_R}\geq0$ and $\normsup{a_R}\geq 0$. Now, $\normop{\tilde a}=\normsup{\tilde a}=0$ if and only if $\tilde a=0$, namely, reminding definition~\ref{def:equiveff}, $a_R=0_\emptyset$.
2. For every $a_R\in\Cntset{\rA_G}_{L\Reals}$, by definitions~\ref{def:supnormeff}, \ref{def:equiveff}, and~\ref{def:opnormeff}, one straightforwardly has that, for every $\mu\in\Reals$,
\begin{align*}
&\normop{\mu a_R }=\normop{\mu\tilde a}=|\mu|\normop{\tilde a}=|\mu|\normop{a_R},\\
&\normsup{\mu a_R}=\normsup{\mu\tilde a}=|\mu|\normsup{\tilde a}=|\mu|\normsup{a_R}.
\end{align*}
3. Let now $c_{R\cup S}=a\otimes e_{S\setminus R}+e_{R\setminus S}\otimes b$ for $a\in\Cntset{\rA_R}_\Reals$ and $b\in\Cntset{\rA_S}_\Reals$, as from definition~\ref{def:equiveff}. Then, by lemma~\ref{lem:normotimese}, proposition~\ref{prop:eqnormop} and by the triangle inequality, we have
\begin{align*}
&\normop{a_R+b_S}=\normop{c_{R\cup S}}=\normop c\\
&\quad\leq\normop a+\normop b=\normop{a_R}+\normop{b_S},\\
&\normsup{a_R+b_S}=\normsup{c_{R\cup S}}=\normsup c\\
&\quad\leq\normsup a+\normsup b=\normsup{a_R}+\normsup{b_S}.&&\qedhere
\end{align*}
\end{proof}

We remark that, in a theory that satisfies assumption~\eqref{eq:assfullcone}, one has $\normop \cdot\equiv\normsup \cdot$ on $\Cntset{\rA}_\Reals$, and thus $\normop\cdot=\normsup\cdot$ on $\Cntset{\rA_G}_\Reals$ (the proof can be found in appendix~\ref{app:equinor}). In general, however, the sup-norm is stronger than the operational norm, as we now prove.
\begin{lemma}\label{lem:supnstrongopn}
Let $a_R\in\Cntset{\rA_G}_{L\Reals}$. Then $\normop{a_R}\leq\normsup{a_R}$.
\end{lemma}
\begin{proof}
Let us consider $\mu\in J(a)$. Then for every $\rho\in\Stset{\rA_R}_1$ we have
\begin{align*}
(\mu e_{R}\pm a|\rho)\geq0,
\end{align*}
i.e.~$|(a|\rho)|\leq\mu$. Then, taking the supremum over states on l.h.s.~we obtain $\normop{a_R}\leq\mu$, and finally, taking the infimum over $J(a)$ we obtain the thesis.
\end{proof}

The space of local effects that we constructed so far is a normed real vector space. We now make it into a Banach space, the space of quasi-local effects, by the usual completion procedure for normed spaces. We first introduce the space $\Cntset{\rA_G}_{C\Reals}$ of Cauchy sequences 
\begin{align*}
a:\Nats\to\Cntset{\rA_G}_{L\Reals}::n\mapsto {a_n}_{R_n}. 
\end{align*}
We then define the equivalence relation between Cauchy sequences in $\Cntset{\rA_G}_{C\Reals}$ defined by $a\cong b$ iff $\lim_{n\to\infty}\normsup{{a_n}_{R_n}-{b_n}_{S_n}}=0$. Finally, we define the space $\Cntset{\rA_G}_{Q\Reals}$ of {\em generalised quasi-local effects}, by taking the quotient 
\begin{align}
\Cntset{\rA_G}_{Q\Reals}\coloneqq \Cntset{\rA_G}_{C\Reals}/\cong.
\end{align}
The elements of this space will be denoted by $a=[{a_n}_{R_n}]$. In this context, the class of a constant Cauchy sequence ($R_n=R$ and ${a_n}_{R_n}=a_R$) will be denoted by $a_R\coloneqq [{a_n}_{R_n}]$. Clearly, since $|\normsup{{a_n}_{R_n}}-\normsup{{a_m}_{R_m}}|\leq\normsup{{a_n}_{R_n}-{a_m}_{R_m}}$, for a Cauchy sequence $a=[{a_n}_{R_n}]$ also $\normsup{{a_n}_{R_n}}$ is a Cauchy sequence, and we define $\normsup a\coloneqq \lim_{n\to\infty}\normsup{{a_n}_{R_n}}$. One can easily verify that the sup-norm on $\Cntset{\rA_G}_{Q\Reals}$ thus defined is independent of the specific sequence within an equivalence class.
The space $\Cntset{\rA_G}_{Q\Reals}$ is by construction a real Banach space, and $\Cntset{\rA_G}_{L\Reals}$ can be identified with the dense submanifold containing constant sequences $a_R=[a_R]$. Clearly, in this case $\normsup{a}=\normsup{a_R}$. 
\begin{definition}
Let $a\in\Cntset{\rA_G}_{Q\Reals}$. If there is a Cauchy sequence ${a_n}_{R_n}$ in the class defining $a$, such that, for some $n_0\in\Nats$, for every $n\geq n_0$ one has $\tilde a_n\in\Cntset{\rA_{R_n}}$, then we call $a$ a {\em quasi-local effect}. We denote the set of quasi-local effects by $\Cntset{\rA_G}_Q$. 

The cone $\Cntset{\rA_G}_{Q+}$ contains all elements that are proportional to an element in $\Cntset{\rA_G}_Q$ by a positive constant.

The {\em deterministic quasi-local effect} $e_G$ is the class $e_G\coloneqq 1_\emptyset$. 
\end{definition}

The cone $\Cntset{\rA_G}_{Q+}$ introduces a partial ordering in $\Cntset{\rA_G}_{Q\Reals}$, that we denote by
\begin{align*}
a\succeq b\ \Leftrightarrow\ a-b\in\Cntset{\rA_G}_{Q+}.
\end{align*}
Quasi-local effects can be interpreted as effects that can be arbitrarily well approximated by local observation procedures. The effect $e_G$ plays an important role, as it is the unique deterministic effect in $\Cntset{\rA_G}_Q$. This will be proved shortly.
Let us start providing the first important property of $e_G$.


\begin{lemma}\label{lem:egdomin}
Let $a\in\Cntset{\rA_G}_Q$. Then also $e_G-a\in\Cntset{\rA_G}_Q$.
\end{lemma}

\begin{proof}
Let 
$a=[{a_n}_{R_n}]$. Let ${a'_n}_{R_n}\coloneqq (e-a_n)_{R_n}\in\Cntset{\rA_{R_n}}$. Then one has $(a'_n-a'_m)_{R_n\cup R_m}=(a_m-a_n)_{R_m\cup R_n}$, and thus ${a'_n}_{R_n}$ is a Cauchy sequence whose class we call $a'\in\Cntset{\rA_G}_Q$. Finally, since $a+a'$ is the class $[{a_n}_{R_n}]+[{a'_n}_{R_n}]=[(a_n+a'_n)_{R_n}]=[1_\emptyset]=e_G$, we have that $a+a'=e_G$, namely $e_G-a=a'\in\Cntset{\rA_G}_Q$.
\end{proof}

By the above lemma we know that not only $e_G\succeq a$ for every quasi-local effect $a$, but also that every quasi-local effect $a$ can be complemented by a quasi-local effect $a'$ such that $a+a'=e_G$. In other words $(a,e_G-a)$ is a binary quasi-local observation test.

We can also prove the converse result.

\begin{lemma}\label{lem:cnessef}
Let $a\in\Cntset{\rA_G}_{Q+}$, and $e_G-a\in\Cntset{\rA_G}_{Q+}$. Then $a,e_G-a\in\Cntset{\rA_G}_Q$.
\end{lemma}
\begin{proof}
Indeed, the statement is true for finite systems thanks to the no-restriction hypothesis (see theorem~\ref{th:norest}), and thus it holds for a given Cauchy sequence in the class of $a$, namely if ${a'_n}_{R_n}\coloneqq (e-a_n)_{R_n}\in\Cntset{\rA_{R_n}}_+$ for all $n$, then $a_n,a'_n\in\Cntset{\rA_{R_n}}$. Thus, $a=\lim_{n\to\infty}a_n\in\Cntset{\rA_G}_{Q}$, and $a'=\lim_{n\to\infty}a'_n=e_G-a\in\Cntset{\rA_G}_{Q}$.
\end{proof}

We now prove that $e_G$ lies in the interior of $\Cntset{\rA_G}_{Q+}$, namely there is a ball of radius $r$ around $e_G$ in sup-norm that is contained in $\Cntset{\rA_G}_{Q+}$.

\begin{lemma}\label{lem:balleg}
There exists an open ball $B_r(e_G)$ of radius $r$ in $\Cntset{\rA_G}_{Q\Reals}$ that is fully contained in $\Cntset{\rA_G}_{Q+}$.
\end{lemma}
\begin{proof}
Let $\normsup{a-e_G}<r<1/2$. Then, by definition, there exist a sequence ${a_n}_{R_n}$ such that $\lim_{n\to\infty}a_n=a$, and $n_0\in\Nats$ such that for $n\geq n_0$
\begin{align*}
\normsup{a_n-e_G}\leq\normsup{a_n-a}+\normsup{a-e_G}<2r.
\end{align*}
This implies that 
\begin{align*}
[2r e+ (a_n-e)]_{R_n}=[a_n-(1-2r)e]_{R_n}\succeq0. 
\end{align*}
Now, this implies that for $n\geq n_0$ it is $a_n\succeq(1-2r)e_{R_n}\succeq0$, and thus $a_n\in\Cntset{\rA_G}_{L+}$. Finally, by definition, one obtains $a\in\Cntset{\rA_G}_{Q+}$. Thus, the ball $B_{r}(e_G)$ is contained in $\Cntset{\rA_G}_{Q+}$.
\end{proof}
\begin{corollary}\label{cor:balltrans}
Let $a\in\Cntset{\rA_G}_{Q+}$. Then for $\varepsilon>0$, $B_{\varepsilon r}(\varepsilon e_G+a)\subseteq\Cntset{\rA_G}_{Q+}$ for some $r>0$.
\end{corollary}
\begin{proof}
Let $b\in B_{\varepsilon r}(\varepsilon e_G+a)$. Then
\begin{align*}
&\normsup{b-a-\varepsilon e_G}<\varepsilon r\\
&\Leftrightarrow\ \normsup{(b-a)/\varepsilon-e_G}<r.
\end{align*}
Then $(b-a)/\varepsilon=c\in\Cntset{\rA_G}_{Q+}$, ad thus
\begin{align*}
b=\varepsilon c+a\in\Cntset{\rA_G}_{Q+}.&\qedhere
\end{align*}
\end{proof}

We now prove that the deterministic effect $e_G$ allows for an extension of the defining property of the sup-norm.
\begin{lemma}\label{lem:normsupinfj}
The sup-norm of $a\in\Cntset{\rA_G}_{Q\Reals}$ can be expressed as
\begin{align*}
&\normsup a=\inf J(a),\\
&J(a)\coloneqq \{\mu\geq0\mid \mu e_G\succeq a\succeq-\mu e_G\}.
\end{align*}
\end{lemma}

\begin{proof}
In the case of local effects $a=a_R$ the statement is a trivial recasting of the definition of sup-norm. Let us then consider the class $a\in\Cntset{\rA_G}_{Q\Reals}$ with $a=[{a_n}_{R_n}]$. One has, by definition, $\normsup a=\lim_{n\to\infty}\normsup{a_n}$. Let us consider the sequence $\mu_n\coloneqq \normsup{a_n}+\varepsilon\in J(a_n)$. Clearly, $\lim_{n\to\infty}\mu_n=\normsup a+\varepsilon$. Moreover, since
\begin{align*}
&\normsup{\{(\mu_n-\mu_m)e\pm (a_n-a_m)\}_{R_n\cup R_m}}\\
&\leq|\normsup{a_m}-\normsup{a_n}|+\normsup{a_m-a_n}\leq2\varepsilon,
\end{align*}
the sequence $\{(\mu_n+\varepsilon) e\pm a_n\}_{R_n}$ converges to $(\normsup{a}+\varepsilon)e_G\pm a\in\Cntset{\rA_G}_{Q+}$, thus for every $\varepsilon\geq0$ one has $\normsup a+\varepsilon\in J(a)$. This implies that 
\begin{align*}
\inf J(a)\leq\normsup a.
\end{align*}
On the other hand, let $\mu\in J(a)$, then 
\begin{align*}
\mu e_G\pm a\succeq 0.
\end{align*}
By corollary~\ref{cor:balltrans}, we have that, for $\varepsilon>0$,
\begin{align*}
(\varepsilon+\mu)e_G\pm a\in B_{\varepsilon r}(\varepsilon e_G+\mu e_G\pm a)\subseteq\Cntset{\rA_G}_{Q+}.
\end{align*}
Now, this implies that there are two open balls, centred at $(\mu+\varepsilon)e_G\pm a$, that are fully contained in $\Cntset{\rA_G}_{Q+}$. There exists then $n_0\in\Nats$ such that for $n\geq n_0$ one has
\begin{align*}
(\mu+\varepsilon)e_G\pm a_n\succeq0,
\end{align*}
namely $(\mu+\varepsilon)\in J(a_n)$. Thus, $J(a)\subseteq J(a_n)-\varepsilon$, and
\begin{align}
\inf J(a)\geq\normsup{a_n}-\varepsilon.
\end{align}
Finally, this implies that $\inf J(a)\geq\normsup a$.
\end{proof}
As the cone $\Cntset{\rA_G}_{Q+}$ is closed, we can easily prove that the infimum in the expression of the sup norm is actually a minimum.
\begin{lemma}\label{lem:strictsup}
Let $a\in\Cntset{\rA_G}_{Q\Reals}$. Then
\begin{align}
\normsup a e_G\pm a\succeq0.
\end{align}
\end{lemma}
\begin{proof}
By lemma~\ref{lem:normsupinfj}, for every $\varepsilon>0$ one has
\begin{align*}
(\normsup a+\varepsilon)e_G\pm a\succeq0.
\end{align*}
The sequences $b^\pm_n\coloneqq (\normsup a+1/n)e_G\pm a$ are both Cauchy, and their limits are both in $\Cntset{\rA_G}_{Q+}$, then
\begin{align*}
&\normsup a e_G\pm a\succeq0.&&\qedhere
\end{align*}
\end{proof}

Before concluding, we remark that the operational norm can be extended to $\Cntset{\rA_G}_{Q\Reals}$, by simply defining for $a=[{a_n}_{R_n}]$
\begin{align*}
\normop{a}\coloneqq \lim_{n\to\infty}\normop{{a_n}_{R_n}}.
\end{align*}
Indeed, since
\begin{align*}
|\normop{{a_n}_{R_n}}-\normop{{a_m}_{R_m}}|&\leq\normop{{a_n}_{R_n}-{a_m}_{R_m}}\\
&\leq\normsup{{a_n}_{R_n}-{a_m}_{R_m}},
\end{align*}
the sequence $\normop{{a_n}_{R_n}}$ is Cauchy, and the limit is well defined. Moreover, also on $\Cntset{\rA_G}_{Q\Reals}$ the sup-norm is stronger than the operational norm, as can be straightforwardly concluded from lemma~\ref{lem:supnstrongopn}. However, in theories where there exists $k>0$ such that $\normsup\cdot\leq k\normop\cdot$ in $\Cntset{\rA}_\Reals$ for every finite system $\rA$, the same inequality holds in the limit, and the operational and sup-norm are equivalent. In particular, this is the case under assumption~\eqref{eq:assfullcone}.

We now introduce a diagrammatic notation for effects in $\Cntset{\rA}_{Q\Reals}$ that will make part of the subsequent proofs and arguments more intuitive. First of all, we denote a quasi-local effect $a\in\Cntset{\rA_G}_{Q\Reals}$ by the symbol
\begin{align}
\begin{aligned}
\Qcircuit @C=1em @R=.7em @! R {&\ustick{G}\qw&\measureD{a}}
\end{aligned}\ .
\end{align}
In the case of a local effect $a_R\in\Cntset{\rA_G}_{L\Reals}$, we will draw
\begin{align}
\begin{aligned}
\Qcircuit @C=1em @R=.7em @! R {&\ustick{G}\qw&\measureD{a_R}}
\end{aligned}\ =\     
&\begin{aligned}
\Qcircuit @C=1em @R=.7em @! R {&\ustick{R}\qw&\measureD{a}\\
&\ustick{G\setminus R}\qw&\measureD{e}}
\end{aligned}\ .
\label{eq:atime}
\end{align}
Notice that, since for $(a\otimes e_{R_1})_{R}$ with $a\in\Cntset{\rA_{R_0}}$ and $R_1\coloneqq R\setminus R_0$, it is $(a\otimes e_{R_1})_{R}\sim a_{R_0}$, we have
\begin{align}
&\begin{aligned}
\Qcircuit @C=1em @R=.7em @! R {&\ustick{R_0}\qw&\measureD{a}\\
&\ustick{R_1}\qw&\measureD{e}\\
&\ustick{G\setminus R}\qw&\measureD{e}}
\end{aligned}\ =\  
\begin{aligned}
\Qcircuit @C=1em @R=.7em @! R {&\qw&\ustick{R_0}\qw&\qw&\measureD{a}\\
&\qw&\ustick{(G\setminus R)\cup R_1}\qw&\qw&\measureD{e}}
\end{aligned}\ .
\end{align}
From the above identity, we can intuitively conclude that
\begin{align}
&\begin{aligned}
\Qcircuit @C=1em @R=.7em @! R {
&\qw&\ustick{(G\setminus R)\cup R_1}\qw&\qw&\measureD{e}}
\end{aligned}\ =\  
\begin{aligned}
\Qcircuit @C=1em @R=.7em @! R {&\ustick{R_1}\qw&\measureD{e}\\
&\ustick{G\setminus R}\qw&\measureD{e}}
\end{aligned}\ .
\label{eq:detdet}
\end{align}
Indeed, let $G=R\cup H$, with $|R|<\infty$. Then
$[e_R]=[1_{\emptyset}]=e_G$, which is the equation represented by the diagram in Eq.~\ref{eq:detdet}. Similarly, let $a_R\otimes b_H\coloneqq \lim_{n\to\infty}(a\otimes b_n)$, where $[b_{nS_n}]=b\in\Cntset{\rA_H}_{Q\Reals}$. By corollary~\ref{cor:productsupnorm}, one has $[(a\otimes b_n)_{R\cup S_n}]=(a_R\otimes b_H)$. Thus, in $G=R\cup H$, one has $a_R=[a_R]=[(a_R\otimes e_H)]=a_R\otimes e_H$, which is the meaning of Eq.~\eqref{eq:atime}.

As a final remark, we observe that for every denumerable set $G$, and every subset $R\subseteq G$, including infinite ones, one can define $\Cntset{\rA^{(G)}_R}_{Q\Reals}$ as the closed subspace spanned by those quasi-local effects $a={a_n}_{R_n}$ such that, for every $n\in\Nats$,  $R_n\subseteq R$. If one constructs the system $\rA_R\coloneqq \bigotimes_{g\in R}\rA_g$, it is straightforward to construct an ordered Banach space isomorphism
\begin{align}
\tJ_R^\dag:\Cntset{\rA^{(G)}_R}_{Q\Reals}\to\Cntset{\rA_R}_{Q\Reals}::[{a_n}_{R_n}]_G\mapsto [{a_n}_{R_n}]_R,
\label{eq:isomj}
\end{align}
with the norm coinciding with the sup-norm in both spaces.
The left-inverse of $\tJ_R^\dag$ is
\begin{align}
\tJ_R^{-1\dag}:\Cntset{\rA_R}_{Q\Reals}\to\Cntset{\rA^{(G)}_R}_{Q\Reals}.
\end{align}
$\tJ_R^\dag$ and $\tJ_R^{-1\dag}$ can be diagrammatically denoted as
\begin{align}
&\begin{aligned}
\Qcircuit @C=1em @R=.7em @! R {&\ustick{R}\qw&\multigate{1}{\tJ_R^\dag}&\ustick{R}\qw&\measureD{a}\\
&&\pureghost{\tJ_R^\dag}&\ustick{G\setminus R}\qw&\measureD{e}}
\end{aligned}\ 
\ =\     
\begin{aligned}
\Qcircuit @C=1em @R=.7em @! R {&\ustick{R}\qw&\measureD{a}}
\end{aligned}\ ,\nonumber\\
\\
&
\begin{aligned}
\Qcircuit @C=1em @R=.7em @! R {&\ustick{G}\qw&\gate{\tJ_R^{-1\dag}}&\ustick{R}\qw&\measureD{a}}
\end{aligned}\ 
\ =\     
\begin{aligned}
\Qcircuit @C=1em @R=.7em @! R {&\ustick{R}\qw&\measureD{a}\\
    &\ustick{G\setminus R}\qw&\measureD{e}}
\end{aligned}\ .
\label{eq:restrinv}
\end{align}

\subsection{Extended states}\label{sub:extst}

Now that we defined the space of quasi-local effects, we can define the space of states as the space of bounded linear functionals on $\Cntset{\rA_G}_{Q\Reals}$. 
The set of states is then defined considering those linear functionals that, acting on local effects of an arbitrary finite region $R$, behave as a state in $\Stset{\rA_R}$.
\begin{definition}[Generalised extended states]
The space $\Stset{\rA_G}_\Reals$ of {\em generalised extended states} of $\rA_G$ is the topological dual of $\Cntset{\rA_G}_{Q\Reals}$, i.e.~the Banach space $\Cntset{\rA_G}_{Q\Reals}^*$ of bounded linear functionals on $\Cntset{\rA_G}_{Q\Reals}$, equipped with the norm
\begin{align}
\normst\rho\coloneqq \sup_{\normsup a=1}|(a|\rho)|.
\label{eq:defnorst}
\end{align}
\end{definition}

The operational norm on $\Stset{\rA_G}_\Reals$, defined as
\begin{align}
\normop\rho\coloneqq \sup_{a\in\Cntset{\rA_G}_Q}(2a-e_G|\rho),
\end{align}
coincides with the norm $\normst\cdot$, as we now prove.
\begin{lemma}\label{lem:eqopnormstnorm}
Let $\rho\in\Stset{\rA_G}_\Reals$. Then
\begin{align}
\normst\rho=\sup_{a\in\Cntset{\rA_G}_Q}(2a-e_G|\rho).
\end{align}
\end{lemma}

\begin{proof}
Let $0\leq\varepsilon\leq\normop\rho$.
We then have
\begin{align*}
0\leq\normop\rho-\varepsilon\leq(2a-e_G|\rho),
\end{align*}
for some $a\in\Cntset{\rA_G}_Q$. Invoking lemma \ref{lem:egdomin}, one has
\begin{align*}
&e_G+(2a-e_G)=2a\succeq0,\\
&e_G-(2a-e_G)=2(e_G-a)\succeq0,
\end{align*}
and, by lemma \ref{lem:normsupinfj}, $\normsup{2a-e_G}\leq1$. Then it is
\begin{align*}
\normop\rho-\varepsilon\leq\frac{(2a-e_G|\rho)}{\normsup{2a-e_G}}\leq\normst\rho.
\end{align*}
On the other hand, let us now pick $a\in\Cntset{\rA_G}_\Reals$ with $\normsup a=1$. Then, by lemma~\ref{lem:strictsup}, $e_G\pm a\succeq0$. If we define $a_\pm\coloneqq \frac12(e_G\pm a)$, we have $a_\pm\succeq0$, and
\begin{align}
a_++a_-=e_G,\quad (a_+-a_-)=a,
\end{align}
which by lemma~\ref{lem:cnessef} implies that $a_\pm\in\Cntset{\rA_G}_{Q}$. Then
\begin{align*}
\pm(a|\rho)&=(2a_\pm-e_G|\rho).
\end{align*} 
Thus, $|(a|\rho)|\leq\normop\rho$, for every $\varepsilon>0$, and taking the supremum on the l.h.s.~ we obtain $\normst\rho\leq\normop\rho$.
\end{proof}
Technically speaking, the two norms $\normsup\cdot$ and $\normst\cdot$ are a base and order-unit norm pair \cite{Nagel1974}.

What is missing now is the notion of a convex set of proper states, the {\em extended preparations} of our infinite system $\rA_G$. Let us then give the following definition.

\begin{definition}[Local restriction]\label{def:exlocst}
Given a state $\rho\in\Stset{\rA_G}_\Reals$, the local restriction of $\rho$ to $S\in\reg G $ is the functional $\rho_{\vert S}\in\Stset{\rA_S}_\Reals$ defined through
\begin{align}
(a|\rho_{\vert S})\coloneqq (a_S|\rho),\quad\forall a\in\Cntset{\rA_S}.
\label{eq:exlocst}
\end{align}
\end{definition}

The notion of a restriction can be brought further, considering infinite regions, by invoking the isomorphism $\tJ_R$ defined in Eq.~\eqref{eq:isomj}, as follows.

\begin{definition}[Restriction]\label{def:restr}
Given a state $\rho\in\Stset{\rA_G}_\Reals$, the restriction of $\rho$ to $S\in\infreg G $ is the functional $\rho_{\vert S}\in\Stset{\rA_S}_\Reals$ defined through
\begin{align}
(a|\rho_{\vert S})\coloneqq (\tJ^{-1\dag}_Sa|\rho),\quad\forall a\in\Cntset{\rA_S}.
\label{eq:restr}
\end{align}
\end{definition}
Defining $\hat\tJ_S^{-1}$ by duality as
\begin{align}
(a|\hat\tJ_S^{-1}\rho)\coloneqq (\tJ^{-1\dag}_Sa|\rho),
\end{align}
we can then equivalently express Eq.~\eqref{eq:restr} as
\begin{align}
\rho_{\vert S}=\hat\tJ^{-1}_S\rho.
\end{align}
Representing generalised extended states through diagrams, and reminding eq.~\eqref{eq:restrinv}, the above equations can be recast as
\begin{align*}
\begin{aligned}
    \Qcircuit @C=1em @R=.7em @! R {\prepareC{\rho_{\vert S}}&\ustick{{S}}\qw&\measureD{a}}
\end{aligned}\ &=\ 
\begin{aligned}
    \Qcircuit @C=1em @R=.7em @! R {\prepareC{\rho}&\ustick{R}\qw&\gate{\tJ_S^{-1}}&\ustick{S}\qw&\measureD{a}}
\end{aligned}\\
\\
&=\ 
\begin{aligned}
    \Qcircuit @C=1em @R=.7em @! R {\multiprepareC{1}{\rho}&\qw&\ustick{{S}}\qw&\measureD{a}&\\
    \pureghost{\rho}&\qw&\ustick{{G\setminus S}}\qw&\measureD{e}}
\end{aligned}\ ,
\end{align*}
which can be taken as the definition of the equation
\begin{align}
\begin{aligned}
    \Qcircuit @C=1em @R=.7em @! R {\prepareC{\rho_{\vert S}}&\ustick{{S}}\qw&\qw}
\end{aligned}\coloneqq 
\begin{aligned}
    \Qcircuit @C=1em @R=.7em @! R {\multiprepareC{1}{\rho}&\qw&\ustick{{S}}\qw&\qw&\\
    \pureghost{\rho}&\qw&\ustick{{G\setminus S}}\qw&\qw&\measureD{e}}
\end{aligned}\ .
\end{align}

We can now introduce the set of states $\Stset{\rA_G}$ as the special set of generalised extended states whose restrictions are all states. In other words, an element $\rho$ of the space $\Stset{\rA_G}_\Reals$ is a state if, restricted to any finite region $R\in\reg G $, it defines a state for that region.

\begin{definition}[Extended states]
The set $\Stset{\rA_G}$ of states in the space of generalised extended states $\Stset{\rA_G}_{\Reals}$ is the set of those elements $\rho\in\Stset{\rA_G}_\Reals$ such that for every $R\in\reg G $ the local restriction of $\rho$ to $R$ is a state $\rho_{\vert R}\in\Stset{\rA_R}$. Deterministic states, whose set is denoted by $\Stset{\rA_G}_1$, are those states $\rho\in\Stset{\rA_G}$ with $(e_G|\rho)=1$.
\end{definition}

One can easily verify that the set $\Stset{\rA_G}$ is convex, as a consequence of convexity of $\Stset{\rA}$ for every finite system $\rA$. As we did for finite systems, we also define a positive cone generated by $\Stset{\rA_G}$. Moreover, it is easy to check that the restriciton of a state to an infinite region $S\in\infreg G$ is s state in $\Stset{\rA_S}$.

\begin{definition}[Extended positive cone]
The set $\Stset{\rA_G}_+$ in the space of generalised extended states $\Stset{\rA_G}_{\Reals}$ is the cone of those elements $\rho\in\Stset{\rA_G}_\Reals$ such that there exist $\bar\rho\in\Stset{\rA_G}$ and $\mu\geq0$ such that $\rho=\mu\bar\rho$.
\end{definition}

We can now show an important result about the restriction of states.
\begin{lemma}\label{lem:restst}
The restriction $\hat\tJ_R^{-1}\Stset{\rA_G}$ of the set of states of $\rA_G$ coincides with $\Stset{\rA_R}$.
\end{lemma}
\begin{proof}
It is a straightforward exercise to verify that $\hat\tJ_R^{-1}\Stset{\rA_G}\subseteq\Stset{\rA_R}$. For the converse, let $\rho\in\Stset{\rA_R}$, and let us now construct a state $\sigma\in\Stset{\rA_G}$ such that $\sigma_{\vert R}=\rho$. We define $\bar R\coloneqq G\setminus R$. There are two possible situations: $\bar R\in\reg G $, or $\bar R\in\infreg G $. If $\bar R\in\reg G $, let $a_T\in\Cntset{\rA_G}_{L\Reals}$, with $T\in\reg G $. We define
\begin{align}
(a|\sigma)\coloneqq (a|\rho_{\vert T\cap R}\otimes\nu_{\vert T\cap \bar R}),
\end{align}
for an arbitrary $\nu\in\Stset{\rA_{\bar R}}$, where we introduced the notation $\sigma_{\vert S'}$ for $S'\subseteq S\in\reg G $, and $\sigma$ a state of the composite system $\sigma\in\Stset{\rA_S}=\Stset{\rA_{S'}\rA_{S\setminus S'}}$, with
\begin{equation*}
\begin{aligned}
    \Qcircuit @C=1em @R=.7em @! R {\prepareC{\sigma_{\vert S'}}&\ustick{{S'}}\qw&\qw}
\end{aligned}\coloneqq 
\begin{aligned}
    \Qcircuit @C=1em @R=.7em @! R {\multiprepareC{1}{\sigma}&\qw&\ustick{{S'}}\qw&\qw&\\
    \pureghost{\sigma}&\qw&\ustick{{S\setminus S'}}\qw&\qw&\measureD{e}}
\end{aligned}\ .
\end{equation*}
Clearly, the functional $\sigma$ is bounded, and thus it can be extended to a functional on the full space $\Cntset{\rA_G}_{Q\Reals}$, i.e.~$\sigma\in\Stset{\rA_G}_\Reals$. Moreover one can easily verify that $\sigma _{\vert T}\in\Stset{\rA_T}$ for every $T\in\reg G $, thus $\sigma $ is the desired state in $\Stset{\rA_G}$. A similar construction can be carried out for the case $\bar R\in\infreg G $, taking $\nu\in\Stset{\rA_{\bar R}}$. By construction, also in this case $\sigma \in\Stset{\rA_G}$, as can be straightforwardly checked.
\end{proof}

The next result that we prove is that quasi-local effects $a\in\Cntset{\rA_G}_Q$ are positive on the set $\Stset{\rA_G}$. 
\begin{lemma}\label{lem:poseff}
Let $a\in\Cntset{\rA_G}_Q$, and $\rho\in\Stset{\rA_G}$. Then $(a|\rho)\geq0$
\end{lemma}
\begin{proof}
Let $a\in\Cntset{\rA_G}_Q$ be a local effect $a_R=[a_R]$. Then, by definition, $(a_R|\rho)=(a|\rho_{\vert R})\geq0$. Now, let $a=[{a_n}_{R_n}]$. We have $(a|\rho)=\lim_{n\to\infty}({a_n}_{R_n}|\rho)\geq0$.
\end{proof}

The following result draws a first analogy between the properties of the sup-norm for finite systems and that for infinite parallel compositions.

\begin{lemma}\label{lem:normonpos}
Let $\rho\in\Stset{\rA_G}_+$. Then $\normst\rho=(e_G|\rho)$.
\end{lemma}
\begin{proof}
Let $\rho\in\Stset{\rA_G}_+$. Then, by virtue of lemma~\ref{lem:poseff}, $(a|\rho)\geq0$ for all $a\in\Cntset{\rA_G}$. Let now $\normsup a=1$. By lemma~\ref{lem:strictsup}, one has $e_G\pm a\succeq0$, and thus $(e_G|\rho)\geq|(a|\rho)|$. Then $(e_G|\rho)\geq\normst\rho$. On the other hand, being $e_G$ a legitimate element of $\Cntset{\rA_G}_Q$ with $\normsup{e_G}=1$, one has $\normst\rho=\sup_{\normsup a=1}|(a|\rho)|\geq(e_G|\rho)$.
\end{proof}

Thanks to lemma~\ref{lem:cnessef}, we have the following corollary.

\begin{corollary}\label{cor:domeg}
Let $\rho\in\Stset{\rA_G}$. Then $0\leq(a|\rho)\leq(e_G|\rho)$ for every $a\in\Cntset{\rA_G}_Q$.
\end{corollary}

We can now show that also in the infinite case every state is proportional to a deterministic one. Let us first prove a preliminary lemma.
\begin{lemma}\label{lem:nullst}
Let $\rho\in\Stset{\rA_G}_\Reals$. If $(a|\rho)=0$ for every $a\in\Cntset{\rA_G}_Q$, then $\rho=0$.
\end{lemma}
\begin{proof}
If $(a|\rho)=0$ for every $a\in\Cntset{\rA_G}_Q$ then, using lemma~\ref{lem:egdomin}, we have
\begin{align*}
(2a-e_G|\rho)=(a|\rho)-(e_G-a|\rho)=0,\quad\forall a\in\Cntset{\rA_G}_Q.
\end{align*}
This implies that $\normst\rho=0$, and thus $\rho=0$.
\end{proof}

\begin{proposition} 
Every state $\rho\in\Stset{\rA_G}$ is proportional to a deterministic state $\bar\rho\in\Stset{\rA_G}_1$.
\end{proposition}

\begin{proof}
Let $\rho\in\Stset{\rA_G}$. If $\rho=0$, then for every $\bar\rho\in\Stset{\rA_G}_1$ it is $\rho=0\bar\rho$. Let then $\rho\neq0$. By lemmas~\ref{lem:poseff} and~\ref{lem:nullst}, there must exist $a\in\Cntset{\rA_G}$ such that $(a|\rho)>0$. By corollary~\ref{cor:domeg}, one then has $(e_G|\rho)\geq(a|\rho)>0$. By definition, for every $S\in\reg G $, we have
\begin{align*}
(e_{\rA_S}|\rho_{\vert S})=(e_S|\rho)=(e_G|\rho)>0.
\end{align*}
Thus, if we set $\bar\rho\coloneqq \rho/(e_G|\rho)$, for every $S\in\reg G $ we obtain
\begin{align*}
(e_{\rA_S}|\bar\rho_{\vert S})=(e_G|\bar\rho)=1,
\end{align*}
implying that $\bar\rho_{\vert S}\in\Stset{\rA_S}_1$ for every $S\in\reg G $.
Then, $\bar\rho\in\Stset{\rA_G}_1$, and $\bar\rho$ is such that $\rho=(e_G|\rho)\bar\rho$.
\end{proof}

We now show that the operational norm is equivalently defined on $\Cntset{\rA_G}_{Q\Reals}$ by
\begin{align*}
\normop a\coloneqq \sup_{\rho\in\Stset{\rA_G}}|(a|\rho)|.
\end{align*}
Indeed, let $a=[a_{nR_n}]$. Then 
\begin{align*}
|(a|\rho)|=\lim_{n\to\infty}|(a_{nR_n}|\rho)|=\lim_{n\to\infty}|(a_n|\rho_{\vert R_n})|.
\end{align*}
This implies that 
\begin{align*}
|(a|\rho)|\leq|(a_n|\rho_{\vert R_n})|+\varepsilon\leq\normop{a_{nR_n}}+\varepsilon,
\end{align*}
thus $\sup_{\rho\in\Stset{\rA_G}}|(a|\rho)|\leq\normop a$.
On the other hand, for every $a_{nR_n}$ and every $\varepsilon$, there exists $\rho_\varepsilon\in\Stset{\rA_{R_n}}$ such that $\normop{a_{nR_n}}\leq|(a_n|\rho_\varepsilon)|+\varepsilon$. Now, by lemma~\ref{lem:restst}, this implies that
for every $\varepsilon$ there exists $\tilde\rho_\varepsilon\in\Stset{\rA_G}$ such that
\begin{align*}
\normop{a_{nR_n}}-\varepsilon\leq|(a_{nR_n}|\tilde\rho_\varepsilon)|.
\end{align*}
Finally, this implies that for every $\varepsilon$ there exists $\rho_\varepsilon\in\Stset{\rA_G}$ such that
\begin{align}
\normop a-3\varepsilon&\leq\normop{a_{nR_n}}-2\varepsilon\nonumber\\
&\leq|(a_{nR_n}|\tilde\rho_\varepsilon)|-\varepsilon\nonumber\\
&\leq|(a|\tilde\rho_\varepsilon)|.
\end{align}
Thus, in conclusion, $\normop a\leq\sup_{\rho\in\Stset{\rA_G}}|(a|\rho)|$.

A crucial property of $\Stset{\rA_G}$
is that extended states are separating for $\Cntset{\rA_G}_Q$, namely if for every state two generalised effects give the same value, then they coincide.
\begin{theorem}[States separate effects]\label{th:statesepeff}
Let $a\in\Cntset{\rA_G}_{Q\Reals}$. If $(a|\rho)=0$ for all $\rho\in\Stset{\rA_G}_1$, then $a=0$.
\end{theorem}
\begin{proof}
Let $a=[{a_n}_{R_n}]\in\Cntset{\rA_G}_{Q\Reals}$, and $(a|\rho)=0$ for every $\rho\in\Stset{\rA_G}_1$. Then for every $\rho\in\Stset{\rA_G}_1$ one has 
\begin{align*}
|(a_n|\rho)|=|(a_n-a|\rho)|\leq\normsup{a_n-a},
\end{align*}
and thus for every $\varepsilon>0$ there exists $n_0$ such that, for $n\geq n_0$, $\normop{a_n}\leq\varepsilon$. Thus, we have $\normop a=0$. As a consequence, for every $\varepsilon>0$ there exists $n_0$ such that for $n\geq n_0$ one has 
\begin{align*}
\normsup{a_n}\leq k\normop{a_n}\leq\varepsilon,
\end{align*}
and thus $\normsup a=0$. Now, this implies that $a=0$.
\end{proof}
\begin{remark}
The above result is made possible by requirement~\eqref{eq:equivnor} that the sup-norm and the operational norm for effects of finite systems $\rA$ are bounded as
\begin{align*}
\forall a\in\Cntset{\rA}_\Reals\ \normsup a\leq k\normop a,
\end{align*}
with $k$ independent of the system $\rA$. In particular, this is true under the assumption~\eqref{eq:assfullcone}, i.e.~that $\Cntset{\rA}_+^*\equiv\Stset{\rA}_+$.
\end{remark}

We can now prove the main consequence of uniqueness of the deterministic effect $e_G$ for $\rA_G$, i.e.~that $e_G$ is the unique effect that amounts to 1 on $\Stset{\rA_G}_1$.

\begin{proposition}
Let $a\in\Cntset{\rA_G}_Q$. Then $(a|\rho)=1$ for all $\rho\in\Stset{\rA_G}_1$ if and only if $a=e_G$.
\end{proposition}
\begin{proof}
Notice that, since $e_G\succeq a$ for every $a\in\Cntset{\rA_G}_Q$, and $(b|\rho)\geq0$ for every $b\in\Cntset{\rA_G}_Q$ and $\rho\in\Stset{\rA_G}_1$, we have
\begin{align*}
0\leq(e_G-a|\rho)=1-(a|\rho),
\end{align*}
which implies $(a|\rho)\leq1$ for every $\rho\in\Stset{\rA_G}_1$. Now, if $(a|\rho)=1$ for every $\rho\in\Stset{\rA_G}_1$, we have
\begin{align*}
(e_G-a|\rho)=0, \quad\forall\rho\in\Stset{\rA_G}_1,
\end{align*}
which by theorem~\ref{th:statesepeff} implies $a=e_G$.
\end{proof}

A class of states of particular interest is that of quasi-local states, which are intuitively understood as states whose preparation can be arbitrarily approximated by local procedures. These live in small subspaces of the space $\Stset{\rA_G}_\Reals$. Their construction is similar to that of quasi-local effects, but differs form it in a relevant respect, that is the necessity of defining arbitrary reference states. The 
construction is inspired by pioneering works on infinite tensor products by von Neumann and Murray~\cite{vonNeumann:1939vh}, and can be found in appendix~\ref{app:qlst}.

\subsection{Quasi-local transformations}

Now we are going to define the quasi-local algebra $\Trnset{\rA_G\to\rA_G}_{Q\Reals}$ of 
transformations. In usual approaches to quantum infinite systems, one usually introduces
the {\em effect algebra}, that in the OPT case would correspond to the space of 
quasi-local effects. However, one can easily understand that, unlike effects in a general
OPT, the quantum effect algebra contains far more information than the mere structure of
the space of effects. Indeed, in the quantum effect algebra one can find every operator on
the Hilbert space, and in turn, the operational role of a linear operator is to provide a
Kraus representation of a transformation. In this perspective, one can also understand
why multiplication of effects has a meaning at all: there is no point in multiplying 
effects, but multiplication of Kraus operators is the mathematical representation of 
sequential composition. Thus, the algebraic structure of effects conveniently summarises
information about every kind of event in the theory: effects, with their coarse-graining 
represented by the sum, and transformations, with sequential composition represented 
by multiplication. 

Cellular automata are defined in quantum theory by specifying their action on the effect
algebra. Such an action thus provides information about how both effects and 
transformations are transformed by the cellular automaton. For a general OPT, however, 
such a compact algebraic structure embodying all the relevant information is absent, and 
the definition of a CA on the space of effects is not sufficient: we need to specify how
the CA transforms transformations. 

For these reasons we construct now the Banach algebra of quasi-local transformations, 
in a way that is very closely reminiscent of the construction of quasi-local effects. 
Here, in order to define a local transformation, we recur to the notion of a 
transformation that acts non-trivially on finitely many systems, while it acts as the 
identity on the remaining ones. In other words, the role that is played by the 
deterministic effect in the case of local effects is played here by the identity 
transformation. Also in this case, the topological closure will be taken in the sup-norm, 
and this choice is of great relevance to ensure a Banach algebra structure for the 
closure.

Let us then introduce the Banach algebra of quasi-local transformations.

\begin{definition}
Let $R\subseteq G$ be an arbitrary finite region of $G$. We define {\em local transformation} $\tA_R$ of $\rA_G$ any pair $(\tA,R)$, where $\tA\in\Trnset{\rA_R\to\rA_R}_\Reals$.
The action $\tA^\dag_R$ of $\tA_R$ on $\set{Pre}\Cntset{\rA_G}_{L\Reals}$ is defined as follows 
\begin{align}
\begin{aligned}
    \Qcircuit @C=1em @R=.7em @! R {&\ustick{{R\setminus S}}\qw&\multimeasureD{2}{\tA_R^\dag a_S}\\
&\ustick{{R\cap S}}\qw&\ghost{\tA_R^\dag a_S}\\    
&\ustick{{S\setminus R}}\qw&\ghost{\tA_R^\dag a_S}}
\end{aligned}\ \coloneqq \ 
\begin{aligned}
    \Qcircuit @C=1em @R=.7em @! R {&\ustick{{R\setminus S}}\qw&\multigate{1}{\tA_R}&\ustick{{R\setminus S}}\qw&\measureD{e}\\
&\ustick{{R\cap S}}\qw&\ghost{\tA_R}&\ustick{{R\cap S}}\qw&\multimeasureD{1}{a_S}\\    
&\ustick{{S\setminus R}}\qw&\qw&\qw&\ghost{a_S}}
\end{aligned}\ .
\label{eq:loctronloceff}
\end{align}
We also define $\set{Pre}\Trnset{\rA_G\to\rA_G}_{L\Reals}$ as the set of local transformations $\tA_R$. 
\end{definition}

\begin{lemma}\label{lem:locteqloceff}
Let $a_S\sim b_T$. Then $\tA^\dag_R a_S\sim \tA^\dag_R b_T$.
\end{lemma}
\begin{proof}
By hypothesis, there exists $c\in\Cntset{\rA_{S\cap T}}_\Reals$ such that
\begin{align*}
a=c\otimes e_{S\setminus T},\quad b=c\otimes e_{T\setminus S}.
\end{align*}
Then we have
\begin{align*}
&(\tA^\dag_R a_S|=(c\otimes e_{S\setminus T}\otimes e_{R\setminus S}|(\tA\otimes\tI_{S\setminus R}),\\
&(\tA^\dag_R b_T|=(c\otimes e_{T\setminus S}\otimes e_{R\setminus T}|(\tA\otimes\tI_{T\setminus R}).
\end{align*}
By Eq.~\eqref{eq:equiveff} the thesis follows.
\end{proof}



We can now define an equivalence relation between local transformations as follows
\begin{align}
\tA_S\sim\tB_T\ \Leftrightarrow\ \left\{
\begin{aligned}
&\tA=\tC\otimes\tI_{S\setminus T},\\
&\tB=\tC\otimes\tI_{T\setminus S},
\label{eq:equivtr}
\end{aligned}
\right.
\end{align}
for some $\tC\in\Trnset{\rA_{S\cap T}}_\Reals$.
The set of local transformations is then defined as
\begin{align*}
\Trnset{\rA_G\to\rA_G}_{L\Reals}\coloneqq \set{Pre}\Trnset{\rA_G\to\rA_G}_{L\Reals}/\sim
\end{align*}

\begin{lemma}\label{lem:exttr}
Let $\tA_R\in\set{Pre}\Trnset{\rA_G\to\rA_G}_{L\Reals}$. Then for every finite region $H\in\reg G $ such that $H\cap R=\emptyset$, one has $(\tA\otimes\tI_{\rA_H})_{R\cup H}\in\set{Pre}\Trnset{\rA_G\to\rA_G}_{L\Reals}$, and $(\tA\otimes\tI_{\rA_H})_{R\cup H}\sim\tA_R$.
\end{lemma}
\begin{proof} It is straightforward to verify that $(\tA\otimes\tI_{\rA_H})_{R\cup H}\sim\tA_R$ by direct inspection of the defining equation~\eqref{eq:equivtr}.\end{proof}

We can now provide a way to identify a canonical representative of the equivalence class of $\tA_R$, which is defined in analogy to the case of effects as follows.
\begin{definition}
The \emph{minimal representative} of the equivalence class $\tA_R$, denoted as $\tilde\tA_{R_\tA}$, is defined through
\begin{align}
&R_\tA\coloneqq \bigcap_{S\in\set R_{(\tA,R)}}S,\quad\tilde\tA_{R_\tA}\sim\tA_R,
\label{eq:minimalreploctr}
\end{align}
where $\set R_{(\tA,R)}$ is the set of all those finite regions $S\subseteq G$ for which there exists $\tB\in\Trnset{\rA_S\to\rA_S}_{\Reals}$ such that $\tB_S\sim\tA_R$.
\end{definition}

\begin{lemma}\label{lem:exunmintr}
The minimal representative exists and is unique. 
\end{lemma}
The proof follows step by step that of lemma~\ref{lem:exuniminst}. 
%

The first result that we need to prove is the following.
\begin{lemma}\label{lem:eqloctloceff}
Let $\tA_R\sim\tB_S$. Then for every 
effect $a_T\in\set{Pre}\Cntset{\rA_G}_{L\Reals}$, one has
\begin{align}
\tA^\dag_Ra_T\sim\tB^\dag_Sa_T.
\end{align}
\end{lemma}
\begin{proof}
If $(\tilde\tC)_{R_\tC}$ is the minimal representative of $\tA_R\sim\tB_S$, by Eq.~\eqref{eq:equivtr}, one has
\begin{align*}
&\tA=\tilde\tC\otimes\tI_{R\setminus R_\tC},\\
&\tB=\tilde\tC\otimes\tI_{S\setminus R_\tC}.
\end{align*}
Now, by the defining equation~\ref{eq:loctronloceff}, we have that
\begin{align*}
&(\tA^\dag_R a_T|=(\tilde a\otimes e_{(T\cup R)\setminus R_a}|(\tilde\tC\otimes\tI_{(T\cup R)\setminus R_\tC})\\
=&(\tilde a\otimes e_{[T\cup(R\cap S)]\setminus R_a}|(\tilde\tC\otimes\tI_{[T\cup (R\cap S)]\setminus R_\tC})\\
&\otimes (e_{R\setminus S}|,\\
&(\tB^\dag_S a_T|=(\tilde a\otimes e_{(T\cup S)\setminus R_a}|(\tilde\tC\otimes\tI_{(T\cup S)\setminus R_\tC})\\
=&(\tilde a\otimes e_{[T\cup (R\cap S)]\setminus R_a}|(\tilde\tC\otimes\tI_{[T\cup (R\cap S)]\setminus R_\tC})\\
&\otimes(e_{S\setminus R}|,
\end{align*}
and finally by Eq.~\eqref{eq:equivlocst} this implies the thesis.
\end{proof}

By virtue of lemmas~\ref{lem:locteqloceff} and~\ref{lem:eqloctloceff}, we can define the action of $\tA_R\in\Trnset{\rA_G\to\rA_G}_{L\Reals}$ on local effects $\Cntset{\rA_G}_{\Reals}$ as
\begin{align*}
\tA^\dag_R[a_T]\coloneqq [\tA^\dag_R a_T],
\end{align*}
where we leave the square braces to denote equivalence classes for $\sim$ in $\set{Pre}\Cntset{\rA_G}_{L\Reals}$, for the sake of clarity.
We can now make local transformations into a vector space as we did for local effects, as follows.

\begin{definition}
Let $\tA_R,\tB_S\in\Trnset{\rA_G\to\rA_G}_{L\Reals}$, and $h\in\Reals$. Then we define
\begin{align*}
&h\tA_R\coloneqq \left\{
\begin{aligned}
&(h\tA)_R&h\neq0,\\
&0_\emptyset&h=0,
\end{aligned}
\right.
\\
&\tA_R+\tB_S\coloneqq \tC_{R\cup S}\\
&\tC\coloneqq \tA\otimes\tI_{S\setminus R}+\tB\otimes\tI_{R\setminus S}.
\end{align*}
\end{definition}

Moreover, the following operation makes the set of local transformations into an algebra.

\begin{definition}
Let $\tA_R,\tB_S\in\Trnset{\rA_G\to\rA_G}_{L\Reals}$. Then we define
\begin{align*}
&\tA_R\tB_S\coloneqq (\{\tA\otimes\tI_{S\setminus R}\}\{\tB\otimes\tI_{R\setminus S}\})_{R\cup S}.\\
\end{align*}
\end{definition}

We omit the straightforward proof that the above definition is well defined, i.e.~independent of the choice of representatives in the classes of $\tA_R$ and $\tB_S$.

\begin{definition}
The \emph{algebra of generalized local transformations} is the unital algebra $\Trnset{\rA_G\to\rA_G}_{L\Reals}$ of finite real combinations of local transformations, with unit $1_\emptyset$ and null element $0_\emptyset$. 
\end{definition}

In order to close the algebra of local operations we introduce the topology given by the \emph{sup-norm}, given in the following, in analogy to the case of effects. We also discuss the interplay of the sup-norm topology with that given by the {\em operational norm}, that we define right away. 
\begin{definition}
The \emph{operational norm} $\normop{\tA_R}$ of $\tilde\tA_{R_\tA}=\tA_R\in\Trnset{\rA\to\rA}_{L\Reals}$ is defined by the following expression
\begin{align}
&\normop{\tA_R}\coloneqq \normop{\tilde\tA}.
\end{align}
\end{definition}

\begin{proposition}\label{prop:eqnormoptr}
Let $\tA_R\in\Trnset{\rA\to\rA}_{L\Reals}$. Then $\normop{\tA_R}=\normop{\tA}$.
\end{proposition}
\begin{proof}
Let $\tA_R\sim\tilde\tA_{R_\tA}$, and let $R=R_\tA\cup R'$, with $R_\tA\cap R'=\emptyset$. Then by Eq.~\eqref{eq:equivtr} we have
\begin{align*}
\tA=\tilde\tA\otimes\tI_{\rA_{R'}},
\end{align*}
Now, by definition of $\normop{\cdot}$ (see \cite{PhysRevA.81.062348}) it straightforwardly follows that $\normop{\tB\otimes\tI_\rC}=\normop{\tB}$. Thus, $\normop{\tA_R}=\normop{\tilde\tA}=\normop{\tA}$.
\end{proof}

Unfortunately, the operational norm does not enjoy the basic property that would make the algebra of transformations into a Banach algebra, i.e.~it is not true that $\normop{\tA\tB}\leq\normop{\tA}\normop{\tB}$. For this reason, in analogy with the case of quasi-local effects, we introduce a second norm, the sup-norm, whose interplay with the operational norm will make it possible to define the closed Banach algebra of quasi-local transformations.


This is obtained extending the definition of sup-norm to the algebra of local transformations of $\rA_G$.

\begin{definition}\label{def:normsupqltr}
The \emph{sup-norm} $\normsup{\tA_R}$ of $\tA_R\in\Trnset{\rA_G\to\rA_G}_{L\Reals}$ is defined as
\begin{align}
&\normsup{\tA_R}\coloneqq \inf J(\tilde\tA)=\normsup{\tilde\tA}.
\end{align}
\end{definition}

\begin{proposition}\label{prop:eqnorsup}
Let $\tA_R\in\Trnset{\rA_G\to\rA_G}_{L\Reals}$. Then $\normsup{\tA_R}=\normsup{\tA}$.
\end{proposition}
\begin{proof}
Let $\tA_R\sim\tilde\tA_{R_\tA}$, and let $R=R_\tA\cup R'$, with $R_\tA\cap R'=\emptyset$. Then by Eq.~\eqref{eq:equivtr} we have
\begin{align*}
\tA=\tilde\tA\otimes\tI_{\rA_{R'}},
\end{align*}
and by corollary~\ref{cor:supnormtid}, $\normsup{\tA}=\normsup{\tilde\tA}=\normsup{\tA_R}$.
\end{proof}

\begin{corollary}
Let $\tA,\tB\in\Trnset{\rA_G\to\rA_G}_{L\Reals}$. Then $\normsup{\tA\tB}\leq\normsup{\tA}\normsup{\tB}$.
\end{corollary}
\begin{proof}
The result is a straightforward consequence of propositions~\ref{prop:eqnorsup} and~\ref{prop:banachsupnorm}.
\end{proof}

\begin{corollary}
On $\Trnset{\rA_G\to\rA_G}_{L\Reals}$, one has $\normop{\tA_R}\leq\normsup{\tA_R}$.
\end{corollary}
\begin{proof}
The result is a straightforward consequence of propositions~\ref{prop:eqnormoptr} and~\ref{prop:eqnorsup}, and corollary~\ref{cor:supstrongerop}.
\end{proof}

The above results allow us to extend the local algebra of events into the \emph{quasi-local} algebra, which is a Banach algebra. The construction is analogous to that of quasi-local states, and proceeds as follows. First we define the algebra $\Trnset{\rA_G\to\rA_G}_{C\Reals}$ of sup-Cauchy sequences of local transformations, i.e. its elements are sequences $\tA:\Nats\to\Trnset{\rA_G\to\rA_G}_{L\Reals}::n\mapsto{\tA_n}_{R_n}$ such that for every $\varepsilon>0$ there exists $n_0\in\Nats$ such that for all $m,n\geq n_0$, one has $\normsup{\tA_n-\tA_m}<\varepsilon$. Now we define the equivalence relation between elements of $\Trnset{\rA_G\to\rA_G}_{C\Reals}$: $\tA\cong\tB$ if for every $\varepsilon>0$ there exists $n_0\in\Nats$ such that for every $n\geq n_0$ one has $\normsup{\tA_n-\tB_n}<\varepsilon$. Finally, we define the quasi-local algebra as follows.

\begin{definition}
The {\em space of quasi-local transformations} of $\rA$ is defined as 
\begin{align*}
\Trnset{\rA_G\to\rA_G}_{Q\Reals}\coloneqq \Trnset{\rA_G\to\rA_G}_{C\Reals}/\cong.
\end{align*}
\end{definition}

\begin{definition}\label{def:qloctr}
An element $\tA$ in $\Trnset{\rA_G\to\rA_G}_{Q\Reals}$ is an \emph{event} if  $\tA=[\tA_{nR_{n}}]$ for a sequence $\tA_{nR_{n}}$ such that $\tA_n\in\Trnset{\rA_{R_{n}}\to \rA_{R_{n}}}$ for all $n\in\Nats$. The set of events will be denoted by $\Trnset{\rA_G\to\rA_G}_{Q}$. 

An element $\tA\in\Trnset{\rA_G\to\rA_G}_{Q}$ is a \emph{channel} if $\tA=[{\tA_n}_{R_{n}}]$ for a sequence ${\tA_n}_{R_{n}}$ such that $\tA_n\in\Trnset{\rA_{R_{n}}\to \rA_{R_{n}}}_1$ for all $n\in\Nats$. The set of channels will be denoted by $\Trnset{\rA_G\to\rA_G}_{Q1}$.

The subset of $\Trnset{\rA_G\to\rA_G}_{Q\Reals}$ containing elements $\tA=[{\tA_n}_{R_{n}}]$ 
for a sequence ${\tA_n}_{R_{n}}$ such that $\tA_n\in\Trnset{\rA_{R_{n}}\to\rA_{R_{n}}}_+$ for all $n\in\Nats$ will be denoted by $\Trnset{\rA_G\to\rA_G}_{Q+}$. Equivalently, we will write $\tA\succeq0$ for $\tA\in\Trnset{\rA_G\to\rA_G}_{Q+}$
\end{definition}

In the following, in analogy with the finite case, we will write $\tA\succeq\tB$ if $\tA-\tB\succeq0$. 
The algebra of quasi-local transformations $\Trnset{\rA_G\to\rA_G}_{Q\Reals}$ is defined as follows.
\begin{definition}
Let $\tA=[{\tA_n}_{R_n}]$, $\tB=[{\tB_n}_{S_n}]$. We define $\tA\tB\coloneqq [{\tA_n}_{R_n}{\tB_n}_{S_n}]$.
\end{definition}

The product $\tA\tB$ is well defined. Indeed: 
\begin{enumerate}
\item ${\tA_n}_{R_n}{\tB_n}_{S_n}$ is a Cauchy sequence.\label{it:cau} 
\item For every ${\tA'_n}_{R'_n}$, ${\tB'_n}_{S'_n}$ in the equivalence classes defining $\tA$ and $\tB$, respectively, one has
${\tA_n}_{R_n}{\tB_n}_{ S_n}\cong{\tA'_n}_{R'_n}{\tB'_n}_{S'_n}$.\label{it:procau}
\end{enumerate}
Item~\ref{it:cau} can be easily proved since we have
\begin{align*}
&\normsup{\tA_n\tB_n-\tA_m\tB_m}\\
=&\normsup{\tA_n\tB_n-\tA_m\tB_n+\tA_m\tB_n-\tA_m\tB_m}\\
\leq&\normsup{\tA_n-\tA_m}\normsup{\tB_n}\\
&+\normsup{\tA_m}\normsup{\tB_n-\tB_m}.
\end{align*}
Similarly, for item~\ref{it:procau} we have
\begin{align*}
&\normsup{\tA_n\tB_n-\tA'_n\tB'_n}\\
=&\normsup{\tA_n\tB_n-\tA'_n\tB_n+\tA'_n\tB_n-\tA'_n\tB'_n}\\
\leq&\normsup{\tA_n-\tA'_n}\normsup{\tB_n}\\
&+\normsup{\tA'_n}\normsup{\tB_n-\tB'_n}.
\end{align*}

The quasi-local algebra is actually a real Banach algebra if equipped with the norm $\normsup{\cdot}$, as we now prove.

\begin{proposition}
The algebra $\Trnset{\rA_G\to\rA_G}_{Q\Reals}$ equipped with the norm $\normsup{\cdot}$ is a Banach algebra.
\end{proposition}
\begin{proof}
We only need to prove that $\normsup{\tA\tB}\leq\normsup\tA\normsup\tB$. Let $\tA=[{\tA_n}_{R_n}]$ and $\tB=[{\tB_n}_{S_n}]$, respectively. Now, for every ${\tA_n}_{R_n}$ and ${\tB_n}_{S_n}$ one has
\begin{align*}
&\normsup{(\tA_n\otimes\tI_{S_n\setminus R_n})(\tB_n\otimes\tI_{R_n\setminus S_n})}\\
&\leq\normsup{\tA_n\otimes\tI_{S_n\setminus R_n}}\normsup{\tB_n\otimes\tI_{R_n\setminus S_n}},
\end{align*}
and by proposition~\ref{prop:eqnorsup} $\normsup{\tA\tB}\leq\normsup\tA\normsup\tB$.
\end{proof}

The subsets $\Trnset{\rA_G\to\rA_G}_{Q}$ and $\Trnset{\rA_G\to\rA_G}_{Q1}$ are convex, and $\Trnset{\rA_G\to\rA_G}_{Q+}$ is a convex cone. Moreover, they are closed in the sup-norm.
\begin{proposition}\label{prop:closcon}
The sets $\Trnset{\rA_G\to\rA_G}_{Q}$, $\Trnset{\rA_G\to\rA_G}_{Q1}$, and $\Trnset{\rA_G\to\rA_G}_{Q+}$ are closed in the sup-norm.
\end{proposition}
\begin{proof}
The argument is the same in both cases. Let $\{\tA_m\}_{m\in\Nats}$ be a Cauchy sequence with $\tA_m\in\Trnset{\rA_G\to\rA_G}_{Q*}$ for all $m$, with $*=\mbox{``nothing''},1,+$. By definition, $\tA_m=[\tA_{mnR_{mn}}]$, with $\tA_{mn}\in\Trnset{\rA_{R_{mn}}\to\rA_{R_{mn}}}_*$. Then one easily proves that the sequence $\{\tA_{mmR_{mm}}\}_{m\in\Nats}$ is Cauchy, and its limit is $\tA\coloneqq \lim_{m\to\infty}\tA_{mmR_{mm}}$, which by definition is in $\Trnset{\rA_G\to\rA_G}_{Q*}$.
\end{proof}


\begin{proposition}
Let $\tA\in\Trnset{\rA_G\to\rA_G}_{Q\Reals}$. Then
\begin{align}
\normop{\tA}\leq\normsup{\tA}.
\end{align}
\end{proposition}
\begin{proof}
By definition $\normop{\tA}=\lim_{n\to\infty}\normop{\tA_n}$ and $\normsup{\tA}=\lim_{n\to\infty}\normsup{\tA_n}$, where $\{{\tA_n}_{R_n}\}_{n\in\Nats}\subseteq\Trnset{\rA_G\to\rA_G}_{L\Reals}$ is a Cauchy sequence converging to $\tA$, thus by proposition~\ref{prop:opleqsupop} one has
\begin{align*}
\normop{\tA_n}\leq\normsup{\tA_n},\ \forall n\in\Nats,
\end{align*}
which implies the inequality in the limit.
\end{proof}

Now, we can prove the following lemma.
\begin{lemma}\label{lem:loctrinvcauch}
Let $a\in\Cntset{\rA_G}_{C\Reals}$ be a Cauchy sequence, and let $\tA\in\Trnset{\rA_G\to\rA_G}_{L\Reals}$. We then have
\begin{align}
\tA^\dag a\in\Cntset{\rA_G}_{C\Reals}.
\end{align}
\end{lemma}
\begin{proof} We just need to evaluate $\normsup{\tA^\dag a_m-\tA^\dag a_n}$, and using proposition~\ref{prop:banachsupnorm} one has
\begin{align*}
\normsup{\tA^\dag(a_m-a_n)}\leq\normsup{a_m-a_n}\normsup\tA.
\end{align*}
Since the sequence $a$ is Cauchy, also $\tA^\dag a$ is Cauchy.
\end{proof}

\begin{lemma}\label{lem:loctrinvqlef}
Let $a,b\in\Cntset{\rA_G}_{C\Reals}$ be equivalent Cauchy sequences, and let $\tA\in\Trnset{\rA_G\to\rA_G}_{L\Reals}$. We then have
\begin{align}
[\tA^\dag a_{nR_n}]=[\tA^\dag b_{nS_n}].
\end{align}
\end{lemma}
\begin{proof} We can bound $\normsup{\tA^\dag a_n-\tA^\dag b_n}$ using proposition~\ref{prop:banachsupnorm}, obtaining
\begin{align*}
\normsup{\tA^\dag(a_n-b_n)}\leq\normsup{a_n-b_n}\normsup\tA.
\end{align*}
Since $[a_{nR_n}]=[b_{nS_n}]$, the thesis follows.
\end{proof}

As a consequence of the above results, all local transformations leave the space $\Cntset{\rA_G}_{Q\Reals}$ invariant. We now prove a further important result: the above statement can be extended to all the quasi-local algebra.

\begin{theorem}
The quasi-local algebra $\Trnset{\rA_G\to\rA_G}_{Q\Reals}$ leaves the space $\Cntset{\rA_G}_{Q\Reals}$ invariant, and for every $\tA\in\Trnset{\rA_G\to\rA_G}_{Q\Reals}$ and every $a\in\Cntset{\rA_G}_{Q\Reals}$, it is
\begin{align}
\normsup{\tA^\dag a}\leq\normsup a\normsup\tA.
\label{eq:boundedoneff}
\end{align}
\end{theorem}
\begin{proof}
By proposition~\ref{prop:banachsupnorm} and lemma~\ref{lem:loctrinvcauch}, the thesis is true for $\tA\in\Trnset{\rA_G\to\rA_G}_{L\Reals}$ and $a\in\Cntset{\rA_G}_{L\Reals}$. Let now $a=[a_{pR_p}]$ and $\tA\in\Trnset{\rA_G\to\rA_G}_{L\Reals}$. Then
\begin{align*}
\normsup{\tA^\dag a_p}\leq\normsup{a_p}\normsup{\tA},
\end{align*}
and taking the limit for $p\to\infty$ we obtain the thesis. Finally, let $\tA=[\tA_{nR_n}]$. In this case, by lemma~\ref{lem:loctrinvcauch} $\tA^\dag_n a\in\Cntset{\rA_G}_{Q\Reals}$ for every $n\in\Nats$. Moreover, $\{\tA^\dag_n a\}_{n\in\Nats}\in\Cntset{\rA_G}_{C\Reals}$. Indeed, by the result we just proved
\begin{align*}
\normsup{(\tA^\dag_n-\tA^\dag_m) a}\leq\normsup{a}\normsup{\tA_n-\tA_m}.
\end{align*}
Finally, by lemma~\ref{lem:loctrinvcauch}, we have
\begin{align*}
\normsup{\tA^\dag_n a}\leq\normsup a\normsup{\tA_n},
\end{align*}
and taking the limit for $n\to\infty$ we obtain the thesis.
%
\end{proof}

A remarkable result that will be important in the following is given by the following lemma.

\begin{lemma}
Let $\tA\in\Trnset{\rA_G\to\rA_G}_{Q}$. Then for every $a\in\Cntset{\rA_G}_Q$ one has $\tA^\dag a\in\Cntset{\rA_G}_Q$.
\end{lemma}
\begin{proof}
By hypothesis, 
$\tA=[\tA_{nR_n}]$ and 
$a=[a_{mS_m}]$,  
with $\tA_n\in\Trnset{\rA_{R_n}\to\rA_{R_n}}$ and $a_m\in\Cntset{\rA_{S_m}}$. Thus $\tA^\dag_n a_n\in\Cntset{\rA_{R_n\cup S_n}}$. Now, one can easily verify that $\lim_{n\to\infty}\tA^\dag_n a_n=\tA^\dag a$, which proves the statement.
\end{proof}

In particular, one has the following condition.
\begin{lemma}\label{lem:qltransqleff}
Let $\tA\in\Trnset{\rA_G\to\rA_G}_{Q+}$. Then $\tA\in\Trnset{\rA_G\to\rA_G}_Q$ iff $\tA^\dag e_G\in\Cntset{\rA_G}_Q$.
\end{lemma}
\begin{proof}
By definition, one has $\tA\in\Trnset{\rA_G\to\rA_G}_{Q}$ if and only if $\tA=[\tA_{nR_n}]$ with $\tA_{n}\in\Trnset{\rA_{R_n}\to\rA_{R_n}}$, and by our assumptions the latter is equivalent to $a_n\coloneqq \tA_{n}^\dag e_{\rA_{R_n}}\in\Cntset{\rA_{R_n}}$. Now, since $\normsup{a_{nR_n}-a_{mR_m}}\leq\normsup{\tA_{nR_n}-\tA_{mR_m}}$, one has that $\tA\in\Trnset{\rA_G\to\rA_G}_Q$ if and only if $\tA^\dag e_G=[a_{nR_n}]\in\Cntset{\rA_G}_Q$.
\end{proof}

\begin{lemma}\label{lem:*id}
Let $\tA\in\Trnset{\rA\to\rB}_{Q+}$, and $a\coloneqq \tA^\dag e_G$. Then $\normsup{\tA}=\normsup{a}$.
\end{lemma}
\begin{proof}
By definition, $\tA=[\tA_{nR_n}]$, with $\tA_n\in\Trnset{\rA_{R_n}\to\rA_{R_n}}_+$. Now, $\normsup{{\tA_n}_{R_n}}=\normsup{{\tA_n}}$, and 
by lemma~\ref{lem:***}, $\normsup{\tA_n}=\normsup{{\tA^\dag_n}e_{R_n}}=\normsup{\tA^\dag_{nR_n}e_G}$. Then,
\begin{align*}
\normsup{\tA}=\lim_{n\to\infty}\normsup{{\tA_n^\dag}_{R_n} e_G}=\normsup{\tA^\dag e_G}.&&\qedhere
\end{align*}
\end{proof}

On the same line, we can provide the following condition for $\tA\in\Trnset{\rA_G\to\rA_G}_{Q+}$ to be a channel.
\begin{lemma}\label{lem:cnesschan}
Let $\tA\in\Trnset{\rA_G\to\rA_G}_{Q+}$. Then $\tA\in\Trnset{\rA_G\to\rA_G}_{Q1}$ iff $\tA^\dag e_G=e_G$.
\end{lemma}
\begin{proof}
First, let $\tA$ be a channel. By definition it must be $\tA=\lim_{n\to\infty}\tA_{nR_n}$, where $\tA_{nR_n}$ are local channels. Thus, by theorem~\ref{th:charchan}, $\tA^\dag_{nR_n}e_G=e_G$. It follows that $\tA^\dag e_G=\lim_{n\to\infty}\tA^\dag_{nR_n}e_G=e_G$. Now, for the converse, by definition $\tA\in\Trnset{\rA_G\to\rA_G}_{Q+}$ iff $\tA=[{\tA_n}_{R_n}]$ with ${\tA_n}_{R_n}\in\Trnset{\rA_{R_n}\to\rA_{R_n}}_{+}$ for all $n\in\Nats$. Let us take $n\geq n_0$ so that $\normsup{{\tA_n}_{R_n}-\tA}\leq\varepsilon$. Then we have
\begin{align*}
|\normsup{{\tA_n}_{R_n}}-\normsup{\tA}|\leq\varepsilon.
\end{align*}
Now, by lemma~\ref{lem:*id}, since $\tA^\dag e_G=e_G$, we have $\normsup\tA=1$. Let us define ${\tA'_n}_{R_n}\coloneqq {\tA_n}_{R_n}/\normsup{{\tA_n}_{R_n}}$. Clearly,
\begin{align*}
\normsup{{\tA'_n}_{R_n}-{\tA_n}_{R_n}}=\normsup{{\tA_n}_{R_n}}|1-1/\normsup{{\tA_n}_{R_n}}|\leq\varepsilon.
\end{align*}
This implies that ${\tA'_n}_{R_n}$ is a sequence converging to $\tA$, with $\normsup{{\tA'_n}_{R_n}}=1$. As a consequence, there exist channels $\tC_{nR_n}$ such that $(\tC_n-\tA'_n)_{R_n}\succeq0$. Again by lemma~\ref{lem:***},
\begin{align*}
\normsup{(\tC_n-\tA'_n)_{R_n}}&=\normsup{(\tC_n-\tA'_n)^\dag_{R_n}e_{R_n}}\\
&=\normsup{e_G-{\tA'_n}^\dag_{R_n}e_G}\\
&=\normsup{(\tA-\tA'_{nR_n})^\dag e_G}\\
&\leq\normsup{(\tA-{\tA'_n}_{R_n})}\leq\varepsilon.
\end{align*}
This implies that $\tA=[\tA'_{nR_n}]=[{\tC_n}_{R_n}]$, and then $\tA\in\Trnset{\rA_G\to\rA_G}_1$.
\end{proof}

The following weak version of the converse of lemma~\ref{lem:qltransqleff} can be easily proved.
\begin{lemma}\label{lem:loceffloctr}
Let $a\in\Cntset{\rA_G}_{L\Reals}$. Then there exists $\tA\in\Trnset{\rA_G}_{L\Reals}$ such that
\begin{align}
a=\tA^\dag e_G.
\label{eq:effop}
\end{align}
Moreover, for $a\in\Cntset{\rA_G}_L$, Eq.~\eqref{eq:effop} is satisfied with $\tA\in\Trnset{\rA_G\to\rA_G}_{L}$.
\end{lemma}
\begin{proof}
Let $a=a_R\in\Cntset{\rA_G}_{L\Reals}$. Let us define $\tA_R\coloneqq |\sigma)(a|$, where $\sigma\in\Stset{\rA_R}_1$ is an arbitrary deterministic state. Then clearly
\begin{align*}
\tA^\dag_Re_G=a_R.
\end{align*}
It is also clear that, for $a=a_R\in\Cntset{\rA_G}_L$, the above defined transformation $\tA_R$ is in $\Trnset{\rA_G}_L$.
\end{proof}

One can take a further step, and prove the following result that will be of crucial importance in the remainder.
\begin{corollary}\label{cor:succqltr}
Let $a\in\Cntset{\rA_G}_{Q\Reals}$. Then there exists a sequence of quasi-local transformations $\tA_{nR_n}$ such that $\lim_{n\to\infty}\tA_{nR_n}^\dag e_G=a$. If $a\in\Cntset{\rA_G}_Q$ the sequence can be found in $\Trnset{\rA_G\to\rA_G}_Q$.
\end{corollary}

Notice that the sequence ${\tA_n}_{R_n}$ is not necessarily Cauchy, thus one cannot conclude that for every $a\in\Cntset{\rA_G}_{Q\Reals}$ there is $\tA\in\Trnset{\rA_G\to\rA_G}_{Q\Reals}$ such that $a=\tA^\dag e_G$. However, the following lemma completes the picture.
\begin{proposition}\label{prop:qlefftrans}
Let $a\in\Cntset{\rA_G}_{Q\Reals}$. Then there exists $\tA\in\Trnset{\rA_G\to\rA_G}_{Q\Reals}$ such that
\begin{align}
a=\tA^\dag e_G.
\end{align}
Moreover, for $a\in\Cntset{\rA_G}_{Q*}$ with $*\in\{\mbox{``nothing'',1,+}\}$ one can find the above $\tA$ in the set $\Trnset{\rA_G\to\rA_G}_{Q*}$.
\end{proposition}
\begin{proof}
Let us define the sets 
\begin{align*}
\set T^\varepsilon_a\coloneqq \{\tA\in\Trnset{\rA_G\to\rA_G}_{Q\Reals}\mid\normsup{\tA^\dag e_G-a}\leq\varepsilon\}.
\end{align*}
The sets $\set T^\varepsilon_a$ are not empty for every $a$ and every $\varepsilon>0$, as a consequence of corollary~\ref{cor:succqltr}. Moreover, they are closed. Indeed, let $\{\tA_n\}_{n\in\Nats}$ be a Cauchy sequence in $\set T^\varepsilon_a$. This implies that by definition $a_n\coloneqq \tA^\dag_n e_G\in\bar B_\varepsilon(a)$. Since the ball $\bar B_\varepsilon(a)$ is closed, also $\lim_{n\to\infty}a_n=(\lim_{n\to\infty}\tA_n)^\dag e_G\in\bar B_\varepsilon(a)$, namely $\lim_{n\to\infty}\tA_n\in\set T^\varepsilon_a$. Finally, one clearly has that for $\varepsilon<\delta$ it is $\set T^\varepsilon_a\subseteq\set T^\delta_a$, and the diameter $d_\varepsilon\coloneqq \sup\{\normsup{\tA-\tB}\mid\tA,\tB\in\set T^\varepsilon_a\}$ is a non-decreasing function of $\varepsilon$. 
Let then
\begin{align*}
\set T^0_a\coloneqq \bigcap_{\varepsilon>0}\set T^\varepsilon_a.
\end{align*}
Either $\lim_{\varepsilon\to 0^+} d_\varepsilon>0$, in which case $\set T^0_a$ is non empty, and each of its elements satisfies $\tA^\dag e_g=a$, or the limit is $0$, in which case, by Cantor's intersection theorem for complete metric spaces, $\set T^0_a=\{\tA\}$ is a singleton, with $\tA^\dag e_G=a$. The same argument can be applied in each of the cases where $a\in\Cntset{\rA_G}_{Q*}$, considering the closed sets $\set T^{\varepsilon*}_a\coloneqq \set T^\varepsilon_a\cap\Trnset{\rA_G\to\rA_G}_{Q*}$.
\end{proof}

We now prove a result that is analogous to lemma~\ref{lem:cnessef}.
\begin{lemma}\label{lem:cnestr}
Let $\tA\in\Trnset{\rA_G\to\rA_G}_{Q+}$, and $\tC-\tA\in\Cntset{\rA_G\to\rA_G}_{Q+}$ for some $\tC\in\Trnset{\rA_G\to\rA_G}_{Q1}$. Then $\tA\in\Trnset{\rA_G\to\rA_G}_Q$.
\end{lemma}
\begin{proof}
By definition, there are three sequences $\tA_{nR_n}$, $\tA'_{nS_n}$ in $\Trnset{\rA_G\to\rA_G}_L$ and $\tC_{nT_n}$ in $\Trnset{\rA_G\to\rA_G}_{L1}$ such that
$\lim_{n\to\infty}\tA_n=\tA$, $\lim_{n\to\infty}\tA'_n=\tC-\tA$, and $\lim_{n\to\infty}\tC_n=\tC$. This implies that
\begin{align*}
\normsup{\tC_n-(\tA_n+\tA'_n)}\leq\varepsilon,
\end{align*}
and consequently there is a sequence of channels $\tD_n$ in $\Trnset{\rA_G\to\rA_G}_{L1}$ such that
\begin{align*}
\varepsilon\tD_n+\tC_n-(\tA_n+\tA'_n)\succeq0.
\end{align*}
By the above equation, it is easy to check that
\begin{align*}
\tF_n-\frac1{1+\varepsilon}\tA_n\succeq0,\quad\tF_n-\frac1{1+\varepsilon}\tA'_n\succeq0,
\end{align*}
where $\tF_n$ is the channel $\varepsilon/(1+\varepsilon)\tD_n+1/(1+\varepsilon)\tC_n$. Thus, by theorem~\ref{th:norest}, both $\tA_n/(1+\varepsilon)$ and $\tA'_n/(1+\varepsilon)$ belong to $\Trnset{\rA_G\to\rA_G}_L$. Let $\varepsilon_m$ be a real Cauchy sequence converging to 0, and consider the corresponding sequences $\tB_m\coloneqq\tA_{n_m}/(1+\varepsilon_m)$, $\tB'_m\coloneqq\tA'_{n_m}/(1+\varepsilon_m)$. Since
\begin{align*}
&\normsup{\tB_m-\tA_{n_m}}=\left|\frac{\varepsilon_m}{1+\varepsilon_m}\right|\normsup{\tA_{n_m}},\\
&\normsup{\tB'_m-\tA'_{n_m}}=\left|\frac{\varepsilon_m}{1+\varepsilon_m}\right|\normsup{\tA'_{n_m}},
\end{align*}
one has $\tA=[\tA_{nR_n}]=[\tB_{mR_m}]$ and $\tC-\tA=[\tA'_{nS_n}]=[\tB'_{mS_m}]$, with $\tB_n$ and $\tB_n'$ beloning to $\Trnset{\rA_G\to\rA_G}_L$. Thus, $\tA,\tC-\tA\in\Trnset{\rA_G\to\rA_G}_Q$.
\end{proof}

We will now define the action of the quasi-local algebra of transformations on quasi-local states. The action is defined by duality as follows.

\begin{definition}[Dual action of the quasi-local algebra]
Let $\tA\in\Trnset{\rA_G\to\rA_G}_{Q\Reals}$. We define the map $\hat\tA:\Stset{\rA_G}_{\Reals}\to\Stset{\rA_G}_{\Reals}$ as follows
\begin{align}
(a|\hat\tA\rho)\coloneqq (\tA^\dag a|\rho),\quad\forall a\in\Cntset{\rA_G}_{Q\Reals}.
\end{align}
\end{definition}
The first thing we need to prove is that the map $\hat\tA:\Stset{\rA_G}_{\Reals}\to\Stset{\rA_G}_{\Reals}$ is linear and bounded.
\begin{proposition}\label{prop:boundloct}
For every $\tA\in\Trnset{\rA_G\to\rA_G}_{Q\Reals}$ the map $\hat\tA:\Stset{\rA_G}_{\Reals}\to\Stset{\rA_G}_{\Reals}$ is linear and bounded.
\end{proposition}
\begin{proof}
As to linearity, it is sufficient to consider that
\begin{align*}
(a|\hat\tA [x\rho_1+y\rho_2])&=(\tA^\dag a|x\rho_1+y\rho_2)\\
&=x(\tA^\dag a|\rho_1)+y(\tA^\dag a|\rho_2)\\
&=(a|x\hat\tA\rho_1)+(a|y\hat\tA\rho_2),
\end{align*}
which by definition implies 
\begin{align*}
&\hat\tA (x\rho_1+y\rho_2)=x\hat\tA\rho_1+y\hat\tA\rho_2.
\end{align*}
In order to prove boundedness, it is sufficient to consider Eq.~\eqref{eq:boundedoneff}, to conclude that
\begin{align*}
|(a|\hat\tA\rho)|&=|(\tA^\dag a|\rho)|\leq\normsup{\tA^\dag a}\normst\rho\\
&\leq\normsup a\normsup\tA\normst\rho,
\end{align*}
which implies that
\begin{align*}
&\normst{\hat\tA\rho}\leq\normsup\tA\normst\rho.&&\qedhere
\end{align*}
\end{proof}

We now prove that $\Stset{\rA_G}_{Q\Reals}$ is invariant under the quasi-local algebra, and more precisely, every space $\Stset{\rA_G}_{{\rho_0Q\Reals}}^{(B)}$ is invariant.
\begin{theorem}
Let $\tA\in\Trnset{\rA_G\to\rA_G}_{Q\Reals}$. Then
\begin{align*}
&\hat\tA\Stset{\rA_G}_{Q\Reals}\subseteq\Stset{\rA_G}_{Q\Reals},
\end{align*}
and for every $(B,\rho_0)$, 
\begin{align*}
&\hat\tA\Stset{\rA_G}_{{\rho_0Q\Reals}}^{(B)}\subseteq\Stset{\rA_G}_{{\rho_0Q\Reals}}^{(B)}.
\end{align*}
\end{theorem}
\begin{proof}
We start observing that, by Eq.~\eqref{eq:loctronloceff}, when we apply $\tA^\dag_R a_S$ to a local state $\tau_T$ in $\Stset{\rA_G}_{{\rho_0Q\Reals}}^{(B)}$, we obtain
\begin{align*}
&\begin{aligned}
\begin{tikzpicture}
	\begin{pgfonlayer}{nodelayer}
		\node [style=four wire box] (0) at (-3.25, 5.5) {$\tA_R$};
		\node [style=none] (1) at (-4, 7) {};
		\node [style=none] (2) at (-7, 7) {};
		\node [style=none] (3) at (-2.5, 7) {};
		\node [style=none] (5) at (-7, 6) {};
		\node [style=none] (6) at (-4, 6) {};
		\node [style=none] (7) at (-2.5, 6) {};
		\node [style=none] (9) at (-8.5, 5) {};
		\node [style=none] (10) at (-4, 5) {};
		\node [style=none] (11) at (-2.5, 5) {};
		\node [style=none] (12) at (-1.5, 5) {};
		\node [style=none] (13) at (-8.5, 4) {};
		\node [style=none] (14) at (-4, 4) {};
		\node [style=none] (15) at (-2.5, 4) {};
		\node [style=none] (16) at (-1.5, 4) {};
		\node [style=none] (22) at (0.25, 6) {};
		\node [style=none] (37) at (0.25, 3) {};
		\node [style=none] (38) at (-1.5, 3) {};
		\node [style=none] (39) at (0.25, 4) {};
		\node [style=none] (42) at (-1.5, 6) {};
		\node [style=none] (47) at (2, 5) {};
		\node [style=none] (48) at (0.25, 5) {};
		\node [style=none] (49) at (2, 4) {};
		\node [style=none] (51) at (2, 7) {};
		\node [style=none] (52) at (2, 6) {};
		\node [style=none] (53) at (2.5, 3) {};
		\node [style=none] (54) at (-8.5, 3) {};
		\node [style=none] (55) at (-7, 8) {};
		\node [style=none] (56) at (2, 8) {};
		\node [style=none] (59) at (-8.5, 2) {};
		\node [style=none] (60) at (2.5, 2) {};
		\node [style=none] (61) at (-5.5, 7.5) {$\scriptstyle{\rA_{(R\cap S)\setminus T}}$};
		\node [style=none] (62) at (-5.5, 8.5) {$\scriptstyle{\rA_{S\setminus(R\cup T)}}$};
		\node [style=none] (63) at (-5.5, 6.5) {$\scriptstyle{\rA_{R\setminus(S\cup T)}}$};
		\node [style=none] (64) at (-5.5, 5.5) {$\scriptstyle{\rA_{R\cap S\cap T}}$};
		\node [style=none] (65) at (-5.5, 4.5) {$\scriptstyle{\rA_{(R\cap T)\setminus S}}$};
		\node [style=none] (67) at (-5.5, 2.5) {$\scriptstyle{\rA_{T\setminus(R\cup S)}}$};
		\node [style=none] (68) at (-5.5, 3.5) {$\scriptstyle{\rA_{(T\cap S)\setminus R}}$};
		\node [style=none] (69) at (-7, 8.25) {};
		\node [style=none] (70) at (-7, 5.75) {};
		\node [style=none] (71) at (-10.5, 6.25) {};
		\node [style=none] (72) at (-10.5, 7.75) {};
		\node [style=none] (73) at (-10, 5.75) {};
		\node [style=none] (74) at (-10, 8.25) {};
		\node [style=none] (75) at (-8.75, 7) {$\rho_{0(R\cup S)\setminus T}$};
		\node [style=none] (76) at (-8.5, 5.25) {};
		\node [style=none] (77) at (-8.5, 1.75) {};
		\node [style=none] (78) at (-9.5, 5.25) {};
		\node [style=none] (79) at (-9.5, 1.75) {};
		\node [style=none] (80) at (-10, 2.25) {};
		\node [style=none] (81) at (-10, 4.75) {};
		\node [style=none] (82) at (-9.25, 3.5) {$\tau_T$};
		\node [style=none] (83) at (2.5, 4) {};
		\node [style=none] (86) at (2, 4.75) {};
		\node [style=none] (87) at (2, 8.25) {};
		\node [style=none] (88) at (2.5, 4.25) {};
		\node [style=none] (89) at (2.5, 3.75) {};
		\node [style=none] (90) at (3, 4.25) {};
		\node [style=none] (91) at (3, 3.75) {};
		\node [style=none] (96) at (3, 4.75) {};
		\node [style=none] (97) at (3.5, 5.25) {};
		\node [style=none] (98) at (3.5, 7.75) {};
		\node [style=none] (99) at (3.25, 8.25) {};
		\node [style=none] (100) at (2.5, 3.25) {};
		\node [style=none] (101) at (2.5, 2.75) {};
		\node [style=none] (102) at (3, 3.25) {};
		\node [style=none] (103) at (3, 2.75) {};
		\node [style=none] (104) at (2.5, 2.25) {};
		\node [style=none] (105) at (2.5, 1.75) {};
		\node [style=none] (106) at (3, 2.25) {};
		\node [style=none] (107) at (3, 1.75) {};
		\node [style=none] (108) at (2.75, 6.5) {$a_S$};
		\node [style=none] (109) at (2.75, 4) {$e$};
		\node [style=none] (110) at (2.75, 3) {$e$};
		\node [style=none] (111) at (2.75, 2) {$ e$};
	\end{pgfonlayer}
	\begin{pgfonlayer}{edgelayer}
		\draw (2.center) to (1.center);
		\draw (5.center) to (6.center);
		\draw (9.center) to (10.center);
		\draw (13.center) to (14.center);
		\draw (15.center) to (16.center);
		\draw (11.center) to (12.center);
		\draw [in=180, out=0] (12.center) to (22.center);
		\draw [in=-180, out=0] (38.center) to (39.center);
		\draw [in=180, out=0] (48.center) to (49.center);
		\draw [in=180, out=0] (42.center) to (48.center);
		\draw (54.center) to (38.center);
		\draw (37.center) to (53.center);
		\draw (59.center) to (60.center);
		\draw (55.center) to (56.center);
		\draw (3.center) to (51.center);
		\draw (74.center) to (69.center);
		\draw (69.center) to (70.center);
		\draw (70.center) to (73.center);
		\draw (71.center) to (72.center);
		\draw [in=-180, out=90] (72.center) to (74.center);
		\draw [in=-180, out=-90] (71.center) to (73.center);
		\draw (76.center) to (77.center);
		\draw (77.center) to (79.center);
		\draw [in=-90, out=180] (79.center) to (80.center);
		\draw (80.center) to (81.center);
		\draw [in=-180, out=105] (81.center) to (78.center);
		\draw (78.center) to (76.center);
		\draw (87.center) to (86.center);
		\draw (86.center) to (96.center);
		\draw [bend right=45] (96.center) to (97.center);
		\draw (97.center) to (98.center);
		\draw [bend right, looseness=1.25] (98.center) to (99.center);
		\draw (99.center) to (87.center);
		\draw (49.center) to (83.center);
		\draw (88.center) to (89.center);
		\draw (89.center) to (91.center);
		\draw [in=0, out=0, looseness=1.25] (91.center) to (90.center);
		\draw (90.center) to (88.center);
		\draw (100.center) to (101.center);
		\draw (101.center) to (103.center);
		\draw [in=0, out=0, looseness=1.25] (103.center) to (102.center);
		\draw (102.center) to (100.center);
		\draw (104.center) to (105.center);
		\draw (105.center) to (107.center);
		\draw [in=0, out=0, looseness=1.25] (107.center) to (106.center);
		\draw (106.center) to (104.center);
		\draw [in=180, out=0] (16.center) to (37.center);
		\draw (7.center) to (42.center);
		\draw [in=180, out=0] (39.center) to (47.center);
		\draw (22.center) to (52.center);
	\end{pgfonlayer}
\end{tikzpicture}
\end{aligned},
\end{align*}
for every $S$ and every $a_S\in\Cntset{\rA_S}_\Reals$, which implies
\begin{align}
&\begin{aligned}
    \Qcircuit @C=1em @R=.7em @! R {\multiprepareC{2}{\hat\tA_R\tau_T}&\qw&\ustick{\rA_{B(R\setminus T)}}\qw&\qw\\
\pureghost{\hat\tA_R\tau_T}&\qw&\ustick{\rA_{R\cap T}}\qw&\qw\\    
\pureghost{\hat\tA_R\tau_T}&\qw&\ustick{\rA_{S\setminus R}}\qw&\qw}
\end{aligned}\nonumber\\
\label{eq:loctronlocst}\\
&=\  
\begin{aligned}
    \Qcircuit @C=1em @R=.7em @! R {\multiprepareC{1}{\rho_{0B(R\setminus T)}}&\qw&\qw&\qw&\ustick{\rA_{B(R\setminus T)\setminus(R\setminus T)}}\qw&\qw&\qw&\qw\\
    \pureghost{\rho_{0B(R\setminus T)}}&\qw&\ustick{\rA_{R\setminus T}}\qw&\qw&\multigate{1}{\tA_R}&\qw&\ustick{\rA_{R\setminus T}}\qw&\qw\\
\multiprepareC{1}{\tau_T}&\qw&\ustick{\rA_{R\cap T}}\qw&\qw&\ghost{\tA_R}&\qw&\ustick{\rA_{R\cap T}}\qw&\qw\\    
\pureghost{\tau_T}&\qw&\ustick{\rA_{T\setminus R}}\qw&\qw&\qw&\qw&\qw&\qw}
\end{aligned}\ .\nonumber
\end{align}
This means that local states in $\Stset{\rA_G}_{{\rho_0Q\Reals}}^{(B)}$ are mapped to local states in the same sector under $\hat\tA_R$. Now, by proposition~\ref{prop:boundloct}, Cauchy sequences are mapped to Cauchy sequences, so that $\hat\tA_R\Stset{\rA_G}_{{\rho_0Q\Reals}}^{(B)}\subseteq\Stset{\rA_G}_{{\rho_0Q\Reals}}^{(B)}$. Moreover, again by proposition~\ref{prop:boundloct}, also Cauchy sequences of local transformations satisfy the same condition, thus $\hat\tA\Stset{\rA_G}_{{\rho_0Q\Reals}}^{(B)}\subseteq\Stset{\rA_G}_{{\rho_0Q\Reals}}^{(B)}$. Similarly, one can easily prove that $\hat\tA\Stset{\rA_G}_{Q\Reals}\subseteq\Stset{\rA_G}_{Q\Reals}$.
\end{proof}

As a consequence of the above theorem, the action of $\Trnset{\rA\to\rA}_Q$ on the space $\Stset{\rA_G}_{Q\Reals}$ of quasi-local states is decomposed into many irreducible representations, one for every space $\Stset{\rA_G}_{{\rho_0Q}}^{(B)}$.

\begin{remark} 
Notice that, given $\tA,\tB\in\Trnset{\rA_G\to\rA_G}_{Q\Reals}$, one has
\begin{align}
(\tA\tB)^\dag=\tB^\dag\tA^\dag,\quad\widehat{(\tA\tB)}=\hat\tA\hat\tB.
\end{align}
\end{remark} 

Let $R\in\reg G $, and let $\tA\in\Trnset{\rA_G\to\rA_G}_Q$. Suppose that $\tA=[\tA_{mR_m}]$ and for every $m\in \Nats$, $R\cap R_{m}=\emptyset$. This implies that, for $S_m=R_m\cup R$, $\tA=[\tA'_{mS_m}]$, with $\tA'_{mS_m}=(\tA_m\otimes\tI_R)_{S_m}$, for $\tA'_m\in\Trnset{\rA_{S_m\setminus R}}_{\Reals}$. We then write $\tA=\tB\otimes\tI_R$. Notice that in this case, given $\rho\in\Stset{\rA_G}_{\Reals}$, one has 
\begin{align*}
\normop{\rho_{\vert R}-(\hat\tA\rho)_{\vert R}}&=\normop{(\hat\tA_m\rho)_{\vert R}-(\hat\tA \rho)_{\vert R}}\nonumber\\
&\leq\normst{(\hat\tA_m-\hat\tA)\rho}\leq\varepsilon\normst\rho,
\end{align*}
and then $(\hat\tA\rho)_{\vert R}=\rho_{\vert R}$.

For $R\in\reg G$, there is a straightforward ordered Banach algebra isomorphism between $\Trnset{\rA_R\to\rA_R}_{Q\Reals}$ and $\Trnset{\rA^{(G)}_R\to\rA^{(G)}_R}_{Q\Reals}$, where the latter is defined as
\begin{align*}
\Trnset{\rA^{(G)}_R\to\rA^{(G)}_R}_{Q\Reals}\coloneqq \{\tA_{R}\mid\tA\in\Trnset{\rA_R\to\rA_R}_{\Reals}\}.
\end{align*}
A similar isomorphism can be found also in the case of infinite regions $R\in\infreg G $, where by definition $\Trnset{\rA^{(G)}_R\to\rA^{(G)}_R}_{Q\Reals}$ is the closed subalgebra of Cauchy classes $[{\tA_n}_{R_n}]$ with $R_n\subseteq R$ for every $n\in\Nats$.
The isomorphism in this case is given by
\begin{align}
\Trnset{\rA_R\to\rA_R}_{Q\Reals}^\dag=\tJ_R^\dag\Trnset{\rA^{(G)}_R\to\rA^{(G)}_R}_{Q\Reals}^\dag{\tJ_R^{-1}}^\dag,
\label{eq:isomqloctr}
\end{align}
where $\tJ_R^\dag:\Cntset{\rA^{(G)}_R}_{Q\Reals}\to\Cntset{\rA_R}_{Q\Reals}$ is defined in Eq.~\eqref{eq:isomj}.
In the following, when we want to specify that a given quasi-local transformation $\tA$ is in the subalgebra $\Trnset{\rA_R\to\rA_R}_{Q\Reals}$, we will write $\tA_R$.

\begin{remark}\label{rem:loctrnotlm}
Notice that every quasi-local transformation in $\Trnset{\rA_R\to\rA_R}_{Q\Reals}$ does not only represent a linear map on $\Cntset{\rA_R}_{Q\Reals}$, but a family of maps that represent the same transformation acting on $\Cntset{\rA_G}$ for every $R\subseteq G$. This is due to the isomorphism of $\Trnset{\rA_R\to\rA_R}_{Q\Reals}$ with the subalgebra $\Trnset{\rA^{(G)}_R\to\rA^{(G)}_R}_{Q\Reals}$ given in Eq.~\eqref{eq:isomqloctr}.
\end{remark}

The following result shows that $\Trnset{\rA^{(G)}_R\to\rA^{(G)}_R}_{Q\Reals}$ preserves the subspace $\Cntset{\rA^{(G)}_R}_{Q\Reals}$.

\begin{lemma}
Let $\tA\in\Trnset{\rA^{(G)}_R\to\rA^{(G)}_R}_{Q\Reals}$. Then
\begin{align*}
\tA^\dag\Cntset{\rA^{(G)}_R}_{Q\Reals}\subseteq\Cntset{\rA^{(G)}_R}_{Q\Reals}.
\end{align*}
\end{lemma}
\begin{proof}
Let $\tA_{nR_n}$ be a sequence in the class of $\tA$, with $R_n\subseteq R$ for every $n$. Let $a_{mS_m}$ be a sequence in the class of $a\in\Cntset{\rA^{(G)}_R}_{Q\Reals}$, with $S_m\subseteq R$ for all $m$. Then  $\tA^\dag a=[(\tA_n^\dag a_n)_{R_n\cup S_n}]$, and for every $n$, $R_n\cup S_n\subseteq R$. This implies that $\tA^\dag a\in\Cntset{\rA^{(G)}_R}_{Q\Reals}$.
\end{proof}
The following intuitive result states that an effect that is localised in the region $R\subseteq G$ is obtained by a quasi-local transformation that is localised in the same region. This has a straightforward proof that will be omitted.
\begin{lemma}\label{lem:qleftr}
Let $a\in\Cntset{\rA^{(G)}_R}_{Q*}$, with $*\in\{\mbox{``nothing''},1,+,\Reals\}$. Then
\begin{align*}
a=\tA^\dag e_G,
\end{align*}
with $\tA\in\Trnset{\rA^{(G)}_R\to\rA^{(G)}_R}_{Q*}$.
\end{lemma}

We conclude the section with a result that confirms, under restrictive hypotheses,  an obvious expectation.
\begin{lemma}\label{lem:norsupql}
Let $\tA\in\Trnset{\rA_G\to\rA_G}_{Q\Reals}$. Then $\normsup \tA=\inf J(\tA)$, where 
\begin{align*}
&J(\tA)\\
&\coloneqq \{\lambda\in\Reals\mid \exists\tC\in\Trnset{\rA_G\to\rA_G}_{Q1},\lambda\tC\pm\tA\succeq0\}.
\end{align*}
\end{lemma}
\begin{proof}
Let $\lambda\in J(\tA)$. Then there exists a quasi-local channel $\tC$ such that $\lambda\tC\pm\tA\succeq0$. By definition, this implies that $\lambda\tC\pm\tA=[\tB^\pm_{nR^\pm_n}]$ with $\tB^\pm\succeq0$. Defining $\tA_n\coloneqq 1/2(\tB^+_n-\tB^-_n)$ and $\tC_n\coloneqq 1/(2\lambda)(\tB^+_n+\tB^-_n)$, by construction one has
\begin{align*}
&[\tC_{nS^+_n}]=\tC,\quad
[\tA_{nS^-_n}]=\tA,
\end{align*}
where the domains $S^\pm_n$ are suitably defined. On the other hand, $\tC=[\tD_{nT_n}]$ for $\tD_{nT_n}\in\Trnset{\rA_{T_n}\to\rA_{T_n}}_1$. Thus, one has $\normsup{\tC_n-\tD_n}\leq\varepsilon$, taking suitable care in  defining the domain $U_n\coloneqq S^+_n\cup T_n$ of $\tC_n-\tD_n$. In other terms, one can find a channel $\tF_n$ such that
\begin{align*}
&\varepsilon\tF_n+\tD_n-\tC_n\succeq0,\\
&\varepsilon\tF_n-\tD_n +\tC_n\succeq0.
\end{align*}
In turn, this implies that
\begin{align*}
&(1+\varepsilon) e_{U_n}-c_n\succeq0,\\
&(\varepsilon-1) e_{U_n}+c_n\succeq0,
\end{align*}
where $c_n\coloneqq \tC_n^\dag e_{U_n}$.
Considering the first of the above relations, we can define
\begin{align*}
\tG_n\coloneqq \frac1{1+\varepsilon} \tC_n+|\rho_n)(d_n|,
\end{align*}
where $d_n\coloneqq e_{U_n}-1/(1+\varepsilon) c_n\succeq0$ and $\rho_n$ is an arbitrary state in $\Stset{\rA_{U_n}}_1$. By our assumptions, since $\tG_n\succeq0$ and $\tG_n^\dag e_{U_n}=e_{U_n}$, $\tG_n$ is a channel. Moreover, by construction we have that
\begin{align*}
\lambda(1+\varepsilon)\tG_n\pm\tA_n=\tB^\pm_n+\lambda(1+\varepsilon)|\rho_n)(d_n|\succeq0,
\end{align*}
thus $\lambda(1+\varepsilon)\in J(\tA_n)$, i.e.~$\lambda(1+\varepsilon)\geq\normsup{\tA_n}$. In the limit, we then have $\lambda\geq\normsup{\tA}$. On the other hand, if $\lambda\not\in J(\tA)$, for every $\tC\in\Trnset{\rA_G\to\rA_G}_{Q1}$ one has
\begin{align}
\lambda\tC+\tA\not\succeq0\ \vee\ \lambda\tC-\tA\not\succeq0.
\end{align}
Since the sequence $\lambda\tC\pm\tA_n$ converges to $\lambda\tC\pm\tA$ for every sequence $\tA_n$ converging to $\tA$, and the positive cone is closed, for every $\tC$ and every $\tA_n$ as above, there must be $n_0$ such that for $n\geq n_0$ 
\begin{align}
\lambda\tC+\tA_n\not\succ0\ \vee\ \lambda\tC-\tA_n\not\succ0.
\end{align}
This holds, in particular, for every local $\tC_n$ with the same domain as $\tA_n$, which implies $\lambda\not\in J(\tA_n)$. As a consequence, $\lambda<\normsup{\tA_n}$, and finally $\lambda\leq\normsup\tA$. It is now easy to conclude that $\inf J(\tA)=\normsup{\tA}$.
\end{proof}

We now prove that the infimum defining the sup norm is actually a minimum, analogously to the case of effects. The proof in this case is less straightforward, and we first need the following lemma.
\begin{lemma}\label{lem:closev}
Given $\tA\in\Trnset{\rA_G\to\rA_G}_{Q\Reals}$, let $\set S^\lambda_\tA\subseteq\Trnset{\rA_G\to\rA_G}_{Q1}$ 
be the set of channels $\tC$ such that  
$\lambda\tC\pm\tA\succeq0$. Then $\set S^\lambda_\tA$ is closed in the sup-norm.
\end{lemma}
\begin{proof}
The empty set is closed, so we will consider the case where $\set S^\lambda_\tA$ is not empty. Let $\{\tC_n\}_{n\in\Nats}$ be a Cauchy sequence in $\set S^\lambda_\tA$. Then $\tD^\pm_n\coloneqq \lambda\tC_n\pm\tA\succeq0$ are both Cauchy sequences, that by definition converge to $\lambda\tC\pm\tA\succeq0$. By proposition~\ref{prop:closcon}, $\tC\in\Trnset{\rA_G\to\rA_G}_{Q1}$.
\end{proof}

\begin{proposition}\label{prop:infmint}
Let $\tA\in\Trnset{\rA_G\to\rA_G}_{Q\Reals}$. Then there exists a quasi local channel $\tC$ such that
\begin{align}
\normsup\tA\tC\pm\tA\succeq0.
\end{align}
\end{proposition}
\begin{proof}
By lemma~\ref{lem:norsupql}, for every $\varepsilon>0$ there exists $\tC_\varepsilon$ such that $(\normsup\tA+\varepsilon)\tC_\varepsilon\pm\tA\succeq0$.
Let us then consider the non empty sets $\set S_\varepsilon\coloneqq \set S^{\normsup\tA+\varepsilon}_\tA$, that are closed by lemma~\ref{lem:closev}.
It is straightforward to verify that for $\delta<\varepsilon$ it is $S_\delta\subseteq\set S_\varepsilon$. Let $d_\varepsilon$ be the diameter of $\set S_\varepsilon$, namely $d_\varepsilon\coloneqq \sup\{\normsup{\tC_1-\tC_2}\mid\tC_1,\tC_2\in\set S_\varepsilon\}$. Notice that $0\leq d_\varepsilon\leq 2$. Being $d_\varepsilon$ a non-decreasing function of $\varepsilon$, $d_0=\lim_{\varepsilon\to0^+}$ exists. The limit $d_0$ is the diameter of
\begin{align*}
\set S_0\coloneqq \bigcap_{\varepsilon>0}\set S_\varepsilon.
\end{align*}
If $d_0>0$, then $\set S_0$ is clearly not empty, and each of its elements is a channel $\tC$ such that $\normsup\tA\tC\pm\tA\succeq0$. If $d_0=0$, then by Cantor's intersection theorem for complete metric spaces 
the set $\set S_0$ is a singleton $\set S_0=\{\tC_\infty\}$, with $\normsup\tA\tC_\infty\pm\tA\succeq0$. 
\end{proof}

We finally prove a crucial result, that is the converse of lemma~\ref{lem:cnestr}. This shows that every quasi-local event is an element of a quasi-local test.

\begin{proposition}\label{prop:chandomeff}
Let $\tA\in\Trnset{\rA_G\to\rA_G}_{Q+}$. Then $\tA\in\Trnset{\rA_G\to\rA_G}_{Q}$ if and only if there exists $\tC\in\Trnset{\rA_G\to\rA_G}_{Q1}$ such that $\tC-\tA\in\Trnset{\rA_G\to\rA_G}_{Q+}$.
\end{proposition}
\begin{proof}
Sufficiency is proved in lemma~\ref{lem:cnestr}. Let us then prove that if $\tA\in\Trnset{\rA_G\to\rA_G}_{Q}$ then there exists $\tC\in\Trnset{\rA_G\to\rA_G}_{Q1}$ such that $\tC-\tA\succeq0$. First of all, notice that $\tA=[\tA_{nR_n}]$ with $\tA_n\in\Trnset{\rA_{R_n}\to\rA_{R_n}}$, and then by
\begin{align*}
\normsup{\tA}&=\lim_{n\to\infty}\normsup{\tA_n}\\
&=\lim_{n\to\infty}\normsup{\tA_{nR_n}^\dag e_G}\leq1.
\end{align*}
By proposition~\ref{prop:infmint}, there exists $\tC\in\Trnset{\rA_G\to\rA_G}_{Q1}$ such that
\begin{align*}
&\tC-\tA\succeq\normsup\tA\tC-\tA\succeq0.&&\qedhere
\end{align*}
\end{proof}

\section{Global update rules}\label{sec:UR}

We have now set the theoretical background needed for the
definition of a cellular automaton. In the present
section we give the definition of an update rule (UR),
that along with the system $\rA_G$---the array of
cells, whose definition has been analyzed in detail in
the previous sections---will provide the backbone of
the notion of a cellular automaton. We then introduce
the notion of a global update rule (GUR), as a family
of update rules satisfying suitable admissibility conditions
in order to represent a local action when extended to composite 
systems $\rA_G\rC$. A GUR thus
describes the evolution occurring when we apply the
global update rule on a joint state of $\rA_G$ and 
$\rC$, leaving $\rC$ unaffected. We then analyse the
main features of GURs, and in particular we focus on
the {\em causal influence} relation given by a global
update rule, and the {\em block decomposition}
that allows one to calculate the action of a GUR on
local effects using only local transformations. 

\begin{remark}\label{rem:nontriv}
In the remainder we will assume that for every $g\in G$ the system $\rA_g$ is not trivial, 
namely $\rA_g\not\simeq\rI$.
\end{remark}

First of all let $\tV^\dag$ be a bounded automorphism of the space $\Cntset{\rA_G}_{Q\Reals}$ of quasi-local effects. Then, with a notation that is reminiscent of the one adopted for quasi-local transformations, we will denote its action as
\begin{align*}
\tV^\dag:\Cntset{\rA_G}_{Q\Reals}\to\Cntset{\rA_G}_{Q\Reals}::a\mapsto\tV^\dag a.
\end{align*}
The action on effects allows us to define the {\em dual map} that acts on the space of extended states, as follows
\begin{align}
(a|\hat\tV\rho)\coloneqq (\tV^\dag a|\rho),\quad\forall a\in\Cntset{\rA_G}_{Q\Reals}.
\label{eq:adjaut}
\end{align}
One can easily verify that $\hat\tV$ is linear and bounded, thanks to boundedness of the map $\tV^\dag$. Unlike the usual notion of a map in quantum theory or theories with local discriminability, however, the action of $\hat\tV$ or $\tV^\dag$ is not sufficient to identify uniquely the corresponding transformation. In particular, those actions are not sufficient to determine that of $\tV\otimes \tI_\rC$, when system $\rA_G$ is considered as a part of the composite system $\rA_G\rC$. For this reason we need to define an automorphic transformation of $\Cntset{\rA_G}_{Q\Reals}$ with care. 
One further delicate issue is the following. When one conjugates a quasi-local 
transformation $\tA$ with an automorphism $\tV^\dag$, obtianing 
$\tA'=\tV^{-1\dag}\tA\tV^\dag$, the latter is a linear bounded map on 
$\Cntset{\rA_G}_{Q\Reals}$, but there is no guarantee that it is still an element of the 
quasi-local algebra $\Trnset{\rA_G\to\rA_G}_{Q\Reals}$, which is surely a desideratum for 
candidate cellular automata.
In the remainder, when we write $\rA_G\rC$ we mean to deal with a system $\rA_{G'}$ with $G'=G\cup h_0$ and $\rA_{h_0}\cong\rC$. In view of the above observations, we introduce the following notion
\begin{definition}[Automorphic family]\label{def:autfam}
An {\em automorphic family} of maps on $\Cntset{\rA_G}_{Q\Reals}$ is a collection of automorphisms $\tV^\dag_\rC$ of $\Cntsetcomp{\rA_{G}}{\rC}_{Q\Reals}$, one for every $\rC\in\Elety(\Theta)$, with the properties i) for every $\tA\in\Trnset{\rA_C\to\rA_C}_\Reals$ one has $\tV^{-1\dag}_\rC\tA_{h_0}^\dag\tV^\dag_\rC=\tA_{h_0}^\dag$; ii) for fixed $\rC_0$ and every choice of system $\rC_1$, setting $\rC=\rC_0\rC_1$ and $\rA_{G'}\coloneqq \rA_G\rC_0$, and for every $\tA\in\Trnset{\rA_{G'}\to\rA_{G'}}_{Q\Reals}$, there exist $\tA',\tA''\in\Trnset{\rA_{G'}\to\rA_{G'}}_{Q\Reals}$ such that
\begin{align}
&\tV_\rC^{-1\dag}\tA_{G'}^\dag\tV_{\rC}^\dag=\tA'^\dag_{G'},\label{eq:injauf}\\
&\tA^\dag_{G'}=\tV_\rC^{\dag}\tA''^\dag_{G'}\tV_{\rC}^{-1\dag},\label{eq:surjauf}
\end{align}
where $\tA'^\dag=(\tV_{\rC_0}^{-1\dag}\tA^\dag\tV_{\rC_0}^\dag)$, and $\tA''^\dag=(\tV_{\rC_0}^{-1\dag}\tA^\dag\tV^{\dag}_{\rC_0})$
%
%
%
%
%
%
%
%
%
%
%
\end{definition}
We remark that, as a special case of item ii) in definition~\ref{def:autfam}, taking $\rC_0\equiv\rI$ and $\rC=\rC_1$, for every choice of system $\rC$, and every $\tA\in\Trnset{\rA_{G}}_{Q\Reals}$, one has
\begin{align*}
&\tV_\rC^{-1\dag}\tA_{G}^\dag\tV_{\rC}^\dag=(\tV_{\rI}^{-1\dag}\tA^\dag_{G}\tV_{\rI}^\dag)_{G},\\
&\tV_\rC^{\dag}\tA_{G}^\dag\tV_{\rC}^{-1\dag}=(\tV_{\rI}^{\dag}\tA^\dag_{G}\tV_{\rI}^{-1\dag})_{G}.
\end{align*}

%
When referring to an automorphic family we will use the shorthand $\tV^\dag$. One has to remind, however, that this symbol refers to a family of automorphisms. In particular, the definition is meant to allow for formulas such as
$\tV\tA\tV^{-1}$, $\tV\Trnset{\rA_G\to\rA_G}_{\$*}\tV^{-1}$, for $\$=Q,L$ and $*=\Reals,+,1$ or nothing. The meaning of the two expressions above is given by the following definitions:
\begin{align}
&\tV\tA\tV^{-1}\coloneqq\{\hat\tV_{\rC_0\rC}(\hat\tA\otimes\hat\tI_\rC)\hat\tV^{-1}_{\rC_0\rC}\mid\rC\in\Elety(\Theta)\},\label{eq:locaufa}\\
&\tV\Trnset{\rA_{G'}\to\rA_{G'}}_{\$*}\tV^{-1}\coloneqq\nonumber\\
&\qquad\{\tV\tA\tV^{-1}\mid\tA\in\Trnset{\rA_{G'}\to\rA_{G'}}_{\$*}\}.
\end{align}
Notice that by definition of an automorphic family and by Eq.~\eqref{eq:locaufa}, one has
\begin{align}
\tV\tA\tV^{-1}=\{(\hat\tV_{\rC_0}\hat\tA\hat\tV_{\rC_0}^{-1})\otimes\hat\tI_\rC\mid\rC\in\Elety(\Theta)\},\label{eq:invauf}
\end{align}
namely an automorphic family maps $\Trnset{\rA_{G'}\to\rA_{G'}}_{Q\Reals}$ onto itself surjectively. In particular, this is true of $\Trnset{\rA_{G}\to\rA_{G}}_{Q\Reals}$. This implies that for $\tA\in\Trnset{\rA_G\rC\to\rA_G\rC}_{Q\Reals}$, $\tV\tA\tV^{-1}$ represents a quasi-local transformation in $\Trnset{\rA_G\rC\to\rA_G\rC}_{Q\Reals}$.

An automorphic family is defined as to satisfy some necessary conditions that we require for the representatives of a transformation of the form $\tV\otimes\tI_\rC$. However, the definition of an automorphic family is not sufficient to ensure consistency. This fact must be taken into account in the next subsection. Sufficient conditions will be only provided through the notion of admissibility. 

\subsection{Update rule}

We can now use the notion of automorphic family to introduce update rules. 
An update rule
is defined as a family of automorphisms, one for each extended system $\rA_G\rC$, with the 
constraint that the rule must act reversibly on the set of states and, by conjugation, on 
the cone of positive local transformations.
\begin{definition}[Update rule]
An {\em update rule} (UR) is a triple $(G,\rA,\tV^\dag)$, where $\tV^\dag$ is an automorphic family of isometric maps on $\Cntset{\rA_G}_{Q\Reals}$ 
such that 
\begin{enumerate}
\item
the maps $\hat\tV_\rC$ leave $\Stset{\rA_{G}\rC}$ invariant, i.e.
\begin{align}
\hat\tV_\rC\Stset{\rA_{G}\rC}=\Stset{\rA_{G}\rC};
\label{eq:inveff}
\end{align}
\item the map ${\tV}\cdot{\tV^{-1}}$ 
leaves the cones $\Trnset{\rA_{G}\rC\to\rA_{G}\rC}_{L+}$ 
invariant, i.e.~for every $\tA\in\Trnset{\rA_{G}\rC\to\rA_{G}\rC}_{L+}$ there are 
$\tA',\tA''\in\Trnset{\rA_{G}\rC\to\rA_{G}\rC}_{L+}$ such that
\begin{align}
&\tV\tA\tV^{-1}=\tA',\label{eq:inj}\\
&\tA=\tV\tA''\tV^{-1}.\label{eq:surj}
\end{align}
\end{enumerate}
\end{definition}

The first result is that 
%
the inverse of a UR is a UR.
\begin{lemma}\label{lem:invauf}
Let $(G,\rA,\tV^\dag)$ be a UR. Then $(G,\rA,\tV^{-1\dag})$ is a UR.
\end{lemma}
\begin{proof}
In the first place, the family $\tV_\rC^{-1}$ is an automorphic family: items i) and ii) in definition~\ref{def:autfam} are trivially proved. 
Now, if $\normsup{\tV^\dag_\rC a}=\normsup a$ for every $a\in\Cntsetcomp{\rA_{G}}{\rC}_{Q\Reals}$, then $\normsup{{\tV^{-1\dag}_\rC} a}=\normsup{\tV_\rC^\dag{\tV_\rC^{-1\dag}} a}=\normsup a$, which proves isometricity of $\tV_\rC^{-1\dag}$. From the definition in Eq.~\eqref{eq:adjaut} one can easily verify that $\hat{(\tV_\rC^{-1})}=(\hat\tV_\rC)^{-1}$.
Now, it is a straightforward observation that by Eq.~\eqref{eq:inveff}, $\hat\tV_\rC^{-1}\Stset{\rA_{G}\rC}=\Stset{\rA_{G}\rC}$. 
Finally, 
multiplying both sides of Eq.~\eqref{eq:inj} by $\tV^{-1}$ to the left and by $\tV$ to the right, for every $\tA\in\Trnset{\rA_{G}\rC\to\rA_{G}\rC}_{L+}$ one has 
\begin{align*}
&\tA=\tV^{-1}\tA'\tV,
\end{align*}
which expresses the same condition as in Eq.~\eqref{eq:surj} for $\tV^{-1}$, and similarly, multiplying both sides of Eq.~\eqref{eq:surj} by $\tV^{-1}$ to the left and by $\tV$ to the right, one obtains the same condition as in Eq.~\eqref{eq:inj} for $\tV^{-1}$.
\end{proof}


In the following we will often use $\tV^\dag$ to denote a UR, when $G$ and $\rA_G$ are clear from the context. The symbols $\hat\tV$ and $\tV^\dag$ will be often used instead of $\hat\tV_\rC$ and $\tV^\dag_\rC$, respectively.
Accordingly, the symbols $a,b,\ldots$ and $\tA,\tB,\ldots$ will denote $a\otimes e_\rC,b\otimes e_\rC,\ldots$ and $\tA\otimes\tI_\rC,\tB\otimes\tI_\rC,\ldots$, respectively. Finally, we will sometimes write $G'$ for $G\cup h_0$ with $\rA_{h_0}\cong\rC$, so that $\rA_{G'}=\rA_G\rC$.

The next result states that an update rule leaves 
the cone $\Trnset{\rA^{(G\cup h_0)}_{G}\to\rA^{(G\cup h_0)}_{G}}_{L+}$ invariant.

\begin{lemma}\label{lem:inveffur}
Let $(G,\rA,\tV^\dag)$ be a UR, and define $\rA_{G'}\coloneqq\rA_G\rC$ for arbitrary $\rC\in\Elety(\Theta)$. Then the map ${\tV}\cdot{\tV}^{-1}$ 
leaves the cone $\Trnset{\rA^{(G')}_{G}\to\rA^{(G')}_{G}}_{L+}$ 
invariant, namely for $\tA\in\Trnset{\rA^{(G')}_G\to\rA^{(G')}_G}_{L+}$ Eqs.~\eqref{eq:inj} and~\eqref{eq:surj} hold with $\tA',\tA''\in\Trnset{\rA^{(G')}_G\to\rA^{(G')}_G}_{L+}$.
%
\end{lemma}
\begin{proof}
Let $\tA\in\Trnset{\rA^{G'}_G\to\rA^{G'}_G}_{L+}\subseteq\Trnset{\rA_{G'}\to\rA_{G'}}_{L+}$. By definition of UR
\begin{align*}
\tV\tA_G\tV^{-1}=\tA'_G\in\Trnset{\rA_{G'}\to\rA_{G'}}_{L+}. 
\end{align*}
Moreover, by definition of an automorphic family of maps and Eq.~\eqref{eq:invauf}, we have 
\begin{align*}
&\tA'\in\Trnset{\rA^{(G')}_G\to\rA^{(G')}_G}_{L\Reals}.
\end{align*}
The same holds exchanging $\tV^{-1}$ and $\tV$.
\end{proof}

As a consequence of isometricity on $\Cntsetcomp{\rA_{G}}{\rC}_{Q\Reals}$, a UR $\tV^\dag$ has a dual $\hat\tV$ that acts isometrically on $\Stset{\rA_{G}\rC}_\Reals$.

\begin{lemma}
Let $\tV^\dag$ be a UR. The maps $\hat\tV_\rC$ on $\Stset{\rA_{G'}}_\Reals$ are isometric.
\end{lemma}
\begin{proof}
Let $\rho\in\Stset{\rA_{G}\rC}_\Reals$, and remind that
\begin{align*}
\normst\rho=\sup_{\normsup a=1}|(a|\rho)|.
\end{align*}
Then we have
\begin{align*}
\normst{\hat\tV_\rC\rho}&=\sup_{\normsup a=1}|(a|\hat\tV_\rC\rho)|=\sup_{\normsup a=1}|(\tV_\rC^\dag a|\rho)|\\
&=\sup_{\normsup {a'}=1}|(a'|\rho)|=\normst\rho,
\end{align*}
thanks to isometricity and invertibility of $\tV_\rC^\dag$. 
\end{proof}
\begin{corollary}\label{cor:detst}
Let $\tV^\dag$ be a UR. Then $\hat\tV_\rC\Stset{\rA_{G}\rC}_1=\Stset{\rA_{G}\rC}_1$
\end{corollary}
\begin{proof}
By eqs.~\eqref{eq:inveff} and ~\eqref{eq:defnorst} and isometricity of $\hat\tV_\rC$, one has $\hat\tV_\rC\Stset{\rA_{G}\rC}_1\subseteq\Stset{\rA_{G}\rC}_1$. The same is true for $\hat\tV^{-1}_\rC$, thus $\hat\tV_\rC\Stset{\rA_{G}\rC}_1=\Stset{\rA_{G}\rC}_1$.
\end{proof}

The next results will allow us to conclude that a UR preserves $\Cntset{\rA_G}_Q$. We start proving that an update rule $\tV^\dag$ preserves the unique deterministic effects $e_{G'}$.
\begin{lemma}\label{lem:veg}
If $\tV^\dag$ is a UR, then $\tV^\dag e_{G'}=e_{G'}$, for $\rA_{G'}\coloneqq\rA_G\rC$ and arbitrary $\rC\in\Elety(\Theta)$.
\end{lemma}
\begin{proof}
By definition of a UR, for every $\rho\in\Stset{\rA_{G'}}$ one has $\hat\tV\rho\in\Stset{\rA_{G'}}$, 
and lemma~\ref{lem:normonpos}, gives for every $\rho\in\Stset{\rA_{G'}}$
\begin{align*}
(\tV^\dag e_{G'}|\rho)=(e_{G'}|\hat\tV\rho)=\normst{\hat\tV\rho}=\normst{\rho}=(e_{G'}|\rho).
\end{align*}
Thanks to theorem~\ref{th:statesepeff}, we then have $\tV^\dag e_{G'}=e_{G'}$. 
\end{proof}

\begin{corollary}
Let $\tV^\dag$ be a UR. Then $\tV^\dag a_{h_0}=a_{h_0}$ for all $a_{h_0}\in\Trnset{\rA^{(G')}_{h_0}\to\rA^{(G')}_{h_0}}_{Q\Reals}$.
\end{corollary}
\begin{proof}
By proposition~\ref{prop:qlefftrans}, $a_{h_0}=\tA^\dag_{h_0}e_{G'}$. Thus, by definition~\ref{def:autfam} and lemmas~\ref{lem:veg} and~\ref{lem:inv}, we have
\begin{align*}
\tV^\dag a_{h_0}&=\tV^\dag\tA^\dag_{h_0}\tV^{-1\dag}e_{G'}\\
&=\tA^\dag_{h_0}e_{G'}=a_{h_0}.&&\qedhere
\end{align*}
\end{proof}

We now prove that the cones of effects $\Cntset{\rA_{G'}}_{L+}$, $\Cntset{\rA_G}_{L+}$, $\Cntset{\rA_{G'}}_{Q+}$ and $\Cntset{\rA_G}_{Q+}$ are invariant under the action of any UR.

\begin{lemma}\label{lem:inveff}
Let $\tV^\dag$ be a UR. Then $\tV^\dag\Cntset{\rA_{G'}}_{L+}=\Cntset{\rA_{G'}}_{L+}$, and $\tV^\dag\Cntset{\rA_{G'}}_{Q+}=\Cntset{\rA_{G'}}_{Q+}$. In particular, for $\rC=\rI$ it is $\tV^\dag\Cntset{\rA_{G}}_{Q+}=\Cntset{\rA_{G}}_{Q+}$
\end{lemma}
\begin{proof}
Let $a\in\Cntset{\rA_{G'}}_{L+}$. By proposition~\ref{prop:qlefftrans}, there exists $\tA\in\Trnset{\rA_{G'}\to\rA_{G'}}_{L+}$ such that $a=\tA^\dag e_{G'}$. Then
\begin{align*}
\tV^\dag a&=\tV^\dag\tA^\dag e_{G'}\\
&=\tV^\dag\tA^\dag\tV^{-1\dag}\tV^\dag e_{G'}\\
&=\tV^\dag\tA^\dag\tV^{-1\dag} e_{G'}\\
&=\tA^{\prime\dag} e_{G'}=a'.
\end{align*}
We observe that, by definition of UR and by lemma~\ref{lem:inveffur}, $\tV\cdot\tV^{-1}$ preserves the cones $\Trnset{\rA_{G'}\to\rA_{G'}}_{L+}$. 
Thus, $a'\in\Cntset{\rA_{G'}}_{L+}$, and $\tV^\dag\Cntset{\rA_{G'}}_{L+}\subseteq\Cntset{\rA_{G'}}_{L+}$. Since also $\tV^{-1\dag}$ is a UR, we have $\tV^\dag\Cntset{\rA_{G'}}_{L+}=\Cntset{\rA_{G'}}_{L+}$. Now, isometricity grants that Cauchy sequences of local effects in $\Cntset{\rA_{G'}}_{L+}$ are mapped to Cauchy sequences in $\Cntset{\rA_{G'}}_{L+}$, and that $\lim_{n\to\infty}\tV^\dag a_n=\tV^\dag\lim_{n\to\infty}a_n$. Finally, by proposition~\ref{prop:closcon}, $\tV^\dag\Cntset{\rA_{G'}}_{Q+}\subseteq\Cntset{\rA_{G'}}_{Q+}$.
%
%
Since also $\tV^{-1\dag} $ is a UR, we conclude that 
${\tV^\dag}\Cntset{\rA_{G'}}_{Q+}=\Cntset{\rA_{G'}}_{Q+}$.
\end{proof}

\begin{proposition}
Let $\tV^\dag$ be a UR. Then $\tV^\dag\Cntset{\rA_{G'}}_Q=\Cntset{\rA_{G'}}_Q$. In particular, for $\rC=\rI$, $\tV^\dag\Cntset{\rA_{G}}_Q=\Cntset{\rA_{G}}_Q$
\end{proposition}
\begin{proof}
Let $a\in\Cntset{\rA_{G'}}_Q$. Then $e_{G'}-a\in\Cntset{\rA_{G'}}_{Q+}$, by lemma~\ref{lem:cnessef}, and thus, by virtue of lemmas~\ref{lem:veg} and~\ref{lem:inveff}, $\tV^\dag a,e_{G'}-\tV^\dag a\in\Cntset{\rA_{G'}}_{Q+}$. Again by lemma~\ref{lem:cnessef}, this implies that $\tV^\dag a\in\Cntset{\rA_{G'}}_{Q}$, and thus $\tV^\dag\Cntset{\rA_{G'}}_{Q}\subseteq\Cntset{\rA_{G'}}_{Q}$. Finally, since also $\tV^{-1\dag} $ is a UR, the thesis follows.
\end{proof}

Let now $\tV^\dag$ be an update rule. By our definition, the map $\tV\cdot\tV^{-1}$ preserves the cones $\Trnset{\rA_{G'}\to\rA_{G'}}_{L+}$. We will now prove that $\tV\cdot\tV^{-1}$ also preserves the cones $\Trnset{\rA_{G'}\to\rA_{G'}}_{Q+}$, as well as the sets of transformations $\Trnset{\rA_{G'}\to\rA_{G'}}_{L}$ and $\Trnset{\rA_{G'}\to\rA_{G'}}_{Q}$.

We start with the following lemma
\begin{lemma}\label{lem:locchan}
Let $\tV^\dag$ be a UR. Then $\tV\cdot\tV^{-1}$ leaves the sets $\Trnset{\rA_{G'}\to\rA_{G'}}_{L1}$ invariant, namely for $\tA\in\Trnset{\rA_{G'}\to\rA_{G'}}_{L1}$ Eqs.~\eqref{eq:inj} and~\eqref{eq:surj} hold, with $\tA',\tA''\in\Trnset{\rA_{G'}\to\rA_{G'}}_{L1}$. In particular, for $\rC=\rI$ the thesis holds with $G'=G$.
\end{lemma}
\begin{proof}
Let $\tC\in\Trnset{\rA_{G'}\to\rA_{G'}}_{L1}$. By definition and by lemma~\ref{lem:inveffur}, $\tV\tC\tV^{-1}=\tC'$, and $\tC=\tV\tC''\tV^{-1}$, for $\tC',\tC''\in\Trnset{\rA_{G'}\to\rA_{G'}}_{L+}$. 
By lemmas~\ref{lem:invauf} and~\ref{lem:veg},
\begin{align*}
&\tC'^\dag e_{G'}=\tV^{\dag}\tC^\dag\tV^{-1\dag}e_{G'}=e_{G'},\\
&\tC''^\dag e_{G'}=\tV^{-1\dag}\tC^\dag\tV^{\dag}e_{G'}=e_{G'},
\end{align*}
and due to lemma~\ref{lem:cnesschan}, $\tC',\tC''\in\Trnset{\rA_{G'}\to\rA_{G'}}_{L1}$. 
\end{proof}

Given a UR $\tV^\dag$, the above result allows us to prove that $\tV\cdot\tV^{-1}$ maps local transformations to local transformations, and is isometric on $\Trnset{\rA_{G'}\to\rA_{G'}}_{L\Reals}$.
\begin{lemma}\label{lem:isomloc}
Let $\tV^\dag$ be a UR, and $\tA\in\Trnset{\rA_{G'}\to\rA_{G'}}_{L\Reals}$. Then $\tV\Trnset{\rA_{G'}\to\rA_{G'}}_{L\Reals}\tV^{-1}=\Trnset{\rA_{G'}\to\rA_{G'}}_{L\Reals}$, and
\begin{align}
\normsup{\tV\tA\tV^{-1}}=\normsup{\tA}.
\end{align}
\end{lemma}
\begin{proof}
By proposition~\ref{prop:infmint}, there exists $\tC\in\Trnset{\rA_{G'}\to\rA_{G'}}_{L1}$ such that
\begin{align*}
\normsup{\tA}\tC\pm\tA\succeq0.
\end{align*}
By definition of a UR, we then have
\begin{align}
\normsup{\tA}\tV\tC\tV^{-1}\pm\tV\tA\tV^{-1}\succeq0,
\label{eq:seq}
\end{align}
and being $\tV\tC\tV^{-1}\in\Trnset{\rA_{G'}\to\rA_{G'}}_{L1}\subseteq\Trnset{\rA_{G'}\to\rA_{G'}}_{L\Reals}$, it follows that $\tV\tA\tV^{-1}\in\Trnset{\rA_{G'}\to\rA_{G'}}_{L\Reals}$. Consequently, we have $\tV\Trnset{\rA_{G'}\to\rA_{G'}}_{L\Reals}\tV^{-1}\subseteq\Trnset{\rA_{G'}\to\rA_{G'}}_{L\Reals}$. Finally, since $\tV^{-1\dag}$ is a UR, we have that $\tV\Trnset{\rA_{G'}\to\rA_{G'}}_{L\Reals}\tV^{-1}=\Trnset{\rA_{G'}\to\rA_{G'}}_{L\Reals}$. Moreover, by lemma~\ref{lem:locchan}, Eq.~\eqref{eq:seq} implies that $\normsup\tA\in J(\tV\tA\tV^{-1})$. Thus $\normsup{\tV\tA\tV^{-1}}\leq\normsup{\tA}$. Now, being $\tV^{-1\dag}$ a UR, the equality $\normsup{\tV\tA\tV^{-1}}=\normsup{\tA}$ follows.
\end{proof}

\begin{lemma}\label{lem:vqloc}
Let $\tV^\dag$ be a UR. Then $\tV\Trnset{\rA_{G'}\to\rA_{G'}}_{Q\Reals}\tV^{-1}=\Trnset{\rA_{G'}\to\rA_{G'}}_{Q\Reals}$, and
\begin{align}
&\lim_{n\to\infty}\tV\tA_n\tV^{-1}=\tV(\lim_{n\to\infty}\tA_n)\tV^{-1}.\label{eq:vqloc}
\end{align}
\end{lemma}
\begin{proof}
Since $\tV\cdot\tV^{-1}$ is isometric and surjective on the dense submanifolds $\Trnset{\rA_{G'}\to\rA_{G'}}_{L\Reals}$, by lemma~\ref{lem:isomloc} it maps Cauchy sequences to Cauchy sequences, and their equivalence classes to equivalence classes, thus it maps $\Trnset{\rA_{G'}\to\rA_{G'}}_{Q\Reals}$ to itself. Surjectivity follows from surjectivity on $\Trnset{\rA_{G'}\to\rA_{G'}}_{L\Reals}$. Eq.~\eqref{eq:vqloc} can be independently checked applying both sides to arbitrary $a\in\Cntset{\rA_{G'}}_{Q\Reals}$.
\end{proof}

\begin{lemma}\label{lem:invcon}
Let $\tV^\dag$ be a UR. Then $\tV\Trnset{\rA_{G'}\to\rA_{G'}}_{Q1}\tV^{-1}=\Trnset{\rA_{G'}\to\rA_{G'}}_{Q1}$.
\end{lemma}

\begin{proof}
Let $\{\tC_n\}_{n\in\Nats}$ be a Cauchy sequence in $\Trnset{\rA_{G'}\to\rA_{G'}}_{L1}$. Then by definition of UR and lemma~\ref{lem:vqloc} we have that $\{\tV\tC_n\tV^{-1}\}_{n\in\Nats}$ is a Cauchy sequence in $\Trnset{\rA_{G'}\to\rA_{G'}}_{L1}$, and 
its limit $\tV\tC\tV^{-1}$ is in $\Trnset{\rA_{G'}\to\rA_{G'}}_{Q1}$.
%
%
%
\end{proof}

The last results finally allow us to prove that for a UR $\tV^\dag$, the map $\tV\cdot\tV^{-1}$ preserves the cones $\Trnset{\rA_{G'}\to\rA_{G'}}_{Q+}$.

\begin{lemma}\label{lem:prescon}
Let $\tV^\dag$ be a UR. Then $\tV\Trnset{\rA_{G'}\to\rA_{G'}}_{Q+}\tV^{-1}=\Trnset{\rA_{G'}\to\rA_{G'}}_{Q+}$.
\end{lemma}
\begin{proof}
By lemmas~\ref{lem:inveffur} and~\ref{lem:vqloc}, along with the definition of a UR, $\tV\cdot\tV^{-1}$ maps Cauchy sequences in $\Trnset{\rA_{G'}\to\rA_{G'}}_{L+}$ to Cauchy sequences in $\Trnset{\rA_{G'}\to\rA_{G'}}_{L+}$, thus $\tV\Trnset{\rA_{G'}\to\rA_{G'}}_{Q+}\tV^{-1}\subseteq \Trnset{\rA_{G'}\to\rA_{G'}}_{Q+}$. Since $\tV^{-1\dag}$ is a UR, the thesis follows.
\end{proof}

Next, we show that for a UR $\tV^\dag$, $\tV^\dag\Cntset{\rA_{G'}}_{L\Reals}=\Cntset{\rA_{G'}}_{L\Reals}$.
\begin{lemma}\label{lem:urloc}
Let $\tV^\dag$ be a UR. Then $\tV^\dag\Cntset{\rA_{G'}}_{L\Reals}=\Cntset{\rA_{G'}}_{L\Reals}$.
\end{lemma}
\begin{proof}
Let $a_R\in\Cntset{\rA_{G'}}_{L\Reals}$. Then by proposition~\ref{prop:qlefftrans}
\begin{align*}
a_R=\tA_R^\dag e_{G'},
\end{align*}
for some $\tA_R\in\Trnset{\rA_{G'}\to\rA_{G'}}_{L\Reals}$. Now, we have 
\begin{align*}
\tV^\dag a_R=\tV^\dag\tA_R^\dag e_{G'}=\tV^\dag\tA_R^\dag\tV^{-1\dag}e_{G'},
\end{align*}
and by lemma~\ref{lem:isomloc} $\tV^\dag\tA_R^\dag\tV^{-1\dag}=\tA'^\dag_{R'}$ with $\tA'_{R'}\in\Trnset{\rA_{G'}\to\rA_{G'}}_{L\Reals}$. Then $\tV^\dag a_R=a'_{R'}$, with $R'\in\reg {G'}$. As to surjectivity, it can be straightforwardly proved exploiting lemma~\ref{lem:invauf}.
\end{proof}

Finally, for a UR $\tV^\dag$, the map $\tV\cdot\tV^{-1}$ preserves the set of quasi-local transformations.
\begin{lemma}
Let $\tV^\dag$ be a UR. Then one has
\begin{align}
&\tV\Trnset{\rA_{G'}\to\rA_{G'}}_{Q}\tV^{-1}=\Trnset{\rA_{G'}\to\rA_{G'}}_{Q},\\
&\tV\Trnset{\rA_{G'}\to\rA_{G'}}_{L}\tV^{-1}=\Trnset{\rA_{G'}\to\rA_{G'}}_{L}.
\end{align}
\end{lemma}
\begin{proof}
Let $\tA\in\Trnset{\rA_{G'}\to\rA_{G'}}_{Q}$. Then $\normsup\tA\leq 1$, and by proposition~\ref{prop:chandomeff}, there is $\tC\in\Trnset{\rA_{G'}\to\rA_{G'}}_{Q1}$ such that
\begin{align*}
\tC-\tA\succeq0.
\end{align*}
By lemma~\ref{lem:prescon}, $\tV\tA\tV^{-1}\in\Trnset{\rA_{G'}\to\rA_{G'}}_{Q+}$. Moreover, by lemma~\ref{lem:locchan},$\tV\tC\tV^{-1}\in\Trnset{\rA_{G'}\to\rA_{G'}}_{Q1}$.
By the above observations, we have
\begin{align*}
&\tV\tA\tV^{-1}\succeq0,\\
&\tV\tC\tV^{-1}-\tV\tA\tV^{-1}\succeq 0. 
\end{align*}
Finally, by lemma~\ref{lem:cnestr}, one has
\begin{align*}
&\tV\tA\tV^{-1},\tV\tC\tV^{-1}-\tV\tA\tV^{-1}\in\Trnset{\rA_{G'}\to\rA_{G'}}_Q. 
\end{align*}
The same argument holds for $\tA \in\Trnset{\rA_{G'}\to\rA_{G'}}_{L}$. Since $\tV^{-1\dag}$ is a UR, surjectivity on both sets is proved.
\end{proof}

Clearly, given a UR $\tV^\dag$, the image of the algebra of local transformations of $\rA_g$ under $\tV\cdot\tV^{-1}$, i.e.
\begin{align*}
\tV\Trnset{\rA^{(G')}_g\to\rA^{(G')}_g}_{Q\Reals}\tV^{-1}&\subseteq\Trnset{\rA^{(G')}_G\to\rA^{(G')}_G}_{Q\Reals},
\end{align*} 
is a subalgebra of $\Trnset{\rA^{(G')}_G\to\rA^{(G')}_G}_{Q\Reals}$ isomorphic to $\Trnset{\rA^{(G')}_g\to\rA^{(G')}_g}_{Q\Reals}$.

\subsection{Admissibility and local action}

As we have seen, independently of the nature of system $\rC$, the map $\tV\cdot\tV^{-1}$ does more than mapping linear maps to linear maps: it sends transformations to transformations. This is an important result, as the definitions of automorphic family and UR are meant to embody necessary conditions for a family of maps to represent $\tV\otimes\tI_\rC$. However, URs are not yet consistent with such a factorisation. 
The reason for this is that local discriminability in our theory is not assumed. Thus,
while knowledge of the linear maps $\tV_\rC^\dag$ is sufficient to determine the action 
of the UR on factorised effects $a_G\otimes b_\rC$ and, by conjugation, on generalised 
transformations $\tA_G\otimes\tI_\rC$, and such actions are consistent with a factorised
action of the form $\tV\otimes\tI_\rC$, still we do not have enough information to
determine the action of $\tV_\rC^\dag$ on those elements of 
$\Cntsetcomp{\rA_G}{\rC}_{Q\Reals}$ that lie outside the subspace 
$\Cntset{\rA_G}_{Q\Reals}\otimes\Cntset{\rC}_{Q\Reals}$ (and analogously for the conjugate 
action on transformations).
For this purpose we introduce in this section the notion of admissibility. 

We remind that $G'$ stands for $G\cup h_0$, with $\rA_{h_0}\cong\rC$ for some system $\rC$. Let $R\in \reg{G'}$ be a finite region. Since $\Cntset{\rA^{(G')}_R}_{Q\Reals}$ is isomorphic to $\Cntset{\rA_R}_\Reals$, whose size is $D_R\coloneqq D_{\rA_R}$, we can find a basis $\{b_{jR}\}_{j=1}^{D_R}\subseteq\Cntset{\rA^{(G')}_R}_{Q\Reals}$. Thus, by lemma~\ref{lem:urloc}, $\tV^\dag\Cntset{\rA^{(G')}_R}_{Q\Reals}\subseteq\Cntset{\rA^{(G')}_{R'}}_{Q\Reals}$ for some finite region $R'\in\reg {G'}$, that can be obtained as
\begin{align*}
R'\coloneqq\bigcup_{j=1}^{D_R} R'_j,\quad\tV^\dag b_{jR}\in\Cntset{\rA^{(G')}_{R'_j}}_{Q\Reals}.
\end{align*}
This implies that, for every finite region $R\in\reg {G'}$, the UR $\tV^\dag$ induces a 
linear map $\tV(R)^\dag:\Cntset{\rA_{R}}_{\Reals}\to\Cntset{\rA_{R'}}_\Reals$.
The following definition of admissible UR is now given in terms of the properties of the induced maps $\tV(R)^\dag$.
\begin{definition}
Let $(G,\rA,\tV^\dag)$ be a UR, and let $\tV(R)^\dag$ be the induced maps on finite regions $R\subseteq G'$, for arbitrary choice of $\rA_{h_0}\cong\rC$. We call the UR {\em admissible} if for every finite $S\in\reg  G$ the family of maps $\tV(S\cup h_0)^\dag$ represents a transformation $\tV_S\in\Trnset{\rA_{S'}\to\rA_S}_{\Reals}$, i.e.
\begin{align}
\tV(S\cup h_0)^\dag=\tV_S^\dag\otimes\tI^\dag_{\rC}.
\end{align}
\end{definition}
Consider now a partition $G=R\cup\bar R$, where $\bar R\coloneqq G\setminus R$. We will write $\rA_{G}=\rA_R\otimes\rA_{\bar R}$. By definition, the space $\Cntset{\rA_{G}}_{Q\Reals}$ of quasi-local effects contains the spaces $\Cntset{\rA^{(G)}_{R}}_{Q\Reals}$ and $\Cntset{\rA^{(G)}_{\bar R}}_{Q\Reals}$ as closed subspaces. However, if the theory does not satisfy local discriminability, it is not true in general that $\Cntset{\rA_{G}}_{Q\Reals}=\Cntset{\rA^{(G)}_{R}}_{Q\Reals}\otimes\Cntset{\rA^{(G)}_{\bar R}}_{Q\Reals}$. This makes it difficult to define a UR of the form $\tV\otimes\tW$.

In the remainder of the section we close this gap. For this purpose, we start defining what it means, for a UR $\tV^\dag$ on $\rA_G$, to act locally on $R$---in symbols $\tV^\dag=\tV'^\dag\otimes\tI^\dag_{\bar R}$. For our purpose, we then need to provide a suitable definition of the identity $\tV^\dag=\tV'^\dag\otimes\tI^\dag_{\bar R}$, which satisfies the necessary (but not sufficient) condition 
\begin{align*}
(\tV'^\dag\otimes\tI^\dag_{\bar R})(a_S\otimes b_{T})=(\tV'^\dag a_S)\otimes(b_T),
\end{align*}
for every $S,T\in\reg G$ such that $S\subseteq R$ and $T\subseteq \bar R$.

\begin{definition}
Let $\tV^\dag$ be a UR for $\rA_G$. We say that $\tV^\dag$ {\em acts locally} on the region $R\subseteq G$ if for every $\rC$ and every finite region $S\in\reg {G'}$ with $G'=G\cup h_0$ and $\rA_{h_0}\cong\rC$, the induced map $\tV(S)^\dag$ 
has the form 
\begin{align}
&\tV^\dag(S)=\tW_{S\cap R}^\dag\otimes\tI^\dag_{S\setminus R}.
\end{align}
\end{definition}

As a consequence of the above definition we now prove the following result, for which we do not provide the proof, that is straightforward.
\begin{lemma}\label{lem:bsh}
Let $(G,\rA,\tV^\dag)$ be a UR acting locally on $R$. Then 
\begin{align}
\tV^\dag(a_S\otimes b_T)=(\tW^\dag_Sa_S)\otimes b_T,
\end{align}
for $S\subseteq R$ and $T\subseteq \bar R$.
In particular, $\tV^\dag\Cntset{\rA_R^{(G')}}_{Q\Reals}=\Cntset{\rA_R^{(G')}}_{Q\Reals}$, and $\tV^\dag b=b$ for every $b\in\Cntset{\rA_{\bar R}^{(G')}}_{Q\Reals}$.
\end{lemma}

\begin{lemma}\label{lem:locaction}
Let $(G,\rA,\tV^\dag)$ be a UR acting locally on $R$. Then there exists a UR $(R,\rA,\tV'^\dag)$ such that $ \tV^\dag{\tJ_{R'}^{-1}}^\dag a={\tJ_{R'}^{-1\dag}}\tV'^\dag a$ and $\tV^{-1\dag}\tJ_{R'}^{-1\dag} a=\tJ_{R'}^{-1\dag}\tV'^{-1\dag} a$ for every $a\in\Cntset{\rA_{R'}}_{Q\Reals}$, with $\rA_{R'}\coloneqq \rA_R\rC$. 
\end{lemma}
\begin{proof}
Linearity and isometricity of the restriction $\tV^\dag_{R'}$ of $\tV^\dag$ to $\Cntset{\rA^{(G')}_{R'}}_{Q\Reals}$ are straightforward by lemma~\ref{lem:bsh}. Thus, also $\tV'^\dag\coloneqq {\tJ_{R'}^\dag}\tV^\dag_{R'}\tJ_{R'}^{-1\dag}$ is linear and isometric. Let now $\tA\in\Trnset{\rA_{R'}\to\rA_{R'}}_{Q}$. Then one has
\begin{align*}
\tV^{\prime\dag}\tA^{\dag}\tV'^{-1\dag}&=\tJ_{R'}^\dag
\tV_{{R'}}^\dag\tJ_{R'}^{-1\dag}\tA^\dag\tJ_{R'}^\dag \tV_{R'}^{-1\dag}\tJ_{R'}^{-1\dag}\\
&=\tJ_{R'}^\dag\tV_{R'}^{\dag}\tA_{R'}^\dag\tV_{R'}^{-1\dag}\tJ_{R'}^{-1\dag}\\
&=\tJ_{R'}^{\dag}(\tV^{\dag}\tA_{R'}^{\dag}\tV^{-1\dag})_{R'}\tJ_{R'}^{-1\dag},
\end{align*}
where $\tA_{R'}^\dag\coloneqq \tJ_{R'}^{-1\dag}\tA^\dag\tJ_{R'}^{\dag}$. In the above chain of equalities we used the fact that, since $\tV^\dag$, $\tV^{-1\dag}$ and $\tA_{R'}^\dag$ preserve the space $\Cntset{\rA^{(G')}_{R'}}_{Q\Reals}$, then also $\tV^{\dag}\tA_{R'}^\dag\tV^{-1\dag}$ does, and one can thus define its restriction $(\tV^{\dag}\tA_{R'}^\dag\tV^{-1\dag})_{R'}$. 
Finally, since $\tV^\dag$ is a UR, one has $\tV^{\dag}\tA^\dag_{R'}\tV^{-1\dag}=\tA'^\dag$ with $\tA'\in\Trnset{\rA^{(G')}_{R'}\to\rA^{(G')}_{R'}}_Q$.
Thus,
$\tV'^{-1}\tA\tV'\in\Trnset{\rA_{R'}\to\rA_{R'}}_Q$.
A similar argument leads to the existence of $\tA''\in\Trnset{\rA_{R'}\to\rA_{R'}}_Q$ such that $\tA=\tV'^{-1}\tA''\tV'$.
Now, let $\rho\in\Stset{\rA_G}$. Then $\rho$ defines a state $\rho_{\vert {R'}}$ on $\Cntset{\rA_{R'}}_{Q\Reals}$, by lemma \ref{lem:restst}, with
\begin{align*}
(a|\rho_{\vert {R'}})=(a_{R'}|\rho)=(\tJ_{R'}^{-1\dag} a|\rho),
\end{align*}
for every $a_{R'}\in\Cntset{\rA^{(G')}_{R'}}_{Q\Reals}$. Now, we have
\begin{align*}
(a|&[\hat\tV\rho]_{\vert R'})=(\tJ_{R'}^{-1\dag} a|\hat\tV\rho)=(\tV^\dag \tJ_{R'}^{-1\dag} a|\rho)\\
&=(\tV^\dag_{R'}\tJ_{R'}^{-1\dag}  a|\rho)=({\tJ_{R'}^{-1\dag}}\tJ_{R'}^\dag\tV^\dag_{R'}{\tJ_{R'}^{-1\dag}} a|\rho)\\
&=({\tJ_{R'}^{-1\dag}} \tV'^\dag a|\rho)=(a|\hat\tV'\rho_{\vert {R'}}).
\end{align*}
We then showed that $[\hat\tV\rho]_{\vert {R'}}=\hat\tV'(\rho_{\vert {R'}})$. Moreover, since $\tV$ preserves the set of states, $[\hat\tV\rho]_{\vert {R'}}$ is a state. Thus, $\hat\tV'\Stset{\rA_{R'}}\subseteq\Stset{\rA_{R'}}$. The same argument for $\tV'^{-1}$ brings to the conclusion that $\hat\tV'\Stset{\rA_{R'}}=\Stset{\rA_{R'}}$. Thus, $\tV'^\dag$ is a UR for $\rA_R$. Moreover 
\begin{align*}
\tV^\dag\tJ_{R'}^{-1\dag}  a=\tV^\dag_{ {R'}}{\tJ_{R'}^{-1\dag}} a={\tJ_{R'}^{-1\dag}}\tV'^\dag a,
\end{align*}
and the same holds for $\tV^{-1}$.
\end{proof}
In the following, under the hypothesis of lemma~\ref{lem:locaction}, we will write $\tV^\dag=\tV'^\dag\otimes\tI^\dag_{\bar R}$.

\begin{lemma}\label{lem:commca}
Let $\tV^\dag$ be a UR for $\Cntset{\rA_G}_{\Reals}$ that acts locally on $R$, and $\tW^\dag$ a UR that acts locally on $\bar R$. Then $\tV'^\dag\otimes\tW'^\dag\coloneqq \tV^\dag\tW^\dag=\tW^\dag\tV^\dag$.
\end{lemma}

%


%


We can now define a global update rule as follows.
\begin{definition}[Global update rule]
Let $(G,\rA,\tV^\dag)$ be a UR. We call $(G,\rA,\tV^\dag)$ a {\em Global Update Rule} (GUR), or {\em global rule} for short, if $(G,\rA,\tV^\dag)$ and $(G,\rA,\tV^{-1\dag})$ are admissible.
\end{definition}
\begin{remark}
The set of GURs $(G,\rA,\tV^\dag)$ is a group. The proof of closure under composition and associativity are tedious but straightforward, and we omit them here. Moreover, given $(G,\rA,\tV^\dag)$ and $(F,\rA,\tW^\dag)$ one can define their {\em parallel composition } as $(G\cup F,\rA,\tV^\dag\otimes\tW^\dag)$. The analysis of properties of parallel composition is beyond the scope of the present work, and will be the subject of further studies.
\end{remark}

We now provide an important example of GUR. Let $G=H\times\{0,1\}$, and let $H_i\coloneqq (H,i)$ for $i=0,1$. Let $a_R\in\Cntset{\rA^{(G')}_R}_{Q\Reals}$ with $\rA_{h_0}\cong\rC$. Let $R\cap G=(R_0,0)\cup(R_1,1)$. Then we define $\tilde R\coloneqq R_0\cup R_1\subseteq H$, and $\bar R\coloneqq \tilde R\times\{0,1\}\cup R$. Clearly, $R\subseteq \bar R$. If $(h,i)\in\bar R$, then also $(h,i\oplus1)\in\bar R$. Let then $\bar a_{\bar R}\coloneqq a_R\otimes e_{\bar R\setminus R}$.

Now, for every $T\in\infreg H$ let us define the GUR $(G,\rA,\tS_T^\dag)$ by setting 
\begin{align}
\tS_T^\dag [a_{R}]\coloneqq \left[\left(\tS^\dag_{\tilde R\cap T}\bar a\right)_{\bar R}\right],
\label{eq:swap}
\end{align}
where $\tS_X^\dag\coloneqq \bigotimes_{h\in X}\tS^\dag_h$ for $X\in\reg H$, and  $\tS_h$ is the map that swaps $\rA_{(h,0)}$ and $\rA_{(h,1)}$.
It is straightforward to verify that, for every $T\in\infreg H$, $\tS_T$ is a UR.
Being defined in terms of induced maps, it is immediate to verify that $\tS^\dag_T$ admissible. Moreover, $\tS_T$ is involutive, i.e.~$\tS_T^{-1}=\tS_T$. 
Of particular interest for the following is the GUR $(G,\rA,\tS_H^\dag)$.

Notice that, for a finite disjoint partition $H=\cup_{i=1}^kS_k$, we have $\tS_H^\dag=\bigotimes_{i=1}^k\tS_{S_k}^\dag$. Thus, by lemma~\ref{lem:commca}, for every GUR $(G,\rA,\tV^\dag)$, and for every $i,j$, we have
\begin{align*}
&\tV(\tS_{S_i\cup S_j}\otimes\tI_{H\setminus (S_i\cup S_j)})\tV^{-1}\\
&=\tV(\tS_{S_i}\otimes\tI_{H\setminus S_i})\tV^{-1}\tV(\tS_{S_j}\otimes\tI_{H\setminus S_j})\tV^{-1}\\
&=\tV(\tS_{S_j}\otimes\tI_{H\setminus S_j})\tV^{-1}\tV(\tS_{S_i}\otimes\tI_{H\setminus S_i})\tV^{-1},
\end{align*}
where $\tI_S$ denotes $\tI_{(S,0)}\otimes\tI_{(S,1)}$.

For $\tA\in\Trnset{\rA_R\to\rA_R}$, it is easily verified that
\begin{align}
&\tA_{(R,1)}=\tS_H(\tA_{(R,0)})\tS_H=\tS_R(\tA\otimes\tI_{R})\tS_R,\nonumber\\
&\tA_{(R,0)}=\tS_H(\tA_{(R,1)})\tS_H=\tS_R(\tI_{R}\otimes\tA)\tS_R.
\label{eq:swapconj}
\end{align}
Finally, let us consider a GUR $(G,\rA,\tU^\dag\otimes\tV^\dag)$. One can prove that $\tS_H(\tU\otimes\tV)\tS_H=\tV\otimes\tU$, by the following argument. Let us consider $a\in\Cntset{\rA^{(G')}_R}_{Q\Reals}$. Then
\begin{align*}
\tS_H^\dag&(\tU\otimes\tV)^\dag\tS_H^\dag a_R=\tS_H^\dag(\tU\otimes\tV)^\dag(\tS_{\tilde R}^\dag a)_{\bar R}\\
&=\tS_H^\dag[(\tU(\tilde R,0)^\dag\otimes\tV(\tilde R,1)^\dag)(\tS_{\tilde R}^\dag a)]_{\bar R}\\
&=[\tS_{\tilde R}^\dag(\tU(\tilde R,0)^\dag\otimes\tV(\tilde R,1)^\dag)(\tS_{\tilde R}^\dag a)]_{\bar R}\\
&=[\tV(\tilde R,0)^\dag\otimes\tU(\tilde R,1)^\dag a]_{\bar R}\\
&=(\tV\otimes\tU)^\dag a_R.
\end{align*}

\subsection{Causal influence}

In the present subsection we define the notion of causal influence that establishes, given a GUR, whether the evolution allows interventions on one system to affect other systems after one step, or more generally how many steps are needed for such an  influence to occur. We will often use the notation $\Trnset{\rA^{(G)}_R\rC\to\rA^{(G)}_R\rC}_{**}$ to denote $\Trnset{\rA^{(G')}_R\to\rA^{(G')}_R}_{**}$, with $G'=G\cup h_0$ and $\rA_{h_0}\cong\rC$. In the following we also adopt the convention that for $R\in\infreg G$, $\bar R$ denotes the complementary region $\bar R\coloneqq G\setminus R$. Notice that, consistently with the notation $R=g$, we will write $\bar g$ to denote $G\setminus g$.

\begin{definition}[Causal influence]\label{def:influ}
We say that, according to the global rule $\tV$, system $g$ does not \emph{causally influence} $g'$ if for every external system $\rC$
\begin{align*}
\tV&\Trnset{\rA^{(G)}_{g}\rC\to\rA^{(G)}_{g}\rC}_{Q\Reals} \tV^{-1}
\subseteq\Trnset{\rA^{(G)}_{\bar g'}\rC\to\rA^{(G)}_{\bar g'}\rC}_{Q\Reals}.
\end{align*}
On the other hand, we will say that according to the global rule $\tV$ system $g$ \emph{causally influences} system $g'$, and write $g\rightarrowtail g'$, in the opposite case, i.e.~when
\begin{align*}
&\exists \rC,\tF\in\Trnset{\rA^{(G)}_g\rC\to\rA^{(G)}_g\rC}_{Q\Reals}:\\
&(\tV\otimes\tI_\rC)\tF(\tV^{-1}\otimes\tI_\rC)\not\in\Trnset{\rA^{(G)}_{\bar g'}\rC\to\rA^{(G)}_{\bar g'}\rC}_{Q\Reals}.
\end{align*}
\end{definition}
The above definition gets a clear meaning as a diagram: according to $\tV$ there is causal 
influence from $g$ to $g'$ if there exists $\tF\in\Trnset{\rA_g\rC\to\rA_g\rC}$ such that
\begin{align}\label{eq:influ}
\begin{aligned}
    \Qcircuit @C=1em @R=.7em @! R {&\ustick{\rC} \qw &\qw& \multigate{1}{\tF} &\qw&\ustick{\rC} \qw &\qw\\
    &\ustick{g} \qw &\multigate{1}{\tV^{-1}}& \ghost{\tF} &\multigate{1}{\tV}&\ustick{g} \qw &\qw\\
                 & \ustick{\bar g}\qw &\ghost{\tV^{-1}}&\qw&\ghost{\tV}&\ustick{\bar g}\qw&\qw }
\end{aligned}
\neq
\begin{aligned}
    \Qcircuit @C=1em @R=.7em @! R {&\ustick{\rC}\qw&\multigate{1}{\tF'}&\ustick{\rC} \qw &\qw\\
                 & \ustick{\bar g'}\qw&\ghost{\tF'}&\ustick{\bar g'}\qw&\qw \\
             & \ustick{g'}\qw &\qw&\ustick{g'}\qw&\qw }
\end{aligned}\ ,
\end{align}
and even better in the form
\begin{align*}
\begin{aligned}
    \Qcircuit @C=1em @R=.7em @! R {&\ustick{\rC} \qw &\multigate{1}{\tF}&\qw&\ustick{\rC} \qw &\qw\\
                 &\ustick{g}\qw&\ghost{\tF}&\multigate{1}{\tV}&\ustick{g}\qw&\qw\\
             &\ustick{\bar g}\qw&\qw&\ghost{\tV}&\ustick{\bar g}\qw&\qw }
\end{aligned}
\neq
\begin{aligned}
    \Qcircuit @C=1em @R=.7em @! R {&\ustick{\rC} \qw &\qw&\qw&\multigate{1}{\tF'}&\ustick{\rC} \qw &\qw\\
                 & \ustick{\bar g'}\qw &\multigate{1}{\tV}& \qw&\ghost{\tF'}&\ustick{\bar g'}\qw&\qw \\
             & \ustick{g'}\qw &\ghost{\tV}& \qw &\qw&\ustick{g'}\qw&\qw }
\end{aligned}\ .
\end{align*}

\begin{definition}
We define {\em forward} and {\em backward $g$-neighbourhoods} the sets $N^+_g\coloneqq \{g'|g\rightarrowtail g'\}$ and $N^-_g\coloneqq \{g'|g'\rightarrowtail g\}$, respectively. 
\end{definition}
In terms of diagrams, the forward $g$-neighbourhood $\rA_{N^+_g}$ of $g$ can be understood as the smallest region $K=\bigotimes_{k\in K} A_{g'_k}$ 
 such that for every $k\in K$ there exists $\rC$ and $\tF\in\Trnset{\rA^{(G)}_g\rC\to\rA^{(G)}_g\rC}_{Q\Reals}$ such that
\begin{align}
&\begin{aligned}
    \Qcircuit @C=1em @R=.7em @! R {&\ustick{\rC}\qw &\multigate{1}{\tF}&\qw&\ustick{\rC}\qw &\qw\\
           &\ustick{g}\qw&\ghost{\tF}&\multigate{1}{\tV}&\ustick{g}\qw&\qw \\
           &\ustick{\bar g}\qw&\qw&\ghost{\tV}&\ustick{\bar g}\qw&\qw }
\end{aligned}
\nonumber\\
\label{eq:N+}\\
=\ 
\begin{aligned}
    \Qcircuit @C=1em @R=.7em @! R {&\ustick{\rC}\qw &\qw&\qw&\multigate{1}{\tF'}&\qw& \ustick{\rC}\qw &\qw\\
            & \ustick{N^+_g}\qw &\qw&\multigate{1}{\tV}&\ghost{\tF'}&\qw&\ustick{N^+_g}\qw&\qw \\
        &\ustick{\bar N^+_g}\qw &\qw&\ghost{\tV}&\qw&\qw&\ustick{\bar N^+_g} \qw &\qw}
\end{aligned}\ .
\nonumber
\end{align}
with $\tF'$ acting non trivially on $g'_k$, namely $\tF'\not\in\Trnset{\rA^{(G)}_{\bar g'_k}\rC\to\rA^{(G)}_{\bar g'_k}\rC}_{Q\Reals}$. Notice that $g$ may or may not belong to $N^+_g$. 

Analogously to the definition of $N^\pm_g$, one can give a definition of $N_R^\pm$ for every finite region $R\subseteq G$. 
\begin{definition}
We say that the system $g'$ belongs to the forward neighbourhood of $R$, and write  $g'\in N^+_R$, if
\begin{align*}
&\exists \rC,\tF\in\Trnset{\rA^{(G)}_R\rC\to\rA^{(G)}_R\rC}_{Q\Reals}:\\
&(\tV\otimes\tI_\rC)\tF(\tV^{-1}\otimes\tI_\rC)\not\in\Trnset{\rA^{(G)}_{\bar g'}\rC\to\rA^{(G)}_{\bar g'}\rC}_{Q\Reals}.
\end{align*}
The backward neighbourhood $N^-_R$ of $R$ is defined as 
\begin{align}
N^-_R\coloneqq\{g'\in G\mid N^+_{g'}\cap R\neq\emptyset\}.
\end{align}
\end{definition}
Clearly, one has
\begin{align}\label{eq:incluneighs}
N^+_R\supseteq \bigcup_{g\in R}N^+_g.
\end{align}
We will prove that the latter is actually an equality in the next section.

One might define causal influence without invoking any external system $\rC$, as follows.
\begin{definition}[Individual influence]
We say that the system $g$ does not \emph{individually influence $g'$} if
\begin{align*}
\tV&\Trnset{\rA^{(G)}_{g}\to\rA^{(G)}_{g}}_{Q\Reals} \tV^{-1}\\
&\subseteq\Trnset{\rA^{(G)}_{\bar g'}\to\rA^{(G)}_{\bar g'}}_{Q\Reals}.
\end{align*}
On the other hand, we will say that according to the global rule $\tV$ system $g$ \emph{individually influences} system $g'$, and write $g\rightarrowtail_g g'$, in the opposite case, i.e.~when
\begin{align*}
&\exists \tF\in[\rA^{(G)}_g\to\rA^{(G)}_g]_{Q\Reals}:\\
&\tV\tF\tV^{-1}\not\in\Trnset{\rA^{(G)}_{\bar g'}\to\rA^{(G)}_{\bar g'}}_{Q\Reals}.
\end{align*}
The individual forward neighbourhood of $g$ is
\begin{align*}
N^+_g(g)\coloneqq\{g'\in G\mid g\rightarrowtail_g g'\}.
\end{align*}
\end{definition}
A second definition that can be given without referring to external systems $\rC$ is the following.
\begin{definition}[Cooperative influence]\label{def:clindinflu}
We say that the region $S$ does not \emph{cooperatively influence $g'$} if 
\begin{align*}
\tV\Trnset{\rA^{(G)}_S\to\rA^{(G)}_S}_{Q\Reals}\tV^{-1}\subseteq\Trnset{\rA^{(G)}_{\bar g'}\to\rA^{(G)}_{\bar g'}}_{Q\Reals}.
\end{align*}
In the opposite case we will write $S\rightarrowtail_S g'$. The cooperative forward neighbourhood of a region $S$ is defined as
\begin{align*}
N^{+}_S(S)\coloneqq\{g'\in G\mid S\rightarrowtail_S g'\}.
\end{align*}
\end{definition} 
Notice now that in a theory without local discriminability it may happen that a system $g$ does not individually influence $g'$ but it does cooperatively, namely one can have $g'\in N^+_S(S)$, but $g'\not\in N^+_{g}(g)$ for any $g\in S$. In other words,
$N^+_S(S)\supset\bigcup_{h\in S}N^+_{h}(h)$. We call this phenomenon {\em non-local activation of causal influence}. If such a phenomenon occurs, however, it can be hard to establish which of the ``cooperating'' systems $h\in S$ is responsible for causal influence on $g'\in N^+_S(S)$. This is the reason why we prefer the former definition of causal influence, which provides also this piece of information.

Given a GUR $(G,\rA,\tV^\dag)$, we can now construct a graph $\Gamma(G,E)$ with vertices $g\in G$ and edges $E\coloneqq \{(g,g')\in G\times G|g\rightarrowtail g'\}$. We call such a graph the {\em graph of causal influence} of the GUR $\tV^\dag$. In the next section we will prove that, if we denote by $N^{\pm*}_g$ the neighbourhoods for the GUR $\tV^{-1}$, the following relations hold
\begin{align}
N_f^\pm=N^{\mp*}_f.
\end{align}
We also define the {\em local perspective present} $P^+_R$ of $R\in\reg G $ as the set $P^+_R\coloneqq N^-_{N^+_R}\supseteq R$, namely the set of systems that can causally influence the one-step future $N^+_R$ of $R$. Similarly one can define the {\em local retrospective present} of $R$ as $P^-_R\coloneqq N^+_{N^-_R}\supseteq R$, namely the set of systems that can be influenced by the one-step past $N^-_R$ of $R$. In general, $P^+_R\neq P^-_R$. We will also need to refer to the sets $P^{+*}_R\coloneqq N^{+*}_{N^+_R}$.

\subsubsection{Relation with signalling}

The term {\em influence} that we used above is precisely defined in Eq.~\eqref{eq:influ}.
In the present subsection we will make a clear point that this definition is stronger than the notion of {\em signalling} that is customary in the literature on quantum information theory \cite{Beckman:2001aa,Eggeling:2002aa,Schumacher:2005:LIT:1061551.1061580} or in general operational probabilistic theories \cite{DAriano:2017aa}. In that context, a channel $\tC\in\Trnset{\rA\rB\to\rA'\rB'}$ is called non-signalling (or semicausal) from $\rB$ to $\rA'$ if there exists $\tF\in\Trnset{\rA\to\rA'}$ such that
\begin{align}
\begin{aligned}
    \Qcircuit @C=1em @R=.7em @! R {&\qw &\ustick{\rA}\qw&\multigate{1}{\tC}&\ustick{\rA'}\qw& \qw\\
    &\qw &\ustick{\rB}\qw&\ghost{\tC}&\ustick{\rB'}\qw& \measureD{e}}        
\end{aligned}\ =\ 
\begin{aligned}
    \Qcircuit @C=1em @R=.7em @! R {&\qw &\ustick{\rA}\qw&\gate{\tF}&\ustick{\rA'}\qw& \qw\\
    &\qw &\ustick{\rB}\qw&\measureD{e}&&}        
\end{aligned}\ .
\label{eq:semiloc}
\end{align}
We will now show that, under the hypothesis that $N^{+*}_g$ is finite for every $g\in G$, the condition of Eq.~\ref{eq:N+}, in a suitably defined sense, implies that of Eq.~\eqref{eq:semiloc}. Let us consider the state $\rho\in\Stset{\rA_G}_\Reals$. Then $(a_S|\hat\tA_T\rho)\coloneqq (a'_{S\cup T}|\rho)=(\tA_T^\dag a_S|\rho)$. In the case where $a_S=e_G$, $(\tA_T^\dag e_G|\rho)=(b_T|\rho)$, where $b=\tA^\dag e_T$.
Let us now consider $(\hat\tV\rho)_{\vert R}$, which by definition is the state in $\Stset{\rA_R}_{\Reals}$ such that, for every $a\in\Cntset{\rA_R}_\Reals$
\begin{align*}
(a|[\hat\tV\rho]_{\vert R})=({\tJ_R}^\dag a|\hat\tV\rho)=(a_R|\hat\tV\rho).
\end{align*}
Now, by lemma~\ref{lem:loceffloctr}, one has
\begin{align*}
(a_R|\hat\tV\rho)&=(e_G|\hat\tA_R\hat\tV\rho)\\
&=(e_G|\hat\tV\hat\tV^{-1}\hat\tA_R\hat\tV\rho)\\
&=(e_G|\hat\tV\hat\tA'_{N^{+*}_R}\rho)\\
&=(e_G|\hat\tA'_{N^{+*}_R}\rho)\\
&=(b_{N^{+*}_R}|\rho)=(b|\rho_{\vert N^{+*}_R}).
\end{align*}
In order to determine $[\hat\tV\rho]_{\vert R}$ it is sufficient to know $\rho_{\vert N^{+*}_R}$. Indeed, the linear map $\tV_R:\Stset{\rA_{N^{+*}_R}}_{\Reals}\to\Stset{\rA_{R}}_{\Reals}$ gives 
\begin{align}
(\hat\tV\rho)_{\vert R}=\tV_R(\rho_{\vert N^{+*}_R}).
\label{eq:loctdefeq}
\end{align} 
Diagrammatically, the equality can be recast, with a slight abuse of notation, as in Eq.~\ref{eq:semiloc}, as follows
\begin{align*}
\begin{aligned}
    \Qcircuit @C=1em @R=.7em @! R {&\qw &\ustick{N^{+*}_R}\qw&\multigate{1}{\tV}&\ustick{R}\qw& \qw\\
    &\qw &\ustick{\overline{N^{+*}_R}}\qw&\ghost{\tV}&\ustick{\bar R}\qw& \measureD{e}}        
\end{aligned}\ =\ 
\begin{aligned}
    \Qcircuit @C=1em @R=.7em @! R {&\qw &\ustick{N^{+*}_R}\qw&\gate{\tV_{R}}&\ustick{R}\qw& \qw\\
    &\qw &\ustick{\overline{N^{+*}_R}}\qw&\measureD{e}&&}        
\end{aligned}\ .
\end{align*}
Notice that, by the no-restriction hypothesis, for an admissible $\tV$ the map $\tV_R$ is linear and admissible, and is thus a transformation in $\Trnset{\rA^{(G)}_{N^{+*}_R}\to\rA^{(G)}_{R}}$. Moreover, 
\begin{align}
\tV^\dag a_R=(\tV^\dag_R a)_{N^{+*}_R}.
\label{eq:nosgnl}
\end{align}
Finally, notice that by definition $\tV^\dag_R e_R=e_{N^{+*}_R}$, namely $\tV_R\in\Trnset{\rA^{(G)}_{N^{+*}_R}\to\rA^{(G)}_R}_1$.
The argument can be repeated for the case where $N^{+*}_g$ is not finite, as long as $N^{+*}_g\subset G$.

We remark that in the case of Quantum or Fermionic cellular automata, the condition for no causal influence is equivalent to no-signalling, while in the case of Classical cellular automata no causal influence is strictly stronger than no-signalling. In a different context, a difference between the two notions was pointed out in Ref.~\cite{barrett2019quantum}

\subsection{Block decomposition}\label{s:block}

\begin{figure}[h]
       \includegraphics[width=4cm]{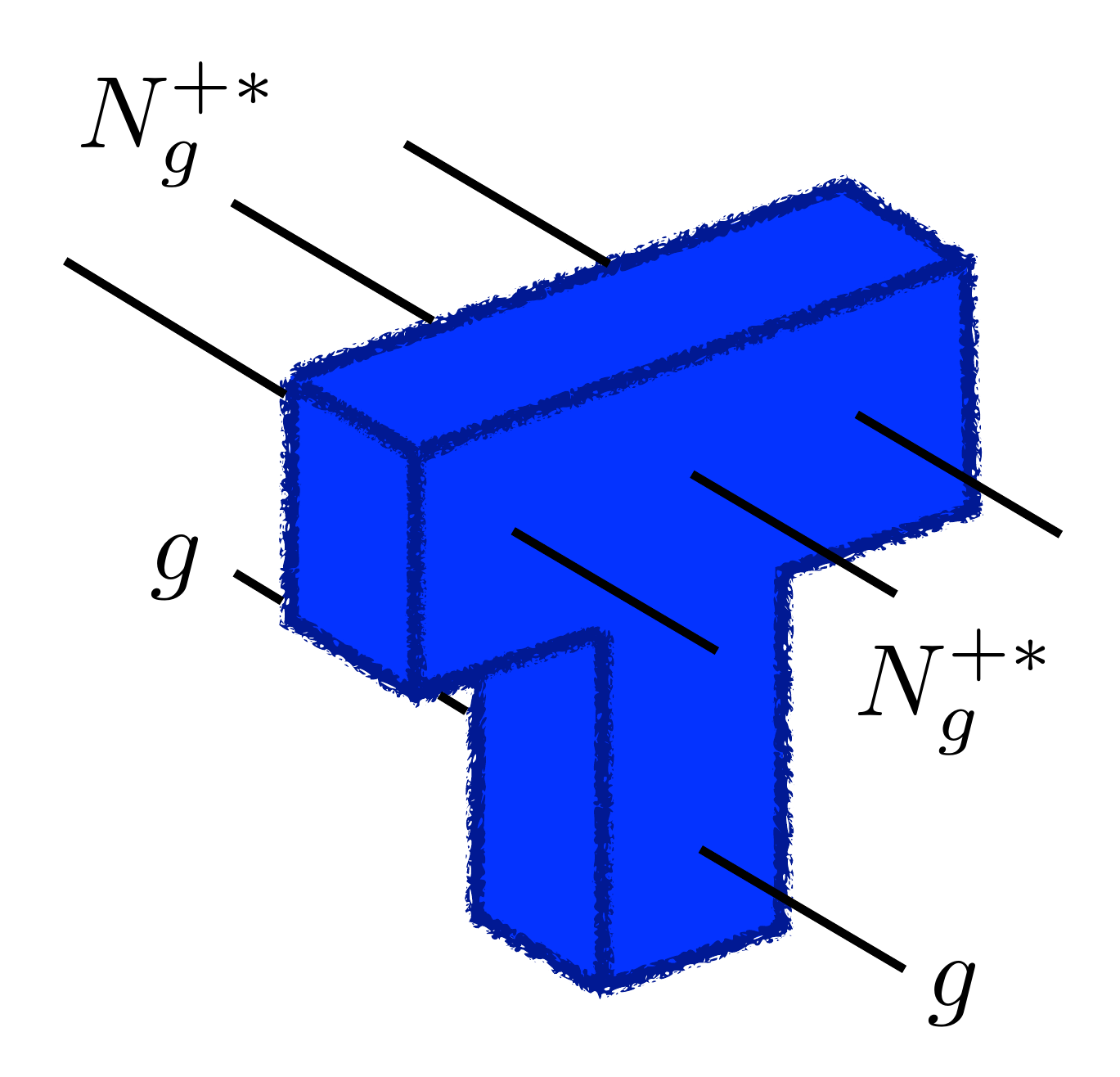}
        \includegraphics[width=8cm]{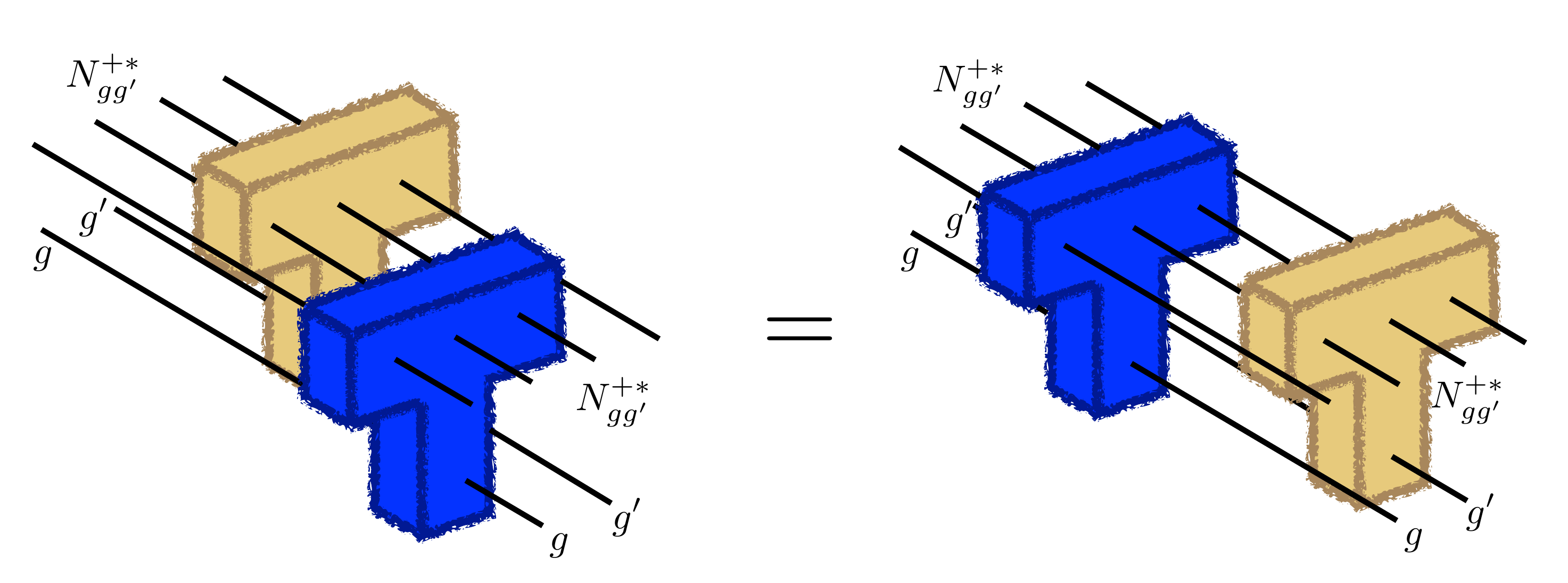}
\caption{In the top picture we provide a graphical representation of the transformation $\tS'_g$, with input wires on the top left and output on the bottom right. In the bottom picture we provide a graphical illustration of the identity $\tS'_g\tS'_{g'}=\tS'_{g'}\tS'_g$.}
\label{f:locru}
\end{figure}

Let us now consider the GUR $(G,\rA,\tW^\dag)$ where we set $G=H\times\{0,1\}$, and $\tW=\tV_{H_0}\otimes\tV^{-1}_{H_1}$, for an arbitrary GUR  $(H,\rA,\tV^\dag)$, $H_i$ denoting a shorthand for $(H,i)$. 
We observe that 
\begin{align*}
\tW=\tV_{H_0}\otimes\tV^{-1}_{H_1}=(\tV_{H_0}\otimes\tI_{H_1})\tS_H(\tV_{H_0}^{-1}\otimes\tI_{H_1})\tS_H.
\end{align*}
Then, by Eq.~\eqref{eq:swap}, given $R\subseteq \rA_G$ and $a\in\Cntsetcomp{\rA^{(G)}_{R}}{\rC}_{Q\Reals}$, with $R\cap G=(R_0,0)\cup(R_1,1)$,
\begin{align*}
&\tW^\dag a_{R\rC}=(\tW^\dag a)_{R'\rC} \\
=&[\tS_H^\dag({\tV^{-1}_{H_0}}\otimes\tI_{H_1})^\dag\tS_H^\dag(\tV_{H_0}\otimes\tI_{H_1})^\dag]a_{R\rC} \\
=&[\tS_H^\dag({\tV^{-1}_{H_0}}\otimes\tI_{H_1})^\dag\tS_H^\dag]a'_{R''\rC}  \\
=&[\tS_H^\dag({\tV^{-1}_{H_0}}\otimes\tI_{H_1})^\dag\tS^\dag_{N^{+*}_{R_0}\cup R_1}]a'_{R''\rC}  \\
=&[\tS_H^\dag({\tV^{-1}_{H_0}}\otimes\tI_{H_1})^\dag\tS^\dag_{N^{+*}_{R_0}\cup R_1}(\tV_{H_0}\otimes\tI_{H_1})^\dag]{a}_{R\rC} ,
\end{align*}
where $a'\coloneqq (\tV_{H_0}\otimes\tI_{H_1})^\dag a$, 
$R'\coloneqq{(N^{+*}_{R_0},0)\cup(N^+_{R_1},1)}$, and $R''\coloneqq(N^{+*}_{R_0},0)\cup(R_1,1)$. Now, defining for $S\in\infreg H$
\begin{align}
&\tS'_S\coloneqq \prod_{h\in S}\tS'_h,\label{eq:prodsp}\\
&\tS'_h\coloneqq (\tV_{H_0}\otimes\tI_{H_1})\tS_h(\tV_{H_0}^{-1}\otimes\tI_{H_1}),\label{eq:sprimh} 
\end{align}
thanks to eq.~\eqref{eq:swapconj} we have
\begin{align*}
({\tV^{-1}_{H_0}}&\otimes\tI_{H_1})^\dag\tS^\dag_{N^{+*}_{R_0}\cup R_1}(\tV_{H_0}\otimes\tI_{H_1})^\dag\\
&=\tS'^\dag_{N^{+*}_{R_0}\cup R_1}.
\end{align*}
We remark that, since $(\tV_{H_0}\otimes\tI_{H_1})\cdot(\tV_{H_0}^{-1}\otimes\tI_{H_1})$ is an automorphism of $\Trnset{\rA_G\to\rA_G}_{Q\Reals}$, one has
\begin{align}
\tS'_f\tS'_g=\tS'_g\tS'_f,\quad\forall f,g\in H.
\label{eq:condone}
\end{align}
Thus, the product $\prod_{h\in N^{+*}_R}{\tS'_{h}}^\dag$ in Eq.~\eqref{eq:prodsp} is well defined.
Notice also that, by definition 
\begin{align}
{\tS'_h}\tS'_h=\tI_{G}. 
\label{eq:condtwo}
\end{align}
Since $\tS'_h\in\Trnset{\rA^{(G)}_{(N_h^+,0)\cup(h,1)}\to\rA^{(G)}_{(N_h^+,0)\cup(h,1)}}_{Q1}$ (see Fig.~\ref{f:locru}), one has that $\tS'_{N^{+*}_{R_0}\cup R_1}\in\Trnset{\rA^{(G)}_{S}\to\rA^{(G)}_{S}}_{Q1}$, with 
\begin{align*}
S\coloneqq (N^{+}_{N^{+*}_{R_0}}\cup N^+_{R_1},0)\cup(N^{+*}_{R_0}\cup R_1,1). 
\end{align*}
Then we can write
\begin{align*}
\tW^\dag a_{R\rC}&=\tS^\dag_H\tS'^\dag_{N^{+*}_{R_0}\cup R_1}a_{R\rC}\\
&=\tS^\dag_{N^+_{N^{+*}_{R_0}}\cup N^+_{R_1}\cup N^{+*}_{R_0}\cup R_1}\tS'^\dag_{N^{+*}_{R_0}\cup R_1}a_{R\rC}.
\end{align*}
However, due to the structure of $\tW^\dag$, one actually has
\begin{align*}
({\tS'_{N^{+*}_{R_0}\cup R_1}}\otimes\tI_\rC)^\dag a_{R\rC}=a''_{(N^+_{R_1},0)\cup(N^{+*}_{R_0},1)\rC},
\end{align*}
and finally this implies that
\begin{align}
\tW^\dag a_{R\rC} &=(\tS_{N^{+*}_{R_0}\cup N^+_{R_1}}^\dag{\tS^{\prime\dag}_{N^{+*}_{R_0}\cup R_1}}\otimes\tI_\rC^\dag) a_{R\rC}.
\label{eq:margo}
\end{align}
We now observe that
\begin{align*}
\tW^{-1}=\tV_{H_0}^{-1}\otimes\tV_{H_1}=\tS_H\tW\tS_H, 
\end{align*}
and thus, by Eq.~\eqref{eq:margo}, it is
\begin{align}
\tW^{-1\dag} a_{R\rC} &=({\tS^{\prime}_{N^{+*}_{R_1}\cup R_0}}\otimes\tI_\rC)^\dag a'_{R'\rC}\nonumber\\
&=({\tS^{\prime\dag}_{N^{+*}_{R_1}\cup R_0}} \tS_{R_0\cup R_1}^\dag\otimes\tI^\dag_\rC)a_{R\rC},
\label{eq:wdaginv}
\end{align}
where $a'_{R'\rC}=[\tS^\dag a]_{(R_1,0)\cup(R_0,1)\rC}$.
Let us then consider $R_1=\emptyset$. In this case for $[a_{(R,0)\rC}]\in\Cntsetcomp{\rA^{(G)}_{(R,0)}}{\rC}$, being $R=(R,0)\cup(\emptyset,1)$, it holds that
\begin{align}
&[(\tV^\dag a)_{(N^{+*}_R,0)\rC}]
=\tS_{N^{+*}_R}^\dag\tS^{\prime\dag}_{N^{+*}_R}\otimes\tI^\dag_\rC [a_{(R,0)\rC}],
\label{eq:blockdir}\\
&[({\tV^{-1}}^\dag a)_{(N^+_R,0)\rC}]
=\tS^{\prime\dag}_{R}\tS_{R}^\dag\otimes\tI_\rC^\dag[a_{(R,0)\rC}].
\label{eq:blockinv}
\end{align}
In view of the above results, we now consider the maps $\tV\tA\tV^{-1}$ for $\tA\in\Trnset{\rA_{(R,0)}\rC\to\rA_{(R,0)}\rC}_{\Reals}$, with $R\in\reg H$. First of all, we observe that
\begin{align*}
(\tV&\tA\tV^{-1})\otimes\tI_{H_1}=\tW(\tA\otimes\tI_{H_1})\tW^{-1}
\end{align*}
is a local transformation in $\in\Trnset{\rA^{(G)}_{(N^+_R,0)}\rC\to\rA^{(G)}_{(N^+_R,0)}\rC}_{Q\Reals}$. Let then $a\in\Cntsetcomp{\rA^{(G)}_{\tilde S}}{\rC}_{\Reals}$ for $S\in\reg{H}$. We have the following (see Appendix~\ref{app:spr})
\begin{align*}
&[(\tV\tA\tV^{-1})\otimes\tI_{H_1}]^\dag a_{\tilde S\rC}\\
=&[\tW(\tA\otimes\tI_{H_1})\tW^{-1}]^\dag a_{\tilde S\rC}\\
=&(\tS^{\prime\dag}_{R}\tS_R^\dag \otimes\tI^\dag_\rC)(\tA\otimes\tI_{H_0})^\dag(\tS_R^\dag\tS^{\prime\dag}_{R}\otimes\tI_\rC^\dag)a_{\tilde S\rC}.
\end{align*}
Finally, since this holds for every $a$, we have that for $\tA\in\Trnset{\rA_{(R,0)}\rC\to\rA_{(R,0)}\rC}_{\Reals}$
\begin{align}
&(\tV\tA\tV^{-1})\otimes\tI_{H_1}\nonumber\\
=&(\tS^{\prime}_{R}\otimes\tI_\rC)(\tI_{H_0}\otimes\tA)(\tS^{\prime}_{R} \otimes\tI_\rC).
\label{eq:condthree}
\end{align}
The above relation, considering Eq.~\eqref{eq:prodsp}, shows that 
\begin{align*}
N^+_R\subseteq \bigcup_{g\in R}N^+_g.
\end{align*}
Along with Eq.~\eqref{eq:incluneighs} the latter proves that actually
\begin{align}\label{eq:npunnp}
N^+_R= \bigcup_{g\in R}N^+_g.
\end{align}
Analogously, since $N^-_R=\{g'\mid \exists g\in R,\ g\in N^+_{g'}\}$, it is
\begin{align}
N^-_R=\{g'\mid \exists g\in R,\ g'\in N^-_{g}\}=\bigcup_{g\in R}N^-_g.
\label{eq:nmrsnmug}
\end{align}
Let us finally consider the maps $(\tV^{-1}\tB\tV)$ for $\tB\in\Trnset{\rA_{(R,1)}\rC\to\rA_{(R,1)}\rC}_{\Reals}$, with $R\in\reg H$. In this case, using Eq.~\eqref{eq:condtwo} and manipulating Eq.~\eqref{eq:condthree} we obtain
\begin{align}
&\tI_{H_0}\otimes(\tV^{-1}\tB\tV)\nonumber\\
&\quad=(\tS'_{N^{+*}_R}\otimes\tI_\rC^\dag)(\tB\otimes\tI_{H_1})({\tS'_{N^{+*}_R}}\otimes\tI_\rC).
\label{eq:condfour}
\end{align}
Since the $\tS'_h$'s that do not commute with $\tB\otimes\tI_{H_1}$ in Eq.~\eqref{eq:condfour} correspond to $h\in N^-_g$ for some $g\in R$, namely $h\in N^-_R$, we can conclude that 
\begin{align}\label{eq:inclunei}
N^{+*}_R\subseteq N^-_R. 
\end{align}
This observation is sufficient to prove that $N^{\pm*}_R=N^\mp_R$, as anticipated in the previous section.
\begin{lemma}
Let $(G,\rA,\tV^\dag)$ be a GUR. Then
\begin{align}
N^{\pm*}_R=N^\mp_R.
\end{align}
\end{lemma}
\begin{proof}
We remind that $g\in N^\pm_f$ iff $f\in N^\mp_g$. Now, considering Eq.~\eqref{eq:inclunei} for $R=g$ we have that
\begin{align*}
g'\in N^{-*}_{g}\Leftrightarrow g\in N^{+*}_{g'}\Rightarrow g\in N^{-}_{g'}\Leftrightarrow g'\in N^{+}_{g}.
\end{align*}
This implies that $N^{-*}_{g}\subseteq N^+_{g}$.
Exchanging $\tV$ and $\tV^{-1}$, we can conclude that $N^{+*}_{g}\subseteq N^-_{g}\subseteq N^{+*}_g$, i.e.~$N^{+*}_{g}= N^-_{g}$. The thesis follows invoking Eqs.~\eqref{eq:npunnp} and~\eqref{eq:nmrsnmug}, which then give us
\begin{align*}
&N^-_R=N^{+*}_R.&&\qedhere
\end{align*}
\end{proof}
In the remainder we will then use $N^\pm_R$ and $P^\pm_R$, and abandon the symbols $N^{\pm*}_R$ and $P^{\pm*}_R$.

We observe that the {\em block decomposition} of $\tV\otimes\tV^{-1}$ that we achieved
through the block transformations $\tS'_g$ is a generalisation of a result that was proved
for quantum cellular automata in Ref.~\cite{arrighi2011unitarity}. Local rules of other 
kinds are used in the literature, with nice properties that the present block 
decomposition lacks, such as composability (see e.g.~Ref~\cite{Arrighi:2013aa}). However, 
for later purposes the block form of Eq.~\eqref{eq:prodsp} is particularly suitable in our 
case, to cope with update rules in theories without local discriminability.

\section{Homogeneity}\label{sec:homog}

The principle of homogeneity, as it is usually formulated in physics, regards equivalence of points in space-time. Here, however, space-time is not the background for occurrence of physical events, on the contrary it emerges as a convenient description of relations between information processing events. We then provide a definition of homogeneity for a  global update rule, that regards the way in which different systems are treated by the evolution. In particular, roughly speaking, every system must be treated in the same way by the GUR. 

It would be easy to state the homogeneity principle if we knew in advance that the set $G$ has some geometric structure that is invariant under the action of a group---technically speaking, if $G$ were a symmetric space---so that we could define pairs of \emph{homologous regions} as those mapped into one another by some group element. In this case, we could provide a criterion for the evolution to treat every system in the same way, consisting in the existence of some representation of the group of symmetries of the set $G$ that connects the evolution of every two homologous regions. The group in this case would provide the two following ingredients that are needed for the definition of homogeneity: i) a notion of ``running the same test on different regions'', and ii) a criterion for equivalence of two homologous regions provided by the requirement that the events of the same test in the two regions have equal probabilities, if the regions are prepared in the same state. 

However, we are in a situation where the structure of $G$ must emerge from the evolution given by the global rule $\tV$. The detailed idea is thus to turn the above argument on its head, and define two regions to be homologous if they are ``treated in the same way" by $\tV$. The precise construction is the following. 

First of all, in order for the dynamics to treat two regions in the same way, the two 
regions need to have the same structure, in such a way that we can make sense of 
``performing the same range of tests in the two different regions''. This concept is 
captured by the extension of the notion of operational equivalence (see 
definition~\ref{def:opeqsys}) to regions. Intuitively, two operationally equivalent 
regions must correspond to operationally equivalent systems. However, they also need 
to``have the same structure'', where the term structure refers to the fact that a system 
$\rA_R$ 
associated with a region $R$ is decomposed in a preferred way into subsystems 
$\rA_{g_i}$ 
corresponding to the cells that compose the region $R=\{g_1,g_2,\ldots,g_k\}$. 
The notion of operational equivalence of regions must then also capture the idea that two 
equivalent regions admit equivalent decompositions.

The definition is thus recursive, starting from elementary regions $R=\{g\}$, and then defining equivalence of $R_1$ and $R_2$ as the equivalence of any {\em strict subregion} $S_1\subset R_1$ with a strict subregion $S_2\subset R_2$, and viceversa.

We recall Remark~\ref{rem:nontriv}, namely that for the remainder every elementary region
$g\in G$ corresponds to some non-trivial system, $\rA_g\not\simeq\rI$.

\begin{definition}\label{def:opeqreg}
Two elementary regions $g$ and $h$ are \emph{operationally equivalent} if $\rA_g\cong\rA_h$. 
Two finite regions $R_1,R_2\subseteq G$ are \emph{operationally equivalent} 
if $\rA_{R_1}\simeq\rA_{R_2}$, and for every $S_1\subset R_1$ there exists $S_2\subset R_2$, and viceversa for every $S_2\subset R_2$ there exists $S_1\subset R_1$, such that 
$\rA_{S_1}$ and $\rA_{S_2}$ are operationally equivalent.
\end{definition}

Two operationally equivalent regions have the same structure, namely they can be decomposed into pairwise equivalent subsystems, as we now prove.

\begin{lemma}\label{lem:elemopeqreg}
An elementary region $R_1=\{g\}$ cannot be operationally equivalent to a non-elementary one $R_2=\{h_1,\ldots, h_k\}$.
\end{lemma}
\begin{proof}
The statement is trivial, since $R_1=\{g_1\}$ does not have proper subregions, while any non-elementary region $R_2$ does, so operational equivalence cannot hold.
\end{proof}

\begin{lemma}\label{lem:cardopeq}
The regions $R_1$ and $R_2$ are operationally equivalent if and only if $R_1=\{g_1,g_2,\ldots,g_k\}$,  $R_2=\{h_1,h_2,\ldots h_k\}$, and $\rA_{g_i}\cong\rA_{h_i}$.
\end{lemma}
\begin{proof}
The condition is clearly sufficient, as one can immediately realise by considering the reversible transformation $\tT_{S}\in\Trnset{\rA_{S_1}\to\rA_{S_2}}_1$ defined as $\tT_{S}\coloneqq \bigotimes_{g_j\in S_1}\tT_j$, with $\tT_j\in\Trnset{\rA_{g_j}\to\rA_{h_j}}_1$ reversible for every $j=1,\ldots,k$. 
We now prove that the condition is necessary. Let $R_1$ and $R_2$ be operationally equivalent. By definition, and by lemma~\ref{lem:elemopeqreg}, for every $g_l\in R_1$ there is $h_m\in R_2$ such that $\rA_{g_l}\cong\rA_{h_m}$, and viceversa. We now introduce the equivalence relation
\begin{align*}
&g_i\sim g_j\ \Leftrightarrow\ \rA_{g_i}\cong\rA_{g_j}.
\end{align*}
We can then partition $R_1$ and $R_2$ into equivalence classes $[g_i]$ and $[h_j]$, respectively, obtaining $R_1=\bigcup_{g_{i_l}}[g_{i_l}]$ and $R_2=\bigcup_{h_{i_l}}[h_{i_l}]$. Let us start considering the case where $R_1$ contains a unique equivalence class: $R_1=[g_1]$. Then $R_2$ must contain at least one element $h_1$ with $\rA_{h_1}\cong\rA_{g_1}$. Moreover, $R_2$ cannot contain $h_0$ such that $h_0\not\sim h_1$, because $h_0$ must be equivalent to an elementary subregion of $R_1$, by lemma \ref{lem:elemopeqreg}. Then, $R_2=[h_1]$. Finally, since $\rA_{g_1}^{\otimes h}\cong\rA_{R_2}\cong\rA_{R_1}\cong\rA_{g_1}^{\otimes k}$, by lemma \ref{lem:compi} it must be $h=k$. In the general case where $R_1$ contains more than one equivalence class, one can simply consider the subregions $S_{1_l}\coloneqq [g_{i_l}]$. Clearly, for each $S_{1_l}$ there must be $S_{2_l}=[h_{j_l}]$, with $|S_{2_l}|=|S_{1_l}|$.
\end{proof}

\begin{remark} Notice that, as a consequence of the above result, two operationally equivalent regions $R_1,R_2\in\reg G $ have the same cardinality $|R_1|=|R_2|$. 
\end{remark}

Given two operationally equivalent regions $R_1,R_2\subseteq G$, we now have a way of defining the notion of ``running the same test'', that resorts to the identification of a canonical isomorphism of the two regions.
\begin{definition}
Given two operationally equivalent regions $R_1=\{g_1,g_2,\ldots,g_k\},R_2=\{h_1,h_2,\ldots,h_k\}\in\reg G $, let us choose a special reversible transformation $\tU\in\Trnset{\rA_{R_1}\to\rA_{R_2}}$ such that $\tU=\bigotimes_{i=1}^{k}\tU_{i}$, where $\tU_i\in\Trnset{\rA_{g_i}\to\rA_{h_i}}$ is reversible for every $i$. We then say that $R_1$ is operationally equivalent to $R_2$ through $\tU$, and that $\{\tB^{(2)}_l\}_{l=1}^n\subseteq\Trnset{\rA_{R_2}\rC\to\rA_{R_2}\rC}$ {\em represents the same test as} $\{\tB^{(1)}_l\}_{l=1}^n\subseteq\Trnset{\rA_{R_1}\rC\to\rA_{R_1}\rC}$ if for every $1\leq l\leq n$ one has 
\begin{align}
\tB^{(2)}_l=(\tU\otimes\tI_\rC)\tB^{(1)}_l(\tU^{-1}\otimes\tI_\rC).
\label{eq:sametest}
\end{align}
\label{def:sametest}
\end{definition}

In order to discriminate the way in which two systems $\rA_{g_1}$ and $\rA_{g_2}$ evolve, one needs to perform a testing-scheme, consisting of a stimulus-test $\{\tA^{(x)}_k\}_{k=1}^m\subseteq\Trnset{\rA_{R_x}\rC\to\rA_{R_x}\rC}_Q$ on the region $R_x\ni g_x$, possibly involving an ancillary system $\rC$, and, after the evolution step $\tV$, a control-test $\{\tB^{(x)}_l\}_{l=1}^n\subseteq\Trnset{\rA_{N^+_{R_x}}\rC\to\rA_{N^+_{R_x}}\rC}_Q$ on the region $N^+_{R_x}$ (where $x\in\{1,2\}$) and the ancilla $\rC$, in order to detect a ``response''.



Two operationally equivalent regions are homologous if there is a GUR $\tT$ that interchanges them while preserving equivalence of local tests of finite regions. In rigorous words, we have the following definition.

\begin{definition}[Homologous regions]\label{def:hom}
Let $(G,\rA,\tV^\dag)$ be a GUR, and let $R_1,R_2\in \reg G$ be operationally equivalent regions through $\tT_i\in\Trnset{\rA_{R_1}\to\rA_{R_2}}$, such that $N^+_{R_1}$ and $N^+_{R_2}$ are operationally equivalent through $\tT_o\in\Trnset{\rA_{R'_1}\to\rA_{R'_2}}$. We say that $R_2$ is \emph{homologous} to $R_1$ if there exists an admissible GUR $(G,\rA,\tT^\dag)$, such that 
for all schemes $(\{\tA^{(i)}_k\}_{k=1}^m,\{\tB^{(i)}_l\}_{l=1}^n)\subseteq\Trnset{\rA_{R_i}\rC\to\rA_{R_i}\rC}_Q\times\Trnset{\rA_{N^+_{R_i}}\rC\to\rA_{N^+_{R_i}}\rC}_Q$
representing the same pair of tests for $i=1,2$, 
one has $\tB^{(1)}_l\tV\tA^{(1)}_k=\tT^{-1}\tB^{(2)}_l\tV\tA^{(2)}_k\tT$.
We denote this relation as $R_2\bowtie R_1$. 
\end{definition}

We remark that, while the relation $\bowtie$ defined above clearly depends on $\tV$, we will generally not make this fact explicit in the notation, as we will always consider contexts where a given $\tV$ is considered, and there will not be room for ambiguity. On the other hand, when we want to specify the GUR $\tT$ that makes $R_1$ homologous to $R_2$, we write $R_2\bowtie_\tT R_1$.

\begin{lemma}\label{lem:invhom}
If $R_2\bowtie_\tT R_1$ then $R_1\bowtie_{\tT^{-1}} R_2$.
\end{lemma}
\begin{proof}
For every scheme $(\{\tA^{(i)}_k\}_{k=1}^m,\{\tB^{(i)}_l\}_{l=1}^n)$ with $\tA^{(1)}_k=(\tT_i^{-1}\otimes\tI_\rC)\tA^{(2)}_k(\tT_i\otimes\tI_\rC)$ and $\tB^{(1)}_l=(\tT_o^{-1}\otimes\tI_\rC)\tB^{(2)}_l(\tT_o\otimes\tI_\rC)$, one has 
$\tB^{(1)}_l\tV\tA^{(1)}_k=\tT^{-1}\tB^{(2)}_l\tV\tA^{(2)}_k\tT$. Now, inverting $\tT$ on the right and $\tT^{-1}$ on the left, one obtains $\tB^{(2)}_l\tV\tA^{(2)}_k=\tT\tB^{(1)}_l\tV\tA^{(1)}_k\tT^{-1}$.
\end{proof}

\begin{lemma}\label{lem:inv}
Let $\tV$ be a global rule, and let the region $R_2$ be homologous to $R_1$ through $\tT$. Then $\tT^{-1}\tV\tT=\tV$.
\end{lemma}
\begin{proof} It is sufficient to consider the special case where the test scheme is $(\{\tA^{(1)}_0\},\{\tB^{(1)}_0\})$, with $\tA^{(1)}_0=\tI_{R_1\rC}$ and $\tB^{(1)}_0=\tI_{N^+_{R_1}\rC}$. Then, we have $\tA^{(2)}_0=\tI_{R_2\rC}$, $\tB^{(2)}_0=\tI_{N^+_{R_2}\rC}$, and $\tV=\tT^{-1}\tV\tT$.
\end{proof}

\begin{lemma}\label{lem:equival}
Let $\tV$ be a global rule, and let $R_2\bowtie R_1$ for some suitable $\tT$. Then 
\begin{align*}
&\tT\Trnset{\rA^{(G)}_{R_1}\rC\to\rA^{(G)}_{R_1}\rC}\tT^{-1}=\Trnset{\rA^{(G)}_{R_2}\rC\to\rA^{(G)}_{R_2}\rC},\\
&\tT\Trnset{\rA^{(G)}_{N^+_{R_1}}\rC\to\rA^{(G)}_{N^+_{R_1}}\rC}\tT^{-1}=\Trnset{\rA^{(G)}_{N^+_{R_2}}\rC\to\rA^{(G)}_{N^+_{R_2}}\rC}.
\end{align*}
\end{lemma}
\begin{proof} 
Let us choose the test scheme as $(\{\tA^{(1)}_k\}_{k=1}^m,\{\tB^{(1)}_0\})$, with $\tB^{(1)}_0=\tI_{R'_1\rC}$. Then, we have $\tB^{(2)}_0=\tI_{R'_2\rC}$, and 
\begin{align*}
&\tV\tA^{(1)}_k=\tT^{-1}\tV\tA^{(2)}_k\tT\\
=&\tT^{-1}\tV(\tT_i\otimes\tI_\rC)\tA^{(1)}_k(\tT_i^{-1}\otimes\tI_\rC)\tT\\
=&\tT^{-1}\tV\tT\tT^{-1}(\tT_i\otimes\tI_\rC)\tA^{(1)}_k(\tT_i^{-1}\otimes\tI_\rC)\tT\\
=&\tV\tT^{-1}(\tT_i\otimes\tI_\rC)\tA^{(1)}_k(\tT_i^{-1}\otimes\tI_\rC)\tT,
\end{align*}
where in the last equality we used the result of lemma~\ref{lem:inv}. If we now invert $\tV$ to the left on both sides, we obtain
\begin{align*}
\tA^{(1)}_k&=\tT^{-1}(\tT_i\otimes\tI_\rC)\tA^{(1)}_k(\tT_i^{-1}\otimes\tI_\rC)\tT,
\end{align*}
and finally
\begin{align*}
&\tT\tA^{(1)}_k\tT^{-1}\\
=&(\tT_i\otimes\tI_\rC)\tA^{(1)}_k(\tT_i^{-1}\otimes\tI_\rC)=\tA^{(2)}_k.
\end{align*}
A similar argument leads us to conclude that
\begin{align*}
&\tT\tB^{(1)}_k\tT^{-1}\\
=&(\tT_o\otimes\tI_\rC)\tB^{(1)}_k(\tT_o^{-1}\otimes\tI_\rC)
=\tB^{(2)}_k.\qedhere
\end{align*}
\end{proof}

We remark that, straightforwardly,
\begin{align}
(R_2\bowtie R_1)\  \Rightarrow\  (\rA_{R_1}\cong \rA_{R_2}).\label{eq:hom->eq}
\end{align}

Now, the discrimination of systems $g_1$ and $g_2$ is successful if there exists a region $R_1\ni g_1$ such that for any operationally equivalent region $R_2\ni g_2$, the region $R_2$ is not homologous to $R_1$. In precise terms, we have the following definition.

\begin{definition}[Absolute discrimination]\label{def:absdiscr}
The evolution given by the global rule $\tV$ discriminates two systems $g_1,g_2\in G$ if for any GUR $\tT$ there exists a region $R_1$, containing $g_1$, for which there is no region $R_2$, containing $g_2$, that is homologous to $R_1$ through $\tT$, with $g_1\bowtie_\tT g_2$.
\end{definition}

In simple words, the evolution given by the global rule $\tV$ allows for discrimination of the two systems $g_1,g_2\in G$ if there exists a test that gives different probabilities depending on whether it is applied at $g_1$ or at $g_2$, or if the tests that one can run on $R_1$ cannot be performed on any region $R_2$ containing $g_2$, or viceversa.

Now, 
the notion of discrimination of systems $g_1$ and $g_2$ with respect to a third system $e$ is given as follows.

\begin{definition}[Relative discrimination]\label{def:reldiscr}
The evolution $\tV$ discriminates two systems $g_1,g_2\in G$ relatively to a reference system $e$, with $e\in G$, if for every $\tT$ there exists a region $R_1$, containing $\{g_1,e\}$, for which there is no region $R_2$, containing $\{g_2,e\}$, homologous to $R_1$ through $\tT$ with $e\bowtie_\tT e$ and $g_1\bowtie_\tT g_2$.
\end{definition}

%
%
%

We can now require the evolution described by a global rule to be homogeneous by a precise statement.
\begin{principle}[Homogeneity]\label{princ:hom}
The global rule $\tV$ discriminates any two systems $g_1\neq g_2\in G$, relatively to an arbitrary reference system $e\in G$, but not absolutely.
\end{principle}

Before providing a convenient restatement of the principle, we analyse some of its aspects and consequences. In the first place, the principle says that if we do not choose a reference element $e\in G$, every two systems ${g_1}$ and ${g_2}$ cannot be discriminated. If we take the negation of definition~\ref{def:absdiscr}, we obtain a straightforward consequence of homogeneity.
\begin{lemma}
Let $\tV$ satisfy the homogeneity principle~\ref{princ:hom}. For every pair $g_1,g_2\in G$, there exists $\tT$ such that for every choice of region $R_1\ni g_1$, there exists an operationally equivalent region $R_2\ni g_2$ through $\tT_i\in\Trnset{\rA_{R_1}\to\rA_{R_2}}$ such that also $N^+_{R_1}$ and $N^+_{R_2}$ are operationally equivalent through $\tT_o\in\Trnset{\rA_{R'_1}\to\rA_{R'_2}}$, and for all 
schemes $(\{\tA^{(i)}_k\}_{k=1}^m,\{\tB^{(i)}_l\}_{l=1}^n)$ representing the same tests for $i=1,2$, 
one has 
$\tB^{(1)}_l\tV\tA^{(1)}_k=\tT\tB^{(2)}_l\tV\tA^{(2)}_k\tT^{-1}$.
\end{lemma}
\begin{proof}
The statement follows negating definition~\ref{def:absdiscr}.
\end{proof}

We can now draw a first non trivial consequence from the statement of principle~\ref{princ:hom}.

\begin{lemma}\label{lem:equiv}
For a homogeneous GUR $(G,\rA,\tV^\dag)$, all local systems $\rA_g$ for $g\in G$ are operationally equivalent.
\end{lemma}
\begin{proof} Let us consider two elements $g_1,g_2\in G$. As a consequence of the homogeneity principle~\ref{princ:hom}, for the special choice $R_1=\{g_1\}$, there exists $R_2\ni g_2$ such that $R_2\bowtie\{g_1\}$ for a suitable $\tT$. Now, by lemma~\ref{lem:cardopeq}, it must be $|R_2|=|\{g_1\}|=1$, i.e.~$R_2=\{g_2\}$.
Thus, we have
\begin{align}
&\rA_{g_2}\cong\rA_{g_1}.&&\qedhere
\end{align}
\end{proof}
\begin{corollary}
If the region $R_2$ containing $g_2$ is homologous to $\{g_1\}$ through $\tT$, then $R_2=\{g_2\}$, and 
\begin{align}
\tT^{-1}\Trnset{\rA^{(G)}_{g_1}\rC\to\rA^{(G)}_{g_1}\rC}\tT&=\Trnset{\rA^{(G)}_{g_2}\rC\to\rA^{(G)}_{g_2}\rC}.
\end{align}
\end{corollary}
\begin{proof} By the proof of lemma~\ref{lem:equiv} one has that ${R_2}={g_2}$.
The thesis then follows from lemma~\ref{lem:equival}.
\end{proof}

The second result that we prove is a technical lemma that will play a crucial role in the following results. 
\begin{lemma}\label{lem:tech}
Let $(G,\rA,\tV^\dag)$ satisfy the homogeneity requirement. Then for every region $R$ one has $\rA_R=\bigotimes_{x\in R}\rA_{g_x}\cong \rA_{0}^{\otimes|R|}$, where $\rA_0\cong\rA_g$ for every $g\in G$. 
\end{lemma}
\begin{proof} 
By lemma~\ref{lem:equiv} we know that for every $g\in G$ one has $\rA_g\cong\rA_0$. The remaining part of the statement thus follows straightforwardly.
\end{proof}

\begin{lemma}\label{lem:homolequi}
Let $\tV$ be non trivial and satisfy the homogeneity requirement, and $R_2\bowtie R_1$. Then $|R_1|=|R_2|$. 
\end{lemma}
\begin{proof} 
This follows from lemmas~\ref{lem:cardopeq} and~\ref{lem:tech}, since $R_1$ and $R_2$ are operationally equivalent by definition.
\end{proof}


Let now $g_1$ and $g_2$ be an arbitrary pair of systems in $G$, and $\tV$ a homogeneous GUR. By the homogeneity principle~\ref{princ:hom}, there exists $\tT$ such that for every region $R_1\ni g_1$ there is a region $R_2\ni g_2$ homologous through $\tT$ to $R_1$.
The set of these transformations is denoted as $T_\tV$, i.e.
\begin{align*}
T_\tV\coloneqq \{\tT\mid \exists &g_1,g_2\in G
\nonumber\\
&:\forall R_1\ni g_1\,\exists R_2\ni g_2,\,R_2\bowtie_\tT R_1\}.
\end{align*}

\begin{lemma}\label{lem:perm}
Let $\tV$ satisfy homogeneity, and let $\tT\in T_\tV$. Then there exists a permutation $\pi:G\to G$ such that if $R_2\bowtie_\tT R_1$ it is
\begin{align}
R_2=\pi(R_1)
\end{align}
\end{lemma}
\begin{proof}
By definition, if $\tT\in T_\tV$  there exist $g_1,g_2\in G$ such that for every $S_1\ni g_1$ there is $S_2\ni g_2$ such that $S_2\bowtie_\tT S_1$. Let now $R_2\bowtie_\tT R_1$, and $f_1\in R_1$. Consider the finite region $V_1\coloneqq \{f_1,g_1\}$. Then there is a region $V_2\ni g_2$ such that $V_2\bowtie_\tT V_1$, and  by lemma~\ref{lem:homolequi}, one has $|V_2|=|V_1|=2$. Consequently, it must be $V_2=\{f_2,g_2\}$. Moreover, since we know that $g_2\bowtie_\tT g_1$, by definitions~\ref{def:opeqreg}, \ref{def:sametest}, and~\ref{def:hom}, it is $f_2\bowtie_\tT f_1$. Thus, for every $f_1\in R_1$ there is a unique $f_2\in R_2$ such that $f_2\bowtie_\tT f_1$. We then define $\pi:G\to G$ by setting $\pi(f_1)\coloneqq f_2$. By lemma~\ref{lem:invhom} the map $\pi$ is both injective and surjective, and thus it is a permutation of $G$.
\end{proof}

In the following, given $\tT\in T_\tV$, we will denote $\tT=\tT_\pi$ where $\pi$ is the permutation of $G$ in lemma~\ref{lem:perm}.

\begin{lemma}
Let $(G,\rA,\tV^\dag)$ be a homogeneous GUR, and $\pi\in\Pi_\tV$. Then the GUR $\tT_\pi$ is unique.
\end{lemma}
\begin{proof}
Let $\tT_\pi,\tS_\pi\in T_\tV$ correspond to the same permutation $\pi$. Let $a\in\Cntsetcomp{\rA_G}{\rC}_{Q\Reals}$. Then by lemma~\ref{lem:qltransqleff} one has $a=\tA^\dag e_{G'}$, with $\tA\in\Trnset{\rA_G\rC\to\rA_G\rC}_{Q\Reals}$. 
Thus
\begin{align*}
(\tS_\pi&\tT_\pi^{-1})^\dag a=[(\tS_\pi\tT_\pi^{-1})^\dag\otimes\tI_\rC^\dag] \tA^\dag e_{G'}\\
&=(\tT_\pi^{-1\dag}\tS^\dag_\pi\otimes\tI_\rC^\dag) \tA^\dag (\tS^{-1\dag}_\pi\tT_\pi^{\dag}\otimes\tI_\rC^\dag) e_{G'},
\end{align*}
where $G'=G\cup h_0$ and $\rA_{h_0}\cong \rC$. Now, by definition 
\begin{align*}
(\tS^\dag_\pi\otimes\tI_\rC^\dag) &\tA^\dag (\tS^{-1\dag}_\pi\otimes\tI_\rC^\dag)\\
&=(\tT^\dag_\pi\otimes\tI_\rC^\dag) \tA^\dag (\tT^{-1\dag}_\pi\otimes\tI_\rC^\dag)\\
&=(\tT^\dag_i\otimes\tI_\rC^\dag) \tA^\dag (\tT^{-1\dag}_i\otimes\tI_\rC^\dag), 
\end{align*}
which implies
\begin{align*}
(\tS_\pi\tT_\pi^{-1})^\dag a&=a.
\end{align*}
Since the latter holds for every $\rC$ and every $a$, we conclude that $\tS_\pi=\tT_\pi$.
\end{proof}

\begin{corollary}
The set $T_\tV=\{\tT_\pi\}$ is a representation of a group $\Pi_\tV=\{\pi\mid\tT_\pi\in T_\tV\}$ of permutations of $G$.
\end{corollary}
We can summarise the above results as follows. For every pair $g_1,g_2$ there exists a reversible map $\pi:G\to G$ such that $\pi(g_1)=g_2$ and a reversible transformation $\tT_\pi$ of $\Stset{\rA_G}_Q$ such that $\tT_\pi^{-1}\cdot\tT_\pi$ is an automorphism of $\Trnset{\rA_G\to\rA_G}_Q$ and for every $R\subset G$ and every system $\rC$ 
\begin{align*}
\Trnset{\rA^{(G)}_{R}\rC\to&\rA^{(G)}_R\rC}_{Q*}\\
&=\tT_\pi^{-1}\Trnset{\rA^{(G)}_{\pi(R)}\rC\to\rA^{(G)}_{\pi(R)}\rC}_{Q*}\tT_\pi, 
\end{align*}
and 
$\tT_\pi$ leaves the GUR $\tV$ invariant, i.e.
\begin{align*}
\tT^{-1}_\pi \tV\tT_{\pi}=\tV.
\end{align*}
The transformations $\pi$ clearly form a group $\Pi_\tV$ whose action $\Pi_\tV\times G\to G$ is {\em transitive}.


We now proceed considering the second part of the homogeneity principle, i.e.~the statement that discrimination of every pair $g_1\neq g_2$ is possible relative to some third element $e\in G$.  The main consequence is that the action $\Pi_\tV\times G\to G$ is {\em free}.

\begin{lemma}\label{lem:freeact}
Let $\tV$ satisfy the homogeneity principle~\ref{princ:hom}. If $e\neq \pi\in 
\Pi_\tV$, there is no element $g$ of $G$ such that $\pi(g)=g$. 
\end{lemma}
\begin{proof}
Let $\tT_\pi\in T_\tV$. Suppose that there is $g\in G$ such that $\pi(g)=g$. Then for every region $R_1\ni g$, the region $R_2\coloneqq \pi(R_1)$ contains $g$ and one has $R_2\bowtie_\tT R_1$, with $g\bowtie_\tT g$. This is in contradiction with definition~\ref{def:reldiscr}, i.e.~with the homogeneity principle~\ref{princ:hom}.
\end{proof}

\begin{corollary}\label{cor:regularact}
Let $\Pi_\tV$ be the group of permutations of $G$ such that $T_\tV$ is a representation of $\Pi_\tV$. The action of $\Pi_\tV$ on $G$ is regular.
\end{corollary}

We can now summarise all the results that we drew from homogeneity as follows:
for every $g_1,g_2\in G$ there exists a permutation $\pi$ of $G$ such that $\pi(g_1)=g_2$, and a GUR $\tT_\pi$ of $\Cntset{\rA_G}_{Q\Reals}$ such that for every $R\in\reg {G'}$ and $\rC$
\begin{align}
\Trnset{\rA^{(G)}_{R}\rC&\to\rA^{(G)}_R\rC}_{Q*}\nonumber\\
&=\tT_\pi^{-1}\Trnset{\rA^{(G)}_{\pi(R)}\rC\to\rA^{(G)}_{\pi(R)}\rC}_{Q*}\tT_\pi, 
\label{eq:tpi}
\end{align}
and 
$\tT_\pi$ leaves the evolution invariant, i.e.
\begin{align*}
\tT^{-1}_\pi \tV\tT_{\pi}=\tT.
\end{align*}
The transformations $\pi$ form a group $\Pi_\tV$ that acts regularly (i.e.~transitively and freely) on $G$.



Now, it is clear that, chosen $e\in G$, the elements of $G$ can be identified with the elements of $\Pi_\tV$, as the action of $\Pi_\tV$ on $G$ is regular. In other therms, $G$ is a principal homogeneous space for $\Pi_\tV$, and there is a bijection $\Pi_\tV\leftrightarrow G$. However, the homogeneity principle implies an even stronger result: the set $G$ can be seen as the set of vertices of a special Cayley graph of $\Pi_\tV$. The following arguments, largely inspired to Refs.~\cite{PhysRevA.90.062106,Ariano2016}, make the mentioned result rigorous.

As a consequence of homogeneity, one has
\begin{align}
\pi(g)\rightarrowtail\pi(g')\ \Leftrightarrow\ g\rightarrowtail g'.
\label{eq:necsuffggp}
\end{align}
In other words, $N^\pm_{\pi(g)}=\pi(N^\pm_g)$ for $\pi\in\Pi_\tV$, as can be checked starting from the definition of $N^\pm_g$. 
The proof is provided in Appendix~\ref{app:sig}.

Let now $g\in G$ and $\pi\in \Pi_\tV$ such that $\pi(e)=g$. One can then label the elements $g'\in N^\pm_g$ by $\pi^{-1}(g')\in N^{\pm}_e=\{h_i^{\pm 1}\}_{i=1}^k$. Notice that for the moment we have no reason to claim that $k<\infty$. We will associate all the pairs $(g,g')\in E$ with the {\em color} $h_i$ if $g=\pi(e)$ and $g'=\pi(h_i)$ for some $\pi\in\Pi_\tV$ \footnote{In principle, the colour of an edge $(g,g')$ might be ambiguously defined if there existed two permutations $\pi,\sigma\in\Pi_\tV$ such that $\pi^{-1}(g)=\sigma^{-1}(g)=e$ and $\pi^{-1}(g'),\sigma^{-1}(g')\in N^\pm_e$, but $\pi^{-1}(g')\neq\sigma^{-1}(g')$. However, due to lemma~\ref{lem:freeact}, this is never the case, since it would imply that $\sigma\pi^{-1}(g')\neq g'$, and thus $\sigma\pi^{-1}\neq e$, but $[\sigma\pi^{-1}](g)=\sigma(e)=g$.}. This construction enriches the graph $\Gamma(G,E)$, which becomes a vertex-transitive coloured directed graph, with colours corresponding to the labels $h_i\in N^+_e$ 

Given two elements $g,g'\in G$, a {\em path} $p$ from $g$ to $g'$ is an array of elements $p=(g_1, g_2,\ldots, g_k)\in G^{\times k}$ such that, setting $g_0\coloneqq g$ and $g_{k+1}\coloneqq g'$, one has
\begin{align*}
\forall 0\leq i\leq k\quad (g_i\rightarrowtail g_{i+1})\vee(g_{i+1}\rightarrowtail g_i).
\end{align*}
The {\em lenght} of a path $p$, denoted $\ell(p)$ is $k+1$ if $p\in G^{\times k}$. In the following we will restrict attention to the case where $\Gamma(G,E)$ is {\em connected}, i.e.~for every two elements $g,g'\in G$ there exists a path from $g$ to $g'$. The set of paths from $g$ to $g'$ will be denoted by $\set p^+_{g,g'}$. We can now introduce the graph metric of $\Gamma(G,E)$ as follows. Let $g,g'\in G$. Then we define
\begin{align*}
d_\Gamma(g,g')\coloneqq \min_{p\in \set p^+_{g,g'}}\ell (p)
\end{align*}
We remark that $d_\Gamma(g,g')=d_\Gamma[\pi(g),\pi(g')]$ for all $\pi\in\Pi_\tV$, since $g_1,\ldots,g_k$ is a path from $g$ to $g'$ if and only if $\pi(g_1),\ldots,\pi(g_k)$ is a path from $\pi(g)$ to $\pi(g')$ due to condition~\eqref{eq:necsuffggp}.

We now use the alphabet $N^+_e\cup N^-_e$ to form arbitrary words, obtaining a free group F: composition corresponds to word juxtaposition, with the empty word $\lambda$ representing the identity, and the formal rule $h_{i} h_i^{-1} = h_i^{-1}h_i = \lambda$. An element $w = h_{i_1}^{p_1}h_{i_2}^{p_2}\ldots h_{i_n}^{p_n}$ of $F$---with $p_j \in\{-1, 1\}$---thus corresponds to a path on the graph, where the symbol $h_i^{-1}\in N^-_e$ denotes a backwards step along the arrow $h_i$. For every $h_{i_1}^{p_1}h_{i_2}^{p_2}\ldots h_{i_m}^{p_m}=w\in F$, one has $w^{-1}=h_{i_m}^{-p_m}\ldots h_{i_2}^{-p_2}h_{i_1}^{-p_1}$. 

The action $M:F\times G\to G$ of words $w=h_{i_1}^{p_1}\ldots h_{i_{m-1}}^{p_{m-1}}h_{i_m}^{p_m}\in F$ on elements $g\in G$ can be defined as follows
\begin{align*}
M(w,g)&=M(h_{i_1}^{p_1} h_{i_{2}}^{p_{2}}\ldots h_{i_m}^{p_m},g)\\
&\coloneqq M[h_{i_2}^{p_2}\ldots h_{i_{m}}^{p_{m}},M(h_{i_1}^{p_1},g)],
\end{align*}
where the action of $h_i\in S$ on the elements $g\in G$ is defined as $M(h_i,g)=g'$ if there exists $\pi\in \Pi_\tV$ such that $\pi^{-1}(g)=e$ and $\pi^{-1}(g')=h_i$, while $M(h_i^{-1},g)=g'$ if there exists $\pi\in \Pi_\tV$ such that $\pi^{-1}(g)=h_i$ and $\pi^{-1}(g')=e$ \footnote{The action is well defined. Suppose indeed that there are two elements $f,f'$ both satisfying the definition of $M(h_i,g)$. This means that there exist $\pi,\sigma$ such that $\pi^{-1}(g)=\sigma^{-1}(g)=e$, while $\pi^{-1}(f)=\sigma^{-1}(f')=h_i$. However, according to lemma~\ref{lem:freeact}, the first condition implies that $\sigma=\pi$. By the same lemma, the second condition then implies that $f=f'$. A similar argument shows that $M(h_i^{-1},g)$ is well defined.}. In the following we will use the shortcut
\begin{align*}
gw\coloneqq M(w,g).
\end{align*}
This notation is particularly handy, since
\begin{align*}
g(h_{i_1}^{p_1} h_{i_{2}}^{p_{2}}\ldots h_{i_m}^{p_m})=(gh_{i_1}^{p_1}) h_{i_{2}}^{p_{2}}\ldots h_{i_m}^{p_m}.
\end{align*}

For every $w\in F$, and for every $g\in G$ and for every permutation $\pi\in\Pi_\tV$, we now show that $\pi(gw)=\pi(g)w$. The first step consists in proving the result for $w=h_i^{\pm1}$. Let $g'=gh_i$, namely by definition $\sigma^{-1}(g)=e$ and $\sigma^{-1}(g')=h_i$ for some (unique) $\sigma\in \pi_\tV$. By Eq.~\eqref{eq:necsuffggp} it is $\pi(g)\rightarrowtail\pi(g')$. Moreover, if we define $f\coloneqq \pi(g)$ and $f'\coloneqq \pi(g')$ we have
\begin{align*}
&h_i=\sigma^{-1}(g')=\sigma^{-1}\pi^{-1}(f'),\\
&e= \sigma^{-1}(g)=\sigma^{-1}\pi^{-1}(f),
\end{align*}
namely $f'=fh_i$. This implies
\begin{align}\label{eq:piact1}
\pi(gh_i)=\pi(g')=\pi(g)h_i.
\end{align}
Similarly, if $g'=gh_i^{-1}$ by definition $\sigma^{-1}(g)=h_i$ and $\sigma^{-1}(g')=e$ for some (unique) $\sigma\in \pi_\tV$. Thus, setting again $f\coloneqq \pi(g)$ and $f'\coloneqq \pi(g')$, we have
\begin{align*}
&h_i=\sigma^{-1}(g)=\sigma^{-1}\pi^{-1}(f),\\
&e= \sigma^{-1}(g')=\sigma^{-1}\pi^{-1}(f'),
\end{align*}
namely $f'=fh_i^{-1}$, and
\begin{align}\label{eq:piact2}
\pi(gh_i^{-1})=\pi(g')=\pi(g)h_i^{-1}.
\end{align}
Notice that the last result implies that for every pair $f,g\in G$, given $\pi\in\Pi_\tV$ such that $\pi(f)=g$, there is a bijection $N^\pm_{f}\leftrightarrow N^\pm_g$, given by 
\begin{align}
N^\pm_{\pi(g)}=\pi(N^\pm_g). \label{eq:npiug}
\end{align}
One can now prove that $\pi(fw)=\pi(f)w$ by induction on the length $l(w)$ of the word $w$. Indeed, we know that it is true for $l(w)=1$. Suppose now that for $l(w)=n-1$ one has $\pi(fw)=\pi(f)w$, and consider $w'$ with $l(w')=n$. Then $w'=wh^{p}_i$ with $l(w)=n-1$ and $p=\pm1$. We have
\begin{align*}
\pi(fw')&=\pi(fwh_{i}^{p})=\pi[(fw)h_{i}^{p}]\\
&=\pi(fw)h_{i}^{p}=\pi(f)wh_{i}^{p}=\pi(f)w',
\end{align*}
where the induction hypothesis is used in the fourth equality.

Let us now suppose that for some $f\in G$ and some word $w\in F$ one has $fw=f$. Then for every $f'\in G$ one can take $\pi$ such that $\pi(f)=f'$, thus obtaining
\begin{align*}
f'w=\pi(f)w=\pi(fw)=\pi(f)=f'.
\end{align*}
Thus, if a path $w\in F$ is closed starting from $f\in G$, then it is closed also starting from any other $g\in G$.

We can now easily see that the subset $R$ of $F$ corresponding to words $r$ such that $gr=g$ for all $g\in G$ is a normal subgroup. Indeed, $R$ is a subgroup because the juxtaposition of two words $s,s'\in R$ is again  a word $ss'\in R$, and for every word $s\in R$ also $s^{-1}\in R$. To prove that $R$ is normal in $F$ we just show that it coincides with its normal closure, i.e.~for every $w\in F$ and every $r\in R$, we have $wrw^{-1}\in R$. Indeed, defining for arbitrary $g$ the element $g'\coloneqq gw$, we have $g'w^{-1}=g$, and thus $gwrw^{-1}=g'rw^{-1}=g'w^{-1}=g$, namely $wrw^{-1}\in R$. 

We thus identified a normal subgroup $R$ containing all the words $r$ corresponding to closed paths. If one takes the quotient $F/R$, one obtains a group whose elements are equivalence classes of words in $F$. If we label an arbitrary element of $G$ by $e$, it is clear that the elements of $G$ are in one-to-one correspondence with the vertices of $G$, since for every $g\in G$ there is one and only one class in $F/R$ whose elements lead from $e$ to $g$. We can then write $g=w$ for every $w\in F$ such that $w$ represents a path leading from $e$ to $g$. Notice that the elements of $F/R$, i.e.~equivalence classes of words $[w]_R$, correspond to elements of $G$, i.e.~for every $g\in G$ there is a unique class $[w]_R$ such that $g=[w]_R$. The criterion for class membership of words is very simple: $w\in g$ iff $w$ connects the vertex $e\in G$ to the vertex $g\in G$.

As a consequence of the above arguments, we can now show that for every $\pi\in \Pi_\tV$ and for every $g=[w]_R\in G$, one has $\pi(g)=[sw]_R$ for a fixed word $s\in F$. Indeed, consider $g_0=[w_0]_R$. Let $[w']_R=\pi(g_0)$. Then $w'=w'w_0^{-1}w_0=sw_0$, with $s\coloneqq w'w_0^{-1}$, and $\pi(g_0)=[sw_0]_R$. Let now $f\in G$. We can always find $t\in F$ so that $f=[w_0t]_R$, e.g.~by setting $t\coloneqq w_0^{-1}z$, $f=[z]_R$. Then we have $\pi(f)=\pi([w_0t]_R)=\pi(g_0t)=\pi(g_0)t=[sw_0t]_R=[sz]_R$, where we use the definition $\pi(k)r\coloneqq [xr]_R$ with $\pi(k)=[x]_R$. This definition clearly makes sense, since when $[a]_R=[b]_R$ and $[r]_R=[u]_R$, also $[ar]_R=[bu]_R$.  

In technical terms, the graph $\Gamma(G,E)=\Gamma(G,S)$ is the Cayley graph of the group $G=F/R$. Homogeneity thus implies that the set $G$ is a group $G$ that can be presented as $G=\<S|R\>$, where $S$ is the set of {\em generators} of $G$ and $R$ is the group of {\em relators}. In the following, if $h_i=h_i^{-1}$ we will draw an undirected edge to represent $h_i$. The presentation can be chosen by arbitrarily dividing $S$ into $S_+\subseteq S$ and $S_-\coloneqq S_+^{-1}$ in such a way that $S_+\cup S_-=S$. The above arbitrariness is inherent the very notion of group presentation and corresponding Cayley graph, and is exploited in the literature, in particular in the definition of isotropy \cite{PhysRevA.90.062106,PhysRevA.96.062101}.

For convenience of the reader we remind the definition of \emph{Cayley
  graph}. Given a group $G$ and a set $S_+$ of generators of the group,
the Cayley graph $\Gamma(G,S_+)$ is defined as the colored directed
graph having vertex set $G$, edge set $\{(g,gh)\mid g\in G, h\in S_+\}$, and
a color assigned to each generator $h\in S_+$. Notice that a Cayley
graph is \emph{regular}---i.e.~each vertex has the same degree---and
\emph{vertex-transitive}---i.e.~all vertices are equivalent, in the sense that
the graph automorphism group acts transitively upon its
vertices. The Cayley graphs of a group $G$ are in one to one
correspondence with its presentations, with $\Gamma(G,S_+)$
corresponding to the presentation $\<S_+|R\>$. 



\section{Locality}\label{sec:local}

If a GUR were to represent a physical law, we would like that, in order to determine it, it is sufficient to test it in a finite region and for a finite amount of time. This possibility is granted by homogeneity in conjunction with a second principle---{\em locality}---to which the present section is devoted. The locality requirement can be phrased in simple words by stipulating that in order to calculate the effects of a physical law, i) it is possible to adopt a reductionist procedure, decomposing an arbitrary system into elementary parts and then calculating its evolution by evolving every part separately, considering only pairwise interactions between parts, and ii) if the reduction is operated, only finite systems will be involved in the calculation for every elementary part. 

In the present section we introduce the precise statement of the locality principle, and analyse its main consequences. In order to make this notion independent of homogeneity, we formulate it for general admissible GURs. In particular, this implies that the mathematical statement of locality alone will not allow for determination of a local GUR by observation of a finite region.

Before giving a formal definition, we provide a few heuristic steps leading to the final formulation. Let us start considering an admissible GUR $(G,\rA,\tV^\dag)$, with uniformly bounded neighbourhood, namely such that there exists $k<\infty$ so that for every $g\in G$ it is $|N^\pm_g|\leq k$. We remark that, since $|N^\pm_g|\leq k$, it is $|P^\pm_g|\leq k^2$. This also implies that the transformations $\tS'_g$ are transformations of finitely many systems.

In the remainder, we will often omit the labels $0,1$ on the wires in diagrams, and adopt the convention that the upper half is labelled $0$ and the lower one $1$.


Now, reminding Eq.~\eqref{eq:blockdir}, given $a\in\Cntset{\rA^{(G)}_R\rC}_{Q\Reals}$, one has
\begin{align*}
[(\tV^\dag a)_{(N^-_R,1)\rC}]&=[\{(\tV_R^\dag\otimes\tI_\rC^\dag) a\}_{(N^-_R,1)\rC}]\\
&=[\{({\tS^{\prime\dag}_{N^-_R}}\otimes\tI_\rC^\dag)a\}_{(N^-_R,1)\rC}].
\end{align*}
Thus, we have 
\begin{align}
&\begin{aligned}
    \Qcircuit @C=1em @R=.7em @! R {&\ustick{N^-_R}\qw &\gate{\tV_R}&\ustick{R}\qw&\qw
    }
\end{aligned}\ =\ 
\begin{aligned}
    \Qcircuit @C=1em @R=.7em @! R {\prepareC{\psi}&\ustick{P^-_R}\qw &\multigate{2}{\tS'_{N^-_R}}&\qw&\ustick{R}\qw&\qw&\qw\\
    &&\pureghost{\tS'_{N^-_R}}&\qw&\ustick{P^-_R\setminus R}\qw&\qw&\measureD{e}\\
    &\ustick{N^-_R}\qw &\ghost{\tS'_{N^-_R}}&\qw&\ustick{N^-_R}\qw&\qw&\measureD{e}}
\end{aligned}\ .
\label{eq:vr}
\end{align}
Notice that, by admissibility of the GUR $\tV$, and thanks to the no-restriction hypothesis, 
\begin{align*}
\tV_R\in\Trnset{\rA_{N^-_R}\to\rA_R}_1. 
\end{align*}

Moreover, the GUR $\tV$ can be equivalently defined starting from its action on $\Cntset{\rA_G}_{L\Reals}$, by setting $\tV^\dag[a_R]\coloneqq [(\tV^\dag_R a)_{N^-_R}]$. Indeed, following this definition, one can prove that $\tV$ is right-invertible, by the argument discussed in Appendix~\ref{app:rinvlinv}. We remark that the construction of the inverse provides a new GUR $\tW$.

As a consequence, we have that $\tW^\dag\tV^\dag=\tV^\dag\tW^\dag=\tI^\dag_G$ for some GUR $\tW$, and thus $\tW=\tV^{-1}$. We want to stress here that in proving this result we only use properties~\eqref{eq:condone}, \eqref{eq:condtwo}, \eqref{eq:condthree} and~\eqref{eq:condfour} of $\tS'_g$, derived in section \ref{s:block}, along with uniform boundedness of the neighbourhoods for the GUR $\tV$ (see Appendix~\ref{app:rinvlinv} for the details). 


Let us now abstract our focus from the situation where a GUR $(G,\rA,\tV^\dag)$ is given. Instead,  we will suppose that, given the pair $(G,\rA)$, a finite neighbourhood $N^+_g\subseteq G$ is defined for every $g\in G$, and for $f\in G$ we set
\begin{align*}
N^-_f\coloneqq \{h\in G\mid f\in N^+_h\}.
\end{align*}
For $X\in\reg G$ we then define $N^\pm_X\coloneqq\bigcup_{g\in X}N^\pm_g$, and the sets $P^\pm_g$ are then defined exactly as for a GUR. Finally, we suppose that a family of maps $\tS'_g$ is defined, with the properties~\eqref{eq:condone} and~\eqref{eq:condtwo}, and an analogue of~\eqref{eq:condthree} and~\eqref{eq:condfour}.

\begin{definition}[Local rule]\label{def:lur}
A $k$-{\em local rule} $(N^+,\rA,\tS')$ on $G$ is i) a map $N^+:G\to\reg G $ that associates $g\in G$ with a finite region $N^+_g\in \reg G $ such that $|N^+_g|,|N^-_g|\leq k$ for every $g\in G$; 
ii) a map $\tS':g\mapsto\tS'_g$, where $\tS'_g$ are maps in $\Trnset{\rA_{(N^+_g,0)}\rA_{(g,1)}\to\rA_{(N^+_g,0)}\rA_{(g,1)}}_1$ with the following properties for every $f,g\in G $ and $R,S\in\reg G$:
\begin{enumerate}[leftmargin=*]
\item \label{it:idemp} $\tS'_g\tS'_g=\tI_{\rA_{N^+_g}\rA_g}$;
\item \label{it:commuz} $\tS'_g\tS'_f=\tS'_f\tS'_g$; 
\item \label{it:mixev}for every $\tC\in\Trnset{\rA_{(R,0)\cup(S,1)}\rC\to\rA_{(R,0)\cup(S,1)}\rC}$
\begin{adjustwidth}{-\leftmargini}{0cm}
\begin{align}
&\begin{aligned}
    \Qcircuit @C=1em @R=.7em @! R {&\ustick{P^-_R\setminus N^+_S}\qw &\qw&\multigate{4}{\tS'_{N^-_R\cup S}}&\qw&\ustick{N^+_{N^-_R\cup S}\setminus R}\qw&\qw&\multigate{4}{\tS'_{N^-_R\cup S}}&\qw&\ustick{P^-_R\setminus N^+_S}\qw&\qw\\
    &\ustick{N^+_S}\qw &\qw&\ghost{\tS'_{N^-_R\cup S}}&
    \ustick{R}\qw&\multigate{2}{\tC}&\ustick{R}\qw&\ghost{\tS'_{N^-_R\cup S}}&\qw&\ustick{N^+_S}\qw&\qw\\
        &&&&\ustick{\rC}&\ghost{\tC}&\ustick{\rC}\qw&&&&\\
    &\ustick{N^-_R}\qw &\qw&\ghost{\tS'_{N^-_R\cup S}}&\ustick{S}\qw&\ghost{\tC}&\ustick{S}\qw&\ghost{\tS'_{N^-_R\cup S}}&\qw&\ustick{N^-_R}\qw&\qw\\
    &\ustick{S\setminus N^-_R}\qw &\qw&\ghost{\tS'_{N^-_R\cup S}}&\qw&\ustick{N^-_{R}\setminus S}\qw&\qw&\ghost{\tS'_{N^-_R\cup S}}&\qw&\ustick{S\setminus N^-_R}\qw&\qw}
\end{aligned}
\nonumber\\
\nonumber\\
&\qquad\qquad=
\begin{aligned}
    \Qcircuit @C=1em @R=.7em @! R {  &\qw&\ustick{P^-_R\setminus N^+_S}\qw&\qw&\qw\\
  &\ustick{N^+_S}\qw &\multigate{2}{\tC'}&\ustick{N^+_S}\qw&\qw\\
    &\ustick{\rC}\qw&\ghost{\tC'}&\ustick{\rC}\qw&\qw\\
     &\ustick{N^-_R}\qw&\ghost{\tC'}&\ustick{N^-_R}\qw&\qw\\
     &\qw&\ustick{S\setminus N^-_R}\qw&\qw&\qw}
\end{aligned}\ ,
\label{eq:mixev}
\end{align}
\end{adjustwidth}
where  $\tS'_X\coloneqq \prod_{g\in X}\tS'_g$ for every $X\subseteq G$.
\end{enumerate}
\end{definition}
In the following, we will often omit the subscript $X$ in $\tS'_X$ in the diagrams, since $X$ corresponds to the label of the lower half wires. Moreover, in order to make the equations lighter, from now on we will always write or draw transformations and effects that do not involve explicitly the external system $\rC$, however it will always be assumed that the conditions hold also locally on extended systems, unless explicitly stated otherwise.

\begin{remark}\label{rem:trivext}
Notice that for a transformation $\tC$ of the form $\tC_0\otimes\tI_{(R\setminus R',0)\cup(S\setminus S',1)}$, due to the definition of $\tS'_{N^-_R\cup S}$ and the property \ref{it:commuz} of $\tS'_g$, one has that
\begin{align*}
&\begin{aligned}
    \Qcircuit @C=1em @R=.7em @! R {  &\qw &\qw&\qw&\ustick{P^-_R\setminus N^+_S}\qw&\qw&\qw&\qw\\
  &\qw&\ustick{N^+_S}\qw &\qw&\multigate{2}{\tC'}&\qw&\ustick{N^+_S}\qw&\qw\\
    &\qw&\ustick{\rC}\qw&\qw&\ghost{\tC'}&\qw&\ustick{\rC}\qw&\qw\\  
    &\qw&\ustick{N^-_R}\qw&\qw&\ghost{\tC'}&\qw&\ustick{N^-_R}\qw&\qw\\
     &\qw &\qw&\qw&\ustick{S\setminus N^-_R}\qw&\qw&\qw&\qw}
     \end{aligned}
\ =\  
     \begin{aligned}
    \Qcircuit @C=1em @R=.7em @! R {  &\qw &\qw&\qw&\ustick{P^-_R\setminus N^+_S}\qw&\qw&\qw&\qw\\
    &\qw &\qw&\qw&\ustick{N^+_S\setminus N^+_{S'}}\qw&\qw&\qw&\qw\\
  &\qw&\ustick{N^+_{S'}}\qw &\qw&\multigate{2}{\tC'_0}&\qw&\ustick{N^+_{S'}}\qw&\qw\\
    &\qw&\ustick{\rC}\qw&\qw&\ghost{\tC'_0}&\qw&\ustick{\rC}\qw&\qw\\
    &\qw&\ustick{N^-_{R'}}\qw&\qw&\ghost{\tC'_0}&\qw&\ustick{N^-_{R'}}\qw&\qw\\
    &\qw &\qw&\qw&\ustick{N^-_R\setminus N^-_{R'}}\qw&\qw&\qw&\qw\\
     &\qw &\qw&\qw&\ustick{S\setminus N^-_R}\qw&\qw&\qw&\qw}
\end{aligned}\\
\\
&= 
\begin{aligned}
    \Qcircuit @C=1em @R=.7em @! R {&\qw &\qw&\qw&\qw&\ustick{[P^-_R\cup N^+_S]\setminus [P^-_{R'}\cup N^+_{S'}]}\qw&\qw&\qw&\qw&\qw&\\\\
    &\ustick{P^-_{R'}\setminus N^+_{S'}}\qw &\qw&\multigate{4}{\tS'}&\qw&\ustick{N^+_{N^-_{R'}\cup {S'}}\setminus {R'}}\qw&\qw&\multigate{4}{\tS'}&\qw&\ustick{P^-_{R'}\setminus N^+_{S'}}\qw&\\
    &\ustick{N^+_{S'}}\qw &\qw&\ghost{\tS'}&\ustick{{R'}}\qw&\multigate{2}{\tC_0}&\ustick{{R'}}\qw&\ghost{\tS'}&\qw&\ustick{N^+_{S'}}\qw&\\
    &&&&\ustick{\rC}&\ghost{\tC_0}&\ustick{\rC}\qw&&&&\\
    &\ustick{N^-_{R'}}\qw &\qw&\ghost{\tS'}&\ustick{{S'}}\qw&\ghost{\tC_0}&\ustick{{S'}}\qw&\ghost{\tS'}&\qw&\ustick{N^-_{R'}}\qw&\\
    &\ustick{S'\setminus N^-_{R'}}\qw &\qw&\ghost{\tS'}&\qw&\ustick{N^-_{R'}\setminus S'}\qw&\qw&\ghost{\tS'}&\qw&\ustick{S'\setminus N^-_{R'}}\qw&\\
    &\qw &\qw&\qw&\qw&\ustick{[N^-_R\cup S]\setminus [N^-_{R'}\cup {S'}]}\qw&\qw&\qw&\qw&\qw&}
\end{aligned}\ .
\end{align*}
\end{remark}

The first property that we prove is the following.
\begin{lemma}\label{lem:lurconds}
Given a local rule $(N^+,\rA,\tS')$ on $G$, the following identities hold:
\begin{enumerate}
\item \label{it:consistlur}
For every $S\in\reg G $, and every $\tA\in\Trnset{\rA_S\rC\to\rA_S\rC}$,
\begin{align}
&\begin{aligned}
    \Qcircuit @C=1em @R=.7em @! R {&\ustick{(N^+_S,0)}\qw &\qw&\multigate{2}{\tS'_{N^-_S}}&\qw&\ustick{(N^+_S,0)}\qw&\qw&\multigate{2}{\tS'_{N^-_S}}&\qw&\ustick{(N^+_S,0)}\qw&\qw\\
    &&&&\ustick{\rC}&\multigate{1}{\tA}&\ustick{\rC}\qw&&&&\\
    &\ustick{(S,1)}\qw &\qw&\ghost{\tS'_{N^-_S}}&\ustick{(S,1)}\qw&\ghost{\tA}&\ustick{(S,1)}\qw&\ghost{\tS'_{N^-_S}}&\qw&\ustick{(S,1)}\qw&\qw}
\end{aligned}
\nonumber\\
\nonumber\\
&\qquad\quad =\  
\begin{aligned}
    \Qcircuit @C=1em @R=.7em @! R {    &\ustick{(N^+_S,0)}\qw &\qw&\multigate{1}{\tA^+}&\qw&\ustick{(N^+_S,0)}\qw&\qw\\
    &&\ustick{\rC}&\ghost{\tA^+}&\ustick{\rC}\qw&&\\
    &\qw&\qw&\ustick{(S,1)}\qw &\qw&\qw&\qw}
\end{aligned}\ ;
\label{eq:forwev}
\end{align}
\item \label{it:consistlur2} for every $R\in\reg G$ and every $\tB\in\Trnset{\rA_R\rC\to\rA_R\rC}$
\begin{align}
&\begin{aligned}
    \Qcircuit @C=1em @R=.7em @! R {&\ustick{(P^-_R,0)}\qw &\qw&\multigate{3}{\tS'_{N^-_R}}&
    \qw&\ustick{(P^-_R\setminus R,0)}\qw&\qw&
    \multigate{3}{\tS'_{N^-_R}}&\qw&\ustick{(P^-_R,0)}\qw&\qw\\
    &&&\pureghost{\tS'_{N^-_R}}&
    \ustick{(R,0)}\qw&\multigate{1}{\tB}&\ustick{(R,0)}\qw&
    \ghost{\tS'_{N^-_R}}&&&\\
    &&&&\ustick{\rC}&\ghost{\tB}&\ustick{\rC}\qw&&&&\\
    &\ustick{(N^-_R,1)}\qw &\qw&\ghost{\tS'_{N^-_R}}&\qw&\ustick{(N^-_R,1)}\qw&\qw&\ghost{\tS'_{N^-_R}}&\qw&\ustick{(N^-_R,1)}\qw&\qw}
\end{aligned}\nonumber\\
\nonumber\\
&\qquad\quad=\ 
\begin{aligned}
    \Qcircuit @C=1em @R=.7em @! R {&\qw&\qw&\ustick{(P^-_R,0)}\qw &\qw&\qw&\qw\\
    &&\ustick{\rC}&\multigate{1}{\tB^-}&\ustick{\rC}\qw&&\\
    &\ustick{(N^-_R,1)}\qw &\qw&\ghost{\tB^-}&\qw&\ustick{(N^-_R,1)}\qw&\qw}
\end{aligned}\ \ .
\label{eq:backwev}
\end{align}
\end{enumerate}
\end{lemma}
\begin{proof}
The equalities follow taking $S=\emptyset$ and $R=\emptyset$ in Eq.~\eqref{eq:mixev}, respectively.
\end{proof}

\begin{remark}
Given a local rule $(N^+,\rA,\tS')$ we define
the {\em forward} and {\em backward} evolution of local transformations $\tF\in\Trnset{\rA_R\rC\to\rA_R\rC}$ for a finite region $R\in\reg G $ through the expressions~\eqref{eq:forwev} and~\eqref{eq:backwev}, as follows
\begin{align}
&[\mathsf V^\pm_L(\tF)]_{N^\pm_R\rC}\coloneqq [\tF^\pm]_{N^\pm_R\rC}.
\label{eq:calcvfloc}
\end{align}
\end{remark}


Now, once we established the notion of a local rule and its application to local transformations, 
we can define the property of a global rule of being reducible to a local rule.

\begin{definition}[Reduction to a local rule]
We say that the GUR $(G,\rA,\tV^\dag)$ is {\em reducible to a local rule} 
if there exists a $k$-local rule $(N^+,\rA,\tS')$ for some $k<\infty$
such that for every $R\in\reg G $ and every $\tF\in\Trnset{\rA_R\rC\to\rA_R\rC}$
\begin{align}
&(\tV\otimes\tI_\rC)\tF(\tV^{-1}\otimes\tI_\rC)
=\mathsf V^-_L(\tF)\otimes\tI_{G\setminus  N^-_R},\nonumber\\
&(\tV^{-1}\otimes\tI_\rC)\tF(\tV\otimes\tI_\rC)
=\mathsf V^+_L(\tF)\otimes\tI_{G\setminus  N^+_R}.\label{eq:locrul}
\end{align}
\end{definition}

We now prove the two main results of this section, which relate local rules and GURs in a one-to-one correspondence.

\begin{theorem}\label{th:lurdefgur}
Every local rule $(N^+,\rA,\tS')$ on $G$ identifies a GUR $(G',\rA,\tW^\dag)$, with $G'=G\times\{0,1\}$, and $\tW^{-1}=\tS_G\tW\tS_G$.
\end{theorem}
\begin{proof}
First of all, one can define the linear maps $\tW_\rC^\dag$ on 
$\Cntsetcomp{\rA_{G}}{\rC}_{L\Reals}$ as follows. For $a_{Z\rC}\in\Cntsetcomp{\rA_{G}}{\rC}_{L\Reals}$, with $Z=(R,0)\cup(S,1)$, we set 
\begin{align*}
\tW^\dag [a_{Z\rC}]\coloneqq [a^-_{(N^-_R,0)\cup(N^+_S,1)\rC}],
\end{align*} 
where
\begin{align*}
\\
&\quad
\begin{aligned}
    \scalebox{0.75}{\Qcircuit @C=.9em @R=1.2em @! R {
    &\ustick{\tilde N^+_{N^-_R\cup S}}\qw&\qw&\measureD{e}\\
    &\ustick{S\setminus N^-_R}\qw&\qw&\measureD{e}\\
    &\ustick{N^-_R}\qw&\qw&\multimeasureD{2}{a^-}\\
    &\ustick{\rC}\qw&\qw&\ghost{a^-}\\
    &\ustick{N^+_S}\qw&\qw&\ghost{a^-}\\
    &\ustick{P^-_R\setminus N^+_S}\qw&\qw&\measureD{e}\\
    &\ustick{\widehat {N^-_R\cup S}}\qw&\qw&\measureD{e}}}
\end{aligned}\ \coloneqq 
\ \begin{aligned}
    \scalebox{0.75}{\Qcircuit @C=.9em @R=1.2em @! R {    &\ustick{\tilde N^+_{N^-_R\cup S}}\qw&\qw&\multigate{6}{\tS}&\qw&\qw&\ustick{\widehat{N^-_R\cup S}}\qw&\qw&\qw&\qw&\qw&\measureD{e}\\
    &\ustick{S\setminus N^-_R}\qw&\qw&\ghost{\tS}&\qw&\ustick{P^-_R\setminus N^+_S}\qw&\qw&\multigate{4}{\tS'}&\qw&\ustick{N^+_{N^-_R\cup S}\setminus R}\qw&\qw&\measureD{e}\\
    &\ustick{N^-_R}\qw&\qw&\ghost{\tS}&\qw&\ustick{N^+_S}\qw&\qw&\ghost{\tS'}&\qw&\ustick{R}\qw&\qw&\multimeasureD{2}{a}\\
    &&&\pureghost{\tS}&&&&\pureghost{\tS'}&&&\ustick{\rC}\qw&\ghost{a}\\
&\ustick{N^+_S}\qw&\qw&\ghost{\tS}&\qw&\ustick{N^-_R}\qw&\qw&\ghost{\tS'}&\qw&\ustick{S}\qw&\qw&\ghost{a}\\
&\ustick{P^-_R\setminus N^+_S}\qw&\qw&\ghost{\tS}&\qw&\ustick{S\setminus N^-_R}\qw&\qw&\ghost{\tS'}&\qw&\ustick{N^-_R\setminus S}\qw&\qw&\measureD{e}\\
&\ustick{\widehat {N^-_R\cup S}}\qw&\qw&\ghost{\tS}&\qw&\qw&\ustick{\tilde N^+_{N^-_R\cup S}}\qw&\qw&\qw&\qw&\qw&\measureD{e}}}
\end{aligned}\ .
\end{align*}
We omitted the subscript for $\tS$ and $\tS'$, as the systems on which the transformations act are clear from the context. Finally, $\tilde N^+_{X}$ is a shorthand for $N^+_{X}\setminus X$, and $\widehat{X}$ for $X\setminus N^+_{X}$. Writing $a=\tA^\dag e$, for $\tA\in\Trnset{\rA_{(R,0)\cup(S,1)}\rC\to\rA_{(R,0)\cup(S,1)}\rC}$, and using Eq.~\eqref{eq:mixev}, one can easily check that the effect $a^-$ is actually in $\Cntsetcomp{\rA_{(N^-_R,0)\cup(N^+_S,1)}}{\rC}$. We remark that the map $\tW^\dag$ is well defined: for $a\otimes e_W\in[a_{Z\rC}]$, with $W=(W_0,0)\cup(W_1,1)$, and $W_0\cap R=W_1\cap S=\emptyset$, one has 
\begin{align*}
\tW^\dag[(a\otimes e_W)_{(R\cup W_0,0)\cup (S\cup W_1,1)\rC}]=\tW^\dag[a_{Z\rC}].
\end{align*}
The above identity is verified exploiting Eq.~\eqref{eq:mixev}, along with the identity in Remark~\ref{rem:trivext}.
As to reversibility, for $[a_{Z\rC}]$ with $Z=(R,0)\cup(S,1)$, one can define 
\begin{align*}
\tZ^\dag[a_{Z\rC}]\coloneqq [a^+_{(N^+_R,0)\cup(N^-_S,1)\rC}],
\end{align*}
where
\begin{align*}
\\
&\ 
\begin{aligned}
    \scalebox{0.7}{\Qcircuit @C=.9em @R=1.2em @! R {
    &\ustick{\widehat{N^+_S\cup R}}\qw&\qw&\measureD{e}\\
    &\ustick{P^-_S\setminus N^+_R}\qw&\qw&\measureD{e}\\
    &\ustick{N^+_R}\qw&\qw&\multimeasureD{2}{a^+}\\
    &\ustick{\rC}\qw&\qw&\ghost{a^+}\\
    &\ustick{N^-_S}\qw&\qw&\ghost{a^+}\\
    &\ustick{R\setminus N^-_S}\qw&\qw&\measureD{e}\\
    &\ustick{\tilde N^+_{N^-_S\cup R}}\qw&\qw&\measureD{e}}}
\end{aligned}\ \coloneqq 
\ 
\begin{aligned}
    \scalebox{0.75}{\Qcircuit @C=.9em @R=1.2em @! R {    
    &\qw&\ustick{\widehat{N^-_R\cup S}}\qw&\qw&\qw&\qw&\qw&\multigate{6}{\tS}&\qw&\ustick{\tilde N^+_{N^-_S\cup R}}\qw&\qw&\measureD{e}\\
    &\ustick{P^-_S\setminus N^+_R}\qw&
    \qw&\multigate{4}{\tS'}&\qw&\ustick{N^+_{N^-_S\cup R}\setminus S}\qw&\qw&
    \ghost{\tS}&\qw&\ustick{N^-_S\setminus R}\qw&\qw&\measureD{e}\\
    &\ustick{N^+_R}\qw&\qw&
    \ghost{\tS'}&\qw&\ustick{S}\qw&\qw&\ghost{\tS}&\qw&\ustick{R}\qw&\qw&\multimeasureD{2}{a}\\
&&&\pureghost{\tS'}&&&&\pureghost{\tS}&&&\ustick{\rC}\qw&\ghost{a}\\
&\ustick{N^-_S}\qw&\qw&\ghost{\tS'}&\qw&\ustick{R}\qw&\qw&\ghost{\tS}&\qw&\ustick{S}\qw&\qw&\ghost{a}\\
&\ustick{R\setminus N^-_S}\qw&\qw&\ghost{\tS'}&\qw&\ustick{N^-_S\setminus R}\qw&\qw&\ghost{\tS}&\qw&\ustick{N^+_{N^-_S\cup R}\setminus S}\qw&\qw&\measureD{e}\\
&\qw&\ustick{\tilde N^+_{N^-_S\cup R}}\qw&\qw&\qw&\qw&\qw&\ghost{\tS}&\qw&\ustick{\widehat{N^-_R\cup S}}\qw&\qw&\measureD{e}}}
\end{aligned}\ ,
\end{align*}
and, similarly to what was done for $\tW$, check that $\tZ^\dag$ is well defined. One can now verify that $\tW^\dag\tZ^\dag=\tZ^\dag\tW^\dag=\tI_{\Cntset{\rA_{G}}_{L\Reals}}$. 
The above observations show that $\tW^\dag$ and $\tZ^\dag=\tW^{-1\dag}$ act isometrically on $\Cntset{\rA_{G}}_{L\Reals}$, since 
\begin{align*}
\normsup{a_{R\rC}}&=\normsup{\tZ^\dag\tW^\dag a_{R\rC}}\\
&\leq\normsup{\tW^\dag a_{R\rC}}\leq\normsup{a_{R\rC}},\\
\normsup{a_{R\rC}}&=\normsup{\tW^\dag\tZ^\dag a_{R\rC}}\\
&\leq\normsup{\tZ^\dag a_{R\rC}}\leq\normsup{a_{R\rC}}.
\end{align*}
Being isometric, both $\tW^\dag$ and $\tW^{-1\dag}$ preserve Cauchy sequences and their equivalence classes. As a consequence, $\tW^\dag$ and $\tW^{-1\dag}$ can be uniquely extended to invertible isometries of $\Cntset{\rA_{G'}}_{Q\Reals}$, with $\tZ^\dag=\tW^{\dag-1}$. 
Let now $\rho\in\Stset{\rA_{G'}\rC}$, and $a_{(R,0)\cup(S,1)\rC}\in\Cntsetcomp{\rA_{G'}}{\rC}_{L\Reals}$. Then
\begin{align*}
&(a_{(R,0)\cup(S,1)\rC}|\hat\tW\rho)=(\tW^\dag a_{(R,0)\cup(S,1)\rC}|\rho)\\
=&(\tS_{N^-_R\cup S\cup N^+_{N^-_R\cup S}}^\dag\tS^{\prime\dag}_{N^-_R\cup S}a_{(S,0)\cup(T,1)\rC}|\rho)\\
=&(a_{(R,0)\cup(S,1)\rC}|\hat\tS^{\prime}_{N^-_R\cup S}\hat\tS_{N^-_R\cup S\cup N^+_{N^-_R\cup S}}\rho)\\
=&(a|[\hat\tS^{\prime}_{N^-_R\cup S}\hat\tS_{N^-_R\cup S\cup N^+_{N^-_R\cup S}}\rho]_{\vert(N^-_R,0)\cup(N^+_S,1)\rC}),
\end{align*}
and clearly this implies that $\hat\tW\rho_{\vert(R,0)\cup(S,1)}$ is a state for every $R,S\in\reg G$. Then $\hat\tW\Stset{\rA_{G}\rC}\subseteq\Stset{\rA_{G}\rC}$. The same argument holds for $\hat\tW^{-1}$, thus $\hat\tW\Stset{\rA_{G}\rC}=\Stset{\rA_{G}\rC}$. Let now $\tA_{T\rC}\in\Trnset{\rA_{G}\rC\to\rA_{G}\rC}_{L}$, with $T={(R,0)\cup(S,1)}$. One can evaluate $\hat\tW^{-1}\tA_{T\rC}\hat\tW$ by dually evaluating $\tW^\dag\tA_{T\rC}^\dag\tW^{-1\dag}$. The calculation is a straightforward application of Eq.~\eqref{eq:mixev}, and shows that $\hat\tW^{-1}\tA_{T\rC}\hat\tW\in\Trnset{\rA_{G}\rC\to\rA_{G}\rC}_L$. 
The same argument holds for $\hat\tW\tA_{T\rC}\hat\tW^{-1}$, and thus
\begin{align*}
\hat\tW^{-1}\Trnset{\rA_{G}\rC\to\rA_{G}\rC}_L\hat\tW=\Trnset{\rA_{G}\rC\to\rA_{G}\rC}_L.
\end{align*}
All the above observations prove that $\tW$ is a UR.
Finally, being constructed from local transformations, one can easily check that the family $\{\tW_\rC\mid\rC\in\Elety\}$ is an admissible UR. This concludes the proof that $(G',\rA,\tW^\dag)$ is a GUR. Now, comparing the equations defining $a^\pm_{(N^\pm_R,0)\cup(N^\mp_S,1)\rC}$, one can easily conclude that $\tZ=\tW^{-1}=\tS_G\tW\tS_G$.
\end{proof}

\begin{theorem}\label{th:luroogur}
Let $(N^+,\rA,\tS')$ be a local rule on $G$. Then the GUR $(G',\rA,\tW^\dag)$, with $G'=G\times\{0,1\}$ in theorem~\ref{th:lurdefgur}, is of the form $\tW=\tV\otimes \tV^{-1}$, where $(G,\rA,\tV^\dag)$ is reducible to $(N^+,\rA,\tS')$.
\end{theorem}
The detailed proof can be found in Appendix~\ref{app:proof}. 
The properties used in the proof are~\eqref{eq:condone}, \eqref{eq:condtwo}, \eqref{eq:condthree}, and~\eqref{eq:condfour}. The first two are required as items~\ref{it:idemp} and~\ref{it:commuz} in the definition~\ref{def:lur} of a local rule, while the remaining two are a consequence of item~\ref{it:mixev} as shown by lemma~\ref{lem:lurconds}.

In the case of a homogeneous update rule, does the existence of a local rule allow one to determine the full $(G,\rA,\tV^\dag)$ by local tests? One would intuitively expect that the answer is positive, since $\tS'_g=\tS'_e$ are the same for every $g\in G$, by virtue of homogeneity, and thus, knowing $\tS'_e$, one could calculate the action of $\tV$ through Eq.~\eqref{eq:backwev}. On the other hand, after giving the question another thought, one can realise that calculating the action of $\tV$ through the maps $\mathsf V^\pm_L$ requires another piece of information besides knowing $\tS'_e$. Actually, determining the cellular automaton with local tests also requires knowledge of the regions $N^+_R$ for every $R\in\reg G $---or, in other words, knowledge of the structure of the graph of influence relations. This information is indeed necessary in order to know how to correctly calculate the transformation $\mathsf V^\pm_L(\tA_{R})$ for a given region $R$. Indeed, it might happen that knowledge of finitely many closed paths from the elements $g\in R$ is not sufficient to reconstruct all the new closed paths that appear as the size of the region $R$ increases. This means that one needs infinitely many rules to identify systems in $N^\pm_{g_1}\cap N^\pm_{g_2}$. However, this is not the case if closed paths in the graph of influence relations can be decomposed into elementary closed paths having a uniformly bounded size, let us say by a constant length $l$. In this case, knowledge of local rules and of the local structure of the graph up to some finite distance $l$ is sufficient to calculate the action of $\tV$ in an arbitrary finite region. This is the reason for the following definition.

\begin{definition}[Decomposability into bounded regions]
We say that a global update rule $(G,\rA,\tV^\dag)$ is {\em decomposable into bounded regions} if the closed paths of the influence graph of $\tV$ can be decomposed into elementary closed paths of uniformly bounded length.
\end{definition}

We are finally in position to formulate the locality principle, which can be thought of as the requirement that finite, local information is needed in order to reconstruct the update rule.

\begin{principle}[Locality]
The GUR $(G,\rA,\tV^\dag)$ is reducible to a local rule and decomposable into bounded regions.
\end{principle}

\section{Cellular Automata}\label{sec:cas}

The principles of homogeneity and locality single out the special class of global rules that we will call {\em cellular automata}.

\begin{definition}
Let $(G,\rA,\tV^\dag)$ be a global update rule obeying the principles of homogeneity and locality. We say that $(G,\rA,\tV^\dag)$ is a {\em cellular automaton}.
\end{definition}

Notice that the influence graph of a cellular automaton is the Cayley graph of a {\em finitely presented} group, i.e.~ a group that is finitely generated and can be presented through finitely many relators. This fact has the very important consequence that every cellular automaton defines a Cayley graph that is quasi-isometric to some smooth Riemannian manifold of dimension $d\leq 4$ (see \cite{de2000topics}, sec. IV pag. 90).

An important remark is in order. Thanks to homogeneity, one can easily show that the local rule $\tS'_g$ for a cellular automaton on the Cayley graph $\Gamma(G,S_+)$ has the property that $\tS'_g=\tS'_{e_G}$ for every $g\in G$. Thus, cellular automata can be identified by their Cayley graph $\Gamma(G,S_+)$ and by a reversible transformation $\tS'_{e_G}\in\Trnset{\rA_{N^+_{e_G}}\rA_{e_G}\to\rA_{N^+_{e_G}}\rA_{e_G}}_1$.
%

\subsection{Results}

We now prove two important theorems regarding cellular automata. The first one shows that, in order to obey the properties of homogeneity and locality, a cellular automaton cannot be in the quasi-local algebra.

\begin{theorem}
Let $|G|=\infty$. The homogeneous and local cellular automaton $\tV$ is an element of $\Trnset{\rA_G\to\rA_G}_{Q1}$ if and only if it is trivial: $\tV=\tI_G$.
\end{theorem}
\begin{proof} 
Let $\tV$ be in the quasi-local algebra, and $\tV_{nR_n}$ be a sequence of local transformations in the class of $\tV$. Then for $\varepsilon>0$ there exists $n_0$ such that for every $n\geq n_0$ one has
\begin{align*}
\normsup{\tV_{nR_n}-\tV}\leq \varepsilon.
\end{align*}
Moreover, for $\pi\in\Pi_\tV$, one has $\tT_\pi\tV\tT_\pi^{-1}=\tV$. Then 
\begin{align*}
\normsup{\tV_{nR_n}-\tT_\pi\tV\tT_\pi^{-1}}\leq \varepsilon.
\end{align*}
Since $\tT_\pi$ is a GUR, it is isometric and thus
\begin{align}
&\normsup{\tT_\pi^{-1}\tV_{nR_n}\tT_\pi-\tV}\nonumber\\
&\qquad=\normsup{\tV_{nR_n}-\tT_\pi\tV\tT_\pi^{-1}}\leq \varepsilon.
\label{eq:parad}
\end{align}
Now, this implies that
\begin{align*}
\normsup{\tV_{nR_n}-\tT_\pi^{-1}\tV_{nR_n}\tT_\pi}\leq 2\varepsilon.
\end{align*}
It is easy to prove that, if $|G|=\infty$, for every $R\in\reg G$ there must exist $\pi\in\Pi_\tV$ so that $R\cap\pi(R)=\emptyset$. Suitably choosing $\pi$, condition~\eqref{eq:parad} then takes the form
\begin{align*}
&\normsup{\tW\otimes\tI-\tI\otimes\tW}\\
&=\normsup{\tW\otimes\tI-\tW\otimes\tW+\tW\otimes\tW-\tI\otimes\tW}\\
&\leq2\normsup{\tW-\tI}\leq \varepsilon.
\end{align*}
Being $\varepsilon$ arbitrary, we must conclude that $\tW=\tI$, namely $\tV_{nR_n}=1_\emptyset$.
\end{proof}

The second result pertains the following types of theory.
\begin{enumerate}
\item\label{ass:loc}
Theories with local discriminability \cite{PhysRevA.81.062348,chiribella2011informational,DAriano:2017aa,hardy2012limited}, or more generally every theory where the algebra of transformations $\tA\in\Trnset{\rA_1\ldots\rA_n\to\rA_1\ldots\rA_n}$ is generated by local transformations, i.e.~transformations of the form 
\begin{align*}
\tA_{i}\in\tI_{\bar\rA_i}\otimes\Trnset{\rA_i\to\rA_i},
\end{align*}
where $\bar\rA_i$ is the composite system made of all $\rA_1\ldots\rA_n$ except from $\rA_i$. This class includes, among others, classical information theory and quantum theory.
\item\label{ass:biloc}
Theories where the algebra of transformations $\tA\in\Trnset{\rA_1\ldots\rA_n\to\rA_1\ldots\rA_n}$ is generated by bipartite transformations, i.e.~transformations of the form 
\begin{align*}
\tA_{i,j}\in\tI_{\bar\rA_i\bar\rA_j}\otimes\Trnset{\rA_i\rA_j\to\rA_i\rA_j},
\end{align*}
where $\bar\rA_i\bar\rA_j$ is the composite system made of all $\rA_1\ldots\rA_n$ except from $\rA_i$ and $\rA_j$. This case includes the Fermionic theory as well as real quantum theory.
\end{enumerate}

In the above cases, for groups $G$ with suitable properties that we will immediately specify, the admissible local rules for a cellular automaton can be sought considering finite-dimensional systems. In the remainder we will then restrict to the families of theories introduced above.

Before proving this result, we need some preliminary definition and lemma.

\begin{definition}[Quotient group] Let $G,H$  be two groups, and suppose that there exists a group homomorphism $\varphi:G\to H$. Then $H$ is a \emph{quotient} of $G$. If $H$ is finite, it is called a {\em finite quotient} of $G$.
\end{definition}
The reason why we are interested in finite quotients is that, under suitable assumptions that 
%
%
will be made rigorous shortly, the same local rule defines an automaton on $G$ and an automaton on any of its finite quotients $H$. This result makes it much easier to provide unitarity conditions in all those cases where the finite quotients of the group $G$ satisfy the mentioned requirements. We now explain the hypotheses in detail, and finally prove the main theorem.

We start remarking that if the group $G$ has a finite presentation 
\begin{align*}
G=\<a_1,\ldots,a_n|r_1,\ldots,r_k\>, 
\end{align*}
it is straightforward to verify that any quotient $H$ of $G$ has a finite presentation of the form 
\begin{align*}
H=\<\varphi(a_1),\ldots,\varphi(a_n)|\varphi(r_1),\ldots,\varphi(r_k),b_1,\ldots,b_j\>,
\end{align*}
where the new relations $b_k$ do not belong to the group $R$, conjugate closure of $\{\varphi(r_1),\ldots,\varphi(r_k)\}$. 
When $H$ is a quotient of $G$, referring to the presentation above, we will denote by $R'$ the conjugate closure of the group generated by $\{\phi(r_l)\}_{l=1}^k\cup\{b_m\}_{m=1}^j$. Moreover, 
the neighbourhoods of $f$ in $\Gamma(H,\phi(S_+))$ will be denoted by $N'^\pm_f$ for $f\in H$.
%
We now state two lemmas that we use in the following. The proofs are provided in Appendix~\ref{app:ped}

\begin{lemma}\label{lem:pedantic}
Let $N^+$ be a neighbourhood system corresponding to the Cayley graph $\Gamma(G,S_+)$ of a a finitely generated group $G$, and let $H=\phi(G)$ be a finite quotient of $G$ such that for every $h_a,h_b,h_c,h_d\in S_+$ one has $\phi(h_ah_b^{-1})\in R'$, if and only if $h_ah_b^{-1}\in R$ and $\phi(h_ah_b^{-1}h_ch_d^{-1})\in R'$ if and only if $h_ah_b^{-1}h_ch_d^{-1}\in R$. 
\begin{enumerate}
\item $N'^\pm_{\phi(g)}=\phi(N^\pm_{g})$, and $|N'^\pm_{\phi(g)}|=|N^\pm_{g}|$.
\item $P'^\pm_{\phi(g)}=\phi(P^\pm_{g})$, and $|P'^\pm_{\phi(g)}|=|P^\pm_{g}|$.
\item Given $f_1,f_2\in H$, one can choose $g_i\in \phi^{-1}(f_i)$ such that $N'^\pm_{f_1}\cap N'^\pm_{f_2}=\phi(N^\pm_{g_1}\cap N^\pm_{g_2})$ and $|N'^\pm_{f_1}\cap N'^\pm_{f_2}|=|N^\pm_{g_1}\cap N^\pm_{g_2}|$.
\item Given $f_1,f_2\in H$, one can choose
$g_i\in \phi^{-1}(f_i)$ such that $P'^\mp_{f_2}\cap N'^\pm_{f_1}=\phi(P^\mp_{g_2}\cap N^\pm_{g_1})$ and $|P'^\mp_{f_2}\cap N'^\pm_{f_1}|=|\phi(P^\mp_{g_2}\cap N^\pm_{g_1})|$.
\item If $g_2\not\in N^\pm_{g_1}$ and $\phi(g_2)\in N'^\pm_{\phi(g_1)}$, it is $P^\pm_{g_1}\cap N^\mp_{g_2}=\emptyset$.
\end{enumerate}
\end{lemma}

\begin{lemma}\label{lem:pedantic2}
In addition to the hypotheses of lemma~\ref{lem:pedantic}, let $\phi(h_ah_b^{-1}h_ch_d^{-1}h_eh_f^{-1})\in R'$ iff $h_ah_b^{-1}h_ch_d^{-1}h_eh_f^{-1}\in R$ for every $h_a,h_b,h_c,h_d,h_e,h_f\in S_+$. 
Then
\begin{enumerate}
\item 
Given $f_1,f_2\in H$ with $f_2\in P'^\pm_{f_1}$, one has $P'^\pm_{f_1}\cap P'^\pm_{f_2}=\phi(P^\pm_{g_1}\cap P^\pm_{g_2})$, for some $g_i\in \phi^{-1}(f_i)$, with $g_2\in P^\pm_{g_1}$, and $|P'^\pm_{f_1}\cap P'^\pm_{f_2}|=|P^\pm_{g_1}\cap P^\pm_{g_2}|$.
\item 
Let $g_2\not\in P^\pm_{g_1}$ and $\phi(g_2)\in P'^\pm_{\phi(g_1)}$. Then $N^\pm_{g_1}\cap N^\pm_{g_2}=P^\pm_{g_1}\cap P^\pm_{g_2}=\emptyset$.
\end{enumerate}
\end{lemma}

\begin{corollary}
Under the hypotheses of lemma~\ref{lem:pedantic}, the homomorphism $\phi$ is invertible on $N^\pm_g$ and $P^\pm_g$ for every $g\in G$.
\end{corollary}

\begin{figure}[h]
\centering
\begin{minipage}{4cm}
\includegraphics[width=4cm]{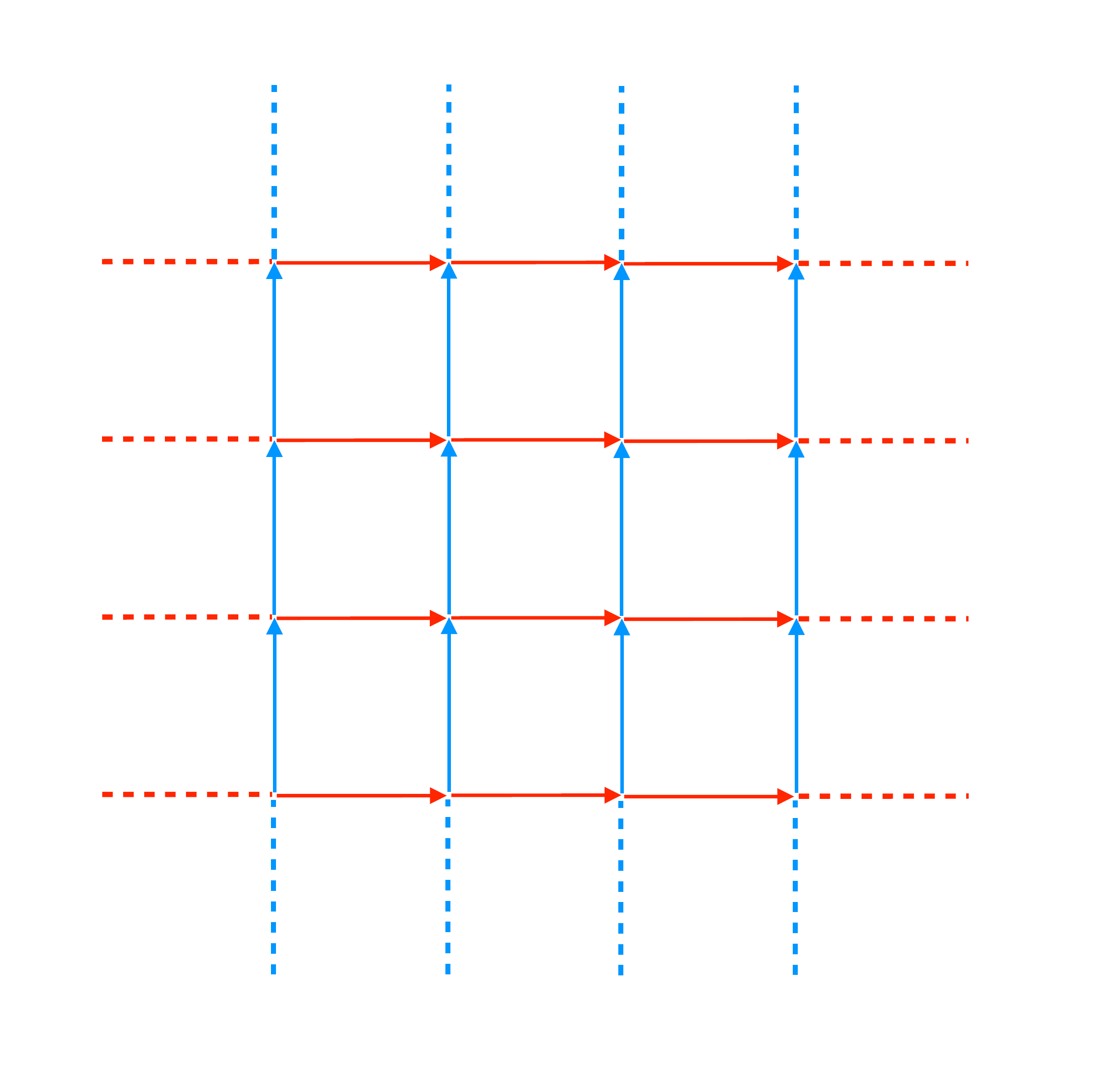}\\(a)
\end{minipage}\\
\begin{minipage}{4cm}
\includegraphics[width=4cm]{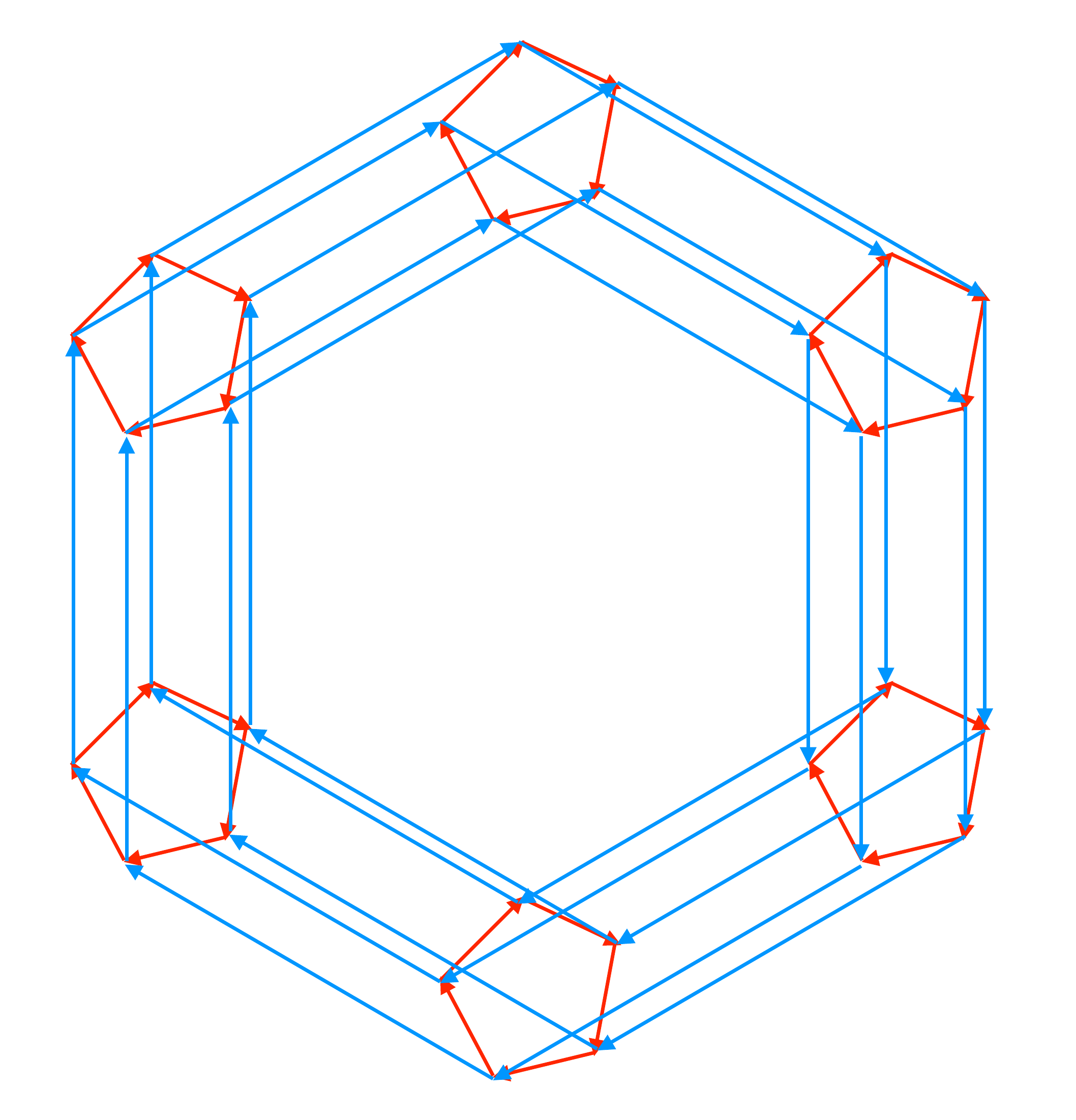}\\
(b)
\end{minipage}
\begin{minipage}{4cm}
\includegraphics[width=4cm]{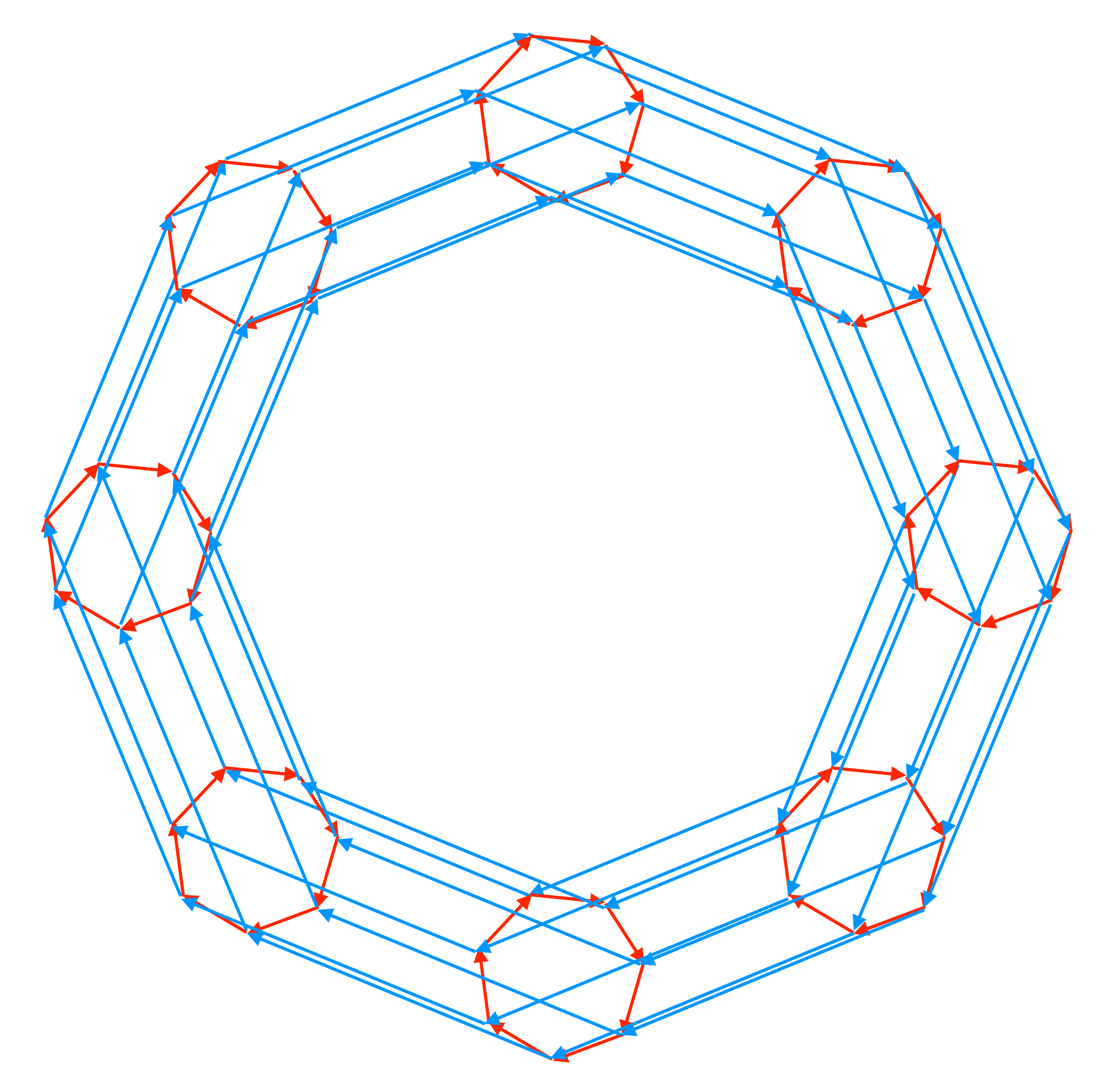}\\
(c)
\end{minipage}
\caption{
(a) The Cayley graph of $\mathbb Z^2$ presented as $\<a,b|aba^{-1}b^{-1}\>$; (b) the Cayley graph of the finite quotient of $\Intg^2$ $\<\phi(a),\phi(b)| \phi(aba^{-1}b^{-1}),\phi(a)^6,\phi(b)^5\>$, satisfying the hypotheses of lemma~\ref{lem:pedantic} but not those of~\ref{lem:pedantic2}; (c) the Cayley graph of the finite quotient of $\Intg^2$ $\<\phi(a),\phi(b)| \phi(aba^{-1}b^{-1}),\phi(a)^8,\phi(b)^7\>$, satisfying the hypotheses of both lemmas~\ref{lem:pedantic} and~\ref{lem:pedantic2}. Notice that we implicitly assume that $S_+$ contains also $a^{-1}$ and $b^{-1}$, but we do not express the presentation accordingly, to keep the formulas readable. In principle there should be two extra generators $c,d$ (and $\phi(c)$, $\phi(d)$ in the finite quotients), along with the relations $ac$ and $bd$ (and $\phi(ac)$, $\phi(bd)$, accordingly).}
\label{fig:wraps}
\end{figure}

In order to make the hypotheses of lemmas~\ref{lem:pedantic} and~\ref{lem:pedantic2} 
clearer, in Fig.~\ref{fig:wraps} we show two examples of finite quotients of the group 
$\Intg^2$, the first one satisfying the hypotheses of lemma~\ref{lem:pedantic} but not
those of lemma~\ref{lem:pedantic2}, and the second one satisfying both.

We can now prove the following results, which are the core of the final theorem at the end of the section.

\begin{lemma}\label{lem:wra}
For theories of type~\ref{ass:loc}, under the hypotheses of lemma~\ref{lem:pedantic}, given a homogeneous local rule $(N^+,\rA,\tS')$ with neighbourhood system $N^+$ such that $|N^+_{e_G}|=|S_+|$, the local rule $(N'^+,\rA,\tilde{\tS}^\prime)$, with $\tilde\tS^\prime_{e_H}\coloneqq \tS^\prime_{e_G}$ is well-defined on the Cayley graph $\Gamma(H,\phi(S_+))$, with $N'^+_{\phi(g)}=\phi(N^+_g)$. Viceversa, given a local rule $(N'^+,\rA,\tilde{\tS}^\prime)$ whose neighbourhood system $N'^+$ corresponds to a Cayley graph $\Gamma(H,\phi(S_+))$, the local rule $(N^+,\rA,\tS')$, with $\tS^\prime_{e}\coloneqq \tilde\tS^\prime_{e}$ is well-defined on the Cayley graph $\Gamma(G,S_+)$, with $\phi(N^+_{g})=N'^+_{\phi(g)}$
\end{lemma}
\begin{proof}
We remark that, by definition of the maps $\tS_g'$ (see Definition \ref{def:lur}) a given choice of $\tS'_{e_H}$ is in principle compatible with every Cayley graph with $|N'^\pm_{e_H}|=|N^\pm_{e_G}|$, i.e.~the same number of generators (we remind that $\tS'_g=\tS'_{e_H}\in\Trnset{\rA_{N^+_{e_G}}\rA_{e_G}\to \rA_{N^+_{e_G}}\rA_{e_G}}_1$). 
By lemma~\ref{lem:pedantic}, the cardinality of $N'^+_{e_H}$ is the same as that of $N^+_{e_G}$. Then $\tilde\tS'_{e_H}=\tS'_{e_G}$ is well defined on $\Gamma(H,\phi(S_+))$ iff it is well defined on $\Gamma(G,S_+)$. What remains to be proved is that the map $\tilde\tS^\prime_{e_H}=\tS'_{e_G}$ identifies a local rule on the system of neighbourhoods $N'^+$---namely that Eq.~\eqref{eq:mixev} as well as items~\ref{it:idemp} and~\ref{it:commuz} in definition~\ref{def:lur} hold---iff it does on the system of neighbourhoods $N^+$. Item~\ref{it:idemp} is satisfied by $\tilde\tS^\prime=\tS'$ by definition in both cases. As to item~\ref{it:commuz}, it is trivial to verify that the only important feature in order to decide whether it holds in any neighbourhood system $N'^+_{e_H}$ is the cardinality of $N'^+_{e_H}\cap N'^+_{f}$ (or $N^+_{e_G}\cap N^+_{g}$) for every $f\in P'^+_{e_H}$ (or $g\in P^-_{e_G}$), since this identifies the systems where both $\tilde\tS^\prime_{e_H}=\tS'_{e_G}$ and $\tilde\tS^\prime_f=\tS'_g$ act non-trivially, namely those where the products $\tilde\tS^\prime_{e_H}\tilde\tS^\prime_f$ or $\tS'_{e_G}\tS'_g$ might not be commutative. Lemma~\ref{lem:pedantic} thus ensures that item~\ref{it:commuz} is satisfied by $\tilde\tS^\prime$ iff it is by $\tS$. Let us then focus on Eq.~\eqref{eq:mixev}. For theories of type~\ref{ass:loc}, it is sufficient to verify that Eq.~\eqref{eq:mixev} holds for $|R|=1$ and $|S|=0$ or $|R|=0$ and $|S|=1$. Both conditions easily follow if we consider lemma~\ref{lem:pedantic}.
\end{proof}

\begin{lemma}\label{lem:wra2}
For theories of type~\ref{ass:biloc}, under the hypotheses of lemma~\ref{lem:pedantic2}, given a homogeneous local rule $(N^+,\rA,\tS')$ with neighbourhood system $N^+$ such that $|N^+_{e_G}|=|S_+|$, the local rule $(N'^+,\rA,\tilde{\tS}^\prime)$, with $\tilde\tS^\prime_{e_H}\coloneqq \tS^\prime_{e_G}$ is well-defined on the Cayley graph $\Gamma(H,\phi(S_+))$, with $N'^+_{\phi(g)}=\phi(N^+_g)$. Viceversa, given a local rule $(N'^+,\rA,\tilde{\tS}^\prime)$ whose neighbourhood system $N'^+$ corresponds to a Cayley graph $\Gamma(H,\phi(S_+))$, the local rule $(N^+,\rA,\tS')$, with $\tS^\prime_{e}\coloneqq \tilde\tS^\prime_{e}$ is well-defined on the Cayley graph $\Gamma(G,S_+)$, with $\phi(N^+_{g})=N'^+_{\phi(g)}$
\end{lemma}

\begin{proof}
The first part of the proof proceeds exactly as that of lemma~\ref{lem:wra}. The non trivial part of the proof regards the equivalence of Eq.~\eqref{eq:mixev} for both $\tS'$ on $G$ and $\tilde\tS'$ on $H$.
Suppose that $\tS'$ defines a local rule for the neighbourhood system $N^+_g$. In theories of type~\ref{ass:biloc}, it is sufficient to verify that Eq.~\eqref{eq:mixev} holds for regions with $|R|=2$, $|S|=0$, or $|R|=1$, $|S|=1$, or $|R|=0$, $|S|=2$.
Let then $R,S\subseteq\Gamma(H,\phi(S_+))$, with $R=\phi(\tilde R)$ and $S=\phi(\tilde S)$.
In the first case we have $R=\{f_1,f_2\}$, with $\tilde R=\{g_1,g_2\}$ and $f_i=\phi(g_i)$, while $S=\emptyset$. The relevant informations in order to verify Eq.~\eqref{eq:mixev} are the cardinalities i) $|N'^-_{f_1}\cap N'^-_{f_2}|$ and ii) $|P'^-_{f_1}\cap P'^-_{f_2}|$. If $f_2\in P'^-_{f_1}$, by lemma~\ref{lem:pedantic2} we know that we can choose $g_1$ and $g_2$ so that $|N'^-_{f_1}\cap N'^-_{f_2}|=|N^-_{g_1}\cap N^-_{g_2}|$ and $|P'^-_{f_1}\cap P'^-_{f_2}|=|P^-_{g_1}\cap P^-_{g_2}|$. 
Thus, since Eq.~\eqref{eq:mixev} holds for $\tC\in\Trnset{\rA^{(G)}_{\tilde R}\to\rA^{(G)}_{\tilde R}}$, and 
\begin{align*}
\rA^{(G)}_{\tilde R}\cong\rA^{(H)}_R,\quad\tilde\tS'=\tS',
\end{align*}
we conclude that Eq.~\eqref{eq:mixev} holds also for $\tC\in\Trnset{\rA^{(H)}_R\to\rA^{(H)}_R}$. 
On the other hand, if $f_2\not\in P'^-_{f_1}$, we can calculate $\tilde\tS'_{N'^-_{f_1,f_2}}\tC\tilde\tS'_{N'^-_{f_1,f_2}}$ in two steps. First we apply $\tilde\tS'_{N'^-_{f_1}}=\tS'_{N^-_{g_1}}$, treating the system $f_2$ as an external system. Reminding that $f_2\not\in P'^-_{f_1}$, it is $N'^-_{f_1}\cap N'^-_{f_2}=\emptyset$. By Eq.~\eqref{eq:mixev}, that holds for $\tS'_{N^-_{g_1}}$ we then have 
\begin{align*}
&\begin{aligned}
    \Qcircuit @C=1em @R=.7em @! R {&\qw &\qw&
    \qw&\qw&    \ustick{(P'^-_{f_2}\setminus [P'^-_{f_1}\cup f_2],0)}\qw&
    \qw&\qw&\qw&\qw&\qw\\    
    &\qw &\qw&
    \ustick{(f_2,0)}\qw&\qw&\multigate{1}{\tC}&
    \qw&\ustick{(f_2,0)}\qw&\qw&\qw&\qw\\
    &\ustick{(P'^-_{f_1},0)}\qw &\qw&\multigate{3}{\tilde\tS'_{N'^-_{f_1}}}&
    \ustick{(f_1,0)}\qw&\ghost{\tC}&\ustick{(f_1,0)}\qw&
    \multigate{3}{\tilde\tS'_{N^-_{f_1}}}&\qw&\ustick{(P'^-_{f_1},0)}\qw&\qw\\
        &&&\pureghost{\tilde\tS'_{N'^-_{f_1}}}&
    &&&
    \pureghost{\tilde\tS'_{N'^-_{f_1}}}&&&\\
    &&&\pureghost{\tilde\tS'_{N'^-_{f_1}}}&
    \qw&\ustick{(P'^-_{f_1}\setminus f_1,0)}\qw&\qw&
    \ghost{\tilde\tS'_{N'^-_{f_1}}}&&&\\
    &\ustick{(N'^-_{f_1},1)}\qw &\qw&\ghost{\tilde\tS'_{N'^-_{f_1}}}&\qw&\ustick{(N^-_{f_1},1)}\qw&\qw&\ghost{\tilde\tS'_{N'^-_{f_1}}}&\qw&\ustick{(N'^-_{f_1},1)}\qw&\qw\\
    &\qw &\qw&
    \qw&\qw&    \ustick{(N'^-_{f_2},1)}\qw&
    \qw&\qw&\qw&\qw&\qw}
\end{aligned}\\
\\
&\qquad\quad=\ 
\begin{aligned}
    \Qcircuit @C=1em @R=.7em @! R {&\qw&\qw&\ustick{(P'^-_{f_1}\cup P'^-_{f_2}\setminus f_2,0)}\qw &\qw&\qw&\qw\\
    &\ustick{(f_2,0)}\qw &\qw&\multigate{1}{\tC'}&\qw&\ustick{(f_2,0)}\qw&\qw\\
    &\ustick{(N'^-_{f_1},1)}\qw &\qw&\ghost{\tC'}&\qw&\ustick{(N'^-_{f_1},1)}\qw&\qw\\\\
&\qw&\qw&\ustick{(N'^-_{f_2},1)}\qw &\qw&\qw&\qw    }
\end{aligned}\ \ .
\end{align*}
Then, we apply $\tilde\tS'_{N'^-_{f_2}}$, obtaining
\begin{align}
&\begin{aligned}
    \Qcircuit @C=1em @R=.7em @! R {  
    &\qw &\qw&
    \ustick{(N'^-_{f_1},1)}\qw&\qw&\multigate{1}{\tC'}&
    \qw&\ustick{(N'^-_{f_1},1)}\qw&\qw&\qw&\qw\\
        &\ustick{(N'^-_{f_2},1)}\qw &\qw&\multigate{3}{\tilde\tS'_{N'^-_{f_2}}}&\ustick{(f_2,0)}\qw&\ghost{\tC'}&\ustick{(f_2,0)}\qw&\multigate{3}{\tilde\tS'_{N'^-_{f_2}}}&\qw&\ustick{(N'^-_{f_2},1)}\qw&\qw\\
                &&&\pureghost{\tilde\tS'_{N'^-_{f_1}}}&
    &&&
    \pureghost{\tilde\tS'_{N'^-_{f_1}}}&&&\\
          &&&\pureghost{\tilde\tS'_{N'^-_{f_2}}}&
    \qw&\ustick{(N'^-_{f_2},1)}\qw&\qw&
    \ghost{\tilde\tS'_{N'^-_{f_2}}}&&&\\
        &\ustick{(P'^-_{f_2},0)}\qw &\qw&\ghost{\tilde\tS'_{N'^-_{f_2}}}&
    \qw&\ustick{(P'^-_{f_2}\setminus f_2,0)}\qw&\qw&
    \ghost{\tilde\tS'_{N'^-_{f_2}}}&\qw&\ustick{(P'^-_{f_2},0)}\qw&\qw\\
     &\qw &\qw&
    \qw&\qw&\ustick{(P'^-_{f_1},0)}\qw&
    \qw&\qw&\qw&\qw&\qw\\
&\qw &\qw&
    \qw&\qw&    \ustick{(P'^-_{f_2}\setminus [P'^-_{f_1}\cup f_2],0)}\qw&
    \qw&\qw&\qw&\qw&\qw    
    }
\end{aligned}\nonumber\\
\nonumber\\
&\qquad\quad=\ 
\begin{aligned}
    \Qcircuit @C=1em @R=.7em @! R {&\qw&\qw&\ustick{(P'^-_{f_1}\cup P'^-_{f_2},0)}\qw &\qw&\qw&\qw\\
    &\ustick{(N'^-_{f_1},1)}\qw &\qw&\multigate{1}{\tC''}&\qw&\ustick{(N'^-_{f_1},1)}\qw&\qw\\
    &\ustick{(N'^-_{f_2},1)}\qw &\qw&\ghost{\tC''}&\qw&\ustick{(N'^-_{f_2},1)}\qw&\qw}
\end{aligned}\ \ ,
\label{eq:extf}
\end{align}
where on the l.h.s.~we reversed the ordering of the wires for the sake of simplicity of the diagram. Overall, considering that $\tilde\tS'_{N'^-_{f_1}\cup N'^-_{f_2}}=\tilde\tS'_{N'^-_{f_1}}\tilde\tS'_{N'^-_{f_2}}=\tilde\tS'_{N'^-_{f_2}}\tilde\tS'_{N'^-_{f_1}}$, we then proved Eq.~\eqref{eq:mixev} for the local rule $\tilde\tS'_f$ on the neighbourhood scheme given by the graph $\Gamma(H,\phi(S_+))$.
In the second case, let $R=f_1=\phi(g_1)$, $S=f_2=\phi(g_2)$. The informations that matter are whether one can choose $g_1,g_2$ so that $f_2\in N'^-_{f_1}$ iff $g_2\in N^-_{g_1}$, and $|P'^-_{f_1}\cap N'^+_{f_2}|=|P^-_{g_1}\cap N^+_{g_2}|$. Again by lemma~\ref{lem:pedantic}, we know that 
the answer is positive.
Then, also for $|R|=|S|=1$, since Eq.~\eqref{eq:mixev} holds for $\tC\in\Trnset{\rA^{(G)}_{(\tilde R,0)\cup(\tilde S,1)}\to\rA^{(G)}_{(\tilde R,0)\cup(\tilde S,1)}}$, and 
\begin{align*}
\rA^{(G)}_{(\tilde R,0)\cup(\tilde S,1)}\cong\rA^{(H)}_{(R,0)\cup(S,1)},\quad\tilde\tS'=\tS',
\end{align*}
we conclude that Eq.~\eqref{eq:mixev} holds also for $\tC\in\Trnset{\rA^{(H)}_{(R,0)\cup(S,1)}\to\rA^{(H)}_{(R,0)\cup(S,1)}}$. The argument for $|R|=0$, $|S|=2$ is the same as for $|R|=2$, $|S|=0$. 
For the converse, one can follow the same argument as for the direct statement. The only differences are that in the case $|R|=2$ and $|S|=0$ one might have $g_2\not\in P^-_{g_1}$ while $f_2=\phi(g_2)\in\phi(P^-_{g_1})=P'^-_{f_2}$, and in the case $|R|=|S|=1$ it might happen that $g_2\not\in N^-_{g_1}$ while $f_2\in N'^-_{f_1}$. By lemma~\ref{lem:pedantic2}, the first case can occur only if $N^+_{g_1}\cap N^+_{g_2}=P^-_{g_1}\cap P^-_{g_2}
=\emptyset$, while by lemma~\ref{lem:pedantic} the second case occurs only if $P^-_{g_1}\cap N^+_{g_2}=\emptyset$. 
In both cases, the different structure of the neighbourhoods between $G$ and $H$ does not preclude the derivation of Eq.~\eqref{eq:mixev} in the case of $G$ from the same identity for the case of $H$, as one can immediately realise by direct inspection of Eq.~\eqref{eq:mixev}.
\end{proof}

\begin{theorem}[Wrapping lemma]
For a theory of type~\ref{ass:loc}, under the hypotheses of lemma~\ref{lem:pedantic}, the local rule $(N^+,\rA,\tS')$ defines a cellular automaton $(G,\rA,\tV)$ on the Cayley graph $\Gamma(G,S_+)$ if and only if the same local rule defines a cellular automaton $(H,\rA,\tW)$ on the Cayley graph $\Gamma(H,\phi(S_+))$.
The same is true of a theory of type~\ref{ass:biloc}, under the hypotheses of lemma~\ref{lem:pedantic2}.
\end{theorem}

\begin{proof}
The thesis now easily follows from lemmas~\ref{lem:wra} and~\ref{lem:wra2}, and theorem~\ref{th:luroogur}.
\end{proof}

The above result allows us, at least in theories of the two types considered so far, and
for CAs on Cayley graphs of groups with finite quotients satisfying the hypotheses of 
lemmas~\ref{lem:pedantic} or~\ref{lem:pedantic2}, to reduce the classification of CAs on
graphs of infinite groups to that of CAs on finite groups, which is in principle a much
easier task. Moreover, once a suitable finite quotient is found, one can simulate the 
evolution of a finite region and for a finite number of steps considering the same 
evolution on the finite {\em wrapped} graph instead of having to consider the actual 
infinite graph.

%
%

\section{Examples}\label{sec:exa}

\subsection{Classical case}

Classical systems $\rm d$ can be classified in terms of the number $d$ of pure states  $\rho_y$, $y=1,\ldots,d$ which coincides with the dimension of the state space $d=\dim\Stset{\rm d}_\Reals$. All the pure states of a classical system are jointly discriminable by atomic effects $a_x$, $x=1,\ldots,d$, i.e.~$(a_{x}|\rho_y)=\delta_{xy}$. For the sake of simplicity, we will restrict attention to classical {\em computation}, i.e.~the sub-theory of classical theory where there is only one kind of non-trivial elementary system, the {\em bit} having $d=2$, and all other systems are composite systems of $n$ bits, having dimension $d=2^n$. Every effect in $\Cntset{\rA_G}_Q$ for an infinite system $|G|=\infty$ can be decomposed as a series of effects corresponding to finite, arbitrarily large bit strings $b_R$, with
\begin{align*}
b_R\coloneqq [a_R],\quad a=a_{b_{r_1}}\otimes a_{b_{r_2}}\otimes\ldots\otimes a_{b_{r_{|R|}}},
\end{align*}
with $R=\{r_1,r_2,\ldots,r_{|R|}\}$. In this case, let $b_R$ denote a local effect. A global rule $\tV$ can be specified by the input/output relation $\tV^\dag b_\rR=b'_{R'}$. However, in the extensive literature about classical cellular automata---see e.g.~\cite{Hadeler2017} for reference, though we cannot even try to provide an exhaustive account of the relevant bibliography---it is customary to express $\tV$ through its local rule, which is specified by its action on local states. The only point we want to discuss here is the apparent clash of our general wrapping lemma with results in the literature. In particular, in Ref.~\cite{schumacher2004reversible} a counterexample to the classical version of the wrapping lemma is provided, referring to \cite{inomizo05}. The local rule for the counterexample is the following. Let $G=\mathbb Z_r$ with $r\not \in3\mathbb Z$. The rule defined by
\begin{align}
b'_j=b_{j\ominus_r1}\oplus b_{j}\oplus b_{j\oplus_r 1}
\label{eq:classstr}
\end{align}
What we claim here is that, according to our notion of neighbourhood based on definition~\ref{def:influ}, the above automaton has a neighbourhood system where $N^+_i$ is actually much larger than $\{{j\ominus_r1},{j},{j\oplus_r 1}\}$. Indeed, if one considers the action of $\tV$ on the algebra of transformations instead of its action on states, one can easily realise that $N^+_i=G$ for every $i$. However, by 
checking the conditions for no-signalling, it is easy to realise that the rule in Eq.~\eqref{eq:classstr} only signals to sites $i\ominus 1$, $i$, and $i\oplus1$. Thus, while no-signalling is necessary for no causal influence, the converse is not true. 

We remark that the definition of causal influence that we give, based on transformations, 
is inspired by the quantum definition, which indeed involves the generators of the local 
algebra, and thus the Kraus representatives of transformations rather than the mere set of 
local states.
In this respect, the notion of neighbourhood used in the literature on CA is different 
from the quantum one, and we point to this relevant difference as the responsible for the 
apparent incompatibility  discussed in Ref.~\cite{schumacher2004reversible}.
The subject was extensively studied in Ref.~\cite{DBLP:conf/jac/ArrighiN10} and later in 
Ref.\cite{PhysRevA.95.012331}, where the 
authors maintain the notion of neighbourhood for classical CA based on the signalling 
condition, and prove a bound on the mismatch between the 
classical neighbourhood of a given CA and the neighbourhood of the quantum version of the 
same CA.

\subsection{Quantum case}

Quantum cellular automata represent now a widely studied field of research.
After the very general analysis of Refs.~\cite{schumacher2004reversible,arrighi2011unitarity,Gross2012,PhysRevA.95.012331}, there are various results in the literature about quantum cellular automata. Here we cite the very general classification of Ref.~\cite{Freedman:2020aa,Freedman:2019uj}, while for extensive reviews we refer to Refs.~\cite{Farrelly:2019vr,Arrighi:2019wq}.

\subsection{Fermionic case}

For the presentation of fermionic theory as an OPT we refer to~\cite{D_Ariano_2014,doi:10.1142/S0217751X14300257}. The Fermionic systems are fully specified by the algebra $\mathcal A$ generated by local field operators $\psi_i$. In fact, $\mathcal A$ is the algebra of field operator polynomials. This algebra is $\Intg_2$-graded, with two modules $\mathcal A_E$ and $\mathcal A_O$, given by the span of even and odd polynomials, respectively. No combination of even and odd polynomials is allowed. As in all quantum theories, $\mathcal A$ provides a simplified representation in terms of Kraus operators of the algebra of transformations, which is not graded, but rather reducible to a direct sum. Linear combinations of transformations indeed do not correspond to linear combinations of the corresponding Kraus operators. The local systems for a site $g$ of a cellular automaton in this case are registers made of one or more local Fermionic modes. Local field operators are then labelled $\psi_{i,g}$, where $g\in G$ and $i\in J_g=\{1,\ldots,n_g\}$, $|J_g|$ representing the size of the register at site $g$. Since the local algebra $\mathcal A_g$ is finitely generated, the locality requirement implies that $\tV$ is given by a local rule, which in turn is completely specified by its action on the local generators $\psi_{i,g}$ for every $g\in G$. Thus, one has
\begin{align*}
\tV&(\psi_{i,f})\\
&=\sum_{\substack{\bvec g\in (N^+_f)^{\times k}\\ \bvec j\in J_{\bvec g}\\ \bvec s,\bvec t\in\{0,1\}^{*}}}T_{i,f}^{(\bvec g,\bvec j,\bvec s,\bvec t)}\psi_{j_1,g_1}^{s_{j_1}}\psi_{j_1,g_1}^{\dag t_{j_1}}\ldots\psi_{j_k,g_k}^{s_k}\psi_{j_k,g_k}^{\dag t_{i_k}},
\end{align*}
where $J_\bvec g\coloneqq \bigtimes_{g\in N^+_f}J_{g}$.
Most of the automata studied so far in the literature are {\em linear}, namely
\begin{align*}
\tV(\psi_{i,f})=\sum_{\substack{g\in N^+_f\\ j\in J_{g}}}T_{i,f}^{j,g'}\psi_{j,g}.
\end{align*}
Remarkable exceptions are represented by Refs.~\cite{PhysRevA.97.032132,Bisio_2018,PhysRevA.98.052337}. In particular, in Ref.~\cite{PhysRevA.98.052337} the full classification of Fermionic cellular automata with $|J_g|=1$ on Cayley graphs of $\Intg_2\times\Intg_2$ and $\Intg$ was carried out.

As regards Fermionic cellular automata, we would like to remark that, despite Fermionic theory being of type~\ref{ass:biloc}, they satisfy the wrapping lemma in the form of lemma~\ref{lem:wra}, instead of lemma~\ref{lem:wra2}. This is due to the fact that Fermionic transofrmations, like quantum transformations, admit a Kraus decomposition, and can thus be represented through Kraus operators instead of completely positive maps. It turns out that local Fermionic field operators $\psi_{j,g}$ and $\psi_{j,g}^\dag$ provide a basis for the full algebra of Fermionic operators. Thus, in the Fermionic case, if Eq.~\eqref{eq:mixev} is satisfied for local transformations, it is always satisfied. Real quantum theory, which shares many features with Fermionic theory, is different in this respect, and we conjecture that a counterexample to the thesis of lemma~\ref{lem:wra} can be found in the latter case.

\section{Conclusion}\label{sec:conc}

Summarising the content of the manuscript, we defined the composition of denumerably many 
systems from an OPT, starting from the space of quasi-local effects, 
and defining quasi-local transformations. We then define states of such system in terms 
of suitable functionals on the space of effects. Among these, there are some 
that can be interpreted as the result of local preparations, and that we deem 
quasi-local states.

We defined global update rules for finite or infinite systems, 
and proved a generalisation of a block-decomposition
theorem for GURs. We then analysed those GURs that are homogeneous. The
definition of homogeneity is a big chapter on its own, and deserves a careful treatment. 
As a consequence of the definition, we proved that for homogenous GURs the memory array
is organised as the Cayley graph of a group. We then defined locality for GURs, 
and studied its consequences in detail. 

In the last section we used homogeneity and locality to define cellular automata, and 
proved the wrapping lemma under very general assumptions.

The theory developed here is of fundamental importance for an approach to physical laws as information-processing algorithms, and will be used for the construction of statistical mechanics in post-quantum theories, as well as the reconstruction of dynamical laws in space-time beyond the now established approach based on quantum walks, thus encompassing interacting quantum and post-quantum field theories.

\acknowledgments

This publication was made possible through the support of a grant from the John Templeton Foundation under the project ID\#60609 Causal Quantum Structures. The opinions expressed in this publication are those of the authors and do not necessarily reflect the views of the John Templeton Foundation.
The author wishes to thank G.~M.~D'Ariano, A.~Bisio, A.~Tosini, and F.~Buscemi for many inspiring discussions. Suggestions about the terminology by G.~Chiribella are also acknowledged that were useful to sharpen the notion that is now defined as causal influence. A particular thank to J.~Barrett for fruitful conversations about various facets of causal influence and signalling, and to M.~Erba for his help in tuning the definitions and dissipating misconceptions on the same subject.

\begin{appendix}
\section{Identification of the sup- and operational norm for effects}\label{app:equinor}

We show that, under the hypothesis of assumption~\eqref{eq:assfullcone}, one has $\normop a=\normsup a$ for every $a\in\Cntset{\rA}_\Reals$. Indeed, if $\Cntset{\rA}_+=\Stset{\rA}_+^*$, then, by theorem~\ref{th:norest}, one has 
\begin{align}
\Cntset{\rA}=\{a\in\Cntset{\rA}_{\Reals}\mid0\leq(a|\rho),(e-a|\rho),\forall\rho\in\Stset{\rA}\}. 
\label{eq:effsst}
\end{align}
First of all, one can rephrase the above condition as
\begin{align}
\Cntset{\rA}=\{a\in\Cntset{\rA}_{\Reals}\mid0\leq(a|\rho),(e-a|\rho),\forall\rho\in\Stset{\rA}_1\}. 
\label{eq:effsnost}
\end{align}
Indeed, since $\Stset{\rA}_1\subseteq\Stset{\rA}$, the set on the l.h.s.~in Eq.~\eqref{eq:effsst} is clearly a subset of that on the l.h.s.~in Eq.~\eqref{eq:effsnost}. Moreover, since every state $\rho\in\Stset{\rA}$ is proportional by a constant $\kappa\in[0,1]$ to a deterministic one $\rho_0\in\Stset{\rA}_1$ (see remark~\ref{rem:norestcau}), if $0\leq(a|\sigma),(e-a|\sigma)$ for all $\sigma\in\Stset{\rA}_1$, then $0\leq(a|\rho),(e-a|\rho)$ for all $\rho\in\Stset{\rA}$. Then, by the identity in Eq.~\eqref{eq:effsnost}, one can identify the set $\Cntset{\rA}$ as the subset of $\Cntset{\rA}_\Reals$ containing only those functionals $a$ on $\Stset{\rA}_\Reals$ such that $0\leq(a|\rho)$ and $0\leq1-(a|\rho)$ for $\rho\in\Stset{\rA}_1$,
and, in turn, the above condition is equivalent to
\begin{align*}
\Cntset{\rA}=\{a\in\Cntset{\rA}_{\Reals}\mid0\leq(a|\rho)\leq1,\forall\rho\in\Stset{\rA}\}.
\end{align*}
Now, we remind that $\lambda\in J(a)$ if and only if $\lambda e\pm a\succeq0$. Under our hypotheses, this means that $\lambda\pm(a|\rho)\geq0$ for every $\rho\in\Stset{\rA}$. Thus, $|(a|\rho)|\leq\lambda$ for every $\rho\in\Stset{\rA}$. Finally, we can conclude that $\normop{a}\leq\normsup a$. On the other hand, if $\normop{a}<\normsup a$, then there exists $\varepsilon>0$ such that for every $\rho\in\Stset{\rA}$ one has $|(a|\rho)|\leq\normsup a-\varepsilon$. In particular, this is true for $\rho_1\in\Stset{\rA}_1$. Consequently
\begin{align*}
|(a|\rho)|=\kappa|(a|\rho_1)|\leq\kappa(\normsup a-\varepsilon)\ \forall \rho\in\Stset{\rA},
\end{align*}
where $0\leq\kappa\leq1$ and $\rho=\kappa\rho_1$. Then $(e|\rho)=\kappa$, and
\begin{align*}
(\{\normsup a-\varepsilon\}e\pm a|\rho)\geq0,\ \forall \rho\in\Stset{\rA}.
\end{align*}
Finally, this implies that $\normsup a-\varepsilon\in J(a)$, which is absurd. Then, $\normsup a=\normop a$.

%
%

\section{Quasi-local states}\label{app:qlst}

We start considering arbitrary finite disjoint partitions of the system $\rA_G$. Let $B\coloneqq \{B_i\}_{i=1}^\infty$ denote a disjoint partition of the set $G$ into finite blocks, i.e. $0<|B_i|<\infty$ for every $i\in\Nats$. The set of such partitions (quotiented by the irrelevant ordering of regions) is denoted by $\set B$. 

\begin{definition}
For a given partition $B\in\set B$, we define the \emph{set of finite $B$-regions} of $G$ as 
\begin{align}
\regpart\coloneqq \left\{R\subseteq G\mid R=\bigcup_{j=1}^k B_{i_j};\ 0\leq k<\infty\right\},
\end{align}
and the \emph{set of $B$-regions} of $G$ as 
\begin{align}
\infregpart\coloneqq \left\{R\subseteq G\mid R=\bigcup_{j=1}^k B_{i_j};\ 0\leq k\leq\infty\right\},
\end{align}
where for $k=0$ we define
\begin{align*}
\bigcup_{j=1}^0 B_{i_j}\coloneqq \emptyset.
\end{align*}
Given $R=\bigcup_{j=1}^k B_{i_j}$ in $\infregpart$, let $\mathbb N^{(B)}_R\coloneqq \{i_j\}_{j=1}^k$.
\end{definition}

\begin{remark}
Clearly, $\regpart\subseteq\infregpart$. It is also trivial to prove that $\infregpart$ and $\regpart$ are closed under $\cup,\cap,\setminus$.
\end{remark}

Given a finite region $R\in\reg G $, for any partition $B\in\set B$ it is possible to cover $R$ with regions of $B$.
\begin{definition}
Given $B\in\set B$, the \emph{cover in $B$ of a region $R\subseteq G$}, denoted $B(R)$, is the minimal $B$-region $C=\bigcup_{j=1}^k B_{i_j}$ such that $R\subseteq C$.
\end{definition}

The above notion of cover is well defined, as proved by the following lemma.
\begin{lemma}
Let $R\subseteq G$. The cover $B(R)$ in $B$ of $R$ exists and is unique.
\end{lemma}
\begin{proof} First of all, the set of $B$-regions $S\in\infregpart$  such that $R\subseteq S$ is not empty, since $R\subseteq G\in\infregpart$. We just need to prove that, having defined $\set C(R,B)\coloneqq \{S\in\infregpart\mid R\subseteq S\}$, there is always a unique minimal element $B(R)\in\set C(R,B)$, i.e.~such that no $B$-region strictly contained in $B(R)$ covers $R$. As to existence, let $h\in R$. Then $h\in B_{l_h}$ for some $l_h$. One can easily realise that $R\subseteq C\coloneqq \bigcup_{h\in R}B_{l_h}$. Moreover, if we remove a given $B$-region $B_{l_h}$ from $C$, and $h\in B_{l_h}$, then $h\not\in C\setminus B_{l_h}$, thus $C$ is a minimal cover of $R$ in $B$. As to uniqueness, suppose that there are two different minimal covers of $R$ in $B$, say $C_1$ and $C_2$. In this case, it is not restrictive to suppose that there is $h\in G$ such that $h\in C_1$ and $h\not\in C_2$. Let $B_i\in B$ be the set such that $h\in B_i$. One has $B_i\subseteq C_1$ and $B_i\cap C_2=\emptyset$. Since $R\subseteq C_2$, we can conclude that $R\cap B_i=\emptyset$, and thus $R\subseteq (C_1\setminus B_i)\in \set C(R,B)$. Thus, $C_1$ cannot be minimal in $\set C(R,B)$.\end{proof}

The cover $B(R)$ can also be thought of as the intersection of all covers of $R$ in $B$.

\begin{lemma}
The cover $B(R)$ of a finite region $R$ is finite.
\end{lemma}
\begin{proof}
Every element $g_k\in R$ belongs to some $B_{i_k}\subseteq B$, where $k\mapsto i_k$ is generally not injective. It is then clear that $R\subseteq\bigcup_{k=1}^{|R|}B_{i_k}$, and thus $B(R)\subseteq\bigcup_{k=1}^{|R|}B_{i_k}$.
\end{proof}
%
%
Given one partition $B\in\set B$, we can now define the set of states that differ on finitely many blocks from a given assignment of states of blocks in $B$. First, we define the reference state as follows.

\begin{definition}[Reference local state]
Given a partition $B\in\set B$, a  \emph{reference local state over $B$} is a map 
\begin{align*}
\rho_0:\Nats\to\bigsqcup_{j\in\Nats}\Stset{\rA_{B_j}}_1::j\mapsto\rho_{0j}\in\Stset{\rA_{B_j}}_1.
\end{align*}
\end{definition}

A reference local state allows one to attribute a state to any finite region $R\in\reg G $ by the following procedure. First, let us consider the $B$-cover $B(R)$ of $R$. Then, the state of $B(R)$ is given by
\begin{align*}
\rho_{0}[B(R)]\coloneqq \bigotimes_{i_j\in\Nats^{(B)}_{B(R)}}\rho_{0i_j}.
\end{align*}
Finally, we discard all the extra systems in the region $B(R)\setminus R$.
\begin{align*}
\rho_{0\vert R}\coloneqq \rho_0[B(R)]_{\vert R}.
\end{align*}

The above construction is then formalised in the following definition.

\begin{figure}
\includegraphics[width=8.5cm]{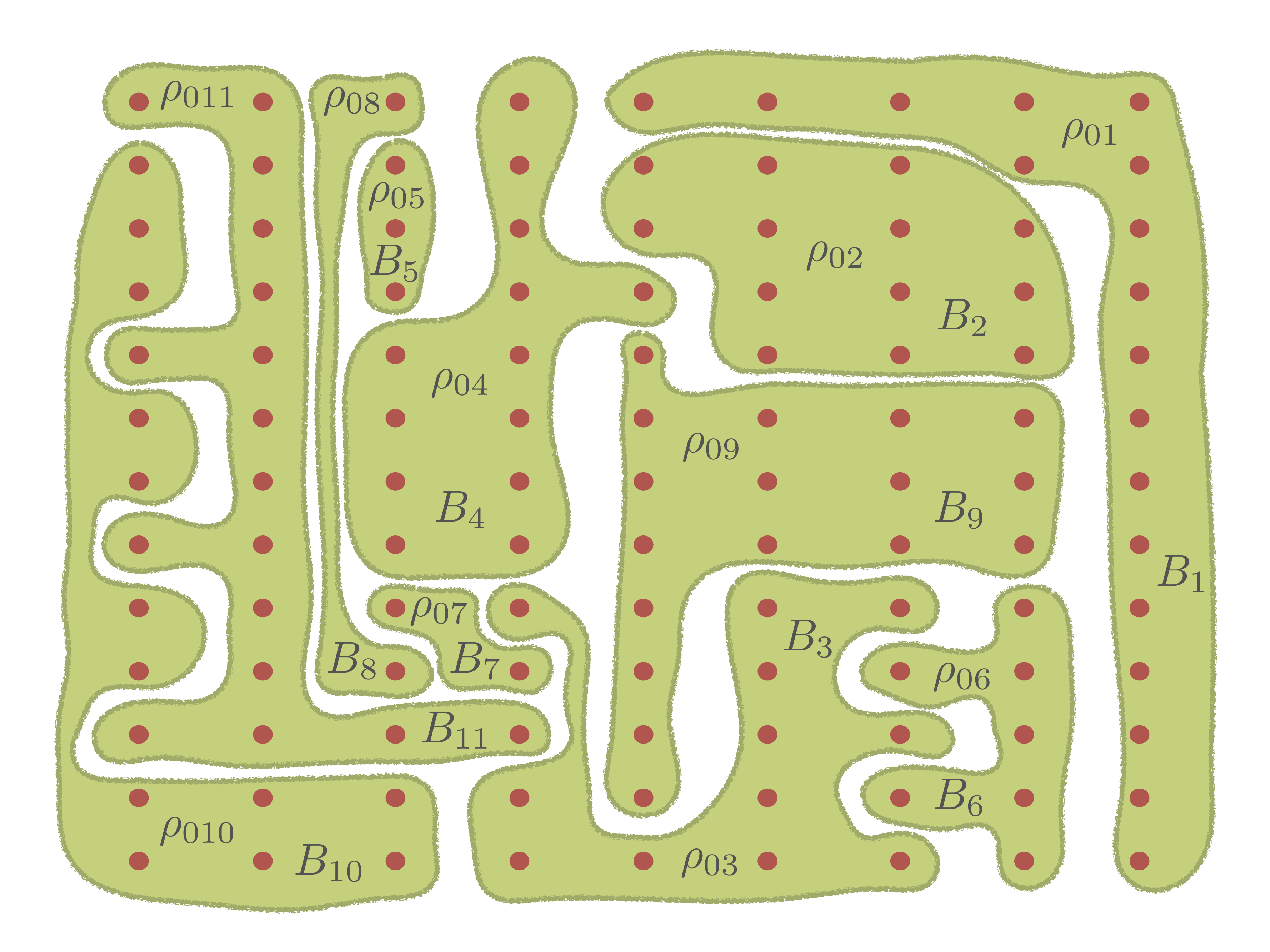}\\
(a)\\
\includegraphics[width=8.5cm]{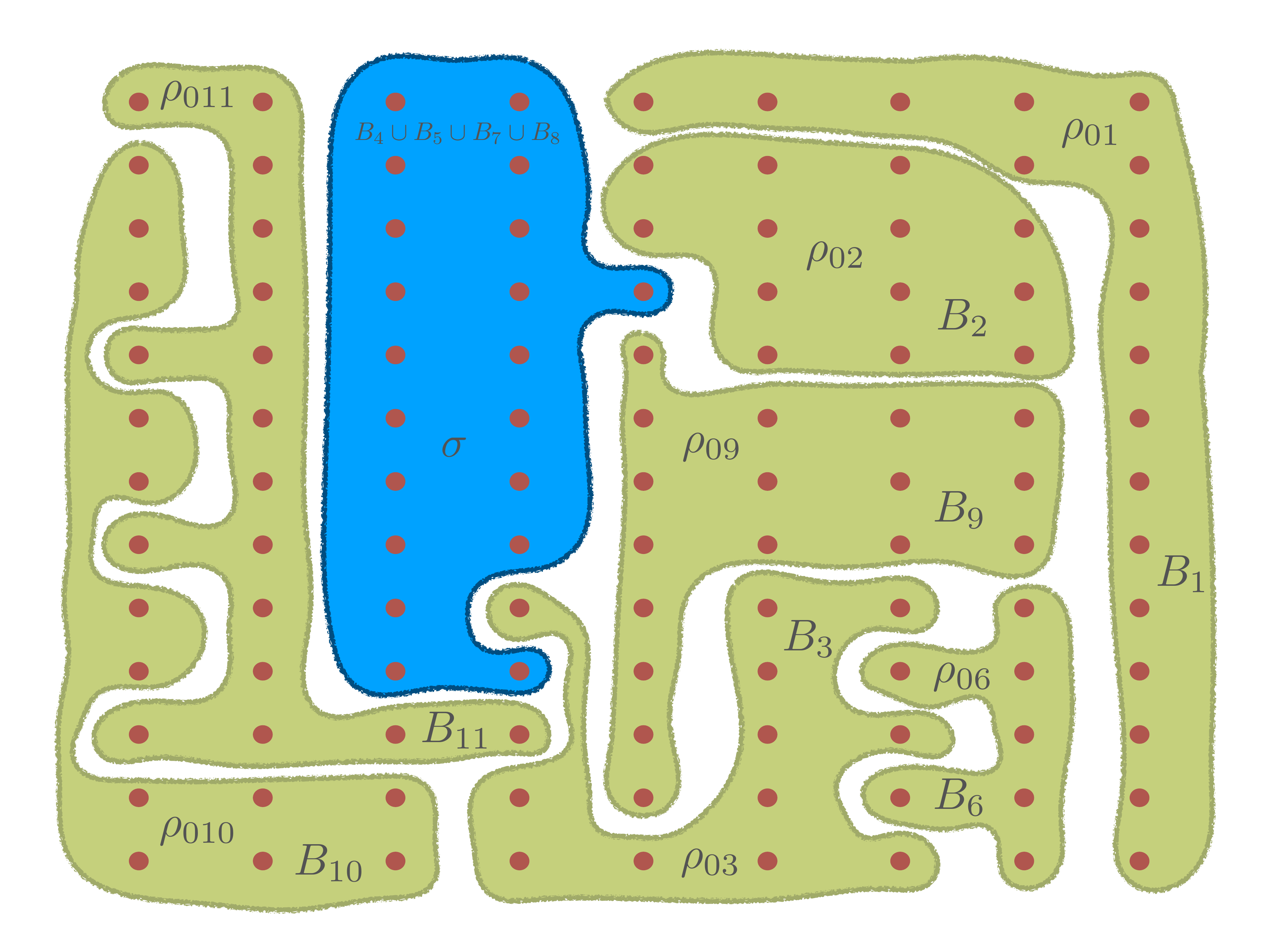}\\
(b)
\caption{\label{f:finitreg}An illustrative example of a reference local state $\rho_0$ (a) for a partition of a finite set $G$ into 11 regions, along with a primitive local state $\sigma_R$ (b) with $R=B_4\cup B_5\cup B_7\cup B_8$. The dots represent the elements of $G$.}
\end{figure}

\begin{definition}
The state of a finite region $R\in\regpart$ in the reference local state $\rho_0$ is defined as
\begin{align*}
&\begin{aligned}
    \Qcircuit @C=1em @R=.7em @! R {\prepareC{\rho_{0\vert R}}&\ustick{R}\qw&\qw}
\end{aligned}\coloneqq 
\begin{aligned}
    \Qcircuit @C=1em @R=.7em @! R {\multiprepareC{1}{\rho_0[B(R)]}&\qw&\ustick{{R}}\qw&\qw&\qw\\
    \pureghost{\rho_0[B(R)]}&\qw&\ustick{{B(R)\setminus R}}\qw&\qw&\measureD{e}}
\end{aligned}\ ,\\
&\rho_{0\vert\emptyset}\coloneqq 1.
\end{align*}
\end{definition}
Given a partition $B$ and a reference local state $\rho_0$, let us now define the set of primitive local states as follows.
\begin{definition}
The set of primitive local states is
\begin{align*}
\set{Pre}\Stset{\rA_G}_{\Reals}^{(B)}\coloneqq \{(\sigma,R)\mid \sigma\in\Stset{\rA_{R}}_\Reals;\ R\in\regpart\}.
\end{align*}
\end{definition}
This set collects all the generalised local preparations of finite regions of $B$. As in the case of effects, we introduce the simplified  notation $\sigma_S\coloneqq (\sigma,S)$.
A choice of reference local state $\rho_0$, allows us to interpret a given element $\sigma_S$ of $\set{Pre}\Stset{\rA_G}_{\Reals}^{(B)}$ as a preparation of the region $S$ in the state $\sigma$. Thus, the element $\sigma_S$ allows one to attribute a generalised state $\tau\in\Stset{\rA_T}_\Reals$ to any region $T\in\reg G $ as follows (see fig.~\ref{f:finitreg})
\begin{align}
\tau\coloneqq \sigma_{\vert{S\cap T}}\otimes \rho_{0\vert(T\setminus S)}.
\label{eq:localstate}
\end{align}

\begin{definition}
Given a reference local state $\rho_0$ over $B$ we define the following equivalence relation in $\set{Pre}\Stset{\rA_G}_{\Reals}^{(B)}$
\begin{align}
	&\sigma_S\sim_{\rho_0}\tau_T\ \Leftrightarrow\ 
	\left\{
	\begin{aligned}
		&\sigma=\nu\otimes \rho_{0\vert(S\setminus T)},\\ 
		&\tau=\nu\otimes \rho_{0\vert(T\setminus S)},
	\end{aligned}\right.
	\label{eq:equivlocst}
\end{align}
for some $\nu\in\Stset{\rA_{S\cap T}}_\Reals$.
\end{definition}
The symbol $\sigma_{S\rho_0}$ will denote the equivalence class of $\sigma_S\in\set{Pre}\Stset{\rA_G}^{(B)}_\Reals$ under the relation $\sim_{\rho_0}$.
Notice that in Eq.~\eqref{eq:equivlocst}, thanks to closure of  $\regpart$ under set difference, one has $(S\setminus T), (T\setminus S)\in\regpart$, and thus the definition of the equivalence relation $\sim_{\rho_0}$ involves exclusively $B$-regions. Moreover, all the regions involved in the definition are finite.
\begin{lemma}\label{lem:extbyb}
Let $\sigma_S\in\set{Pre}\Stset{\rA_G}_{\Reals}^{(B)}$. Then for every $R\in \regpart$ such that $R\cap S=\emptyset$ one has $(\sigma\otimes\rho_{0\vert R})_{S\cup R}\in\set{Pre}\Stset{\rA_G}_{\Reals}^{(B)}$ and $(\sigma\otimes\rho_{0\vert R})_{S\cup R}\sim_{\rho_0}\sigma_S$.
\end{lemma}
\begin{proof} First of all, as $\regpart$ is closed under union, clearly $S\cup R\in\regpart$, and thus $(\sigma\otimes\rho_{0\vert R})_{S\cup R}\in\set{Pre}\Stset{\rA_G}_{\Reals}^{(B)}$. Now, $S\cap(S\cup R)=S$, and thus the condition in Eq.~\eqref{eq:equivlocst} is satisfied for $T=S\cup R$, $T\setminus S=R$, $S\cap T=S$, $\nu=\sigma$, $\rho_{0\vert(T\setminus S)}=\rho_{0\vert R}$, and $\rho_{0\vert(S\setminus T)}=1$.\end{proof}

We now provide a way to identify a canonical representative of the equivalence class $\sigma_{R\rho_0}$ defined as follows.
\begin{definition}
The \emph{minimal representative} $\tilde\sigma_{R_\sigma}$ of the equivalence class $\sigma_{R\rho_0}$ is defined through
\begin{align}
&R_\sigma\coloneqq \bigcap_{S\in\set R_{(\sigma,R)}}S,\quad\tilde\sigma_{R_\sigma}\sim_{\rho_0}\sigma_R,
\label{eq:minimalreplocst}
\end{align}
where 
\begin{align*}
\set R_{(\sigma,R)}\coloneqq \{S\in\regpart\mid\exists\tau\in\Stset{\rA_S}_\Reals:\tau_S\sim_{\rho_0}\sigma_R\}.
\end{align*}
\end{definition}

\begin{lemma}\label{lem:exuniminst}
The minimal representative exists and is unique. 
\end{lemma}

For the proof, see the analogous lemma~\ref{lem:exunmineff}. 

\begin{definition}
The set of \emph{generalised local states over $B$ upon $\rho_0$} is
\begin{align}
\Stset{\rA_G}_{\rho_0L\Reals}^{(B)}\coloneqq \set{Pre}\Stset{\rA_G}_{\Reals}^{(B)}/\sim_{\rho_0}.
\end{align}
Its elements will be denoted by symbols like $\rho_{S\rho_0}$.
\end{definition}
We now define linear combinations of local states.
\begin{definition}\label{def:vecspst}
Let $\sigma_{S\rho_0},\tau_{T\rho_0}\in\Stset{\rA_G}^{(B)}_{\rho_0L\Reals}$, and $a\in\mathbb R$. Then we have
\begin{align*}
&a(\sigma_{S\rho_0})\coloneqq \left\{\begin{aligned}
&(a\sigma)_{S\rho_0}&&a\neq0\\
&0_{\emptyset\rho_0}&&a=0,
\end{aligned}\right.\\
&\sigma_{S\rho_0}+\tau_{T\rho_0}\coloneqq \nu_{S\cup T\rho_0},\\
&\quad \nu\coloneqq \sigma\otimes\rho_{0\vert(T\setminus S)}+\rho_{0\vert(S\setminus T)}\otimes\tau.
\end{align*}
\end{definition}

We will denote by $R_{\sigma+\tau}$ the region where the minimal representative of the class $\sigma_{S\rho_0}+\tau_{T\rho_0}$ differs from $\rho_0$. As in the case of effects, it is not always true that $R_{\sigma+\tau}=R_\sigma\cup R_\tau$. As an example, consider $\sigma=\alpha_{i_1}\otimes\beta_{i_2}$ and $\tau=\alpha_{i_1}\otimes(\rho_{0i_2}-\beta_{i_2})$, with $\alpha_{i_1}\neq\rho_{0i_1}$, $\beta_{i_2}\neq\rho_{0i_2}$, and $R_\sigma=R_\tau=B_{i_1}\cup B_{i_2}$. Then $\sigma+\tau=\alpha_{i_1}\otimes\rho_{0i_2}$, and clearly $R_{\sigma+\tau}=B_{i_1}$, which is strictly included in $R_\sigma=R_\tau=R_\sigma\cup R_\tau$.

\begin{definition}
The \emph{set of local states over $B$ upon $\rho_0$} is
\begin{align*}
\Stset{\rA_G}_{\rho_0L}^{(B)}\coloneqq \{\sigma_{S\rho_0}\in\Stset{\rA_G}_{\rho_0L\Reals}^{(B)}\mid \sigma\in\Stset{\rA_{S}}\}.
\end{align*}
An element $\sigma_{S\rho_0}\in\Stset{\rA_G}_{\rho_0L\Reals}^{(B)}$ is a \emph{local state over $B$ upon $\rho_0$}. A subset of special interest is that of \emph{deterministic local states over $B$ upon $\rho_0$}, defined as
\begin{align*}
\Stset{\rA_G}_{\rho_0L1}^{(B)}\coloneqq \{\sigma_{S\rho_0}\in\Stset{\rA_G}_{\rho_0L}^{(B)}\mid \sigma\in\Stset{\rA_{S}}_1\}.
\end{align*}
\end{definition}

We now characterise the vector space of generalised states as follows.

\begin{lemma}
The set of generalized local states in $B$ upon $\rho_0$ equipped with the operations defined in Def.~\ref{def:vecspst} is a real vector space. It is the space of finite real combinations of deterministic local states in $B$ upon the same state $\rho_0$:
\begin{align}
\Stset{\rA_G}_{\rho_0L\Reals}^{(B)}=\Span_\Reals( \Stset{\rA_G}_{\rho_0L1}^{(B)}),
\end{align}
\end{lemma}
The proof is trivial and we omit it.

%


We now show that quasi-local states are bounded linear functionals on $\Cntset{\rA_G}_{Q\Reals}$. We start considering the pairing of local states $\Stset{\rA_G}_{{\rho_0L\Reals}}^{(B)}$ with local effects 
$\Cntset{\rA_G}_{L\Reals}$, defined as follows.
\begin{definition}
Every $\sigma_{S\rho_0}\in\Stset{\rA_G}_{\rho_0L\Reals}$ identifies a functional on $\Cntset{\rA_G}_{L\Reals}$ as follows. For $a_R\in\Cntset{\rA_G}_{L\Reals}$ let
\begin{align}
(a_R|\sigma_{S\rho_0})\coloneqq 
\begin{aligned}
    \Qcircuit @C=1em @R=.7em @! R {
    \prepareC{\sigma_{\vert R\cap S}}&\ustick{R\cap S}\qw&\multimeasureD{1}{a}\\
    \prepareC{\rho_{0\vert R\setminus S}}&\ustick{R\setminus S}\qw&\ghost{a}}
\end{aligned}\ .
\label{eq:loceffonlocst}
\end{align}
\end{definition}

We now prove that the above pairing is well-defined, namely it does not depend on the representative in the class $\sigma_{S\rho_0}$, nor on the representative in the class $a_R$. This is done in the next two lemmas.
\begin{lemma}
For every $a_R\in\Cntset{\rA_G}_{L\Reals}$ the result of Eq.~\eqref{eq:loceffonlocst} is independent of the choice of representative $\tau_{T\rho_0}$ in the class $\sigma_{S\rho_0}$.
\end{lemma}
\begin{proof}
By hypothesis, there exists $\nu\in\Stset{\rA_{S\cap T}}_\Reals$ such that
\begin{align*}
\sigma=\nu\otimes\rho_{0S\setminus T},\quad\tau=\nu\otimes\rho_{0T\setminus S}.
\end{align*}
Then we have
\begin{align*}
(a_R|\sigma_{S\rho_0})=&(a|\nu_{\vert(S\cap T)\cap R}\otimes\rho_{0\vert[(S\setminus T)\cap R]\cup(R\setminus S)})\\
(a_R|\tau_{T\rho_0})=&(a|\nu_{\vert(S\cap T)\cap R}\otimes\rho_{0\vert[(T\setminus S)\cap R]\cup(R\setminus T)}).
\end{align*}
A straightforward set-theoretic calculation gives that $[(S\setminus T)\cap R]\cup(R\setminus S)=[(T\setminus S)\cap R]\cup(R\setminus T)=R\setminus(S\cap T)$, and thus $(a_R|\sigma_{S\rho_0})=(a_R|\tau_{T\rho_0})$.
\end{proof}

On similar lines, we have the following result.
\begin{lemma}
For every partition $B\in\set B$, every reference local state $\rho_0$ over $B$, and every local state $\tau_{T\rho_0}\in\Stset{\rA_G}_{{\rho_0L\Reals}}^{(B)}$, the result of Eq.~\eqref{eq:loceffonlocst} is independent of the choice of representative $b_S$ in the class $a_R$.
\end{lemma}
\begin{proof}
If $\tilde c_{R_c}$ is the minimal representative of $a_R=b_S$, by Eq.~\eqref{eq:equiveff}, one has
\begin{align*}
&a=\tilde c\otimes e_{R\setminus R_c},\\
&b=\tilde c\otimes e_{S\setminus R_c}.
\end{align*}
Now, by the defining equation~\ref{eq:loceffonlocst}, we have that
\begin{align*}
&(a_R|\tau_{T\rho_0})=(a|\tau_{\vert R\cap T}\otimes\rho_{0\vert(R\setminus T)})\\
=&(\tilde c\otimes e_{R\setminus R_c}|\tau_{\vert R\cap T}\otimes\rho_{0\vert(R\setminus T)})\\
=&(\tilde c|\tau_{\vert R_c\cap T}\otimes\rho_{0\vert R_c\setminus T}),
\end{align*}
and similarly we obtain
\begin{align*}
&(b_S|\tau_{T\rho_0})=
(\tilde c|\tau_{\vert R_c\cap T}\otimes\rho_{0\vert(R_c\setminus T)}).&\qedhere
\end{align*}
\end{proof}

The above results give us a hint that for every $B\in\set B$ and every $\rho_0$, the space $\Stset{\rA_G}_{{\rho_0L\Reals}}^{(B)}$ is a submanifold of the space $\Stset{\rA_G}_\Reals$. We now complete the real vector spaces $\Stset{\rA_G}_{{\rho_0L\Reals}}^{(B)}$ to Banach spaces.
For this purpose, we start equipping all  spaces $\Stset{\rA_G}_{\rho_0L\Reals}^{(B)}$ with a norm.
\begin{definition}
Given an element $\sigma$ of the space $\Stset{\rA_G}_{\rho_0L\Reals}^{(B)}$, its norm is defined as
\begin{align}
\normloc{\sigma}\coloneqq \normop{\tilde\sigma},
\end{align}
where $\normop{\cdot}$ denotes the operational norm on $\Stset{\rA_{R_\sigma}}_\Reals$.
\end{definition}

\begin{lemma}
Given an element $\sigma$ of $\Stset{\rA_G}_{\rho_0L\Reals}^{(B)}$, for $\tau_T$ in the corresponding equivalence class, one has
\begin{align*}
\normloc{\sigma}=\normop{\tau},
\end{align*}
where $\normop{\cdot}$ denotes the operational norm on $\Stset{\rA_T}_\Reals$.
\end{lemma}
\begin{proof} By definition of the relation $\sim_{\rho_0}$, and of minimal representative $\tilde\sigma_{R_\sigma}$ of $\sigma$, one has
\begin{align*}
\tau=\tilde\sigma\otimes\rho_{0(T\setminus R_\sigma)}.
\end{align*}
Thus, $\normop{\tau}=\normop{\tilde\sigma\otimes\rho_{0(T\setminus R_\sigma)}}$. Now, by lemma~\ref{lem:optens}
\begin{align*}
\normop \tau=\normop{\tilde \sigma}=\normloc{\sigma}.&\qedhere
\end{align*}
\end{proof}
We can now show that vectors in $\Stset{\rA_G}_{{\rho_0L\Reals}}^{(B)}$ are bounded linear functionals on $\Cntset{\rA_G}_{Q\Reals}$.
\begin{definition}
Let $a=[{a_n}_{R_n}]\in\Cntset{\rA_G}_{Q\Reals}$, and $\rho=\rho_{S\rho_0}\in\Stset{\rA_G}_{\rho_0L\Reals}$. We define
\begin{align}
(a|\rho)\coloneqq \lim_{n\to\infty}({a_n}_{R_n}|\rho_{S\rho_0}).
\label{eq:defqefflocst}
\end{align}
\end{definition}
With this definition, we can now prove that local states are actually bounded linear functionals on $\Cntset{\rA_G}_{Q\Reals}$.

\begin{lemma}\label{lem:qefflocstbd}
For $a=[{a_n}_{R_n}]\in\Cntset{\rA_G}_{Q\Reals}$, and $\rho_{S\rho_0}\in\Stset{\rA_G}_{\rho_0L\Reals}$, one has 
\begin{align}
|(a|\rho)|\leq\normsup a\normloc{\rho}.
\end{align}
\end{lemma}

\begin{proof}
We start considering the case of a local effect $a_R$. In this case, us define
\begin{align*}
a_0\coloneqq (2\normsup a)^{-1}(\normsup a e_{{R}}+a)\succeq0,\\
a_1\coloneqq (2\normsup a)^{-1}(\normsup a e_{{R}}-a)\succeq0,\\
a_0+a_1=e_{{R}},\quad \normsup a(a_0-a_1)=a,
\end{align*}
we have that
\begin{align*}
|({a}_{R}|\rho_{S\rho_0})|&=|({a}_{R}|\rho_{\vert R\cap S}\otimes\rho_{0\vert R\setminus S})|\\
&=\normsup a|(\{\bar a_0-\bar a_1\}\otimes e_{S\setminus R}|\rho\otimes\rho_{0\vert R\setminus S})|\\
&\leq\normsup a\normloc{\rho_{S\rho_0}},
\end{align*}
where $(\bar a_i|\coloneqq (a_i|\tI_{R\cap S}\otimes|\rho_{0\vert R\setminus S})$.
Let now $a=[{a_n}_{R_n}]$. Then we have
\begin{align*}
|({a_n}_{R_n}&-{a_m}_{R_m}|\rho_{S\rho_0})|\\
&\leq\normsup{{a_n}_{R_n}-{a_m}_{R_m}}\normloc{\rho_{S\rho_0}},
\end{align*}
and thus actually $({a_n}_{R_n}|\rho_{S\rho_0})$ is a Cauchy sequence.
Finally, since for every $n\in\Nats$ one has
\begin{align*}
|({a_n}_{R_n}|\rho_{S\rho_0})|\leq\normsup{a_n}\normloc{\rho_S},
\end{align*}
taking the limit for $n\to\infty$ on both sides we have that
\begin{align*}
|(a|\rho_{S\rho_0})|\leq\normsup a\normloc{\rho_{S\rho_0}}.&&\qedhere
\end{align*}
\end{proof}


We now show that the definition in Eq.~\eqref{eq:defqefflocst}
is independent of the sequence ${a_n}_{R_n}$ in the class of $a$.

\begin{lemma}
Let $a=[{a_n}_{R_n}]=[{b_n}_{R'_n}]\in\Cntset{\rA_G}_{Q\Reals}$, and $\rho_{S\rho_0}\in\Stset{\rA_G}_{\rho_0L\Reals}$. Then
\begin{align}
\lim_{n\to\infty}(a_n|\rho_{S\rho_0})=\lim_{n\to\infty}(b_n|\rho_{S\rho_0}).
\end{align}
\end{lemma}
\begin{proof}
Remind that by definition, for every $\varepsilon$ there exists $n_0$ such that for $n\geq n_0$, it is $\normsup{{a_n}_{R_n}-{b_n}_{R'_n}}\leq\varepsilon$. Now, by lemma~\ref{lem:qefflocstbd}, this implies that
\begin{align*}
|({a_n}_{R_n}-{b_n}_{R'_n}|\rho_{S\rho_0})|\leq\varepsilon\normloc{\rho_{S\rho_0}},
\end{align*}
and thus the thesis follows.
\end{proof}

We finally close the vector space $\Stset{\rA_G}_{\rho_0L\Reals}^{(B)}$ by adding the limits of Cauchy sequences in the norm $\normloc\cdot$, thus obtaining a Banach space that we denote $\Stset{\rA_G}_{\rho_0Q\Reals}^{(B)}$. We start considering Cauchy sequences 
\begin{align*}
\sigma:\Nats\to\Stset{\rA_G}_{\rho_0\Reals}^{(B)}::n\mapsto{\sigma_n}_{R_n\rho_0}\in\Stset{\rA_G}_{\rho_0L\Reals},
\end{align*}
which make a real vector space $\Stset{\rA_G}_{\rho_0C\Reals}^{(B)}$, containing $\Stset{\rA_G}_{\rho_0L\Reals}^{(B)}$ as the subspace of constant sequences. We then quotient $\Stset{\rA_G}^{(B)}_{\rho_0C\Reals}$ by the usual equivalence relation $\cong$, obtaining complete normed real vector space.

\begin{definition}
The space of quasi-local states in $B$ upon $\rho_0$ is defined as the space $\Stset{\rA_G}_{\rho_0Q\Reals}^{(B)}\coloneqq \Stset{\rA_G}_{\rho_0C\Reals}^{(B)}/\cong$.
\end{definition}

Inside the new Banach space that we defined, we can single out a convex set of quasi-local states, corresponding to preparations that can be arbitrarily approximated by local protocols starting from $\rho_0$, along with the cone that they span, and the convex set of deterministic states.

\begin{definition}\label{def:qlstorig}
An element $\sigma$ of $\Stset{\rA_G}_{\rho_0Q\Reals}^{(B)}$ is a \emph{quasi-local state} if there exists ${\sigma_n}_{R_n\rho_0}$ in the class $\sigma$, and $n_0\in\Nats$, such that, for $n\geq n_0$, ${\sigma_n}_{R_n\rho_0}\in\Stset{\rA_G}_{\rho_0L}^{(B)}$. The set of quasi-local states in $B$ upon $\rho_0$ is denoted by $\Stset{\rA_G}_{\rho_0Q}^{(B)}$.

A quasi-local state $\sigma$ in $\Stset{\rA_G}_{\rho_0Q}^{(B)}$ is \emph{deterministic}, if there exists a sequence ${\sigma_n}_{R_n}$ in the class of $\sigma$ and $n_0\in\Nats$ such that, for $n\geq n_0$, ${\sigma_n}_{R_n}\in\Stset{\rA_G}_{\rho_0L1}^{(B)}$. The set of deterministic quasi-local states is denoted by $\Stset{\rA_G}_{\rho_0Q1}^{(B)}$.

We denote the subset of $\Stset{\rA_G}_{\rho_0Q\Reals}^{(B)}$ of elements $\sigma=\lambda\rho$, for $\lambda\geq0$ and $\rho\in\Stset{\rA_G}_{\rho_0Q}^{(B)}$, as $\Stset{\rA_G}_{\rho_0Q+}^{(B)}$. We will equivalently write $\sigma\succeq0$ for $\sigma\in\Stset{\rA_G}_{\rho_0Q+}^{(B)}$
\end{definition}

Analogously to the case of effects, the elements of $\Stset{\rA_G}_{{\rho_0Q\Reals}}^{(B)}$ will be denoted by $\rho=[{\rho_n}_{R_n\rho_0}]$, with the convention that for a constant Cauchy sequence ${\rho_n}_{R_n\rho_0}=\rho_{R\rho_0}$ we will set $\rho_{R\rho_0}\coloneqq [{\rho_n}_{R_n\rho_0}]$. 
We define $\normloc \rho\coloneqq \lim_{n\to\infty}\normloc{{\rho_n}_{R_n\rho_0}}$. 
One can straightforwardly check that the space $\Stset{\rA_G}_{\rho_0Q\Reals}^{(B)}$ of quasi-local states in $B$ upon $\rho_0$ is a Banach space, the sets $\Stset{\rA_G}_{\rho_0Q}^{(B)}$ and $\Stset{\rA_G}_{\rho_0Q1}^{(B)}$ are convex subsets, and $\Stset{\rA_G}_{\rho_0Q+}^{(B)}$ is a convex cone. The space $\Stset{\rA_G}_{{\rho_0L\Reals}}^{(B)}$ can be identified with the dense submanifold containing constant sequences $\sigma_{S\rho_0}$. 

For every partition $B$ and every reference local state $\rho_0$, the Banach space $\Stset{\rA_G}_{\rho_0Q\Reals}^{(B)}$ is separable. Indeed, one can choose a basis in $\Stset{\rA_{S_n}}_\Reals$ for each region $S_n\coloneqq \bigcup_{j=1}^n B_j$ in the increasing sequence $\{S_n\}_{n\in\Nats}$ of regions of $B$. Now, for every $\rho\in \Stset{\rA_G}_{\rho_0Q\Reals}^{(B)}$, and for every $n\in\Nats$, given the class ${\rho_n}_{R_n\rho_0}$ defining $\rho$, one has $R_n\subseteq S_m$ for sufficiently large $m$. Thus, linear combinations of the countable collection of bases for $S_m$ are dense in $\Stset{\rA_G}_{\rho_0Q\Reals}^{(B)}$.

We now show that actually, for every pair $(B,\rho_0)$, the space $\Stset{\rA_G}_{{\rho_0Q\Reals}}^{(B)}$ is a subspace of $\Stset{\rA_G}_\Reals$, and $\normst\cdot\equiv\normloc\cdot$ on $\Stset{\rA_G}_{{\rho_0Q\Reals}}^{(B)}$. 
First of all, we extend the pairing of $\Cntset{\rA_G}_{Q\Reals}$ and $\Stset{\rA_G}_{{\rho_0L\Reals}}^{(B)}$ to $\Cntset{\rA_G}_{Q\Reals}$ and $\Stset{\rA_G}_{{\rho_0Q\Reals}}^{(B)}$.

\begin{definition}
Let $a=[{a_n}_{R_n}]\in\Cntset{\rA_G}_{Q\Reals}$ and $\sigma=[{\sigma_n}_{S_n\rho_0}]\in\Stset{\rA_G}_{{\rho_0Q\Reals}}^{(B)}$. Then we define
\begin{align}
(a|\sigma)\coloneqq \lim_{n\to\infty}(a|{\sigma_n}_{S_m\rho_0}).
\end{align}
\end{definition}
The above definition is well-posed, as shown by the following lemma, whose proof is a straightforward consequence of lemma~\ref{lem:qefflocstbd}.
\begin{lemma}
Let $a\in\Cntset{\rA_G}_{Q\Reals}$, and $\sigma=[{\sigma_n}_{S_n\rho_0}]=[{\sigma'_n}_{S'_n\rho_0}]\in\Stset{\rA_G}_{{\rho_0Q\Reals}}^{(B)}$. Then
\begin{align}
\lim_{n\to\infty}(a|{\sigma_m}_{S_m\rho_0})=\lim_{n\to\infty}(a|{\sigma'_m}_{S'_m\rho_0}).
\end{align}
\end{lemma}

Finally, we use the definition of sup-norm of effects to prove that vectors in the spaces $\Stset{\rA_G}_{\rho_0Q\Reals}^{(B)}$ are bounded linear functionals on $\Cntset{\rA_G}_{Q\Reals}$. Also in this case the proof follows straightforwardly form lemma \ref{lem:qefflocstbd}.

\begin{lemma}\label{lem:boundeffst}
Let $a\in\Cntset{\rA_G}_{Q\Reals}$ and $\rho\in\Stset{\rA_G}_{{\rho_0Q\Reals}}^{(B)}$. Then one has
\begin{align}\label{eq:boundlocn}
|(a|\rho)|\leq\normsup{a}\normloc{\rho}.
\end{align}
\end{lemma}

\begin{lemma}\label{lem:normqlst}
The spaces $\Stset{\rA_G}^{(B)}_{\rho_0Q\Reals}$ are closed subspaces of the extended space of states $\Stset{\rA_G}_\Reals$. On $\Stset{\rA_G}^{(B)}_{\rho_0Q\Reals}$ it holds that $\normloc\cdot\equiv\normst\cdot$.
\end{lemma}

\begin{proof}
We only need to prove that $\normst\cdot\equiv\normq\cdot$, since we already proved that $\Stset{\rA_G}^{(B)}_{\rho_0L\Reals}$ is a linear manifold in the space of bounded linear functionals on $\Cntset{\rA_G}_{Q\Reals}$.
First of all, by eq.~\eqref{eq:boundlocn} one clearly has $\normst\rho\leq\normloc\rho$ for all $\rho\in\Stset{\rA_G}_{Q\Reals}$. For the converse, one can use the same technique as for lemma \ref{lem:eqopnormstnorm}, to prove that $\normloc\rho\leq\normst\rho$ for the case of a local state $\rho\in\Stset{\rA_G}_{{\rho_0L\Reals}}^{(B)}$. Thus $\normst\rho=\normloc\rho$ for local states $\rho\in\Stset{\rA_G}^{(B)}_{\rho_0L\Reals}$. It is then straightforward to conclude that equality holds for every quasi local state $\rho\in\Stset{\rA_G}^{(B)}_{\rho_0Q\Reals}$.
\end{proof}

Notice that two partitions $B$ and $B'$ along with two local reference states $\rho_0$ and $\rho_0'$ might define the same space $\Stset{\rA_G}_{{\rho_0Q\Reals}}^{(B)}\equiv\Stset{\rA_G}_{{\rho_0'Q\Reals}}^{(B')}$. This is the case if for every $\sigma\in\Stset{\rA_G}_{{\rho_0Q\Reals}}^{(B)}$ there exists a state $\sigma'\in\Stset{\rA_G}_{{\rho_0'Q\Reals}}^{(B')}$ such that, $\sigma=\sigma'$, as functionals on $\Cntset{\rA_G}_{Q\Reals}$.
We then provide the following definition.
\begin{definition}
We say that the two reference local states $(B,\rho_0)$ and $(B',\rho'_0)$ are compatible, and denote this relation by the symbol $(B,\rho_0)\sim(B',\rho'_0)$, if for all $\sigma\in\Stset{\rA_G}_{{\rho_0Q\Reals}}^{(B)}$ there exists a state $\sigma'\in\Stset{\rA_G}_{{\rho_0'Q\Reals}}^{(B')}$ such that, for every $a\in\Cntset{\rA_G}_{Q\Reals}$, one has
\begin{align*}
(a|\sigma)=(a|\sigma').
\end{align*}
\end{definition}

%
%
%
The equivalence classes of pairs $(B,\rho_0)$ under the above relation will be denoted by $[(B,\rho_0)]$.

The norm $\normloc{\cdot}$ can be extended to the space of finite linear combinations of generalised quasi-local states. 
\begin{definition}\label{def:finsum}
Let $\sigma_i\in\Stset{\rA_G}^{(B^{(i)})}_{\rho^{(i)}_0Q\Reals}$, for $i=1,\ldots,k$, with $[(B^{(i)},\rho_0^{(i)})]\neq[(B^{(j)},\rho^{(j)}_0)]$. Then we define
\begin{align*}
\left\|\sum_{i=1}^ka_i\sigma_i\right\|_Q\coloneqq \sum_{i=1}^k|a_i|\|{\sigma_i}\|_{B^{(i)},\rho_0^{(i)}}.
\end{align*} 
\end{definition}

The space of finite sums as from definition \ref{def:finsum} is a real vector subspace of $\Stset{\rA_G}_\Reals$. Moreover, as a real vector space, it is equipped with the norm $\normq\cdot$. We can define the space of Cauchy sequences in this space, and complete it by the usual procedure. This leads us to the following definition of the space of quasi-local states.

\begin{definition}\label{def:qlgenst}
The \emph{space of generalised quasi-local states} of $\rA_G$, denoted as $\Stset{\rA_G}_{Q\Reals}$, is 
\begin{align*}
\Stset{\rA_G}_{Q\Reals}\coloneqq \bigoplus_{[B,\rho_0]}\Stset{\rA_G}_{\rho_0Q\Reals}^{(B)},
\end{align*}
where the direct sum denotes closure of the space of finite linear combinations in the norm $\normq\cdot$. 
\end{definition}
The norm introduced above allows us to prove that $\Stset{\rA_G}_{Q\Reals}$ is a space of bounded linear functionals on $\Cntset{\rA_G}_{Q\Reals}$, namely a subspace of $\Stset{\rA_G}_\Reals$.
\begin{lemma}
Every quasi-local state $\rho\in\Stset{\rA_G}_{Q\Reals}$ is a continuous functional on the space of quasi-local effects $a\in\Cntset{\rA_G}_{Q\Reals }$, and
\begin{align}
\label{eq:conqlst}
|(a|\rho)|\leq\normsup a\normq\rho
\end{align}
\end{lemma}
\begin{proof}
Let $\rho\in\Stset{\rA_G}_{\Reals}$ be a finite sum of the form
\begin{align}\label{eq:fins}
\rho=\sum_{i=1}^k a_i\sigma_i,\quad\sigma_i\in\Stset{\rA_G}_{\rho_0Q\Reals}^{(B^{(i)})}. 
\end{align}
Then
\begin{align}
(a|\rho)&\coloneqq \sum_{i=1}^k a_i(a|\sigma_i).
\end{align}
One can easily realise that $|(a|\rho)|\leq\normsup a\normq\rho$. Finally, by the above inequality, also the limit of a Cauchy sequence in $\Stset{\rA_G}_{Q\Reals}$ is a continuous linear functional on $\Cntset{\rA_G}_{Q\Reals}$, and satisfies
\begin{align*}
&|(a|\rho)|\leq\normsup a\normq\rho.&&\qedhere
\end{align*}
\end{proof}

We can now prove the following important result.
\begin{lemma}\label{lem:normqnormst}
The space $\Stset{\rA_G}_{Q\Reals}$ is a closed subspace of the extended space of states $\Stset{\rA_G}_\Reals$. On $\Stset{\rA_G}_{Q\Reals}$ it holds that $\normq\cdot\equiv\normst\cdot$.
\end{lemma}

\begin{proof}
We only need to prove that $\normst\cdot\equiv\normq\cdot$ on finite sums of the form of eq.~\eqref{eq:fins}, since we already proved that $\Stset{\rA_G}_{Q\Reals}$ is obtained closing a linear manifold in the space of bounded linear functionals on $\Cntset{\rA_G}_{Q\Reals}$. The proof follows exactly the same argument as for lemma~\ref{lem:normqlst}, using
eq.~\eqref{eq:conqlst} instead of eq.~\eqref{eq:boundlocn}.
\end{proof}

Notice that all local states allow us to define states of arbitrary finite regions $S\subseteq G$, by Eq.~\eqref{eq:localstate}. We will now extend the definition of local state for all quasi-local states in a fixed sector $\Stset{\rA}_{{\rho_0Q}}^{(B)}$, as  follows. 
First, 
if $\rho$ is quasi local, then $\normloc{\rho_n- \rho_m}\leq\varepsilon$, and $\normop{(\rho_{n}-\rho_{m})\vert_S}\leq\normloc{\rho_n-\rho_m}\leq\varepsilon$. Thus, being the set of states $\Stset{\rA_S}$ compact, $\rho_S\coloneqq \lim_{n\to\infty}\rho_{n}\vert_S\in\Stset{\rA_S}$.

The same is true for finte linear combinations, as well as limits of Cauchy sequences thereof. Indeed, given a finite combination of states $\rho=\sum_ia_i\rho_i$, one has 
\begin{align}
\rho_{\vert S}\coloneqq \sum_ia_i\rho_{i\vert S}\in\Stset{\rA_S}_\Reals. 
\label{eq:finsumlocst}
\end{align}
Finally, if $\tau:\Nats\to\Stset{\rA}_{Q\Reals}$, and $\normq{\tau_n-\tau_m}\leq \varepsilon$, then clearly $\normop{(\tau_{n}-\tau_{m})_{\vert_S}}\leq\varepsilon$, and thus 
\begin{align}
\tau_{\vert{S}}\coloneqq \lim_{n\to\infty}(\tau_n)_{\vert S}\in\Stset{\rA_S}_\Reals.
\label{eq:locgenst}
\end{align}

In view of the above considerations, we can now define the set of states in $\Stset{\rA_G}_{Q\Reals}$.

\begin{definition}\label{def:qlst}
The set $\Stset{\rA_G}_Q$ of quasi-local states is the (convex) set of generalised states $\rho$ for which, for every finite region $S\in\reg G $, the restriction $\rho_{\vert S}$ of $\rho$ to $S$ is a state in $\Stset{\rA_S}$. 
The \emph{positive cone} $\Stset{\rA_G}_{Q+}$ of $\rA_G$ is 
\begin{align*}
&\Stset{\rA_G}_{Q+}\\
&\coloneqq \{\sigma\in\Stset{\rA_G}_{Q\Reals}\mid\exists \lambda\geq0,\rho\in\Stset{\rA_G}_Q:\sigma=\lambda\rho\}.
\end{align*}
The \emph{set of deterministic quasi-local states} of $\rA_G$, denoted as $\Stset{\rA_G}_{Q1}$, is 
\begin{align*}
\Stset{\rA_G}_{Q1}\coloneqq \{\rho\in\Stset{\rA_G}_Q\mid(e_G|\rho)=1\}.
\end{align*}
\end{definition}

\begin{lemma}\label{lem:compdefst}
Definition~\ref{def:qlst} is compatible with definition~\ref{def:exlocst}
\end{lemma}
\begin{proof}
We need to prove that the definition of restriction to the finite region $S$ provided in Eq.~\eqref{eq:exlocst} is equivalent to that of Eq.~\eqref{eq:locgenst}. We start from the case of local states. In this case, since effects are separating for states of finite dimensional systems, one has $(a_S|\rho_T)=(a_S\otimes e_{T\setminus S}|\rho_{\vert T}\otimes\rho_{0\vert S\setminus T})$, and thus $\rho_{\vert S}=\rho_{\vert(T\cap S)}\otimes\rho_{0\vert S\setminus T}$, 
which is precisely the same as in Eq.~\eqref{eq:localstate}. The statement in the case of a general state $\rho\in\Stset{\rA_G}_{Q}$ is then straightforwardly proved, reminding that $|(a_S|\rho-\rho_n)|\leq\normsup{a_S}\normq{\rho-\rho_n}$.
\end{proof}

\begin{corollary}
One has the following inclusions
\begin{align}
\Stset{\rA_G}_Q\subseteq\Stset{\rA_G},\quad\Stset{\rA_G}_{Q1}\subseteq\Stset{\rA_G}_1.
\end{align}
\end{corollary}

\section{Proof of identity~\ref{eq:condthree}}\label{app:spr}
In order to keep the notation lighter, we do the calculation for the case of $\rC\cong\rI$. The proof can then be adapted to the general case by suitably padding all transformations with $\tI_\rC$ and extending the region $\tilde S$ to $\tilde S\rC$.
\begin{align*}
&({\tV}\tA\tV^{-1}\otimes\tI_{H_1})^\dag a_{\tilde S}=\tW^{-1\dag}(\tA\otimes\tI_{H_1})^\dag\tW^\dag a_{\tilde S}\nonumber\\
=&\tW^{-1\dag}(\tA\otimes\tI_{H_1})^\dag\tS_{N^{+*}_S\cup N^+_S}^\dag\tS^{\prime\dag}_{N^{+*}_S\cup S} a_{\tilde S}\nonumber\\
=&\tS^{\prime\dag}_{P^{+*}_S\cup N^{+*}_S\cup R} \tS_{N^{+}_{S}\cup N^{+*}_{S}\cup {R}}^\dag(\tA\otimes\tI_{H_1})^\dag\nonumber\\
&\ \times\tS_{N^{+*}_S\cup N^+_S}^\dag\tS^{\prime\dag}_{N^{+*}_S\cup S} a_{\tilde S}\nonumber\\
=&\tS^{\prime\dag}_{P^{+*}_S\cup N^{+*}_S\cup R} \tS_{N^{+}_{S}\cup N^{+*}_{S}\cup {R}}^\dag(\tA\otimes\tI_{H_1})^\dag\nonumber\\
&\ \times\tS_{N^{+}_{S}\cup N^{+*}_{S}\cup {R}}^\dag\tS^{\prime\dag}_{P^{+*}_S\cup N^{+*}_S\cup R}  a_{\tilde S}\nonumber\\
=&\tS^{\prime\dag}_{{P^{+*}_S}\cup N^{+*}_{S}\cup {R}} \tS_{R}^\dag(\tA\otimes\tI_{H_1})^\dag\tS_{ R}^\dag\tS^{\prime\dag}_{{P^{+*}_S}\cup N^{+*}_{S}\cup {R}} a_{\tilde S}\nonumber\\
=&\tS^{\prime\dag}_{{P^{+*}_S}\cup N^{+*}_{S}\cup {R}} (\tI_{H_0}\otimes\tA)^\dag\tS^{\prime\dag}_{{P^{+*}_S}\cup N^{+*}_{S}\cup {R}} a_{\tilde S}\nonumber\\
=&\tS^{\prime\dag}_{R} (\tI_{H_0}\otimes\tA)^\dag\tS^{\prime\dag}_{R} [a_{\tilde S}]\nonumber\\
=&\tS^{\prime\dag}_{R}\tS_R^\dag (\tA\otimes\tI_{H_0})^\dag\tS_R^\dag\tS^{\prime\dag}_{R} [a_{\tilde S}],
\end{align*}
where in the third identity we used Eq.~\eqref{eq:wdaginv}, observing that $(\tA\otimes\tI_{H_1})^\dag\tS_{N^{+*}_S\cup N^+_S}^\dag\tS^{\prime\dag}_{N^{+*}_S\cup S} [a_{\tilde S}]\in\Cntset{\rA^{(G)}_{(N^{+*}_S\cup R,0}\cup(N^+_S,1)}_{Q\Reals}$. In the fourth identity we used the fact that $a_{\tilde S}$ can be padded with $e_X$ to obtain an effect $\tilde a_X$ on the region $\tilde S\cup X$, and then $\tW^\dag\tilde a_X$ is an effect equivalent to $\tW^\dag a_S$. In the fifth identity we used the fact that $\tS_X$ commutes with $\tY_Y\otimes\tI_{H_1}$ for $Y\cap X=\emptyset$, and in the  seventh identity we used the fact that $\tS'_X$ commutes with $\tI_{H_0}\otimes \tY_Y$ for $Y\cap X=\emptyset$.

\section{Homogeneity and causal influence}\label{app:sig}
This section is devoted to the proof that, for $\pi\in\Pi_\tV$,
\begin{align*}
\pi(g)\rightarrowtail\pi(g')\ \Leftrightarrow\ g\rightarrowtail g'
\end{align*}
By definition, $f\rightarrowtail f'$ if there exists $R\ni f$ and
\begin{align}
&\exists\tF\in\Trnset{\rA^{(G)}_f\rC\to\rA^{(G)}_f\rC}_{Q\Reals},\nonumber\\ 
&(\tV\otimes\tI_\rC)\tF (\tV^{-1}\otimes\tI_\rC)\not\in\Trnset{\rA^{(G)}_{\bar f'}\rC\to\rA^{(G)}_{\bar f'}\rC}_{Q\Reals}.
\label{eq:gnotsiggp}
\end{align}
Let now $f=\pi(g)$ for some $\pi\in\Pi_\tV$. Now, by homogeneity, and in particular by Eq.~\eqref{eq:tpi}, one has that $\tF'\in\Trnset{\rA^{(G)}_{\pi(g)}\rC\to\rA^{(G)}_{\pi(g)}\rC}$
if and only if there exists $\tF\in\Trnset{\rA^{(G)}_{g}\rC\to\rA^{(G)}_{g}\rC}$ such that $\tF'=(\tT_\pi\otimes\tI_\rC)\tF(\tT_\pi^{-1}\otimes\tI_\rC)$.
Then one has
\begin{align}
(\tV\otimes\tI_\rC)&\tF' (\tV^{-1}\otimes\tI_\rC)\nonumber\\
&=(\tV\tT_\pi\otimes\tI_\rC)\tF(\tT_\pi^{-1}\tV^{-1}\otimes\tI_\rC)\nonumber\\
&=(\tT_\pi\tV\otimes\tI_\rC)\tF(\tV^{-1}\tT_\pi^{-1}\otimes\tI_\rC).
\label{eq:relgpig}
\end{align}
As a consequence, again by eq.~\eqref{eq:tpi}, we have that $(\tV\otimes\tI_\rC)\tF'(\tV^{-1}\otimes\tI_\rC)\in\Trnset{\rA^{(G)}_{\bar f'}\rC\to\rA^{(G)}_{\bar f'}\rC}$, for $f'=\pi(g')$, if and only if $(\tV\otimes\tI_\rC)\tF(\tV^{-1}\otimes\tI_\rC)\in\Trnset{\rA^{(G)}_{\bar g'}\rC\to\rA^{(G)}_{\bar g'}\rC}$. Finally, comparing with Eq.~\eqref{eq:gnotsiggp}, this implies that $\pi(g)\rightarrowtail\pi(g')$ if and only if $g\rightarrowtail g'$.

\section{Proof of right and left invertibility of a locally defined GUR}\label{app:rinvlinv}

Here we prove right and left invertibility of $\tV$ defined starting from Eq.~\eqref{eq:vr}.
First, notice that
\begin{align*}
&\begin{aligned}
    \scalebox{.65}{\Qcircuit @C=1em @R=.7em @! R {\prepareC{\psi}&\ustick{P^-_R}\qw &\multigate{3}{\tS'_{N^-_R}}&\qw&\ustick{P^-_R\setminus R}\qw&\qw&\measureD{e}\\
    &&\pureghost{\tS'_{N^-_R}}&\qw&\ustick{R}\qw&\qw&\ghost{a}\\
    &&&&&\ustick{\rC}\qw&\multimeasureD{-1}{a}\\
    &\ustick{N^-_R}\qw &\ghost{\tS'_{N^-_R}}&\qw&\ustick{N^-_R}\qw&\qw&\measureD{e}}}
\end{aligned}\ 
=\ 
\begin{aligned}
    \scalebox{.65}{\Qcircuit @C=1em @R=.7em @! R {\prepareC{\psi}&\ustick{P^-_R}\qw &\multigate{3}{\tS'_{N^-_R}}&\qw&\qw&\ustick{P^-_R\setminus R}\qw&\qw&\measureD{e}\\
        &&\pureghost{\tS'_{N^-_R}}&\qw&\ustick{R}\qw&\ghost{\tA}&\ustick{R}\qw&\measureD{e}\\
        &&\pureghost{\tS'_{N^-_R}}&&\ustick{\rC}\qw&\multigate{-1}{\tA}&\ustick{\rC}\qw&\measureD{e}\\
    &\ustick{N^-_R}\qw &\ghost{\tS'_{N^-_R}}&\qw&\qw&\ustick{N^-_R}\qw&\qw&\measureD{e}}}
\end{aligned}\\
\\
&=\ 
\begin{aligned}
    \scalebox{.7}{\Qcircuit @C=1em @R=.7em @! R {\prepareC{\psi}&\ustick{P^-_R}\qw &\multigate{3}{\tS'_{N^-_R}}&\qw&\qw&\ustick{P^-_R\setminus R}\qw&\qw&\qw&\multigate{3}{\tS'_{N^-_R}}&\ustick{P^-_R}\qw&\measureD{e}\\
    &&\pureghost{\tS'_{N^-_R}}&\qw&\ustick{R}\qw&\ghost{\tA}&\ustick{R}\qw&\qw&\ghost{\tS'_{N^-_R}}&&\\
    &&\pureghost{\tS'_{N^-_R}}&&\ustick{\rC}\qw&\multigate{-1}{\tA}&\ustick{\rC}\qw&\measureD{e}\\
    &\ustick{N^-_R}\qw &\ghost{\tS'_{N^-_R}}&\qw&\qw&\ustick{N^-_R}\qw&\qw&\qw&\ghost{\tS'_{N^-_R}}&\ustick{N^-_R}\qw&\measureD{e}}}
\end{aligned}\\ 
\\
&= 
\begin{aligned}
    \scalebox{.7}{\Qcircuit @C=1em @R=.7em @! R {\prepareC{\psi}&\qw&\ustick{P^-_R}\qw &\qw&\measureD{e}\\
       &\ustick{\rC}\qw &\multigate{1}{\tA'}&\ustick{\rC}\qw&\measureD{e}\\
        &\ustick{N^-_R}\qw &\ghost{\tA'}&\ustick{N^-_R}\qw&\measureD{e}}}
\end{aligned}
\ ,
\end{align*}
where we used Eq.~\eqref{eq:condthree}.

Let us now define $\tW$ by $\tW^\dag b_{S\rC}\coloneqq \{(\tW^\dag_S\otimes\tI_\rC^\dag)b\}_{N^+_S\rC}$, where
\begin{align}
&\begin{aligned}
    \Qcircuit @C=1em @R=.7em @! R {&\ustick{N^+_S}\qw &\gate{\tW_S}&\ustick{S}\qw&\qw
    }
\end{aligned}\ \coloneqq\ 
\begin{aligned}
    \Qcircuit @C=1em @R=.7em @! R {&\ustick{N^+_S}\qw &\multigate{1}{\tS'_{S}}&\qw&\ustick{N^+_S}\qw&\qw&\measureD{e}\\
    \prepareC{\eta}&\ustick{S}\qw &\ghost{\tS'_{S}}&\qw&\ustick{S}\qw&\qw&\qw}
\end{aligned}\ .
\label{eq:ws}
\end{align}
Then one can calculate $\tW^\dag\tV^\dag a_{R\rC}$ by the following diagram
\begin{align*}
&\begin{aligned}
    \scalebox{.75}{\Qcircuit @C=1em @R=.7em @! R {&\ustick{P^-_R\setminus R}\qw&\qw&\multigate{3}{\tS'_{N^-_R}}&\ustick{P^-_R}\qw&\measureD{e}&\prepareC{\psi}&\ustick{P^-_R}\qw &\multigate{3}{\tS'_{N^-_R}}&\qw&\ustick{\!\!\!\! P^-_R\setminus R}\qw&\measureD{e}\\
    &\ustick{R}\qw&\qw&\ghost{\tS'_{N^-_R}}&&&&&\pureghost{\tS'_{N^-_R}}&\ustick{R}\qw&\qw&\multimeasureD{1}{a}\\
       &&&&&&&&&&\ustick{\rC}\qw&\ghost{a}\\
    \prepareC{\eta}&\ustick{N^-_R}\qw&\qw&\ghost{\tS'_{N^-_R}}&\qw&\qw&\ustick{N^-_R}\qw&\qw &\ghost{\tS'_{N^-_R}}&\qw&\ustick{N^-_R}\qw&\measureD{e}}}
\end{aligned}
\\
\\
&
=\ 
\begin{aligned}
    \scalebox{.65}{\Qcircuit @C=1em @R=.7em @! R {&\ustick{\!\!\!\! P^-_R\setminus R}\qw &\multigate{2}{\tS'_{N^-_R}}&\qw&\ustick{P^-_R}\qw&\qw&\measureD{e}\\
    &\ustick{R}\qw&\ghost{\tS'_{N^-_R}}&\ustick{\rC}&\multigate{1}{\tA'}&\ustick{\rC}\qw&\measureD{e}\\
    \prepareC{\eta}&\ustick{N^-_R}\qw&\ghost{\tS'_{N^-_R}}&\ustick{N^-_R}\qw&\ghost{\tA'}&\ustick{N^-_R}\qw&\measureD{e}}}
\end{aligned}\\
\\
&=\ 
\begin{aligned}
    \scalebox{.75}{\Qcircuit @C=1em @R=.7em @! R {&\ustick{\!\!\!\! P^-_R\setminus R}\qw &\multigate{3}{\tS'_{N^-_R}}&\ustick{P^-_R}\qw&\multigate{3}{\tS'_{N^-_R}}&\ustick{R}\qw&\ustick{P^-_R\setminus R}\qw&\qw&\measureD{e}\\
    &\ustick{R}\qw&\ghost{\tS'_{N^-_R}}&&\pureghost{\tS'_{N^-_R}}&\ustick{R}\qw&\multigate{1}{\tA}&\ustick{R}\qw&\measureD{e}\\
    &&&&&\ustick{\rC}&\ghost{\tA}&\ustick{\rC}\qw&\measureD{e}\\
    \prepareC{\eta}&\ustick{N^-_R}\qw&\ghost{\tS'_{N^-_R}}&\ustick{N^-_R}\qw&\ghost{\tS'_{N^-_R}}&\qw&\ustick{N^-_R}\qw&\qw&\measureD{e}}}
\end{aligned}\ 
=\ 
\begin{aligned}
    \scalebox{.65}{\Qcircuit @C=1em @R=.7em @! R {    &\ustick{P^-_R\setminus R}\qw&\qw&\measureD{e}\\
    &\ustick{R}\qw &\qw&\multimeasureD{1}{a}\\
    &\ustick{\rC}\qw &\qw&\ghost{a}}}
\end{aligned}
\ .
\end{align*}
Finally, we observe that $[(a\otimes e_{P^-_R\setminus R})_{P^-_R\rC}]=[a_{R\rC}]$, then $\tW^\dag\tV^\dag=\tI_{G}^\dag$.
Similarly, one can calculate $\tV^\dag\tW^\dag b_{S\rC}$. First we observe that $[b_{S\rC}]=[(b\otimes e_{P^+_S\setminus S})_{P^+_S\rC}]$. Then $\tW^\dag b_{S\rC}=\{(\tW^\dag_{P^+_S}(b\otimes e_{N^+_{P^+_S}\setminus N^+_S})\}_{N^+_{P^+_S\rC}}$, and
\begin{align*}
&\begin{aligned}
    \scalebox{.65}{\Qcircuit @C=1em @R=.7em @! R {&\ustick{N^+_{P^+_S}\setminus N^+_S}\qw &\qw&\multigate{3}{\tS'_{P^+_S}}&\qw&\ustick{N^+_{P^+_S}}\qw&\qw&\measureD{e}\\
    &\ustick{N^+_S}\qw &\qw&\ghost{\tS'_{P^+_S}}&\qw&\ustick{P^+_S\setminus S}\qw&\qw&\measureD{e}\\
    &&&&&&\ustick{\rC}\qw&\multimeasureD{1}{b}\\
    \prepareC{\eta}&\ustick{P^+_S}\qw &\qw&\ghost{\tS'_{P^+_S}}&\qw&\ustick{S}\qw&\qw&\ghost{b}}}
\end{aligned}\ 
=\ 
\begin{aligned}
    \scalebox{.65}{\Qcircuit @C=1em @R=.7em @! R {&\ustick{N^+_{P^+_S}\setminus N^+_S}\qw &\qw&\multigate{3}{\tS'_{P^+_S}}&\qw&\ustick{N^+_{P^+_S}}\qw&\qw&\measureD{e}\\
    &\ustick{N^+_S}\qw&\qw&\ghost{\tS'_{P^+_S}}&\qw&\ustick{P^+_S\setminus S}\qw&\qw&\measureD{e}\\
        &&&&\ustick{\rC}&\multigate{1}{\tB}&\ustick{\rC}\qw&\measureD{e}\\
    \prepareC{\eta}&\ustick{P^+_S}\qw &\qw&\ghost{\tS'_{P^+_S}}&\ustick{S}\qw&\ghost{\tB}&\ustick{S}\qw&\measureD{e}}}
\end{aligned}
\\
\\
=&
\begin{aligned}
    \scalebox{.65}{\Qcircuit @C=1em @R=.7em @! R {&\ustick{N^+_{P^+_S}\setminus{N^+_S}}\qw &\qw&\multigate{3}{\tS'_{P^+_S}}&  \qw&\ustick{N^+_{P^+_S}}\qw&\qw&\qw&\multigate{3}{\tS'_{P^+_S}}&\ustick{N^+_{P^+_S}}\qw&\measureD{e}\\
    &\ustick{N^+_S}\qw&\qw&\ghost{\tS'_{P^+_S}}&\qw&\ustick{P^+_S\setminus S}\qw&\qw&\qw&\ghost{\tS'_{P^+_S}}&&\\
        &&&&\ustick{\rC}&\multigate{1}{\tB}&\ustick{\rC}\qw&\measureD{e}&\pureghost{\tS'_{P^+_S}}&\ustick{P^+_S}\qw&\measureD{e}\\
    \prepareC{\eta}&\ustick{P^+_S}\qw &\qw&\ghost{\tS'_{P^+_S}}&\ustick{S}\qw&\ghost{\tB}&\ustick{S}\qw&\qw&\ghost{\tS'_{P^+_S}}&\ustick{P^+_S}\qw&\measureD{e}}}
\end{aligned}
\\
\\
&=
\begin{aligned}
    \scalebox{.75}{\Qcircuit @C=1em @R=.7em @! R {
    &\qw&\ustick{N^+_{P^+_S}\setminus N^+_S}\qw &\qw&\measureD{e}\\
        &\ustick{N^+_S}\qw &\multigate{1}{\tB''}&\ustick{N^+_S}\qw&\measureD{e}\\
        &\ustick{\rC}\qw &\ghost{\tB''}&\ustick{\rC}\qw&\measureD{e}\\
        \prepareC{\eta}&\qw&\ustick{P^+_S}\qw &\qw&\measureD{e}}}
\end{aligned}
\ ,
\end{align*}
where we used Eq.~\eqref{eq:condfour}.
Now, for $\tV^\dag\tW^\dag[b_{S\rC}]$ we have
\begin{align*}
&\begin{aligned}
    \scalebox{.75}{\Qcircuit @C=1em @R=.7em @! R {
    \prepareC{\psi}&\qw&\ustick{P^-_{N^+_{P^+_R}}}\qw&\multigate{3}{\tS'_{P^+_{P^+_R}}}&
    \qw&\ustick{N^+_{P^+_R}}\qw&\qw&\qw     
    &\multigate{3}{\tS'_{P^+_R}}&\qw&\ustick{N^+_{P^+_R}}\qw&\measureD{e}\\
    &&&\pureghost{\tS'_{P^+_{P^+_R}}}&
    &&&&\pureghost{\tS'_{P^+_R}}&
    &\ustick{\rC}\qw&\multimeasureD{1}{b}\\    
    &&&\pureghost{\tS'_{P^+_{P^+_R}}}&\qw&\ustick{\!\!\! P^-_{N^+_{P^+_R}}\setminus N^+_{P^+_R}}\qw&\measureD{e}&&\pureghost{\tS'_{P^+_R}}&
        \qw&\ustick{R}\qw&\ghost{b}\\
    &\ustick{P^+_{P^+_R}}\qw&\qw&\ghost{\tS'_{P^+_{P^+_R}}}&
    \ustick{{P^+_{P^+_R}}}
    \qw&\measureD{e}&\prepareC{\eta}&\ustick{P^+_R}\qw 
    &\ghost{\tS'_{P^+_R}}&
    \qw&\ustick{\!\!\!\!\!\! P^+_R\setminus R}\qw&\measureD{e}
}}
\end{aligned}
\\
\\
&=
\begin{aligned}
    \scalebox{.75}{\Qcircuit @C=1em @R=.7em @! R {
    \prepareC{\psi}&\qw&\ustick{P^-_{N^+_{P^+_R}}}\qw &\multigate{5}{\tS'_{P^+_{P^+_R}}}&\qw&\ustick{N^+_{P^+_R}\setminus N^+_R}\qw&\qw&\measureD{e}\\
      &&&\pureghost{\tS'_{P^+_{P^+_R}}}&\ustick{N^+_R}\qw&\multigate{1}{\tB''}&\ustick{N^+_R}\qw&\measureD{e}\\
         &&&&\ustick{\rC}&\ghost{\tB''}&\ustick{\rC}\qw&\measureD{e}\\
       &&&\pureghost{\tS'_{P^+_{P^+_R}}}&
    &&&\\
    &&&\pureghost{\tS'_{P^+_{P^+_R}}}&\qw&\ustick{P^-_{N^+_{P^+_R}}\setminus N^+_{P^+_R}}
    \qw&\measureD{e}&\\
    &\qw&\ustick{P^+_{P^+_R}}\qw&\ghost{\tS'_{P^+_{P^+_R}}}&\qw&\ustick{P^+_{P^+_R}}
    \qw&\measureD{e}&}}
\end{aligned}\ 
=\ 
\begin{aligned}
    \scalebox{.65}{\Qcircuit @C=1em @R=.7em @! R {&\ustick{\rC}\qw &\qw&\multimeasureD{1}{b}\\
    &\ustick{R}\qw &\qw&\ghost{b}\\\\
    &\ustick{P^+_{P^+_R}\setminus R}\qw&\qw&\measureD{e}}}
\end{aligned}
\ ,
\end{align*}
where Eq.~\eqref{eq:condfour} was used in the last step. The arguments are straightforwardly generalised to $b\in\Cntsetcomp{\rA^{(G)}_S}{\rC}_{Q\Reals}$.

\section{Proof of theorem~\ref{th:luroogur}}\label{app:proof}

\begin{proof}
In the following, $G_i\coloneqq (G,i)$ for $i=0,1$. The argument in appendix~\ref{app:rinvlinv} shows that 
\begin{align*}
\begin{aligned}
    \Qcircuit @C=.9em @R=1.2em @! R {
    &\ustick{G_0}\qw&\multigate{1}{\tW}&\ustick{G_0}\qw&\measureD{e}\\
    &\ustick{G_1}\qw&\ghost{\tW}&\ustick{G_1}\qw&\qw}
\end{aligned}\ =\ 
\begin{aligned}
    \Qcircuit @C=.9em @R=1.2em @! R {
    &\ustick{G_0}\qw&\measureD{e}\\
    &\ustick{G_1}\qw&\gate{\tV^{-1}}&\ustick{G_0}\qw&\qw}
\end{aligned}\ ,
\end{align*} 
where $(G,\rA,\tV^{-1\dag})$ is a GUR. Moreover, since 
\begin{align*}
\begin{aligned}
    \Qcircuit @C=.9em @R=1.2em @! R {
    &\ustick{G_0}\qw&\multigate{1}{\tS_G}&\ustick{G_1}\qw&\qw\\
    &\ustick{G_1}\qw&\ghost{\tS_G}&\ustick{G_0}\qw&\measureD{e}}
\end{aligned}
\ =\ 
\begin{aligned}
    \Qcircuit @C=.9em @R=1.2em @! R {
    &\ustick{G_0}\qw&\measureD{e}\\
    &\ustick{G_1}\qw&\qw}
    \end{aligned}\ ,
\end{align*}
one has
\begin{align*}
&\begin{aligned}
    \Qcircuit @C=.9em @R=1.2em @! R {
    &\ustick{G_0}\qw&\measureD{e}\\
    &\ustick{G_1}\qw&\qw}
\end{aligned}\ =
\begin{aligned}
    \Qcircuit @C=.9em @R=1.2em @! R {
    &\ustick{G_0}\qw&\multigate{1}{\tW^{-1}}&\ustick{G_0}\qw&\multigate{1}{\tW}&\ustick{G_0}\qw&\measureD{e}\\
    &\ustick{G_1}\qw&\ghost{\tW^{-1}}&\ustick{G_1}\qw&\ghost{\tW}&\ustick{G_1}\qw&\qw}
\end{aligned}\\
\\
=&
\begin{aligned}
    \Qcircuit @C=.9em @R=1.2em @! R {
    &\ustick{G_0}\qw&\multigate{1}{\tS_G}&\ustick{G_1}\qw&\multigate{1}{\tW}&\ustick{G_1}\qw&\multigate{1}{\tS_G}&\ustick{G_0}\qw&\measureD{e}\\
    &\ustick{G_1}\qw&\ghost{\tS_G}&\ustick{G_0}\qw&\ghost{\tW}&\ustick{G_0}\qw&\ghost{\tS_G}&\ustick{G_1}\qw&\gate{\tV^{-1}}&\ustick{G_1}\qw&\qw}
\end{aligned}\\
\\
=&
\begin{aligned}
    \Qcircuit @C=.9em @R=1.2em @! R {
    &\ustick{G_0}\qw&\multigate{1}{\tS_G}&\ustick{G_1}\qw&\multigate{1}{\tW}&\ustick{G_1}\qw&\gate{\tV^{-1}}&\ustick{G_0}\qw&\qw\\
    &\ustick{G_1}\qw&\ghost{\tS_G}&\ustick{G_0}\qw&\ghost{\tW}&\ustick{G_0}\qw&\measureD{e}}
\end{aligned}\ \ ,
\end{align*}
namely
\begin{align*}
&\begin{aligned}
    \Qcircuit @C=.9em @R=1.2em @! R {
    &\ustick{G_1}\qw&\qw\\
    &\ustick{G_0}\qw&\measureD{e}}
\end{aligned}\ =
\begin{aligned}
    \Qcircuit @C=.9em @R=1.2em @! R {
    &\ustick{G_1}\qw&\multigate{1}{\tW}&\ustick{G_1}\qw&\gate{\tV^{-1}}&\ustick{G_0}\qw&\qw\\
    &\ustick{G_0}\qw&\ghost{\tW}&\ustick{G_0}\qw&\measureD{e}}
\end{aligned}\ \ .
\end{align*}
On the other hand, one can write the latter as
\begin{align}\label{eq:parsecw}
&\begin{aligned}
    \Qcircuit @C=.9em @R=1.2em @! R {
    &\ustick{G_1}\qw&\gate{\tV}&\ustick{G_1}\qw&\qw\\
    &\ustick{G_0}\qw&\measureD{e}}
\end{aligned}\ =
\begin{aligned}
    \Qcircuit @C=.9em @R=1.2em @! R {
    &\ustick{G_1}\qw&\multigate{1}{\tW}&\ustick{G_1}\qw&\qw\\
    &\ustick{G_0}\qw&\ghost{\tW}&\ustick{G_0}\qw&\measureD{e}}
\end{aligned}\ \ ,
\end{align}
or equivalently
\begin{align*}
&\begin{aligned}
    \Qcircuit @C=.9em @R=1.2em @! R {
    &\ustick{G_0}\qw&\measureD{e}\\
    &\ustick{G_1}\qw&\gate{\tV}&\ustick{G_1}\qw&\qw}
\end{aligned}\ =
\begin{aligned}
    \Qcircuit @C=.9em @R=1.2em @! R {
    &\ustick{G_0}\qw&\multigate{1}{\tW^{-1}}&\ustick{G_0}\qw&\measureD{e}\\
    &\ustick{G_1}\qw&\ghost{\tW^{-1}}&\ustick{G_1}\qw&\qw}
\end{aligned}\ \ .
\end{align*}
Now, let us consider
\begin{align*}
&\begin{aligned}
    \Qcircuit @C=.9em @R=1.2em @! R {
    &\ustick{G_0}\qw&\multigate{1}{\tilde\tS_G}&\ustick{G_1}\qw&\measureD{e}\\
    &\ustick{G_1}\qw&\ghost{\tilde\tS_G}&\ustick{G_0}\qw&\qw\\
    &\qw&\ustick{G_2}\qw&\qw&\qw}
\end{aligned}\\
\\
&\coloneqq\begin{aligned}
    \Qcircuit @C=.9em @R=1.2em @! R {
    &\ustick{G_0}\qw&\qw&\qw&\multigate{1}{\tS_G}&\qw&\qw&\ustick{G_1}\qw&\measureD{e}\\
    &\ustick{G_1}\qw&\multigate{1}{\tW^{-1}}&\ustick{G_1}\qw&\ghost{\tS_G}&\ustick{G_0}\qw&\multigate{1}{\tW}&\ustick{G_0}\qw&\qw\\
    &\ustick{G_2}\qw&\ghost{\tW^{-1}}&\qw&\ustick{G_2}\qw&\qw&\ghost{\tW}&\ustick{G_2}\qw&\qw}
\end{aligned}\\
\\
&=
\begin{aligned}
    \Qcircuit @C=.9em @R=1.2em @! R {
    &\ustick{G_0}\qw&\qw&\qw&\multigate{1}{\tW}&\ustick{G_0}\qw&\qw\\
    &\ustick{G_2}\qw&\multigate{1}{\tW}&\ustick{G_2}\qw&\ghost{\tW}&\ustick{G_2}\qw&\qw\\   &\ustick{G_1}\qw&\ghost{\tW}&\qw&\ustick{G_1}\qw&\measureD{e}}
\end{aligned}\\
\\
&=
\begin{aligned}
    \Qcircuit @C=.9em @R=1.2em @! R {
        &\qw&\ustick{G_0}\qw&\qw&\multigate{1}{\tW}&\ustick{G_0}\qw&\qw\\   
        &\ustick{G_2}\qw&\gate{\tV}&\ustick{G_2}\qw&\ghost{\tW}&\ustick{G_2}\qw&\qw\\
        &\ustick{G_1}\qw&\measureD{e}}
\end{aligned}\ ,
\end{align*}
from which one concludes that 
\begin{align*}
&\begin{aligned}
    \Qcircuit @C=.9em @R=1.2em @! R {
    &\ustick{G_0}\qw&\multigate{1}{\tilde\tS_G}&\ustick{G_1}\qw&\measureD{e}\\
    &\ustick{G_1}\qw&\ghost{\tS_G}&\ustick{G_0}\qw&\qw}
\end{aligned}\ =\ 
\begin{aligned}
    \Qcircuit @C=.9em @R=1.2em @! R {
    &\ustick{G_1}\qw&\measureD{e}\\
    &\ustick{G_0}\qw&\gate{\tX}&\ustick{G_0}\qw}
\end{aligned}\ ,
\end{align*}
for some $\tX$. Thus
\begin{align*}
\begin{aligned}
    \Qcircuit @C=.9em @R=1.2em @! R {
        &\qw&\ustick{G_0}\qw&\qw&\multigate{1}{\tW}&\ustick{G_0}\qw&\qw\\   
        &\ustick{G_2}\qw&\gate{\tV}&\ustick{G_2}\qw&\ghost{\tW}&\ustick{G_2}\qw&\qw}
\end{aligned}\ =\ 
\begin{aligned}
    \Qcircuit @C=.9em @R=1.2em @! R {
    &\ustick{G_0}\qw&\gate{\tX}&\ustick{G_0}\qw\\
    &\qw&\ustick{G_2}\qw&\qw}
\end{aligned}\ .
\end{align*}
Finally, this implies that $\tW=\tX\otimes\tV^{-1}$, and in particular, by Eq.~\eqref{eq:parsecw}, it must be $\tX=\tV$, namely $\tW=\tV\otimes\tV^{-1}$. Consequently, we have that
\begin{align*}
&\begin{aligned}
    \scalebox{.75}{\Qcircuit @C=.9em @R=1.2em @! R {
    &\ustick{\rC}\qw&\multigate{1}{\mathsf V^-_L(\tA_R)}&\ustick{\rC}\qw&\qw\\
    &\ustick{G_0}\qw&\ghost{\mathsf V^-_L(\tA_R)}&\ustick{G_0}\qw&\qw\\
    &\qw&\ustick{G_1}\qw&\qw}}
\end{aligned}\ =\ 
\begin{aligned}
    \scalebox{.75}{\Qcircuit @C=.9em @R=1.2em @! R {&&&\ustick{\rC}&\multigate{1}{\tA_R}&\ustick{\rC}\qw&&&\\
    &\ustick{G_0}\qw&\multigate{1}{\tW^{-1}}&\ustick{G_0}\qw&\ghost{\tA_R}&\ustick{G_0}\qw&\multigate{1}{\tW}&\ustick{G_0}\qw&\qw\\
    &\ustick{G_1}\qw&\ghost{\tW^{-1}}&\qw&\ustick{G_1}\qw&\qw&\ghost{\tW}&\ustick{G_1}\qw&\qw}}
\end{aligned}\\
\\
&=\ 
\begin{aligned}
    \Qcircuit @C=.9em @R=1.2em @! R {&&&\ustick{\rC}&\multigate{1}{\tA_R}&\ustick{\rC}\qw&&&\\
    &\ustick{G_0}\qw&\gate{\tV^{-1}}&\ustick{G_0}\qw&\ghost{\tA_R}&\ustick{G_0}\qw&\gate{\tV}&\ustick{G_0}\qw&\qw\\
    &\qw&\qw&\qw&\ustick{G_1}\qw&\qw&\qw&\qw&\qw}
\end{aligned}\ ,
\end{align*}
and
\begin{align*}
&\begin{aligned}
    \scalebox{.75}{\Qcircuit @C=.9em @R=1.2em @! R {
    &\qw&\ustick{G_0}\qw&\qw\\
        &\ustick{G_1}\qw&\multigate{1}{\mathsf V^+_L(\tA_R)}&\ustick{G_1}\qw&\qw\\
                &\ustick{\rC}\qw&\ghost{\mathsf V^+_L(\tA_R)}&\ustick{\rC}\qw&\qw}}
\end{aligned}\ =\ 
\begin{aligned}
    \scalebox{.75}{\Qcircuit @C=.9em @R=1.2em @! R {
    &\ustick{G_0}\qw&\multigate{1}{\tW^{-1}}&\qw&\ustick{G_0}\qw&\qw&\multigate{1}{\tW}&\ustick{G_0}\qw&\qw\\
    &\ustick{G_1}\qw&\ghost{\tW^{-1}}&\ustick{G_1}\qw&\multigate{1}{\tA_R}&\ustick{G_1}\qw&\ghost{\tW}&\ustick{G_1}\qw&\qw\\
    &&&\ustick{\rC}&\ghost{\tA_R}&\ustick{\rC}\qw&&&}}
\end{aligned}\\
\\
&=\ 
\begin{aligned}
    \Qcircuit @C=.9em @R=1.2em @! R {
    &\qw&\qw&\qw&\ustick{G_0}\qw&\qw&\qw&\qw&\qw\\
    &\ustick{G_1}\qw&\gate{\tV}&\ustick{G_1}\qw&\multigate{1}{\tA_R}&\ustick{G_1}\qw&\gate{\tV^{-1}}&\ustick{G_1}\qw&\qw\\
    &&&\ustick{\rC}&\ghost{\tA_R}&\ustick{\rC}\qw&&&}
\end{aligned}\ .&\qedhere
\end{align*}
\end{proof}
\section{Proof of lemmas~\ref{lem:pedantic} and~\ref{lem:pedantic2}}\label{app:ped}

We start with the proof of lemma~\ref{lem:pedantic}.
\begin{proof}
(1) Notice that $N'^\pm_{e_H}=\phi(S_\pm)=\phi(N^\pm_{e_G})$. Then $|N'^\pm_{e_H}|\leq|N^\pm_{e_G}|$, with strict inequality if and only if $\phi(h_a)=\phi(h_b)$ for some $h_a\neq h_b\in N^\pm_{e_G}$. However, this is impossible because it would be equivalent to $\phi(h_ah_b^{-1})\in R'$, namely $h_a=h_b$, contrarily to the hypothesis. Thus, $|N'^\pm_{e_H}|=|N^\pm_{e_G}|$. Invariance of neighbourhoods under left-multiplication finally gives the thesis.
\\
\noindent
(2) Let now $f\in P'^\pm_{e_H}=N'^\mp_{N'^\pm_{e_H}}$, namely there exists $h\in N'^\pm_{e_H}$ such that $f\in N'^\mp_h$, or equivalently $h\in N'^\pm_f$. Then, $f\in P'^\pm_{e_H}$ iff there exists $h\in N'^\pm_{e_H}\cap N'^\pm_f$. Equivalently, there exists $h\in H$ such that $h=\phi(h_a)$ and $h=f\phi(h_b)$, namely $f=\phi(h_ah_b^{-1})$. Then $P'^\pm_{e_H}=\{\phi(h_ah_b^{-1})\mid h_a, h_b\in N^\pm_{e_G}\}$. By a similar argument, we have that $P^\pm_{e_G}=\{h_ah_b^{-1}\mid h_a, h_b\in N^\pm_{e_G}\}$. Consequently, $|P'^\pm_{e_H}|\leq|\{h_ah_b^{-1}\mid h_a, h_b\in N^+_{e_G}\}|=|P^\pm_{e_G}|$. Now, strict inequality in the last relation would require that $\phi(h_ah_b^{-1})=\phi(h_dh_c^{-1})$, contrarily to the hypothesis. We conclude that $|P'^\pm_{e_H}|=|P^\pm_{e_G}|$. Also in this case the thesis can be obtained by invariance of neighbourhoods under left-multiplication.
\\
\noindent
(3) Let $N'^\pm_{f_1}\cap N'^\pm_{f_2}=\emptyset$. Then clearly for every $g_i\in \phi^{-1}(f_i)$ it must be $N^\pm_{g_1}\cap N^\pm_{g_2}=\emptyset$. Let then now $h\in N'^\pm_{f_1}\cap N'^\pm_{f_2}$. This implies that $h=f_1\phi(h_a)=f_2\phi(h_b)$ for $h_a,h_b\in N^\pm_{e_G}$. Consequently, $f_2=f_1\phi(h_a h_b^{-1})$. Now, either $f_1=f_2$, and then choosing $g_1=g_2$ the thesis trivially follows, or $\phi(h_a h_b^{-1})\not\in R'$, and then $h_a h_b^{-1}\not\in R$. One can then set $g_2=g_1h_a h_b^{-1}$, and verify that $g_2\in\phi^{-1}(f_2)$, as well as $k=g_1h_a=g_2h_b\in N^\pm_{g_1}\cap N^\pm_{g_2}$. Thus, $h=\phi(k)$, and consequently $  N'^\pm_{f_1}\cap N'^\pm_{f_2}\subseteq\phi(N^\pm_{g_1}\cap N^\pm_{g_2})$.
%
This implies that $|N'^\pm_{f_1}\cap N'^\pm_{f_2}|\leq|N^\pm_{g_1}\cap N^\pm_{g_2}|$, with a strict inequality iff $\phi(k_1)=\phi(k_2)$ for $k_1\neq k_2\in N^\pm_{g_1}\cap N^\pm_{g_2}$. However, form the proof of item (1) we know that if $k_1\neq k_2\in N^\pm_{g}$ it must be $\phi(k_1)\neq\phi(k_2)$. Thus, $ | N'^\pm_{f_1}\cap N'^\pm_{f_2}|=|\phi(N^\pm_{g_1}\cap N^\pm_{g_2})|$, and $  N'^\pm_{f_1}\cap N'^\pm_{f_2}=\phi(N^\pm_{g_1}\cap N^\pm_{g_2})$.
\\
\noindent
(4) Let 
$h\in P'^\mp_{f_2}\cap N'^\pm_{f_1}$. Then $h=f_2\phi(h_c^{-1}h_b)=f_1\phi(h_a)$, with $h_a,h_b,h_c\in N^\pm_{e_G}$. This implies that $f_2=f_1\phi(h_ah_b^{-1}h_c)$. We can now set $g_2\coloneqq g_1h_ah_b^{-1}h_c$ with $g_1\in\phi^{-1}(f_1)$. In this way, one has $f_2=\phi(g_2)$. Let us set $k\coloneqq g_1h_a=g_2h_c^{-1}h_b\in P^\mp_{g_2}\cap N^\pm_{g_1}$. Then, $h=\phi(k)$, and $P'^\mp_{f_1}\cap N'^\pm_{f_1}\subseteq\phi(P^\mp_{g_2}\cap N^\pm_{g_1})$. By the same argument as for item (3), the thesis follows.
\\
\noindent
(5) 
Let $h\in P^\pm_{g_1}\cap N^\mp_{g_2}$. Then
$h=g_1 h_ah_b^{-1},\quad h=g_2 h_c^{-1}$,
for some $h_a,h_b,h_c\in N^\pm_{e_G}$, and thus $g_2=g_1h_ah_b^{-1}h_c$. On the other hand, by hypothesis, for every $h_x\in N^\pm_{e_G}$ one has $g_2\neq g_1 h_x$. Collecting all the informations, we have
$h_ah_b^{-1}h_ch_x^{-1}\not\in R$, i.e.~$\phi(h_ah_b^{-1}h_ch_x^{-1})\not\in R'$,
for every $h_x\in N^\pm_{e_G}$. Then one has
$\phi(g_2)=\phi(g_1)\phi(h_ah_b^{-1}h_c)$. By hypothesis, it is also $\phi(g_2)=\phi(g_1)\phi(h_d)$ for some $h_d\in N^\pm_{e_G}$. This is clearly absurd, thus it must be $P^\pm_{g_1}\cap N^\mp_{g_2}=\emptyset$.
\end{proof}

Now, we provide the proof of lemma~\ref{lem:pedantic2}

\begin{proof}
(1) Let $f_2\in P'^\pm_{f_1}$, namely $f_2=f_1\phi(h_ah_b^{-1})$ for $h_a,h_b\in N^\pm_{e_G}$. If $f_1=f_2$, it is sufficient to choose $g_1=g_2\in\phi^{-1}(f_1)$. Let now $f_1\neq f_2$. Then, let $g_1\in\phi^{-1}(f_1)$, and $g_2\coloneqq g_1h_ah_b^{-1}$. By construction, $\phi(g_2)=f_2$ and $g_2\in P^\pm_{g_1}$. Now, let $h\in P'^\pm_{f_1}\cap P'^\pm_{f_2}$, namely $h=f_1\phi(h_fh_e^{-1})=f_2\phi(h_ch_d^{-1})=f_1\phi(h_ah_b^{-1}h_ch_d^{-1})$. 
Then we have $\phi(h_ah_b^{-1}h_ch_d^{-1}h_eh_f^{-1})\in R'$, and necessarily $h_ah_b^{-1}h_ch_d^{-1}h_eh_f^{-1}\in R$. This implies that $g_1h_fh_e^{-1}=g_1h_ah_b^{-1}h_ch_d^{-1}=g_2h_ch_d^{-1}$. Now, setting $k\coloneqq g_1h_fh_e^{-1}$, we have $k=g_2h_ch_d^{-1}\in P^\pm_{g_1}\cap P^\pm_{g_2}$, and $h=\phi(k)$. Thus, $P'^\pm_{f_1}\cap P'^\pm_{f_2}\subseteq\phi(P^\pm_{g_1}\cap P^\pm_{g_2})$. Finally, by the same argument as for the proof of item (2) of lemma~\ref{lem:pedantic}, we get the thesis. 
\\
\noindent
(2) The hypothesis clearly implies that $N^\pm_{g_1}\cap N^\pm_{g_2}=\emptyset$. Suppose 
now that $h\in P^\pm_{g_1}\cap P^\pm_{g_2}$. Then $h=g_1h_ah_b^{-1}=g_2h_dh_c^{-1}$ for 
$h_a,h_b,h_c,h_d\in N^\pm_{e_G}$. This implies that
$g_2=g_1h_ah_b^{-1}h_ch_d^{-1}$, and consequently $\phi(g_2)=\phi(g_1)\phi(h_ah_b^{-1}
h_ch_d^{-1})$. On the other hand, we proved that for every $h_e,h_f\in N^\pm_{e_G}$ one 
has $g_2\neq g_1h_fh_e^{-1}$, while by hypothesis $\phi(g_2)=\phi(g_1)\phi(h_fh_e^{-1})$ 
for some $h_e,h_f\in N^\pm_{e_G}$. Thus, while 
$h_ah_b^{-1}h_ch_d^{-1}h_eh_f^{-1}\not\in R$, it must be 
$\phi(h_ah_b^{-1}h_ch_d^{-1}h_eh_f^{-1})\not\in R'$. We reached a contradiction, and thus 
it must be $P^\pm_{g_1}\cap P^\pm_{g_2}=\emptyset$. 
\end{proof}

\end{appendix}

\bibliography{biblio-auto}
\bibliographystyle{unsrturl}

\end{document}